\documentclass[review,nopreprintline]{elsarticle}

\usepackage{float}
\usepackage{subcaption}
\usepackage{longtable}
\usepackage[table]{xcolor}
\usepackage{mathtools}
\usepackage{lineno}
\usepackage{booktabs}
\usepackage[graphicx]{realboxes}
\usepackage{multirow}
\usepackage{csquotes}
\modulolinenumbers[5]

\usepackage[%pdfusetitle,
            ]{hyperref}

\usepackage[superscript,biblabel]{cite}
\newcommand{\beginsupplement}{%
	\setcounter{table}{0}
	\renewcommand{\thetable}{S\arabic{table}}%
	\setcounter{figure}{0}
	\renewcommand{\thefigure}{S\arabic{figure}}%
	\setcounter{section}{0}
	\renewcommand{\thesection}{Supplemental \arabic{section}}%
	\setcounter{equation}{0}
	\renewcommand{\theequation}{S\arabic{equation}}
	\setcounter{page}{1}
}

\usepackage{geometry}
\newlength{\mymargin}
\begin{document}

\begin{frontmatter}

\title{Managing CO$_2$ under global and country-specific net-zero emissions targets in Europe}

\author[mpe,novo]{Ricardo Fernandes \corref{cor1}}
\cortext[cor1]{Lead contact and corresponding author, Email: ricardo.fernandes@mpe.au.dk}
\author[mpe]{Martin Greiner}
\author[mpe,novo]{Marta Victoria}

\affiliation[mpe]
{
	organization = Department of Mechanical and Production Engineering, Aarhus University,
	addressline = Katrinebjergvej 89,
	city = Aarhus,
	postcode = 8200,
	country = Denmark
}

\affiliation[novo]
{
	organization = Novo Nordisk Foundation CO$_2$ Research Center,
	addressline = Gustav Wieds Vej 10,
	city = Aarhus,
	postcode = 8000,
	country = Denmark
}

\begin{abstract}
The European Union (EU) aims to reach carbon neutrality by 2050. This requires capturing CO$_2$, eventually transporting it to different regions, and either converting it into valuable products or sequestering it underground. Although the target is set for the entire EU, in practice, most of the governance and strategy to attain it remains in the individual member states. Previous literature modelling how Europe can achieve carbon neutrality has either considered only a global CO$_2$ limit or used coarse spatial and temporal representation without proper network modelling. Here, we use a highly-resolved open model of the European sector-coupled energy system, PyPSA-Eur, to explore the impacts of imposing net-zero emissions globally for the entire EU versus imposing carbon neutrality for each country. Forcing net-zero emissions in every country increases system cost by 1.4\%, demands varied CO$_2$ prices, and triggers higher investment in direct air capture and renewable capacities. Furthermore, in both scenarios, a significant portion of the captured CO$_2$ is transported across Europe, either directly via CO$_2$ pipelines or indirectly via solid biomass or synthetic methane gas, methanol, and oil. Our research enables quantifying the impact of following a collaborative or self-sufficient carbon management strategy to attain carbon neutrality.
\end{abstract}


\end{frontmatter}

\begin{linenumbers}

\nolinenumbers

\clearpage

\section{Introduction}
The European Union (EU) has committed to achieving climate neutrality by 2050 \cite{doi/10.2834/02074}, aiming for an economy with net-zero greenhouse gas (GHG) emissions. Given that carbon dioxide (CO$_2$) constitutes about 75\% of the total GHG emissions globally \cite{Intergovernmental_Panel_on_Climate_Change_(IPCC)_2023}, understanding how this chemical compound could be managed across Europe in a future climate-neutral system is crucial. The EU is implementing a comprehensive policy framework with intermediate steps to reduce GHG emissions by at least 55\% by 2030 and 90\% by 2040 relative to 1990 levels, as key milestones towards the 2050 goal \cite{european_commission_climate_strategies_targets}. This has led the EU Commission to propose, amongst other measures, that at least 50 MtCO$_2$ per year can be sequestered geologically by 2030 \cite{european_carbon_management_strategy_2024}.

Although the Paris Agreement has been signed at the EU level, a significant part of the governance remains at a national level, with each country deciding its own strategy and setting its decarbonisation goals for 2050 and intermediate steps. This distributed decision process is typically absent when modelling future scenarios. Both Energy System Models (ESMs) and Integrated Assessment Models (IAMs) commonly calculate optimal decarbonisation scenarios in which carbon neutrality is imposed at a global level and all the modelled regions contribute to the global objective. This yields non-uniform contributions to emissions mitigation and carbon dioxide removal (CDR) from the different regions raising questions about the fairness of the attained solution.

For the European electricity sector, scenarios with country-level CO$_2$ emissions targets have been explored in Schwenk-Nebbe et al. \cite{SCHWENKNEBBE2021100012} and Pedersen et al. \cite{Pedersen_2023}. In both works, a large variation is found in the CO$_2$ price required in each country for the same CO$_2$ target, highlighting that the mitigation potential, and consequently the need for CO$_2$ price or policy-driven transformations, is radically different amongst countries.

The problem has also been observed on a global scale. Strefler et al. \cite{Strefler_2021} found that scenarios consistent with a 1.5 °C temperature increase involve a highly uneven distribution of CDR technologies across different continents. Furthermore, Bauer et al. \cite{Bauer2020} argue that uniform carbon pricing tend to impose high mitigation costs on developing and emerging economies. In their work, the authors discuss the trade-offs of a system with uniform carbon pricing worldwide and financial transfers, a system with different carbon prices for each region without any financial transfer, and argue that a hybrid combination might be the best solution. The usage and uniformity of different CDR strategies in their model are also affected by the selected carbon pricing scheme.

Coming back to Europe, in Victoria et al. \cite{VICTORIA20221066}, the open sector-coupled model PyPSA-Eur is used to implement myopic transition paths with different carbon budgets, corresponding to temperature increments from 1.5 to 2 °C, and under a globally imposed limit on CO$_2$ emissions for the entire continent. The authors found that the technologies used to capture CO$_2$ are (in order of appearance): (i) point-source capture of industry process emissions, (ii) capture from biomass used for heating in the industry and combusted in combined heat and power (CHP) units, (iii) capture from methane gas used in the industry, and (iv) Direct Air Capture (DAC). Captured CO$_2$ is preferentially sequestered underground and used to produce synthetic oil using the Fischer-Tropsch process.

In Neumann et al. \cite{Neumann2023}, PyPSA-Eur is extended to include an additional option to use CO$_2$ by converting it into synthetic methanol used by the shipping industry. The model is further extended in Hofmann et al. \cite{Hofmann2025} to allow the endogenous built-in of a CO$_2$ network to transport CO$_2$ around the continent.

{
    \renewcommand{\arraystretch}{0.8}% Tighter
    \noindent
    \small
    \label{table_acronyms}
    \begin{tabular}{|*2l|}
        \hline
        \textbf{Acronym} & \textbf{Description}                                              \\
        BECCS            & Bioenergy with carbon capture and storage                         \\
        CC               & Carbon capture                                                    \\
        CCGT             & Combined cycle gas turbine                                        \\
        CCTUS            & Carbon capture, transportation, usage, and storage                \\
        CDR              & Carbon dioxide removal                                            \\
        CHP              & Combined heat and power                                           \\
        CO$_2$           & Carbon dioxide                                                    \\
        DAC              & Direct air capture                                                \\
        DOC              & Direct ocean capture                                              \\
        ENSPRESO         & Energy system potentials for renewable energy sources             \\
        ENTSO-E          & European network of transmission system operators for electricity \\
        ESM              & Energy system model                                               \\
        ETS              & European Union emissions trading system                           \\
        EU               & European Union                                                    \\
        GHG              & Greenhouse gas                                                    \\
        H$_2$            & Hydrogen                                                          \\
        IAM              & Integrated assessment model                                       \\
        JRC-EU-TIMES     & Joint Research Centre European TIMES energy system model          \\
        JRC-IDEES        & Joint Research Centre integrated database European energy sector  \\
        LNG              & Liquefied natural gas                                             \\
        LULUCF           & Land use, land use change, and forestry                           \\
        PHS              & Pump hydro storage                                                \\
        PV               & Photovoltaic                                                      \\
        PyPSA            & Python for power system analysis                                  \\
        PyPSA-Eur        & Python for power system analysis for Europe                       \\
        ROR              & Run-of-river                                                      \\
        SMR              & Steam methane reforming                                           \\
        \hline
        \end{tabular}
}

In the existing literature, there is a gap in modelling climate-neutral systems that include (i) a detailed representation of all the energy sectors, their inter-linkages, and relevant technologies for CO$_2$ capture, transportation, usage, and sequestration, (ii) high spatial and temporal resolution including network representation that allows quantification of the carbon exchange amongst countries, and (iii) country-specific CO$_2$ targets. Closing this research gap is important because detailed modelling is necessary for an accurate representation of the operation of future interconnected systems and because in order to design climate policies that can be perceived as fair by society, a comprehensive understanding of the impacts of local CO$_2$ constraints is needed.

In this work, for the first time, we use a sector-coupled model of Europe including electricity, heating, land transport, aviation, shipping, industry, as well as industrial feedstock, and detailed carbon management to investigate country-wise CO$_2$ targets. We compare a climate-neutral European energy system with net-zero CO$_2$ emissions imposed globally for the whole of Europe versus locally for each country, and investigate the impact of these requirements on the technologies used to capture and convert CO$_2$, their optimal location and operation patterns, and on the underground sequestration of CO$_2$. Moreover, we evaluate the distribution of CO$_2$ emissions amongst countries, the flows of energy and carbon, which quantify the degree of collaboration and exchange, and the required CO$_2$ prices to force more uniform scenarios, which quantify the need for specific policies in the different modelled countries.

\section{Results}
\subsection{Net emitting and net absorbing countries under a global net-zero CO$_2$ emissions constraint}

\begin{figure}[!htb]
    \centering
    \includegraphics[width = 0.81\textwidth]{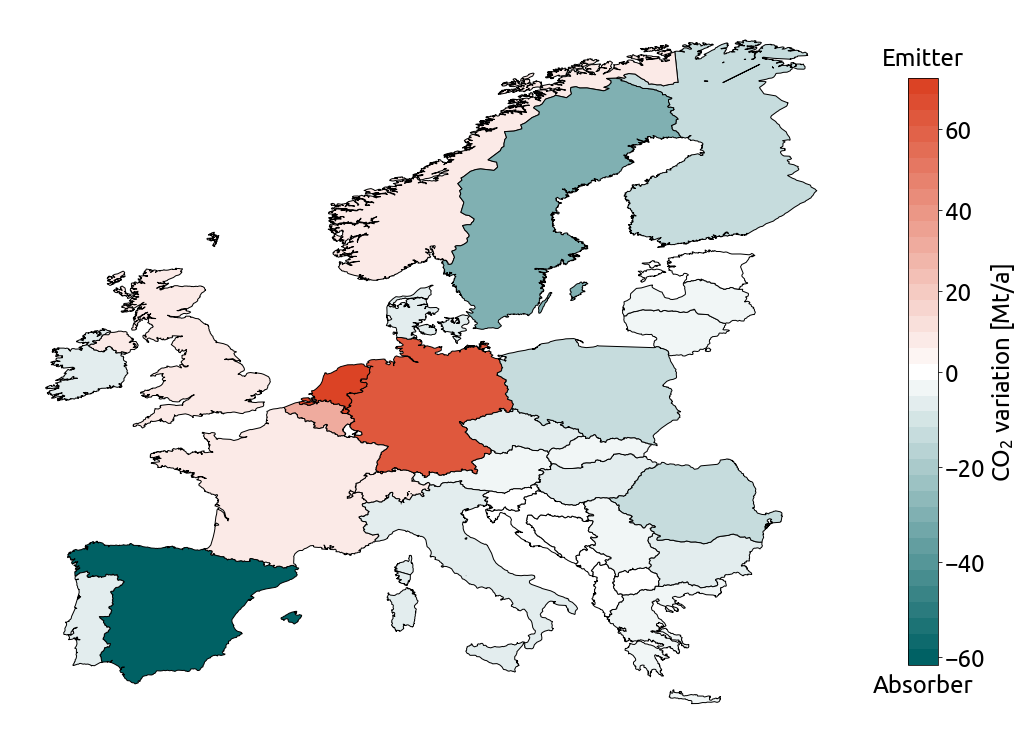}
    \caption{Net CO$_2$ emitter and absorber countries in the global net-zero CO$_2$ emissions scenario. Countries capturing more CO$_2$ than they emit on a yearly average are shown in green, while countries capturing less CO$_2$ than they emit are shown in red. The area corresponding to Kosovo is shown on the map but this country is not modelled. }
    \label{figure_net_co2_emitters_absorbers}
\end{figure}

\begin{figure}[!htb]
    \centering
    \includegraphics[width = 0.8\textwidth]{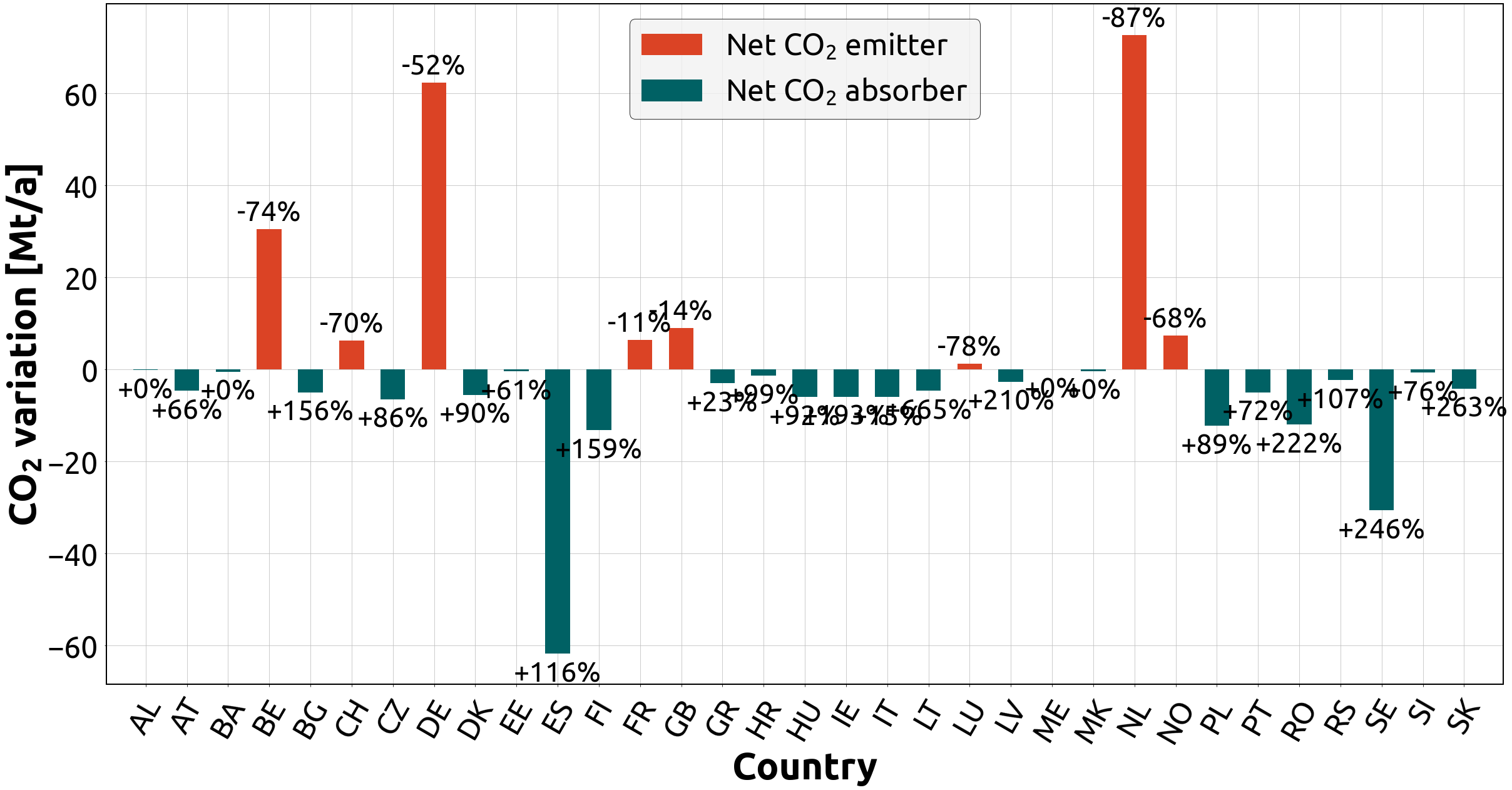}
    \caption{In the global net-zero CO$_2$ emissions scenario, countries that emit more CO$_2$ than they capture on a yearly average are represented by red bars, while countries that capture more CO$_2$ than they emit are represented by green bars. The percentages on top of the red bars indicate how much CO$_2$ net emitter countries do not capture, compared to their own emission values. Conversely, the percentages at the bottom of the green bars indicate how much additional CO$_2$ net absorber countries capture, compared to their own emission values.}
    \label{supplemental:figure_net_co2_emitters_absorbers}
\end{figure}

\begin{figure}[!htb]
    \centering
    \includegraphics[width = 0.8\textwidth]{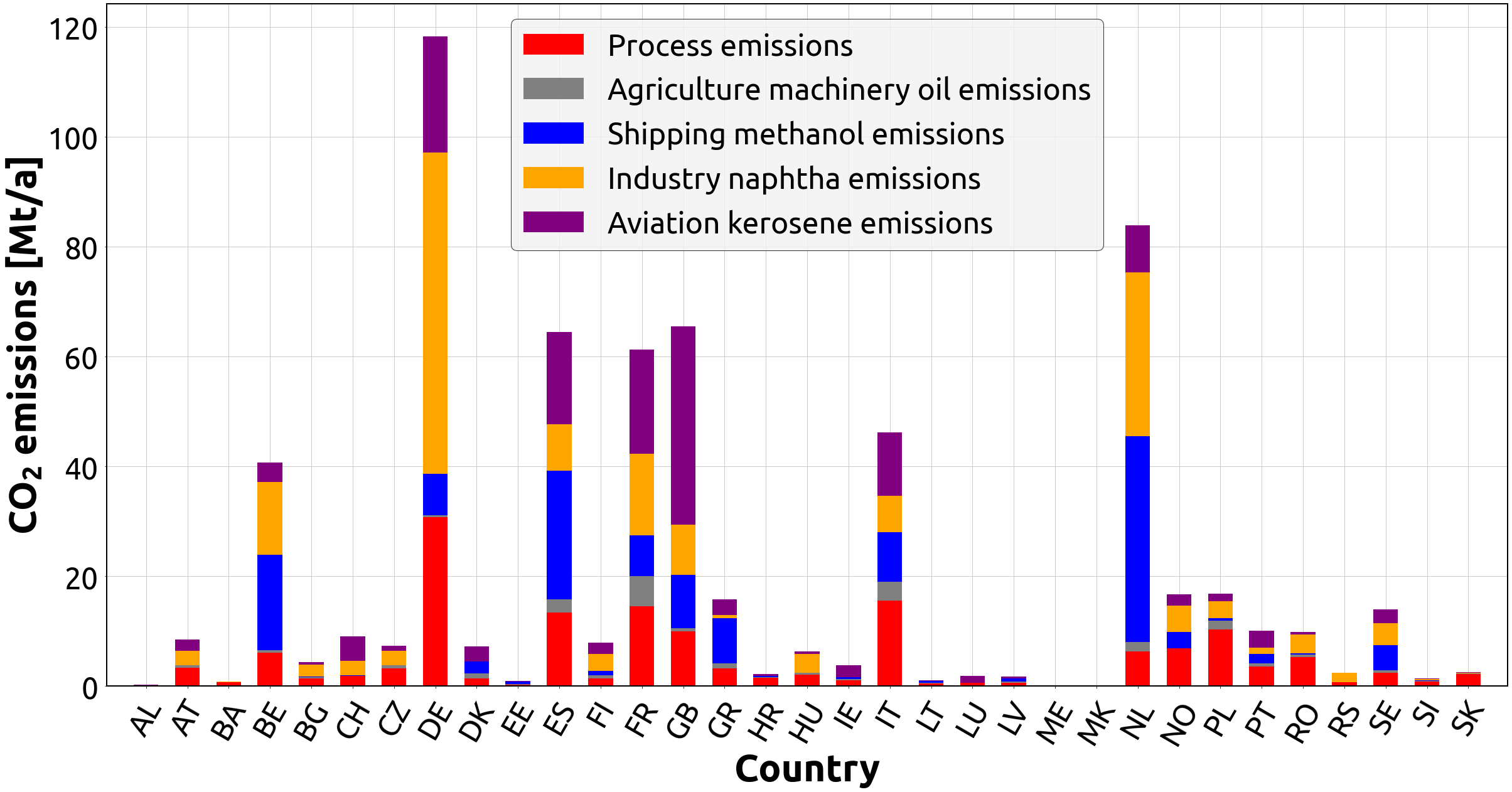}
    \caption{Exogenous CO$_2$ emissions per country and sector in both the global and local net-zero CO$_2$ emissions scenarios. CO$_2$ emissions total 634 Mt/a in Europe, with 153 Mt/a from processes, 23 Mt/a from oil for agriculture machinery, 135 Mt/a from methanol for shipping, 177 Mt/a from naphtha for industry, and 146 Mt/a from kerosene for aviation.}
    \label{supplemental:figure_co2_emissions}
\end{figure}

Using the open model PyPSA-Eur \cite{Neumann2023}, we co-optimised the capacities and dispatch of energy generation, storage, transmission, and conversion in the different sectors, as well as the technologies required to capture, convert, and sequester CO$_2$. Our model represents the sector-coupled European energy system comprising 90 interconnected regions and uses 3-hourly resolution for a full year. Under a global net-zero CO$_2$ emissions constraint, we see non-uniform results, with several countries emitting more CO$_2$ than they capture on average throughout the year, resulting in a total net emission of 195 MtCO$_2$. Specifically, Germany, Belgium, and The Netherlands (hereafter referred to as \enquote{interior countries}) are responsible for 85\% of this total. In contrast, net absorbers such as Spain, Sweden, Finland, Romania, and Poland (hereafter referred to as \enquote{exterior countries}) are collectively responsible for absorbing 66\% of the total net emission (Figures \ref{figure_net_co2_emitters_absorbers} and \ref{supplemental:figure_net_co2_emitters_absorbers}). The lack of uniformity stems from each European nation having (i) different levels of CO$_2$ emissions from process and energy consumption in the industry, agriculture, shipping, and aviation sectors, (ii) disparate potential to sequester CO$_2$ underground, (iii) distinct access to renewable resources such as wind and solar, and (iv) differing solid biomass resources (Figures \ref{supplemental:figure_co2_emissions}, \ref{supplemental:figure_co2_sequestration_potential_usage}, \ref{supplemental:figure_renewable_capacity_factors}, and \ref{supplemental:figure_solid_biomass_potential}). In short, the unique combination of different CO$_2$ emission levels, access to cost-effective energy to capture and convert CO$_2$, and capacity to sequester captured CO$_2$ underground distinctly characterize each country as a net emitter or absorber and by how much. Under a global net-zero CO$_2$ emissions constraint, the total system cost is 878 billion € per year and the main contributors are wind and solar photovoltaic (PV) capacities, technologies to electrify the heating sector, and the production of electrolytic H$_2$ (Figure \ref{figure_total_system_cost_and_variation}).

\subsection{Local net-zero CO$_2$ emissions constraints cause a slight increase in total system cost but dramatic impacts at a country level}
In the locally constrained scenario, the exchange of CO$_2$ with the atmosphere is imposed to be net-zero at every country. However, a country can still exchange CO$_2$ with neighbouring countries in various ways. This can be accomplished either directly through underground CO$_2$ pipelines (built by the model if cost-effective) or indirectly through the exchange of synthetic oil, methane gas, and methanol, all of them produced by combining electrolytic H$_2$ and captured CO$_2$ when economically beneficial. Another way to indirectly exchange CO$_2$ is by transporting solid biomass amongst countries.

\begin{figure}[!htb]
    \centering
    %\subfloat[]{\label{figure_total_system_cost_global}\includegraphics[width = 0.96\linewidth]{./figures/figure_2a.png}}
    \subfloat[]{\label{figure_total_system_cost_global}\includegraphics[width = 0.8\linewidth]{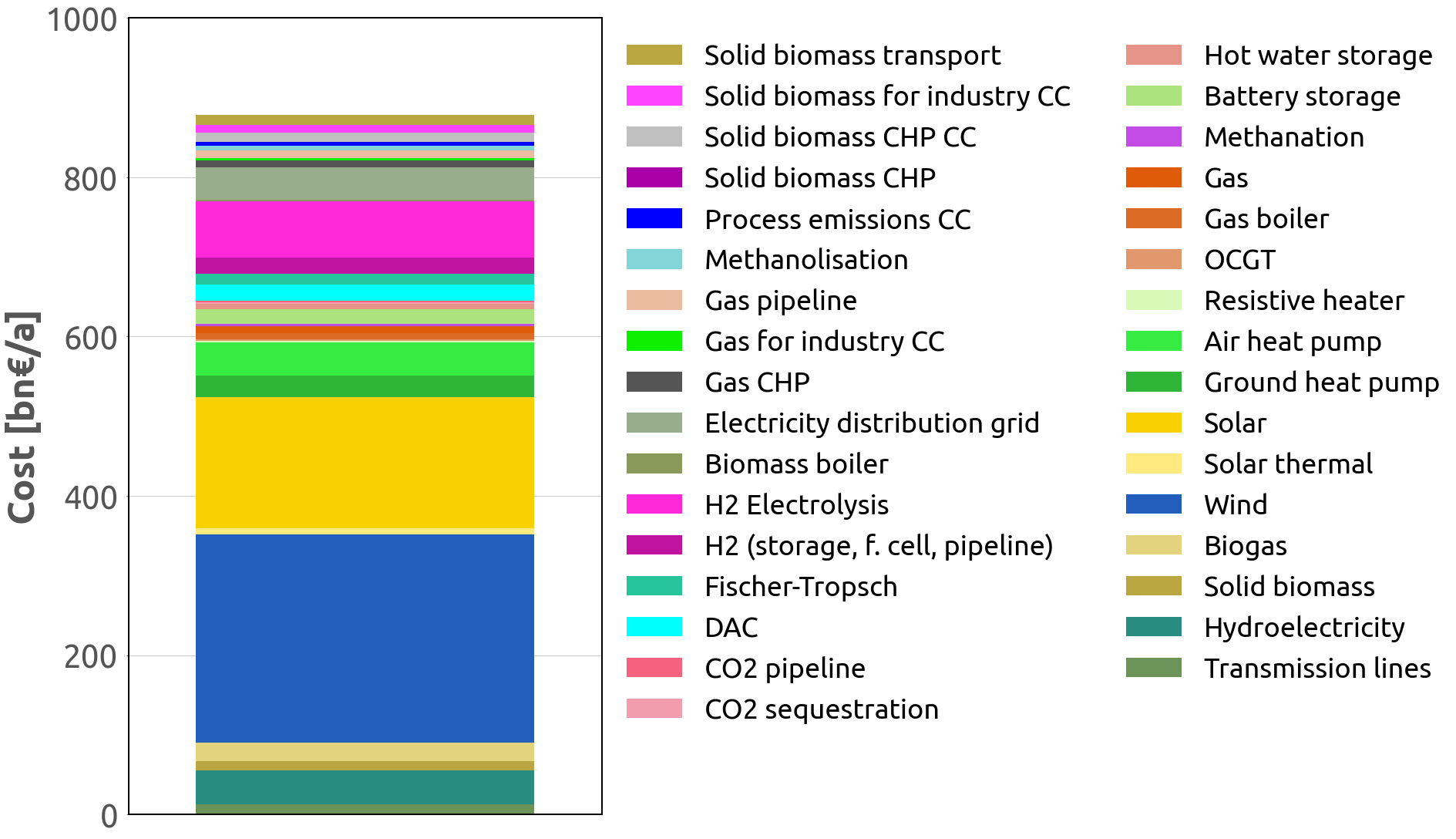}}
    \vspace{20pt}
    %\subfloat[]{\label{figure_cost_variations_per_technology}\includegraphics[width = 0.96\linewidth]{./figures/figure_2b.png}}
    \subfloat[]{\label{figure_cost_variations_per_technology}\includegraphics[width = 0.8\linewidth]{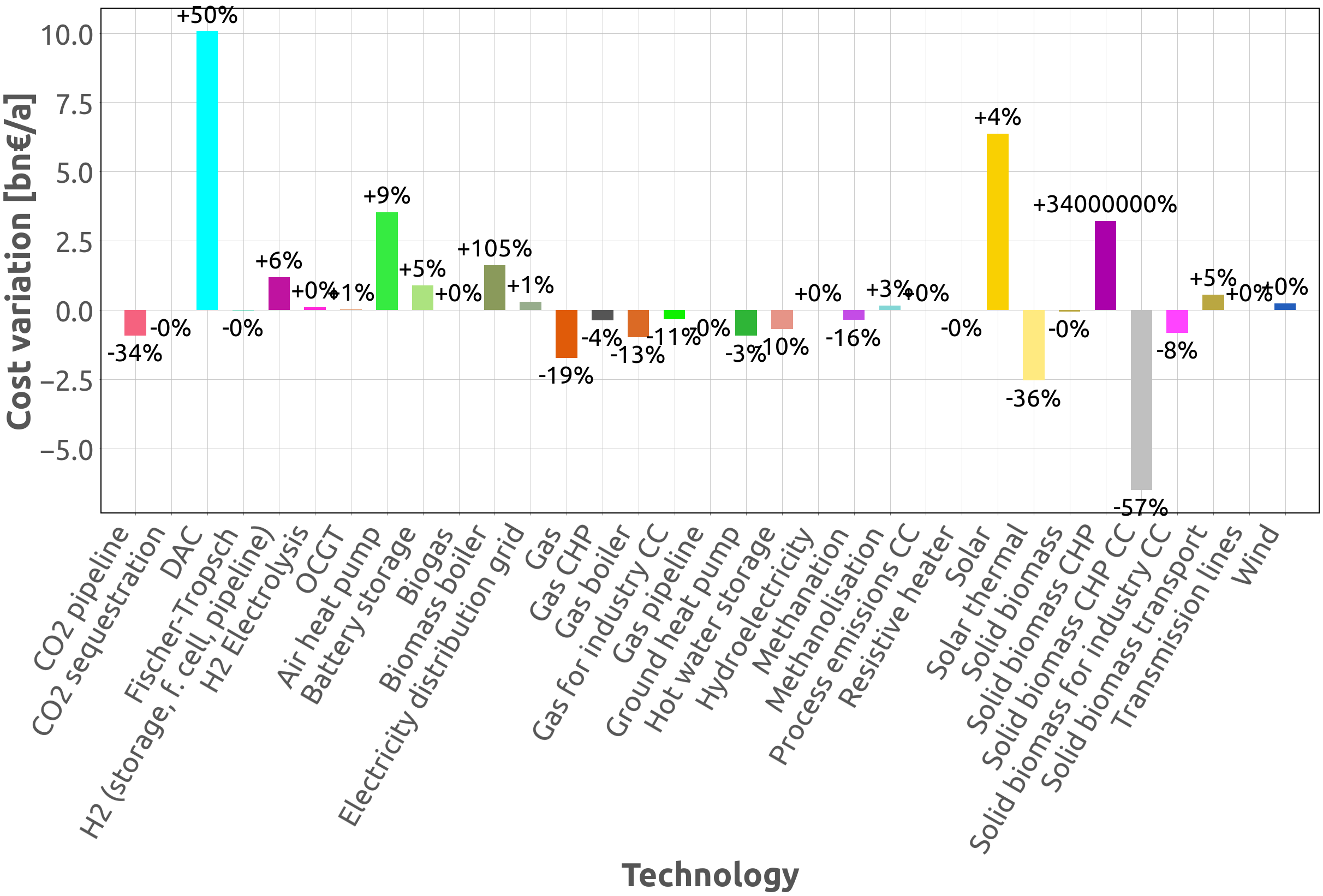}}
    \caption{(a) Total system cost and technology configuration in the global net-zero CO$_2$ emissions scenario and (b) Cost variation per technology between the global net-zero CO$_2$ emissions scenario and the local net-zero CO$_2$ emissions scenario.}
    \label{figure_total_system_cost_and_variation}
\end{figure}

\begin{figure}[!htb]
    \centering
    %\subfloat[]{\label{supplemental:figure_total_system_cost_global_interior_countries}\includegraphics[width = 0.96\linewidth]{./figures/figure_S11a.png}}
    \subfloat[]{\label{supplemental:figure_total_system_cost_global_interior_countries}\includegraphics[width = 0.8\linewidth]{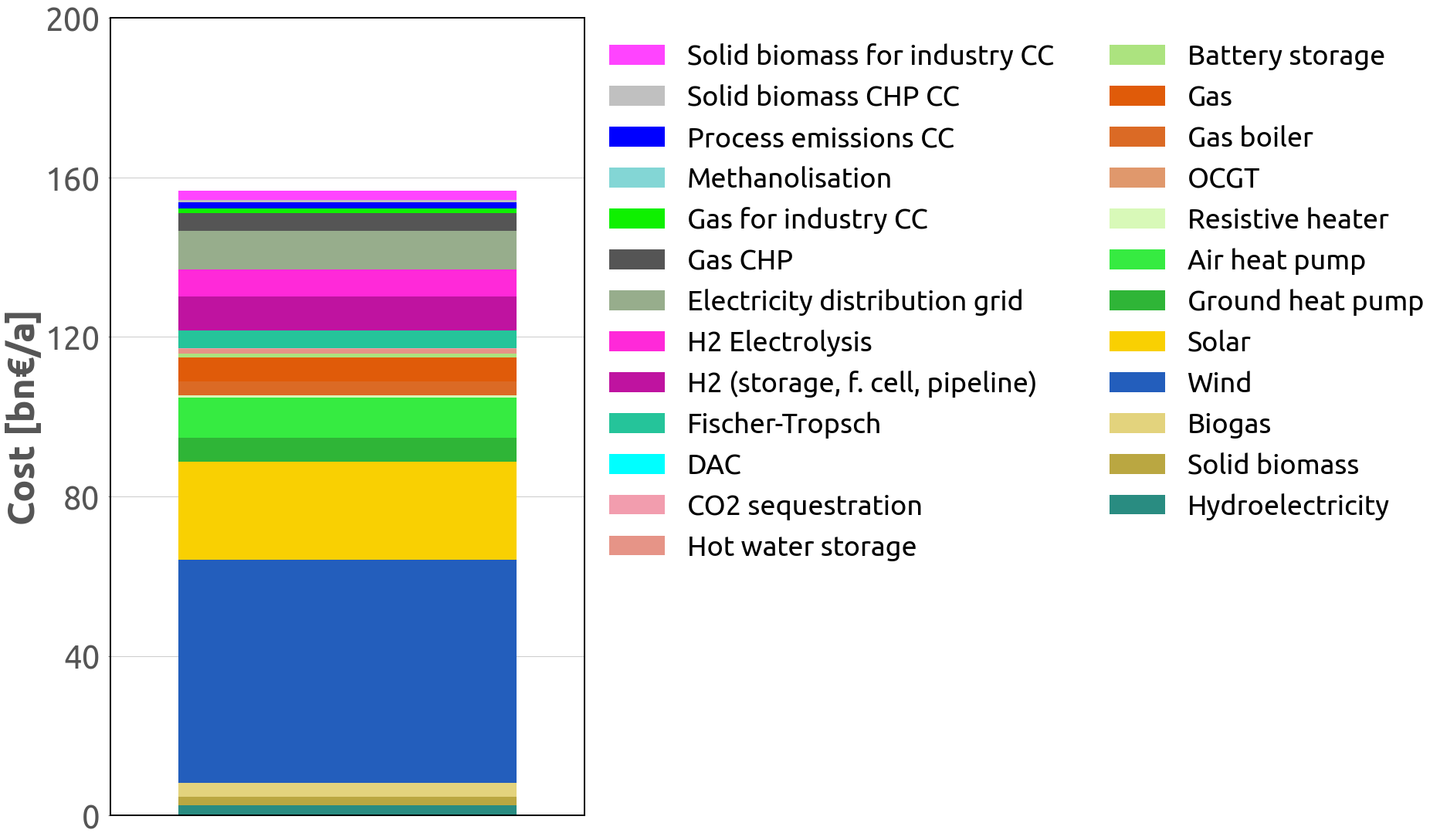}}
    \vspace{20pt}
    %\subfloat[]{\label{supplemental:figure_cost_variations_per_technology_interior_countries}\includegraphics[width = 0.96\linewidth]{./figures/figure_S11b.png}}
    \subfloat[]{\label{supplemental:figure_cost_variations_per_technology_interior_countries}\includegraphics[width = 0.8\linewidth]{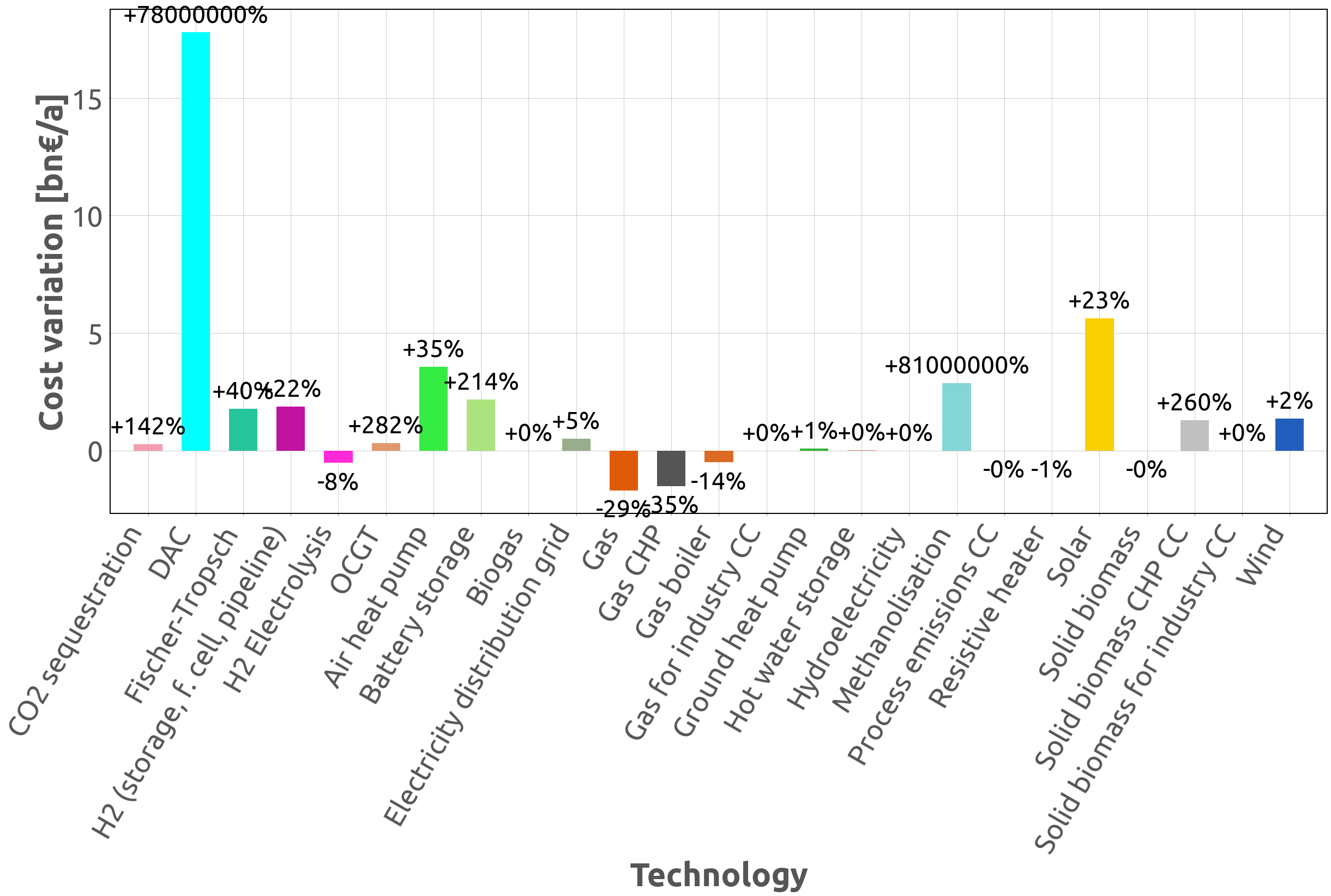}}
    \caption{(a) Total system cost and technology configuration of the interior countries (i.e. Germany, Belgium, and The Netherlands) in the global net-zero CO$_2$ emissions scenario and (b) Cost variation per technology between the global net-zero CO$_2$ emissions scenario and the local net-zero CO$_2$ emissions scenario of the interior countries.}
    \label{supplemental:figure_total_system_cost_and_variation_interior_countries}
\end{figure}

\begin{figure}[!htb]
    \centering
    %\subfloat[]{\label{supplemental:figure_total_system_cost_global_exterior_countries}\includegraphics[width = 0.96\linewidth]{./figures/figure_S12a.png}}
    \subfloat[]{\label{supplemental:figure_total_system_cost_global_exterior_countries}\includegraphics[width = 0.8\linewidth]{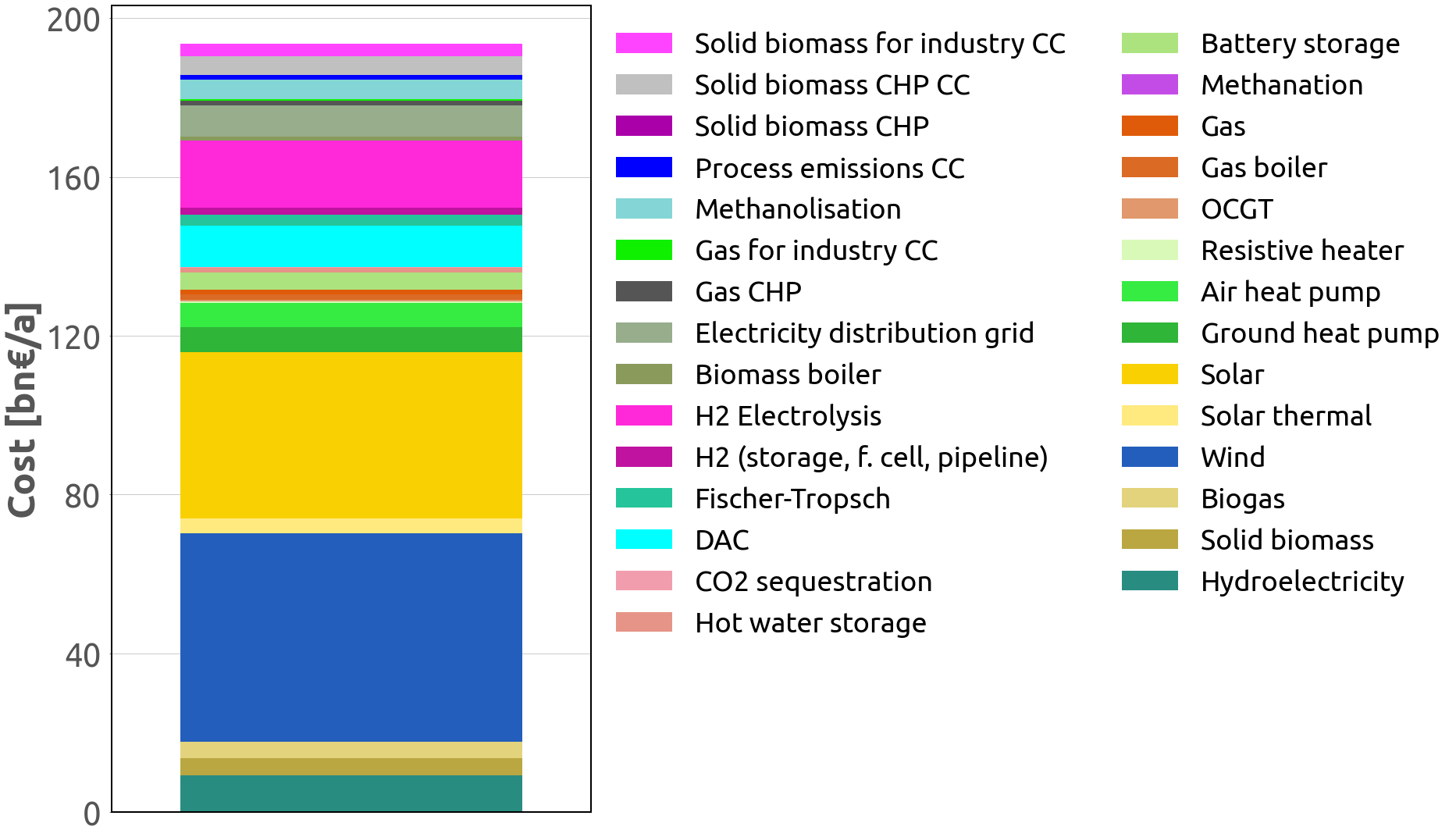}}
    \vspace{20pt}
    %\subfloat[]{\label{supplemental:figure_cost_variations_per_technology_exterior_countries}\includegraphics[width = 0.96\linewidth]{./figures/figure_S12b.png}}
    \subfloat[]{\label{supplemental:figure_cost_variations_per_technology_exterior_countries}\includegraphics[width = 0.8\linewidth]{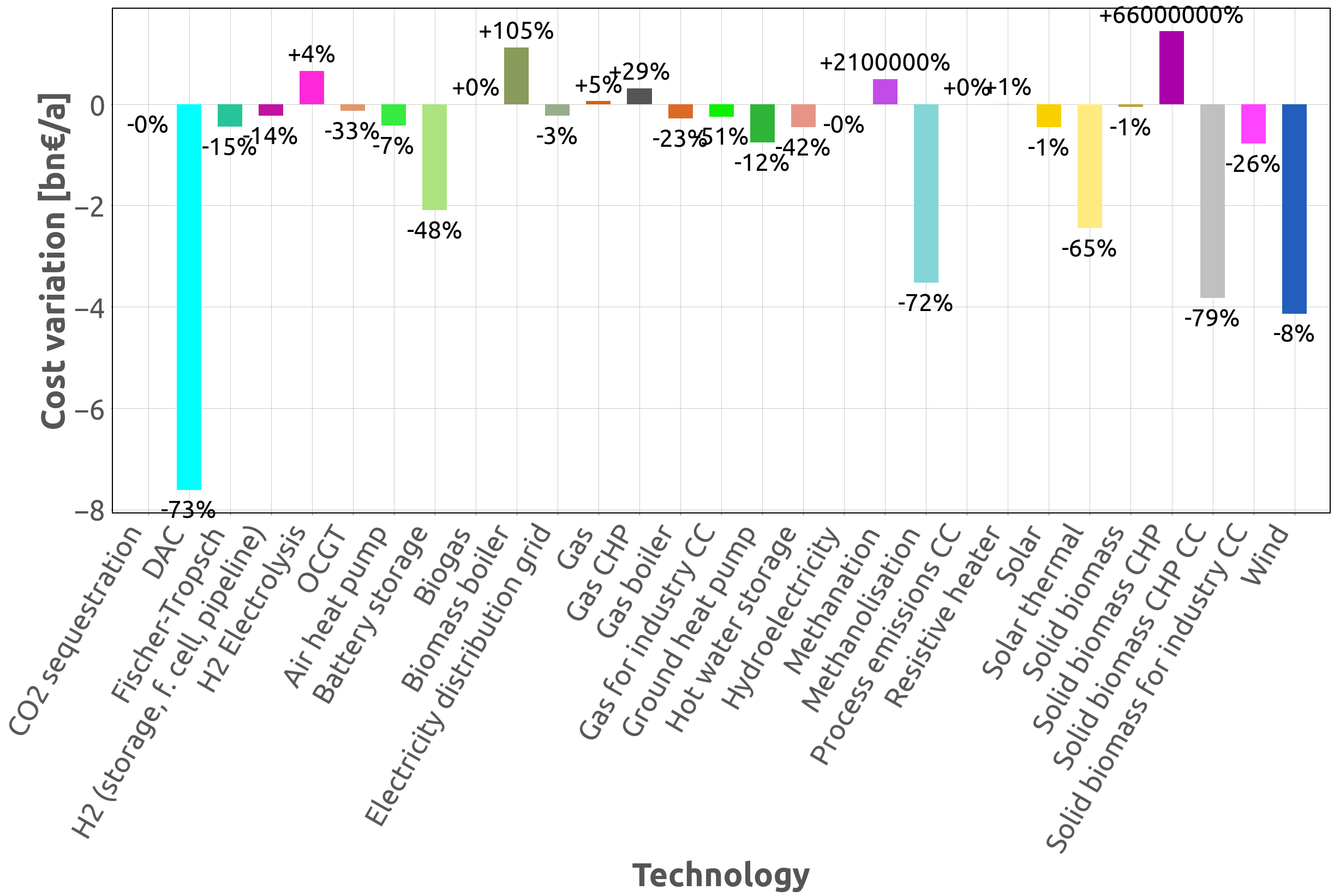}}
    \caption{(a) Total system cost and technology configuration of the exterior countries (i.e. Spain, Sweden, Finland, Poland, and Romania) in the global net-zero CO$_2$ emissions scenario and (b) Cost variation per technology between the global net-zero CO$_2$ emissions scenario and the local net-zero CO$_2$ emissions scenario of the exterior countries.}
    \label{supplemental:figure_total_system_cost_and_variation_exterior_countries}
\end{figure}

Under a local net-zero CO$_2$ emissions scenario, the total system cost increases 1.4\% relative to the system under a global constraint (Figure \ref{figure_total_system_cost_and_variation} and Table \ref{supplemental:table_total_system_cost_per_country}). The main contributors to system cost remain the same, and the cost increase is mainly due to an expansion of DAC and solar PV, especially in the interior countries (Figure \ref{supplemental:figure_total_system_cost_and_variation_interior_countries}), along with higher usage of air heat pumps, biomass boilers, and solid biomass-based CHP units. It is important to note that, while certain technologies increase the total system cost, others decrease it by becoming less relevant under local constraints. These include the reduction of solar thermal and carbon capture from solid biomass-based CHP units in many countries that were net CO$_2$ absorbers under a global constraint since their need to capture CO$_2$ is relaxed. In particular, exterior countries substantially reduce DAC (for the same prior reason) and methanolisation (as less captured CO$_2$ needs to be dealt with), as well as wind to power the two processes (Figure \ref{supplemental:figure_total_system_cost_and_variation_exterior_countries}).

Although the total system cost only increases slightly under local net-zero CO$_2$ emissions, the impact for individual countries can be dramatic. Countries that were net emitters under a global constraint increase their cost: Belgium by 58\%, The Netherlands by 39\%, and Germany by 12\%. Countries that were net absorbers under a global constraint, decrease their cost: Spain by 16\%, Sweden by 15\%, Romania by 13\%, and Finland by 12\%.

The low increase in total system cost is maintained even when reducing the size of regions. We found that a climate-neutral energy system under nodal constraints, where each of the 90 nodes is set with a net-zero CO$_2$ emissions target (Equation \ref{supplemental:material_nodal_co2_constraint}), is only 1.7\% more expensive than under a global constraint.

\subsection{A CO$_2$ transport network is cost-effective but not essential}
\begin{figure}[!htb]
    \centering
    %\subfloat[]{\label{supplemental:figure_total_system_cost_global_without_co2_network}\includegraphics[width = 0.96\linewidth]{./figures/figure_S16a.png}}
    \subfloat[]{\label{supplemental:figure_total_system_cost_global_without_co2_network}\includegraphics[width = 0.8\linewidth]{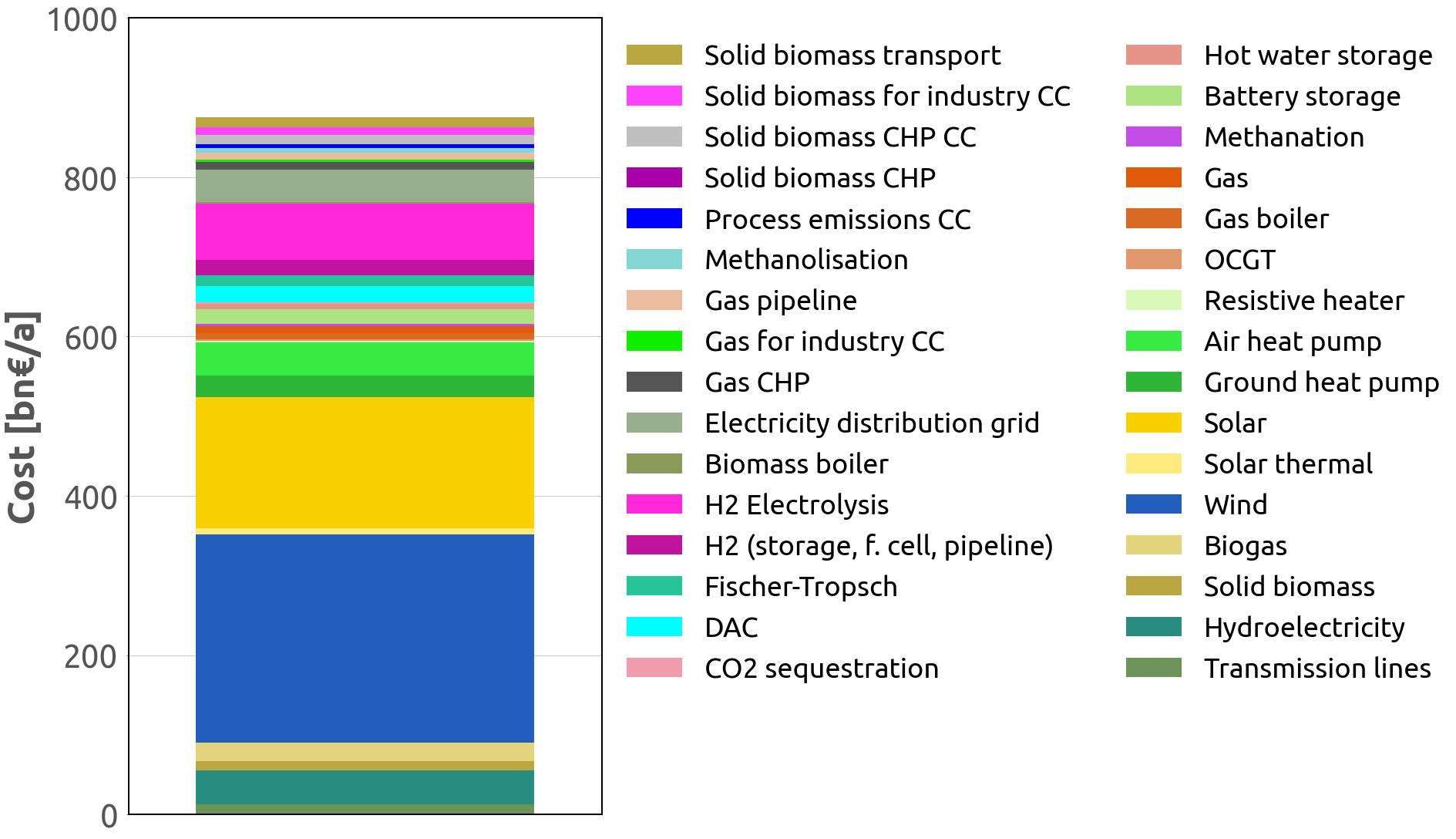}}
    \vspace{20pt}
    %\subfloat[]{\label{supplemental:figure_cost_variations_per_technology_global_without_co2_network}\includegraphics[width = 0.96\linewidth]{./figures/figure_S16b.png}}
    \subfloat[]{\label{supplemental:figure_cost_variations_per_technology_global_without_co2_network}\includegraphics[width = 0.8\linewidth]{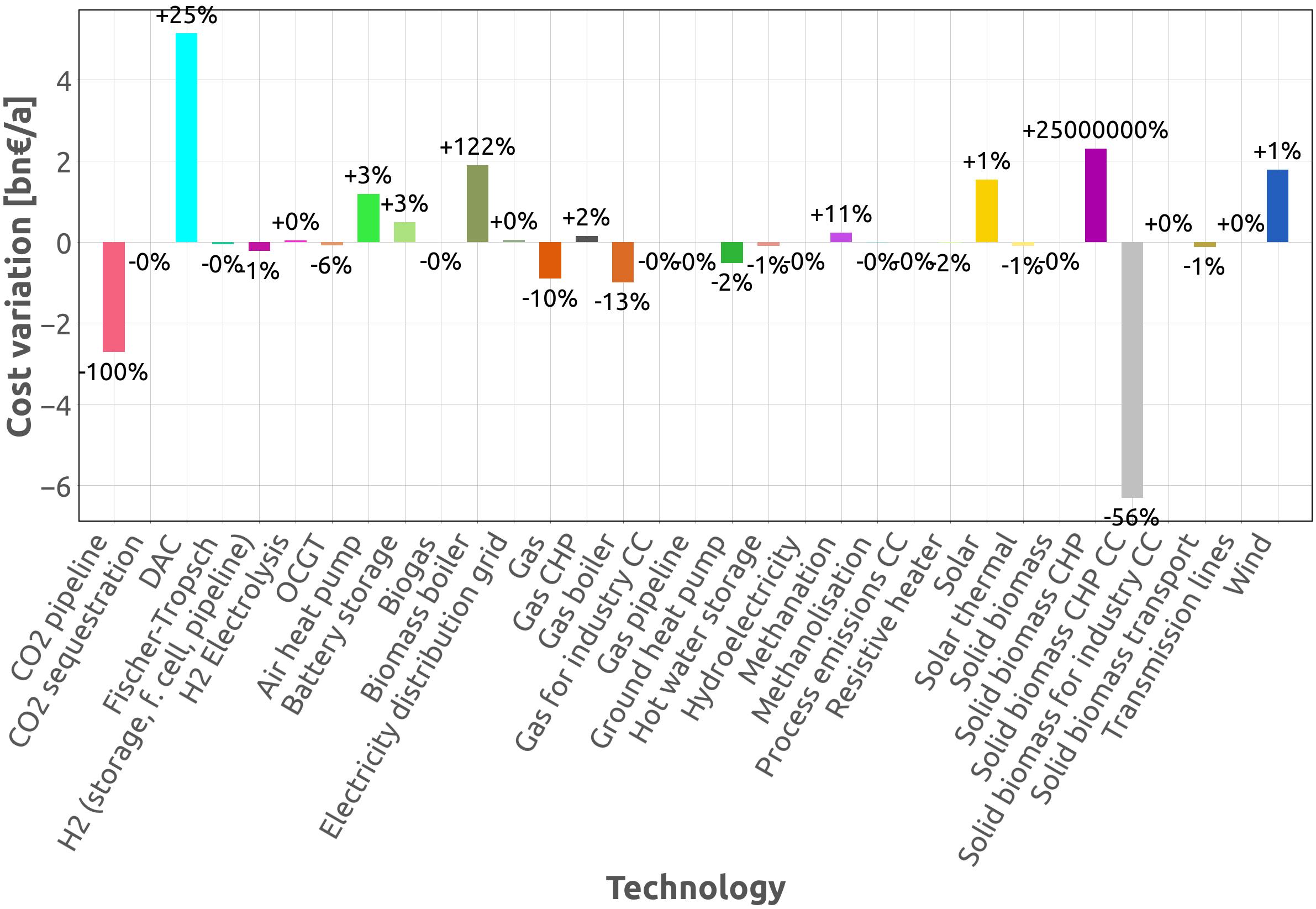}}
    \caption{(a) Total system cost and technology configuration in the global net-zero CO$_2$ emissions scenario without a CO$_2$ network and (b) Cost variation per technology between the global net-zero CO$_2$ emissions scenario with a CO$_2$ network and the global net-zero CO$_2$ emissions scenario without a CO$_2$ network. The total cost increases by 0.3\% when modelling a system without a CO$_2$ network compared to its counterpart equipped with such a network. Other than the differences in deployment levels, the technology configuration is the same in both types of systems.}    \label{supplemental:figure_total_system_cost_and_variation_global_without_co2_network}
\end{figure}

\begin{figure}[!htb]
    \centering
    %\subfloat[]{\label{supplemental:figure_total_system_cost_local_without_co2_network}\includegraphics[width = 0.96\linewidth]{./figures/figure_S17a.png}}
    \subfloat[]{\label{supplemental:figure_total_system_cost_local_without_co2_network}\includegraphics[width = 0.8\linewidth]{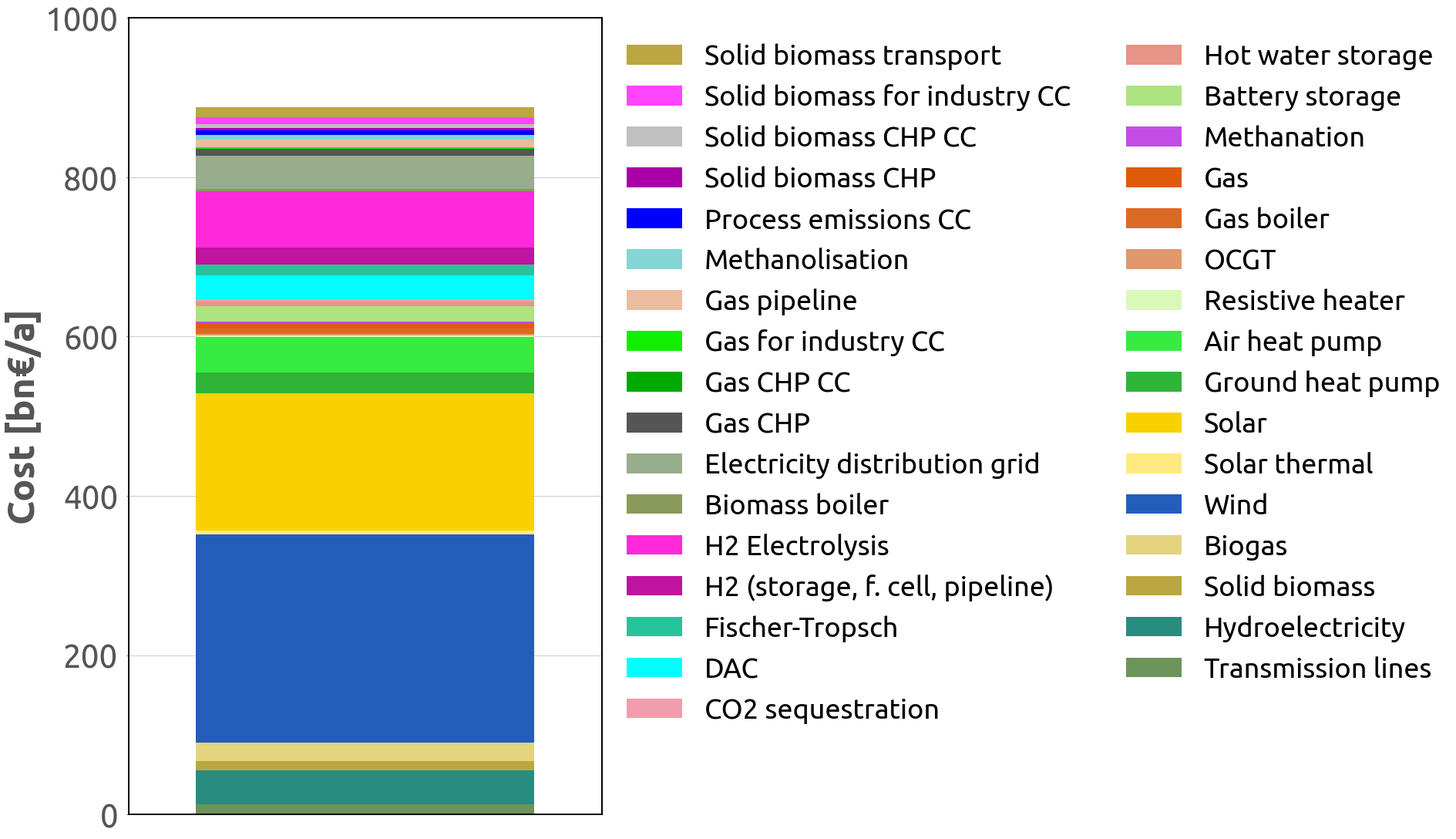}}
    \vspace{20pt}
    %\subfloat[]{\label{supplemental:figure_cost_variations_per_technology_local_without_co2_network}\includegraphics[width = 0.96\linewidth]{./figures/figure_S17b.png}}
    \subfloat[]{\label{supplemental:figure_cost_variations_per_technology_local_without_co2_network}\includegraphics[width = 0.8\linewidth]{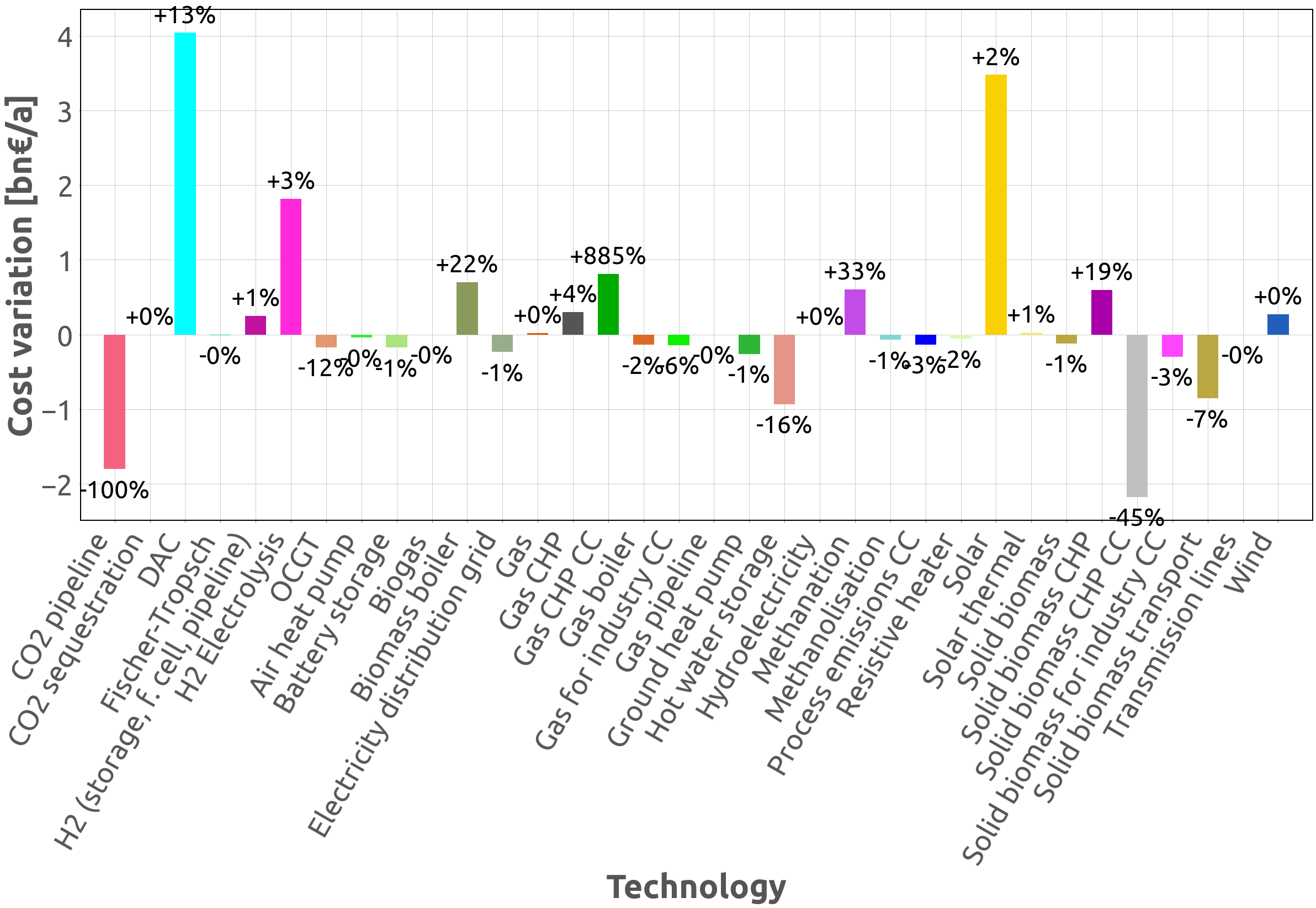}}
    \caption{(a) Total system cost and technology configuration in the local net-zero CO$_2$ emissions scenario without a CO$_2$ network and (b) Cost variation per technology between the local net-zero CO$_2$ emissions scenario with a CO$_2$ network and the local net-zero CO$_2$ emissions scenario without a CO$_2$ network. The total cost increases by 0.6\% when modelling a system without a CO$_2$ network compared to its counterpart equipped with such a network. Other than the differences in deployment levels, the technology configuration is the same in both types of systems.}    \label{supplemental:figure_total_system_cost_and_variation_local_without_co2_network}
\end{figure}

Building a CO$_2$ transport network is found cost effective in all scenarios. In case Europe does not establish a CO$_2$ network amongst its nations, the model predicts that a climate-neutral energy system would experience a 0.3\% cost increase (2.6 billion € per year) in the globally constrained scenario and a 0.6\% cost increase (5.4 billion € per year) in the locally constrained scenario, compared to their counterparts equipped with a CO$_2$ network (Table \ref{table_energy_system_total_cost} and Figures \ref{supplemental:figure_total_system_cost_and_variation_global_without_co2_network} and \ref{supplemental:figure_total_system_cost_and_variation_local_without_co2_network}). The cost increase is due to the model no longer leveraging from a network to manage captured CO$_2$ efficiently by sending it to neighbouring countries for underground sequestration or helping countries to balance their national CO$_2$ demand and supply for manufacturing synthetic products amongst themselves.

\begin{table}[!htb]
    \renewcommand{\arraystretch}{0.8}% Tighter
    \small
    \centering
    \caption{Total cost of a sector-coupled climate-neutral European energy system with and without a network to transport CO$_2$ under a global, local, and nodal net-zero CO$_2$ emissions constraints. Values are in billion € per year and percentage increase with respect to the global scenario with CO$_2$ network.}
    \label{table_energy_system_total_cost}
    \begin{tabular}{*4c}
        \toprule
        \textbf{Energy system} & \textbf{Global} & \textbf{Local} & \textbf{Nodal} \\
        \midrule
        With CO$_2$ network    & 878             & 890 (+1.4\%)   & 893 (+1.7\%)   \\
        Without CO$_2$ network & 881 (+0.3\%)    & 896 (+2.1\%)   & 915 (+4.2\%)   \\
        \bottomrule
    \end{tabular}
\end{table}

\subsection{High CO$_2$ prices needed to attain carbon neutrality}
\begin{figure}[!htb]
    \centering
    %\subfloat[]{\label{supplemental:figure_co2_shadow_price22a}\includegraphics[width = 0.8\linewidth]{./figures/figure_S14a.png}}
    \subfloat[]{\label{supplemental:figure_co2_shadow_price22a}\includegraphics[width = 0.65\linewidth]{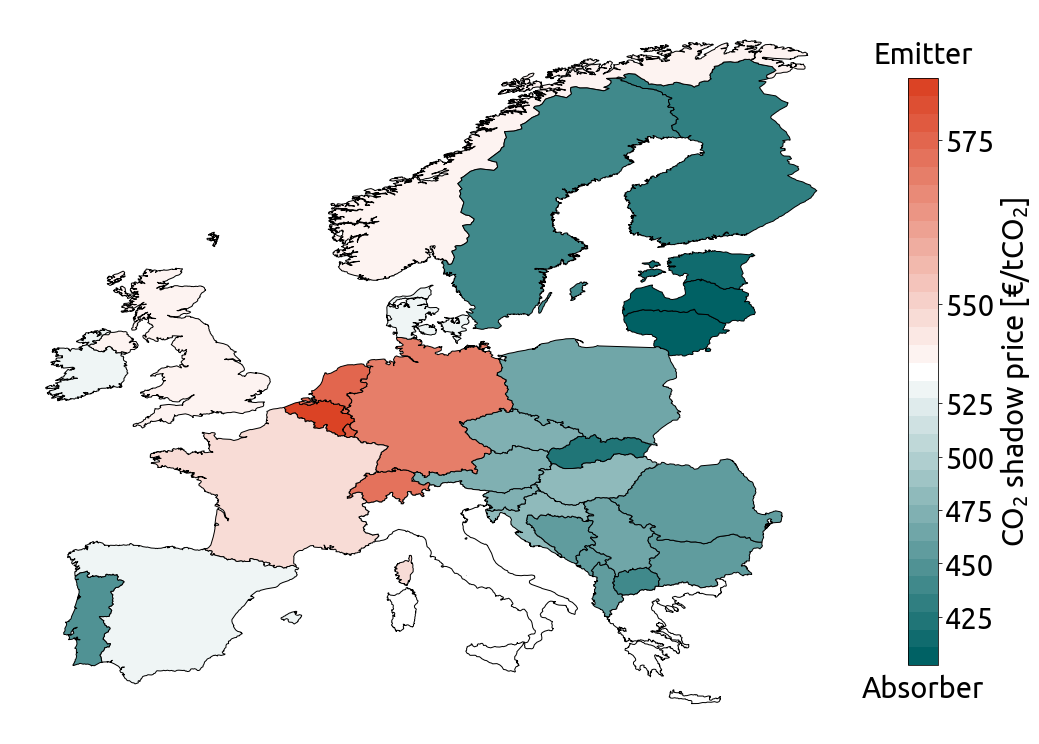}}
    \vspace{20pt}
    %\subfloat[]{\label{supplemental:figure_co2_shadow_price_co2_emissions_map}\includegraphics[width = 0.8\linewidth]{./figures/figure_S14b.png}}
    \subfloat[]{\label{supplemental:figure_co2_shadow_price_co2_emissions_map}\includegraphics[width = 0.65\linewidth]{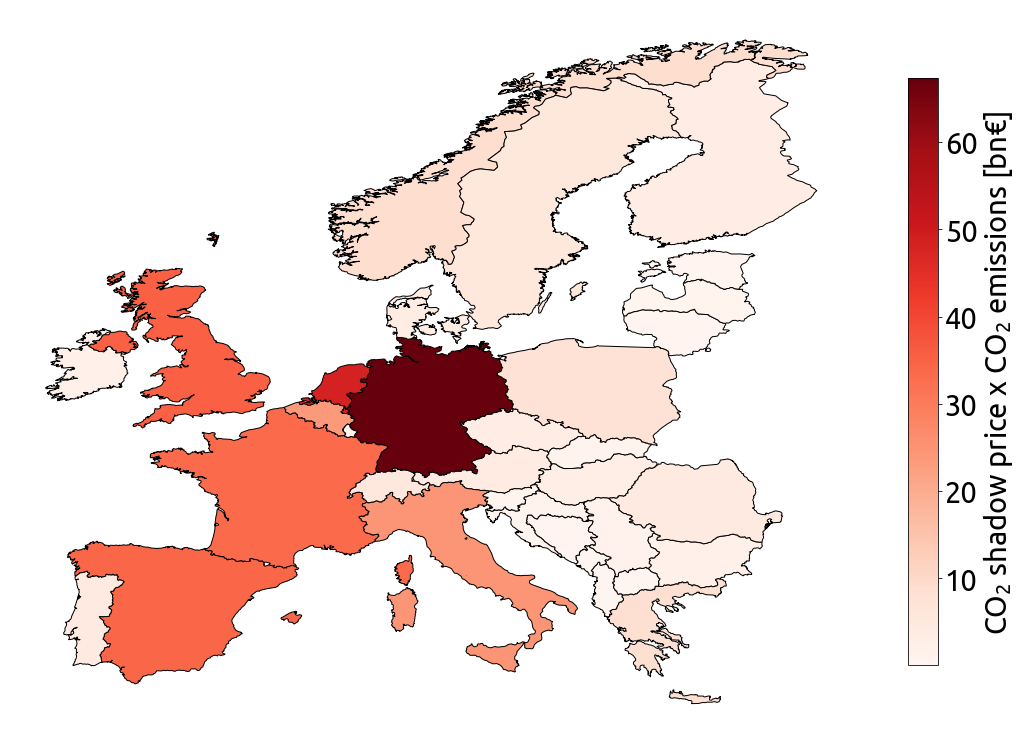}}
    \caption{(a) CO$_2$ shadow price per country across Europe in the local net-zero CO$_2$ emissions scenario and (b) Product of CO$_2$ shadow price and CO$_2$ emissions per country across Europe in the local net-zero CO$_2$ emissions scenario. See Figure \ref{supplemental:figure_co2_emissions} for the amount of CO$_2$ emissions per country. As a reference, the CO$_2$ shadow price for the whole of Europe is set at 540 € per tonne of CO$_2$ in the global net-zero CO$_2$ emissions scenario.}
    \label{supplemental:figure_co2_shadow_price_co2_emissions}
\end{figure}

\begin{figure}[!htb]
    \centering
    %\subfloat[]{\label{supplemental:figure_co2_shadow_price_all_countries}\includegraphics[width = 0.96\linewidth]{./figures/figure_S15a.png}}
    \subfloat[]{\label{supplemental:figure_co2_shadow_price_all_countries}\includegraphics[width = 0.8\linewidth]{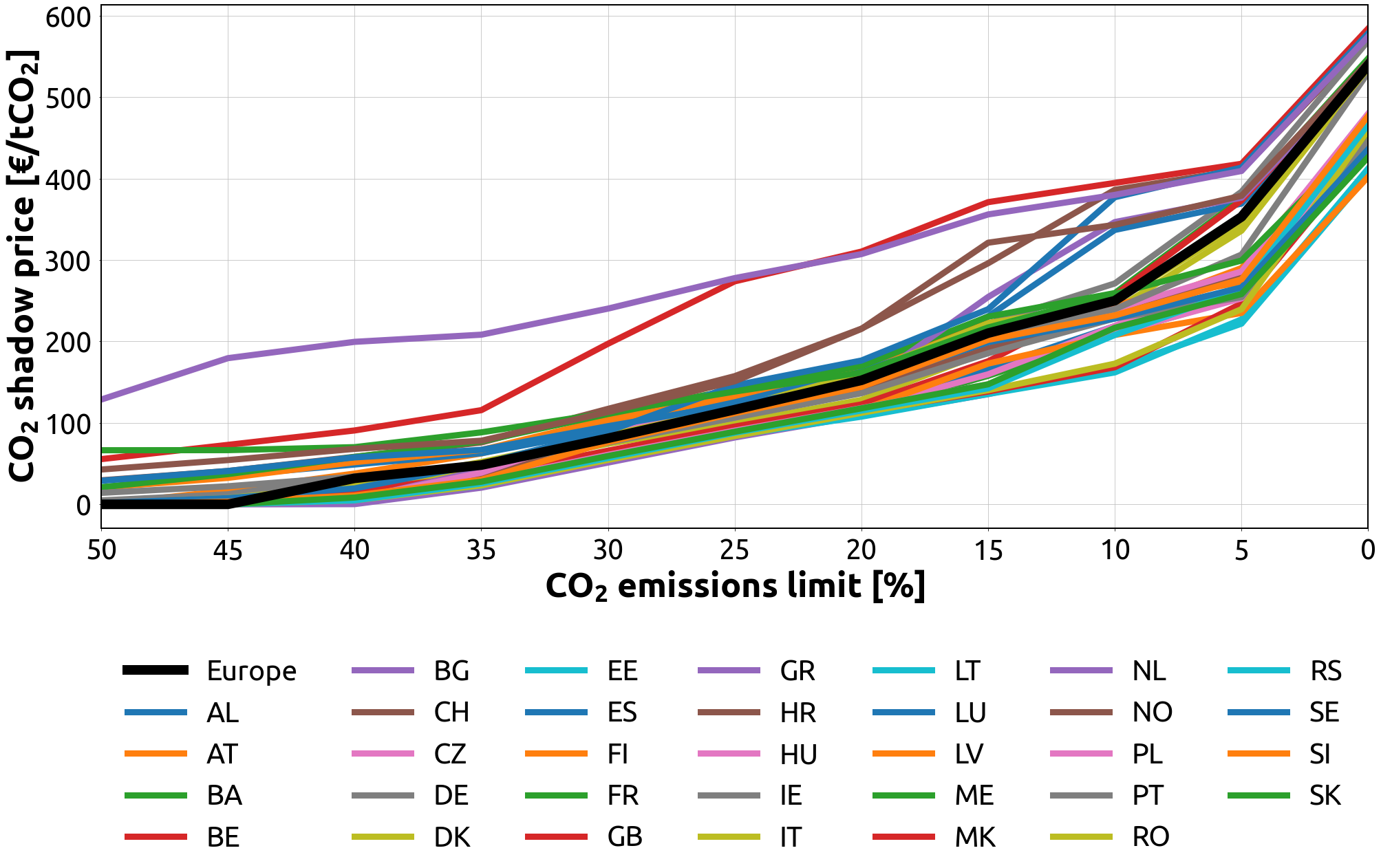}}
    \vspace{20pt}
    %\subfloat[]{\label{supplemental:figure_co2_shadow_price_interior_countries}\includegraphics[width = 0.485\linewidth]{./figures/figure_S15b.png}}\hfill
    \subfloat[]{\label{supplemental:figure_co2_shadow_price_interior_countries}\includegraphics[width = 0.42\linewidth]{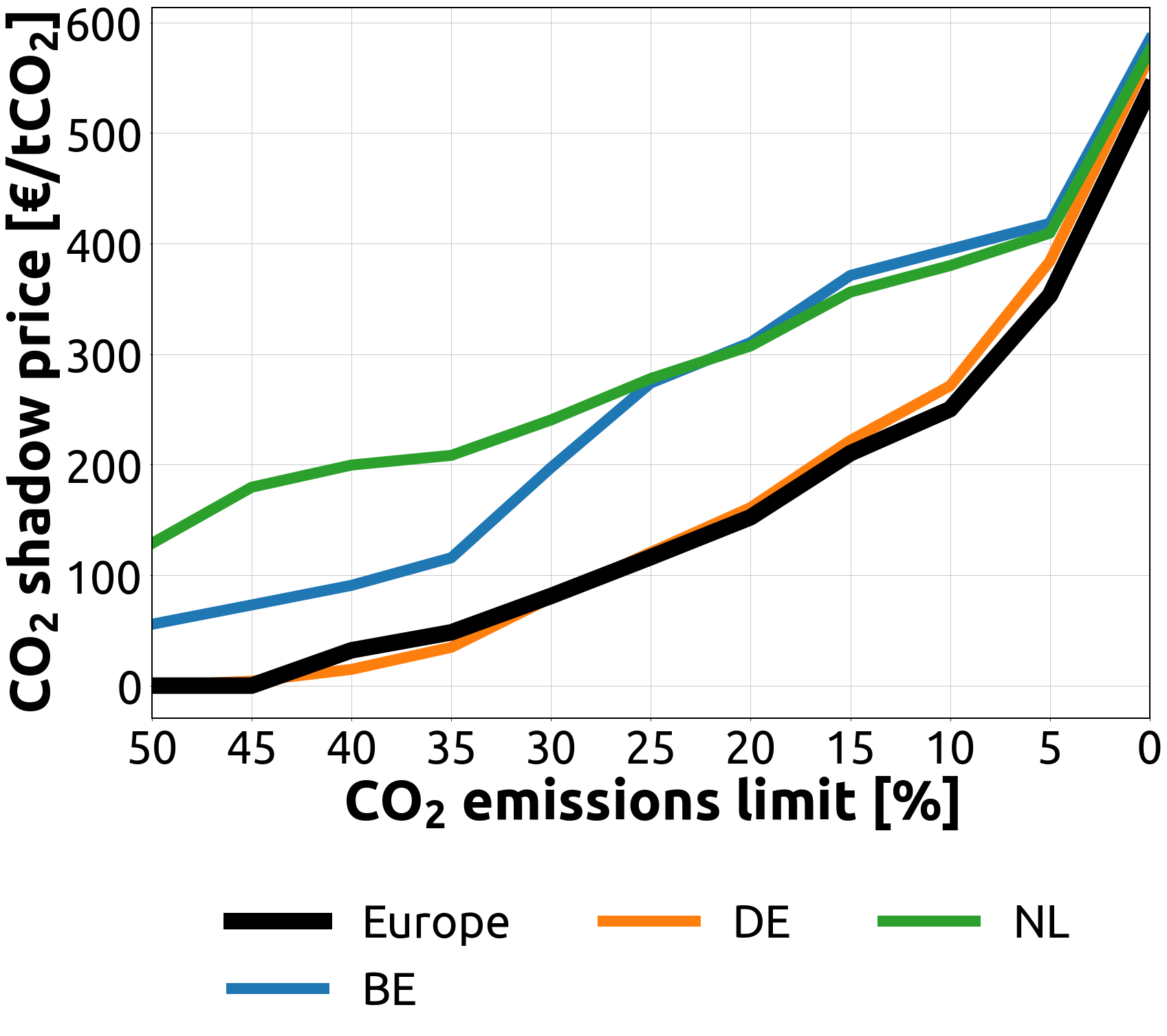}}\hfill
    %\subfloat[]{\label{supplemental:figure_co2_shadow_price_exterior_countries}\includegraphics[width = 0.485\linewidth]{./figures/figure_S15c.png}}\par
    \subfloat[]{\label{supplemental:figure_co2_shadow_price_exterior_countries}\includegraphics[width = 0.42\linewidth]{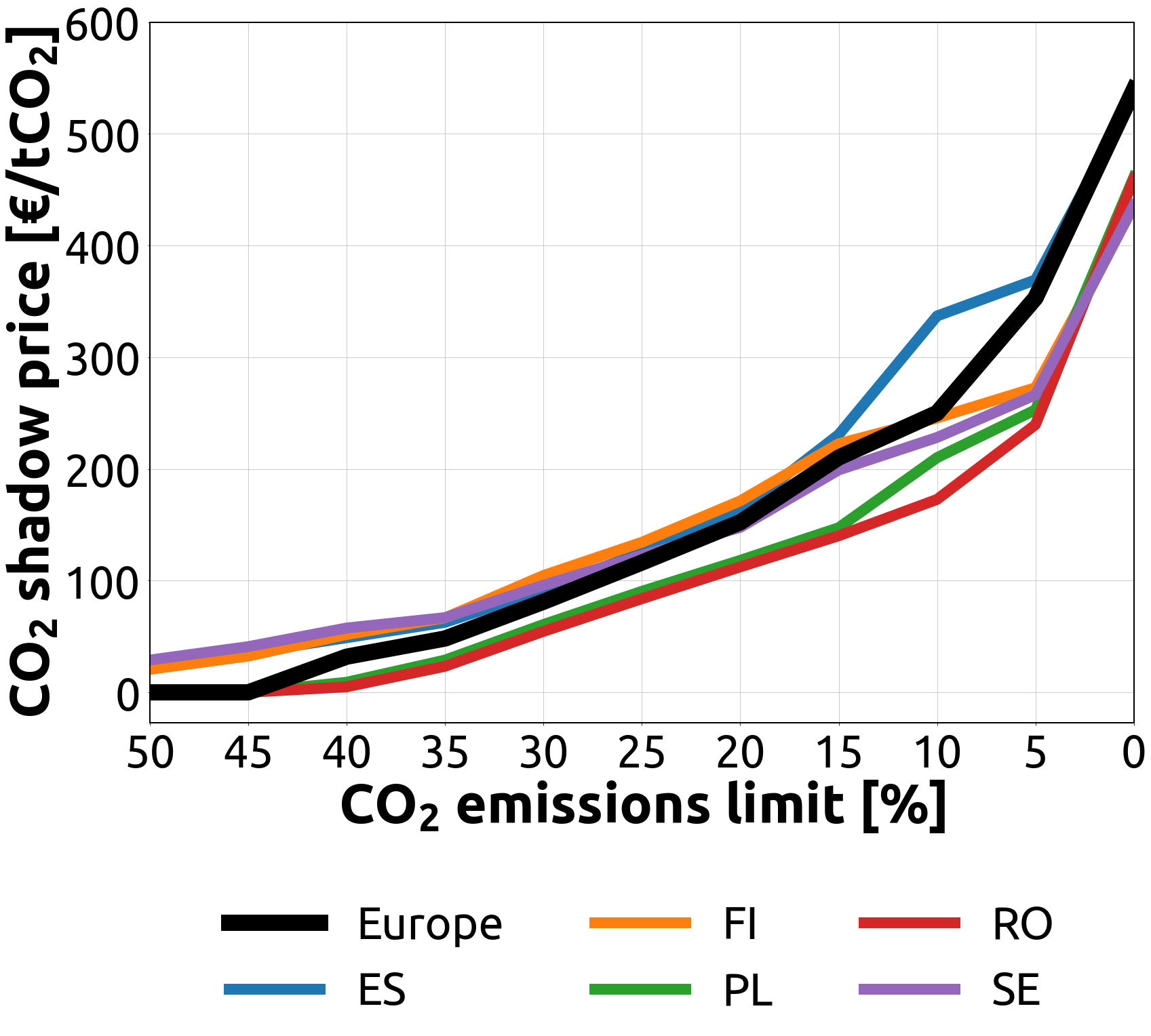}}\par
    \caption{CO$_2$ shadow price sensitivity analysis under different CO$_2$ emissions limits (a) for Europe (global scenario) and all the modelled countries (local scenario), (b) for Europe (global scenario) and the interior countries (local scenario), and (c) for Europe (global scenario) and the exterior countries (local scenario). In a continent fully decarbonised under local constraints, countries that used to emit more CO$_2$ than they absorb annually in the global scenario have local CO$_2$ shadow prices higher than Europe's global CO$_2$ shadow price. Conversely, countries that used to absorb more CO$_2$ than they emit in the global scenario have local CO$_2$ shadow prices lower than Europe's global CO$_2$ shadow price.}
    \label{supplemental:figure_co2_shadow_price}
\end{figure}

The required CO$_2$ price to attain the emissions target is an output of the model and coincides with the Lagrange multiplier associated with the CO$_2$ limit constraint. For a global net-zero emissions constraint, it resulted in 540 €/tCO$_2$. When local net-zero emissions constraints are imposed, the price required to attain climate neutrality varies significantly across countries, ranging from 402 in Latvia to 584 €/tCO$_2$ in Belgium (Figures \ref{supplemental:figure_co2_shadow_price_co2_emissions} and \ref{supplemental:figure_co2_shadow_price} and Table \ref{supplemental:table_co2_shadow_price}). The variation is due to each country having a distinct mix of exogenous CO$_2$ emissions and access to renewable energy resources. As expected, there is a pattern between the price and whether countries are net CO$_2$ emitters or absorbers under a global constraint: previously-emitting countries have higher local prices than the global CO$_2$ price, whereas previously-absorbing countries have lower local prices than the global CO$_2$ price.

\subsection{Carbon capture: DAC collocated with renewables or to boost capture under local CO$_2$ constraint} \label{sec_capture}
\begin{figure}[!htb]
    \centering
    %\subfloat[]{\label{figure_spatial_co2_capture_global}\includegraphics[width = 0.8\linewidth]{./figures/figure_3a.png}}
    \subfloat[]{\label{figure_spatial_co2_capture_global}\includegraphics[width = 0.66\linewidth]{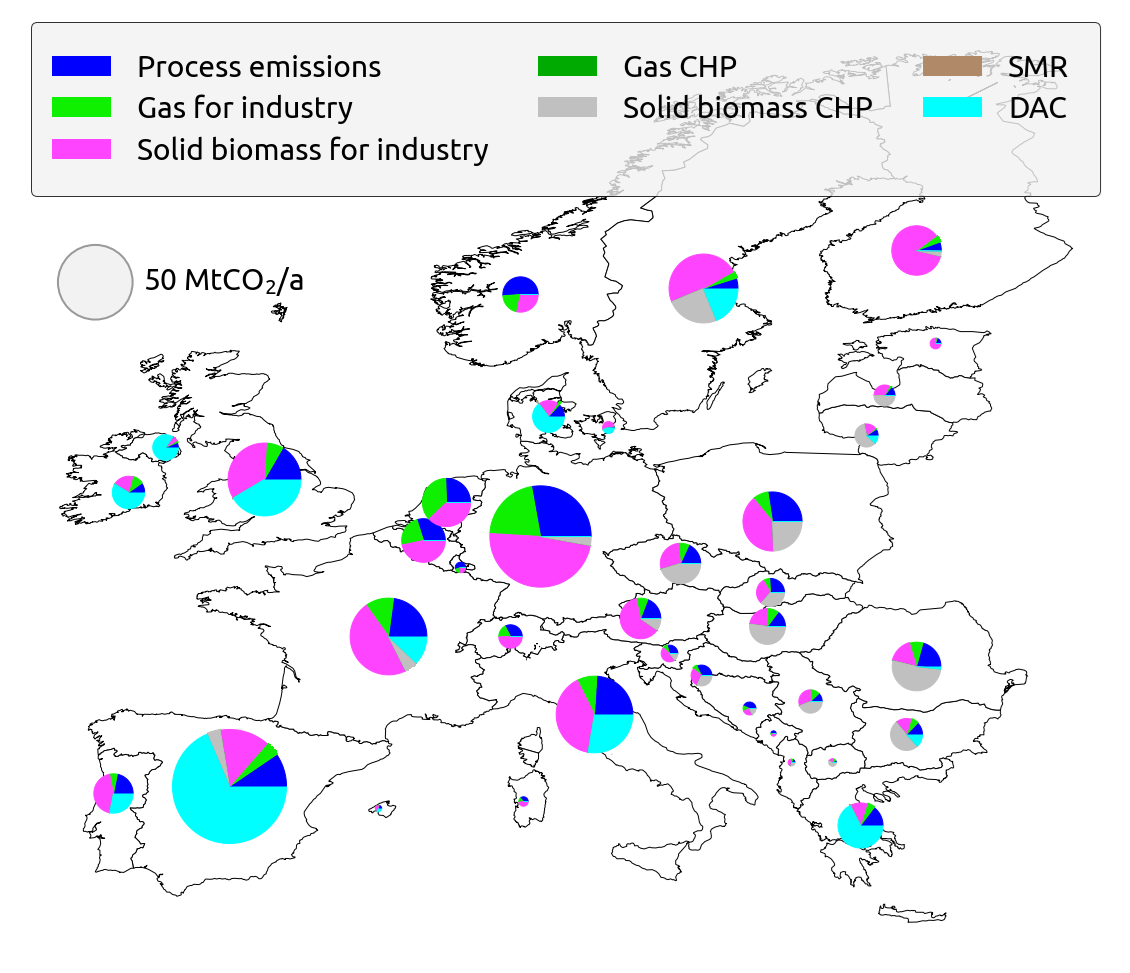}}
    \vspace{20pt}
    %\subfloat[]{\label{figure_spatial_co2_capture_variation}\includegraphics[width = 0.96\linewidth]{./figures/figure_3b.png}}
    \subfloat[]{\label{figure_spatial_co2_capture_variation}\includegraphics[width = 0.8\linewidth]{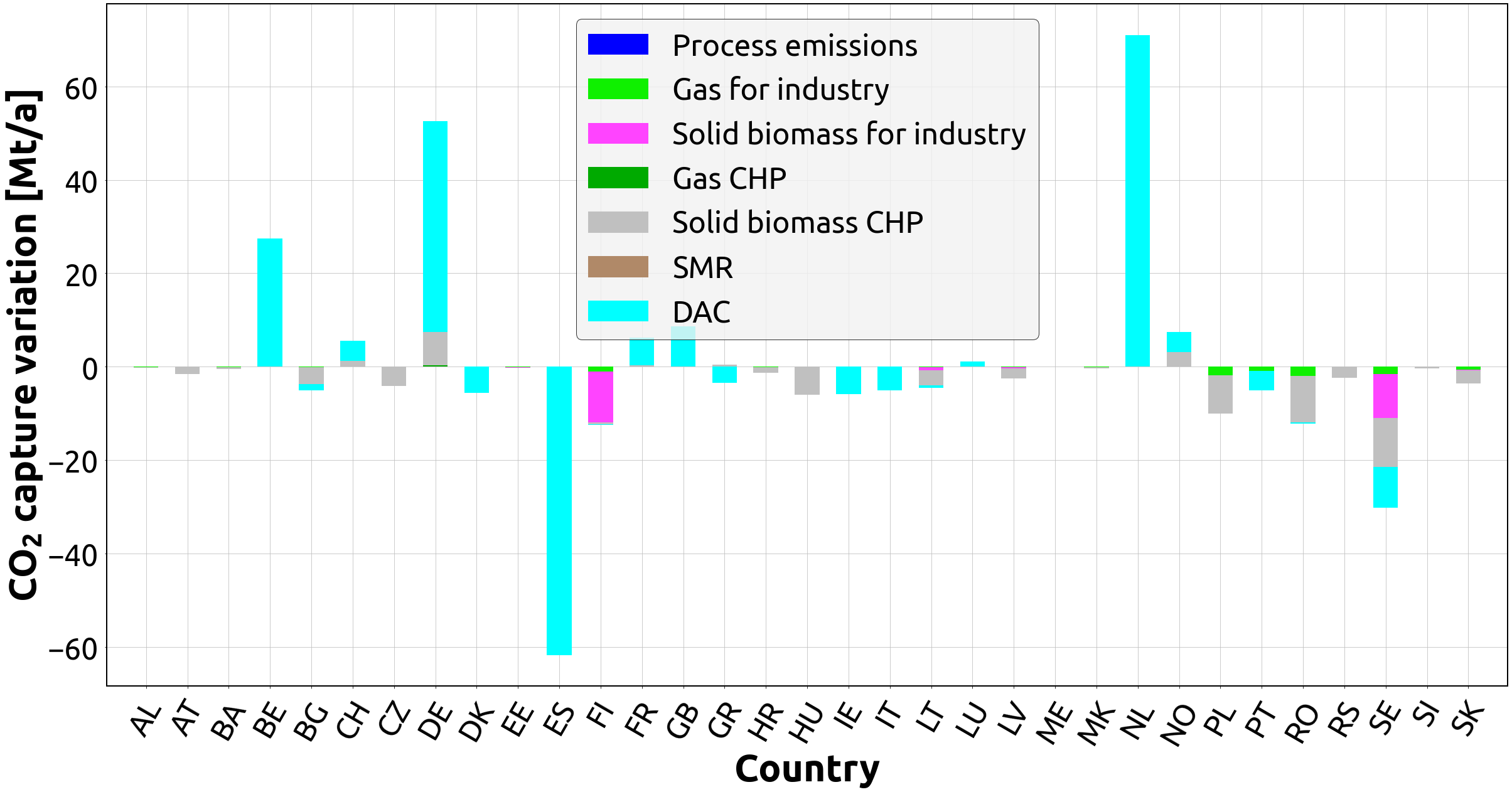}}
    \caption{(a) CO$_2$ capture in the global net-zero CO$_2$ emissions scenario and (b) CO$_2$ capture variation per country and technology between the global net-zero CO$_2$ emissions scenario and the local net-zero CO$_2$ emissions scenario.}
    \label{figure_spatial_co2_capture_and_variation}
\end{figure}

In the model, CO$_2$ is primarily captured from: (i) industry process emissions, (ii) gas and solid biomass used for heating in the industry and combusted in CHP units, and (iii) DAC. At a continental level, both climate-neutral scenarios result in a similar total of 724 MtCO$_2$ emitted and captured annually by these methods. However, at a national level, the amount of CO$_2$ captured and the methods used show significant variations between the two scenarios (Figure \ref{figure_spatial_co2_capture_and_variation}).

The first remarkable change is how DAC shifts from countries that are net CO$_2$ absorber under a global CO$_2$ constraint to net CO$_2$ emitter countries. The former countries have better access to renewable resources (Figure \ref{supplemental:figure_renewable_capacity_factors}) but, under local constraints, do not have such a strong push for capturing CO$_2$, which is transmitted to the latter countries. These install DAC in areas where district heating is available, which gives access to low-cost heat, increasing the competitiveness of DAC.

Solid biomass-based CHP units with carbon capture (CC) decline in most net CO$_2$ absorber countries under local constraints (Figure \ref{supplemental:figure_electricity_production}). In some of these countries, they are replaced by units without CC to reduce costs. In others, this decline is due to fewer solid biomass-based CHP units being deployed since power and heating demand has decreased as a consequence of a reduction in DAC and methanol production (Figure \ref{figure_spatial_co2_conversion_and_variation}), both of which are power-intensive processes, along with an increase in air heat pumps (Figure \ref{figure_cost_variations_per_technology}). CO$_2$ captured from solid biomass used in the industry also decreases in a few of these countries due to no longer being cost-effective under these constraints.

The process emissions and the gas demand in the industry are exogenously fixed, and the model shows that capturing CO$_2$ from both sources is cost-effective in nearly every location, with minimal variation between the two scenarios. The production of blue H$_2$ is not cost-competitive in any region or scenario, and as a result, no CO$_2$ is captured from steam methane reforming (SMR).

\begin{figure}[!htb]
    \centering
    %\subfloat[]{\label{supplemental:figure_temporal_co2_capture_exogenous_global}\includegraphics[width = 0.485\linewidth]{./figures/figure_S24a.png}}\hfill
    \subfloat[]{\label{supplemental:figure_temporal_co2_capture_exogenous_global}\includegraphics[width = 0.415\linewidth]{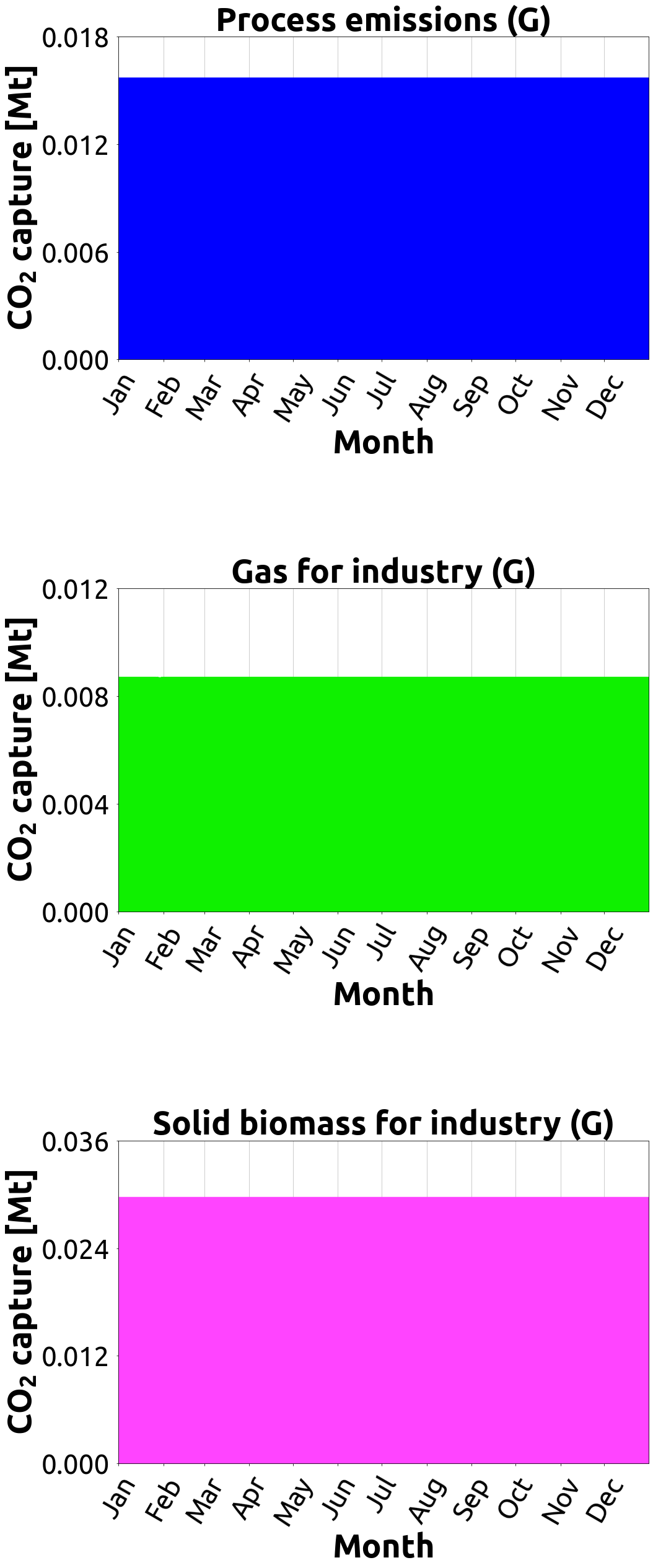}}\hfill
    %\subfloat[]{\label{supplemental:figure_temporal_co2_capture_exogenous_local}\includegraphics[width = 0.485\linewidth]{./figures/figure_S24b.png}}\par
    \subfloat[]{\label{supplemental:figure_temporal_co2_capture_exogenous_local}\includegraphics[width = 0.415\linewidth]{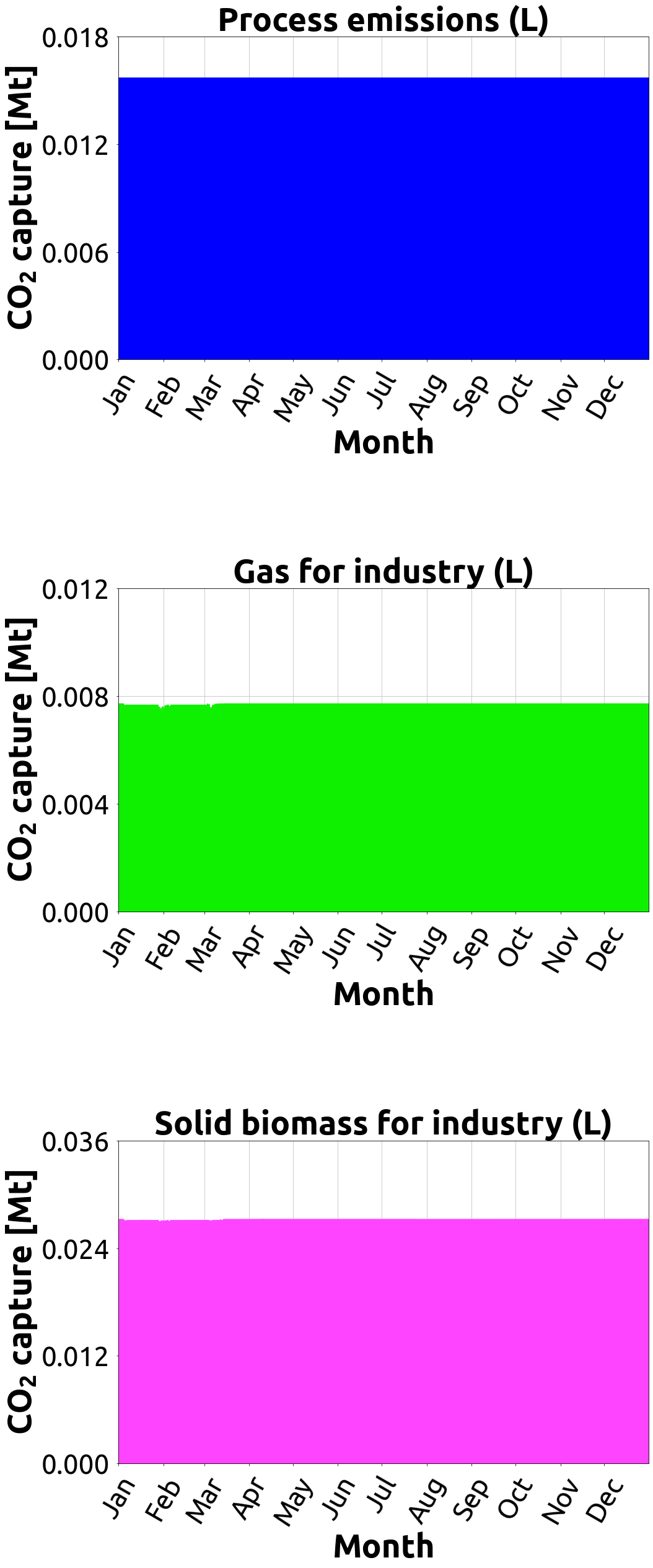}}\par
    \caption{Temporal CO$_2$ capture from exogenously fixed sources across all of Europe based on the weather year 2013 in (a) the global net-zero CO$_2$ emissions scenario and (b) the local net-zero CO$_2$ emissions scenario.}
    \label{supplemental:figure_temporal_co2_capture_exogenous}
\end{figure}

From a temporal perspective, our results show three distinct operational patterns for capturing CO$_2$, present in both scenarios. The first pattern occurs when low-cost electricity is available, typically during the summer, to power DAC. The second pattern occurs when gas and solid biomass are combusted in CHP units to provide heat and power during the winter, using CC. Finally, as the third pattern, since industry is assumed to operate continuously, the amount of CO$_2$ captured from process emissions, as well as gas and solid biomass used in the industry is kept constant throughout the year (Figures \ref{supplemental:figure_temporal_co2_capture_exogenous} and \ref{supplemental:figure_temporal_co2_capture_endogenous}).

\subsection{Carbon conversion and sequestration: Production of carbonaceous fuels shifted to interior countries under local constraints} \label{sec_conversion}
\begin{figure}[!htb]
    \centering
    %\subfloat[]{\label{figure_spatial_co2_conversion_global}\includegraphics[width = 0.8\linewidth]{./figures/figure_4a.png}}
    \subfloat[]{\label{figure_spatial_co2_conversion_global}\includegraphics[width = 0.66\linewidth]{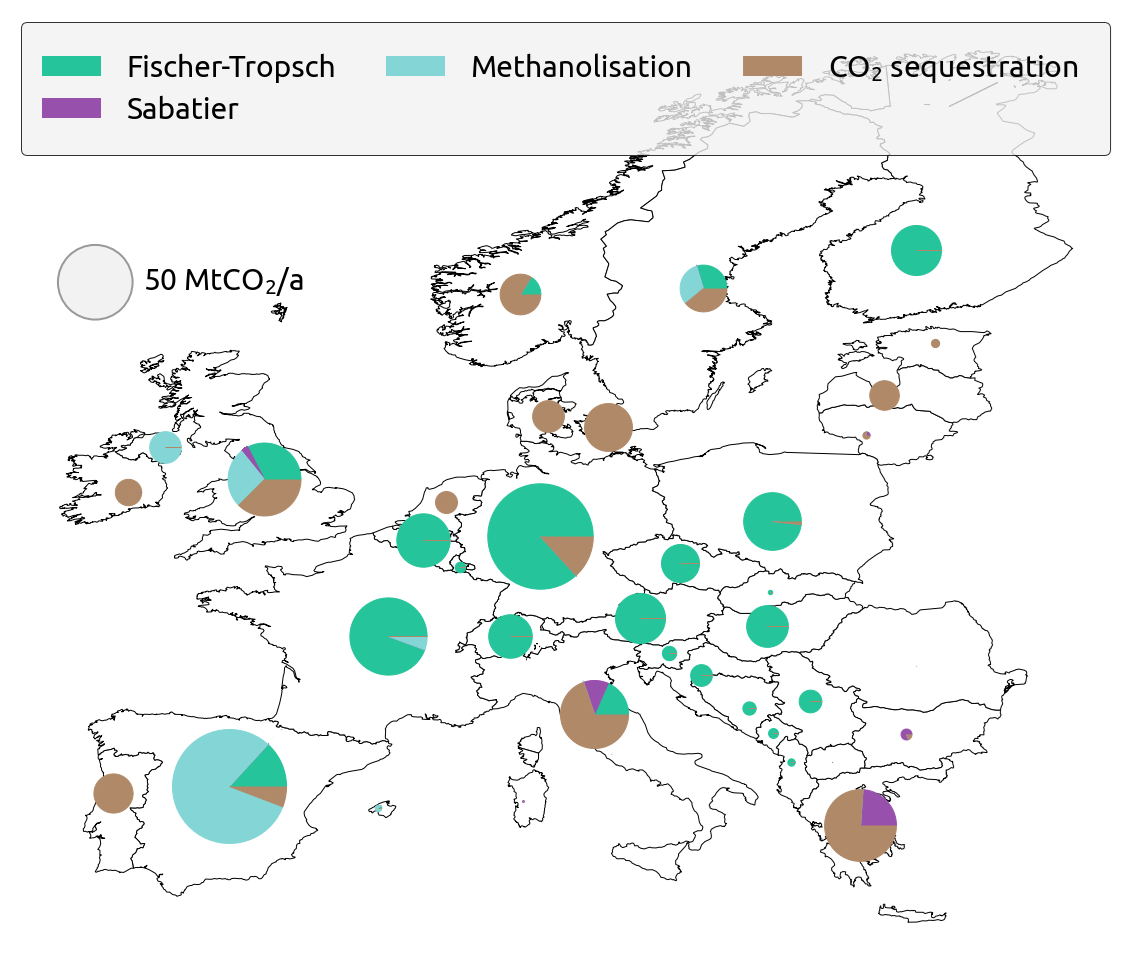}}
    \vspace{20pt}
    %\subfloat[]{\label{figure_spatial_co2_conversion_variation}\includegraphics[width = 0.96\linewidth]{./figures/figure_4b.png}}
    \subfloat[]{\label{figure_spatial_co2_conversion_variation}\includegraphics[width = 0.8\linewidth]{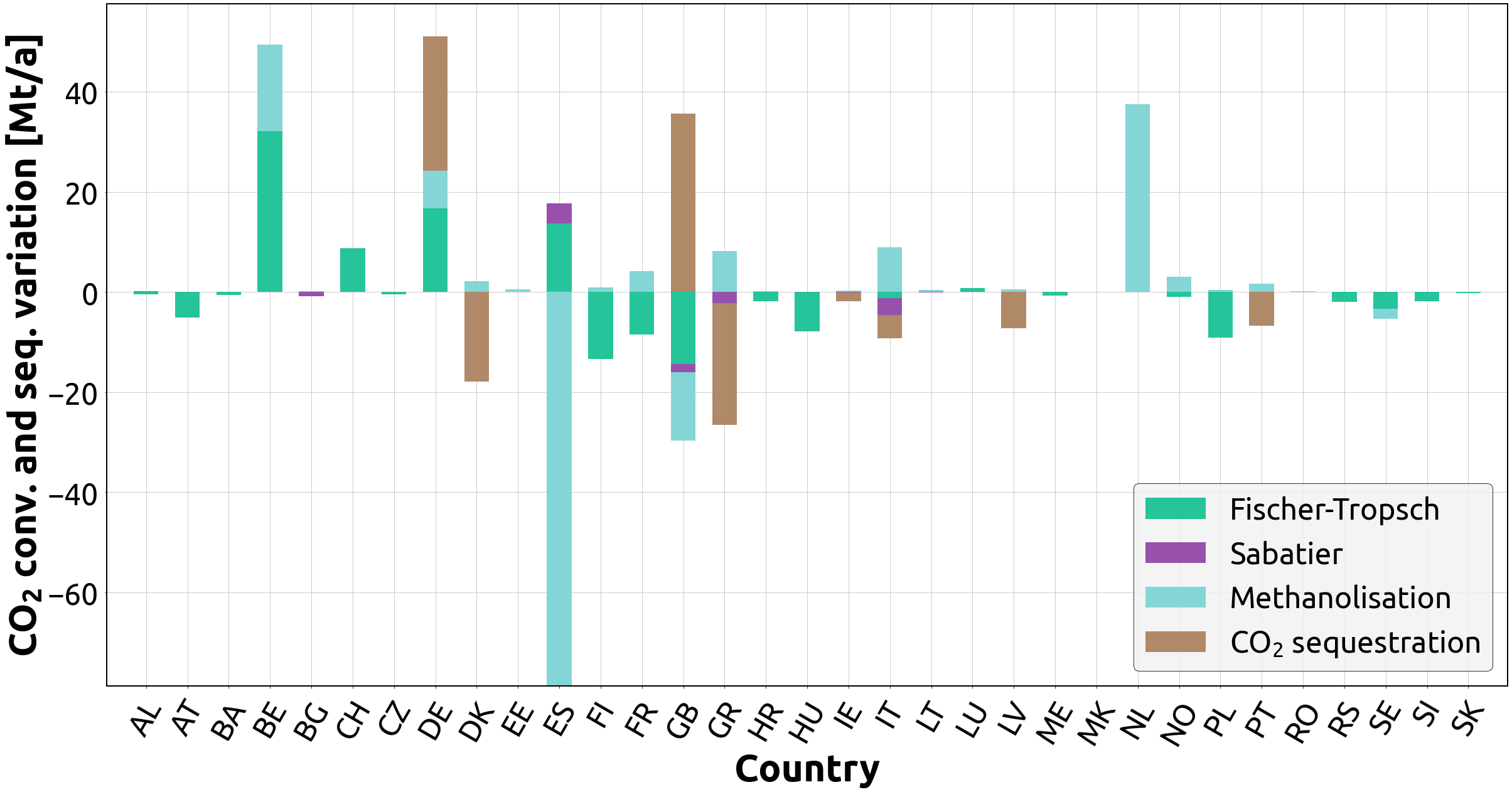}}
    \caption{(a) CO$_2$ conversion and sequestration in the global net-zero CO$_2$ emissions scenario and (b) CO$_2$ conversion and sequestration variation per country and technology between the global net-zero CO$_2$ emissions scenario and the local net-zero CO$_2$ emissions scenario.}
    \label{figure_spatial_co2_conversion_and_variation}
\end{figure}

In the model, captured CO$_2$ is either (i) utilised to produce carbonaceous fuels such as synthetic oil through the Fischer-Tropsch process, synthetic methanol in methanolisation plants, and synthetic methane through the Sabatier reaction, or (ii) sequestered underground. The total CO$_2$ converted and sequestered underground remains essentially the same in both climate-neutral scenarios, 724 Mt per year, at a continental level. However, the amount of converted CO$_2$ and the types of products it is converted into, as well as the levels of CO$_2$ sequestered underground vary significantly in each country between the two scenarios (Figure \ref{figure_spatial_co2_conversion_and_variation}).

\begin{figure}[!htb]
    \centering
    %\subfloat[]{\label{supplemental:figure_synthetic_oil_production_global_map}\includegraphics[width = 0.485\linewidth]{./figures/figure_S22a.png}}\hfill
    \subfloat[]{\label{supplemental:figure_synthetic_oil_production_global_map}\includegraphics[width = 0.475\linewidth]{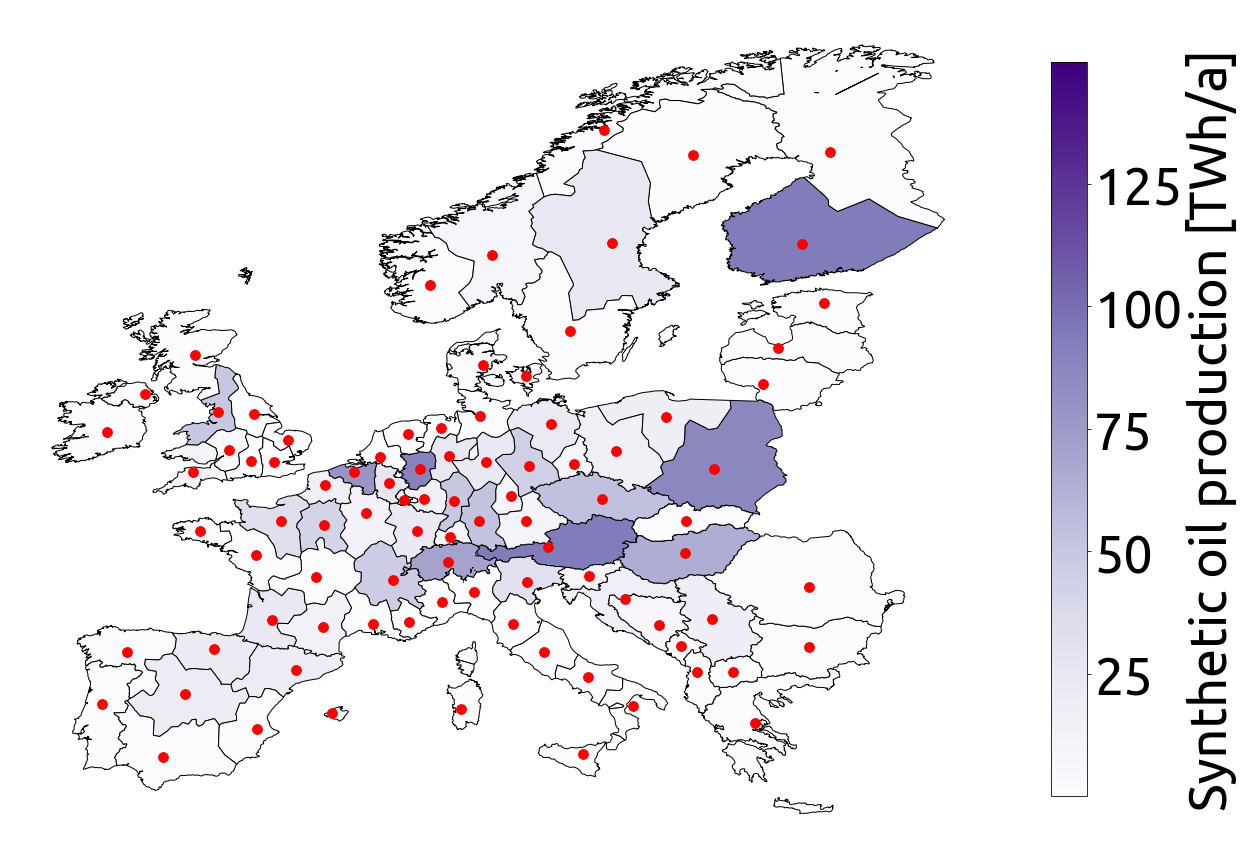}}\hfill
    %\subfloat[]{\label{supplemental:figure_synthetic_oil_production_local_map}\includegraphics[width = 0.485\linewidth]{./figures/figure_S22b.png}}\par
    \subfloat[]{\label{supplemental:figure_synthetic_oil_production_local_map}\includegraphics[width = 0.475\linewidth]{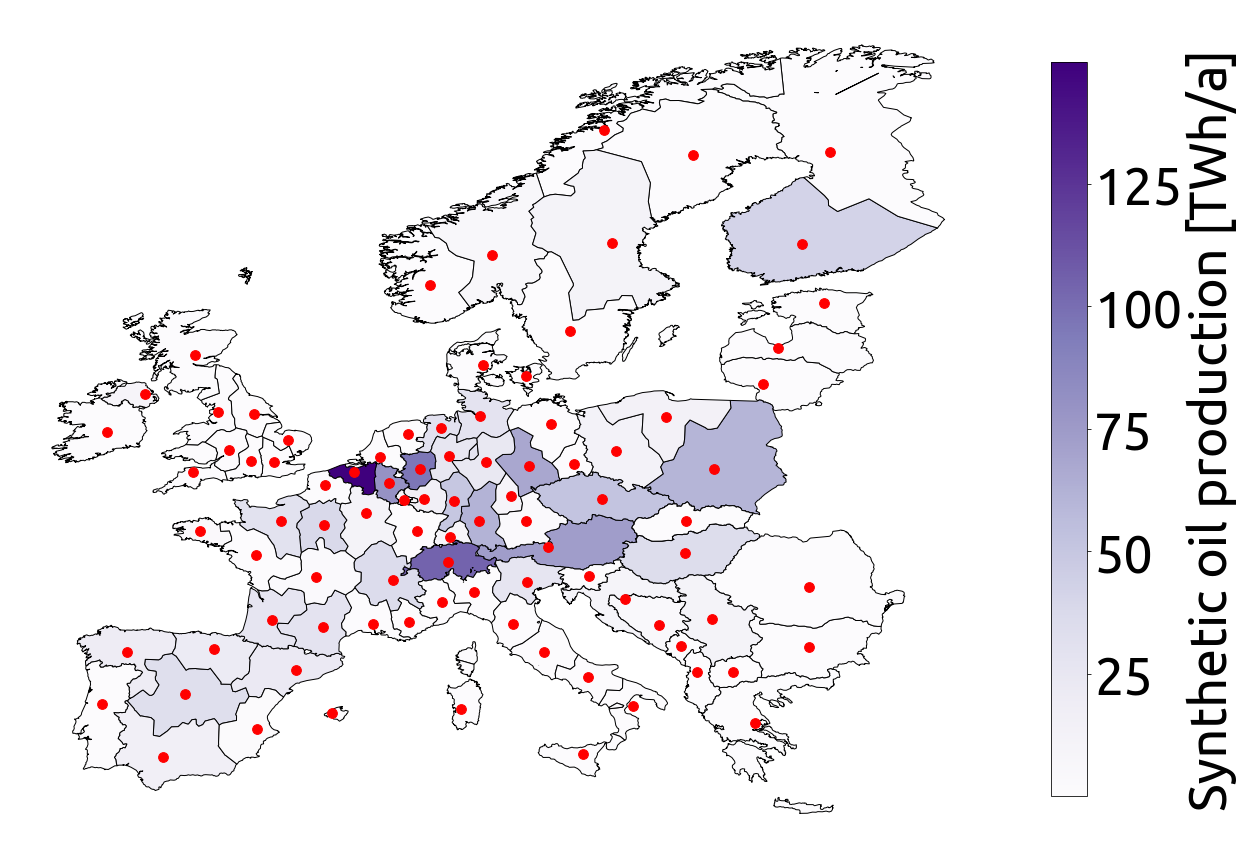}}\par
    \vspace{20pt}
    %\subfloat[]{\label{supplemental:figure_synthetic_oil_producers_consumers_variation}\includegraphics[width = 0.96\textwidth]{./figures/figure_S22c.png}}
    \subfloat[]{\label{supplemental:figure_synthetic_oil_producers_consumers_variation}\includegraphics[width = 0.8\textwidth]{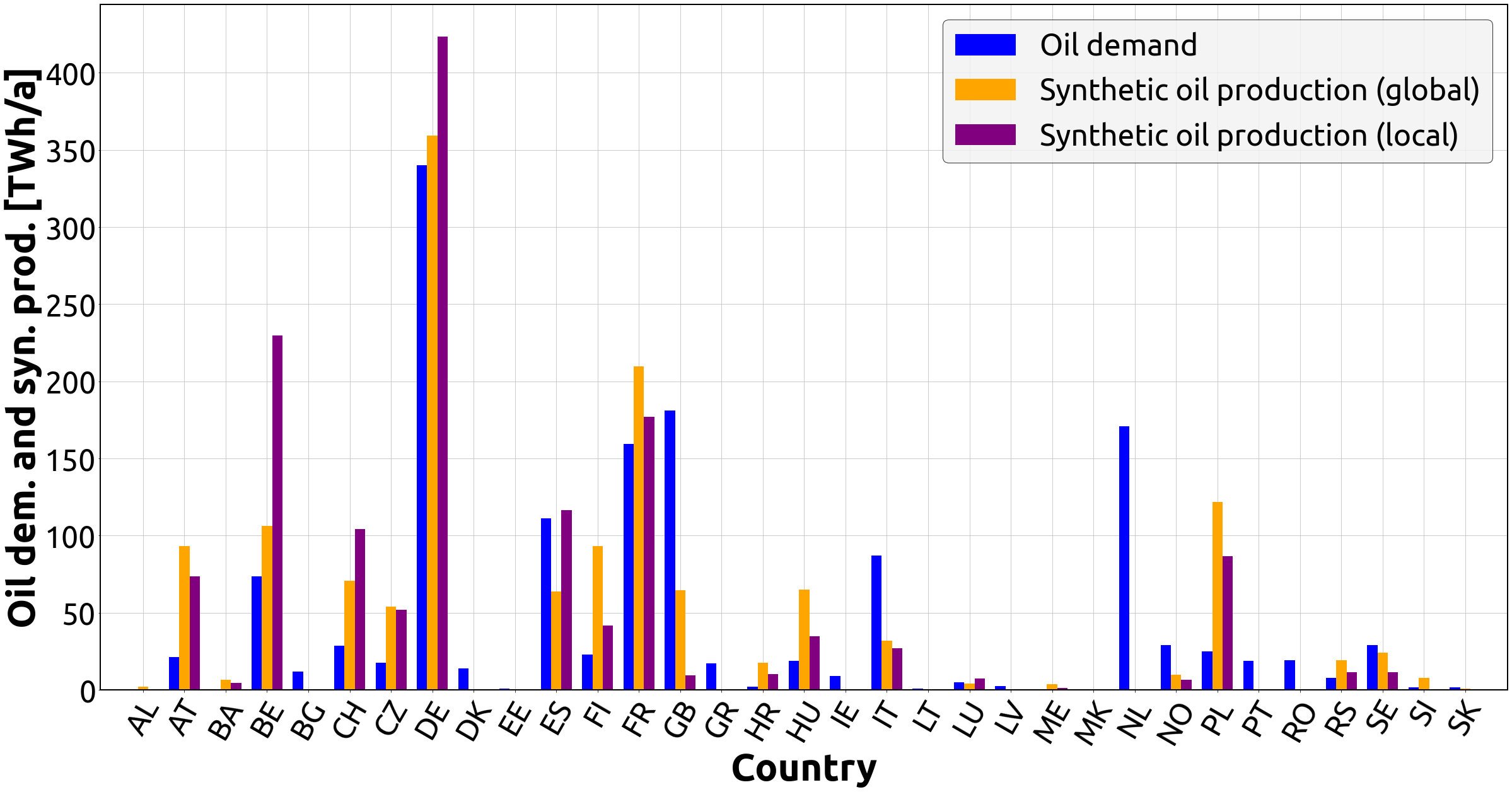}}
    \caption{Synthetic oil production per node in (a) the global net-zero CO$_2$ emissions scenario and (b) the local net-zero CO$_2$ emissions scenario, and (c) Exogenous oil demand and synthetic production per country in the global and local net-zero CO$_2$ emissions scenarios. In both scenarios, all modelled countries exchange (either by sending or receiving) synthetic oil with each other to satisfy their national demands.}
    \label{supplemental:figure_synthetic_oil_producers_consumers_map_variation}
\end{figure}

Under local constraints, Belgium and Germany substantially increase synthetic oil production through Fischer-Tropsch to help meet climate targets (by reusing CO$_2$) and satisfy their high demand for this product (Figure \ref{supplemental:figure_demand}), mainly in the industry and aviation sectors. In contrast, in most countries that were net CO$_2$ absorbers under a global constraint, synthetic oil production decreases given that they do not need to process as much captured CO$_2$ as they would under the former constraints. Furthermore, under both types of constraints, all countries exchange (either by sending or receiving) synthetic oil with each other to satisfy their national demands (Figure \ref{supplemental:figure_synthetic_oil_producers_consumers_map_variation}).

\begin{figure}[!htb]
    \centering
    %\subfloat[]{\label{supplemental:figure_synthetic_methanol_production_global_map}\includegraphics[width = 0.485\linewidth]{./figures/figure_S23a.png}}\hfill
    \subfloat[]{\label{supplemental:figure_synthetic_methanol_production_global_map}\includegraphics[width = 0.475\linewidth]{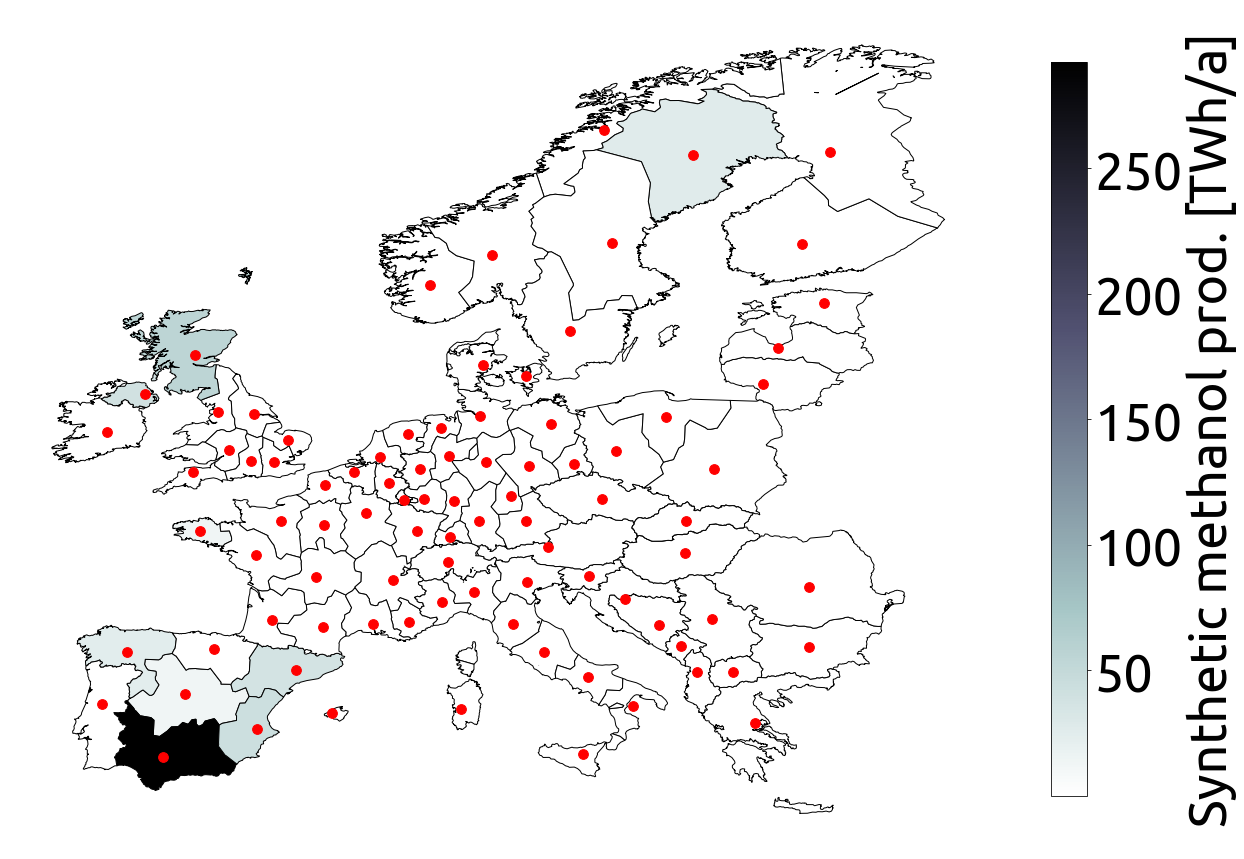}}\hfill
    %\subfloat[]{\label{supplemental:figure_synthetic_methanol_production_local_map}\includegraphics[width = 0.485\linewidth]{./figures/figure_S23b.png}}\par
    \subfloat[]{\label{supplemental:figure_synthetic_methanol_production_local_map}\includegraphics[width = 0.475\linewidth]{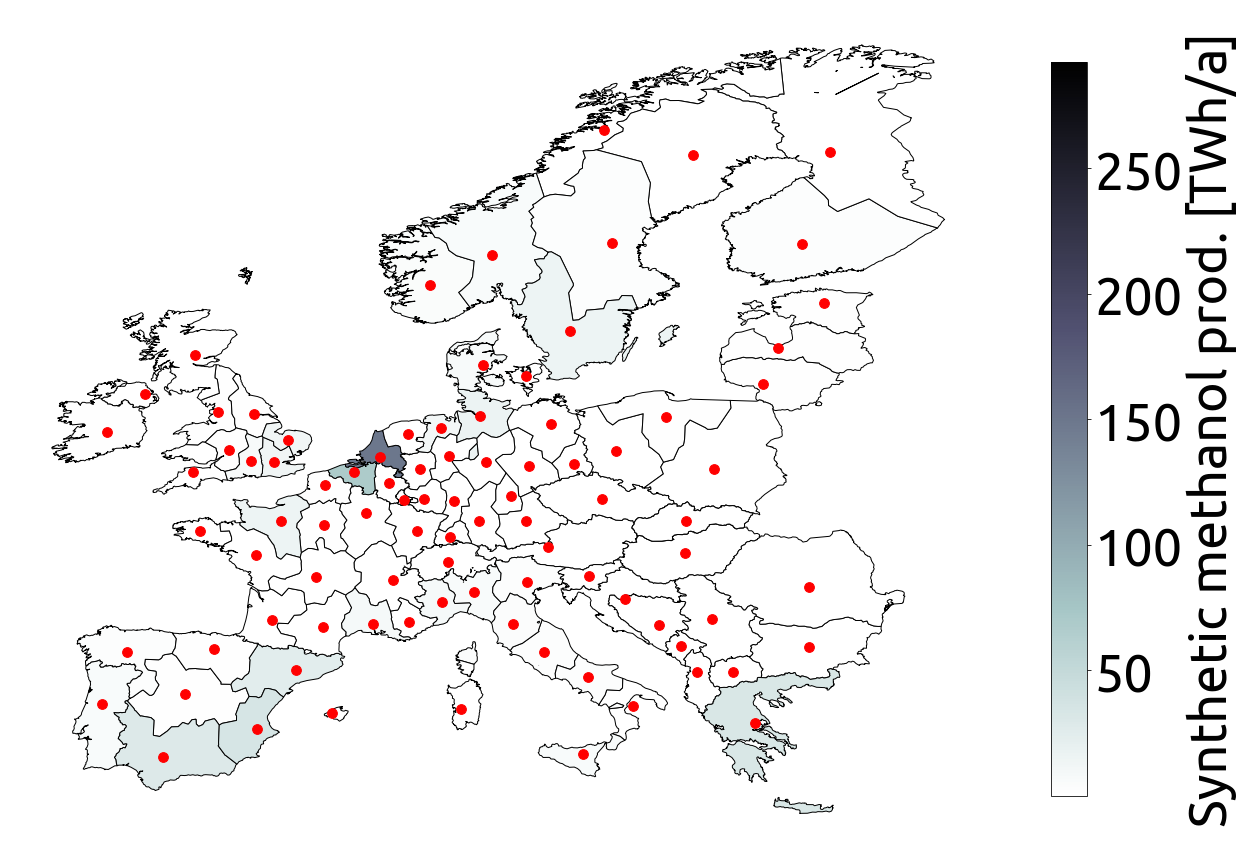}}\par
    \vspace{20pt}
    %\subfloat[]{\label{supplemental:figure_synthetic_methanol_producers_consumers_variation}\includegraphics[width = 0.96\textwidth]{./figures/figure_S23c.png}}
    \subfloat[]{\label{supplemental:figure_synthetic_methanol_producers_consumers_variation}\includegraphics[width = 0.8\textwidth]{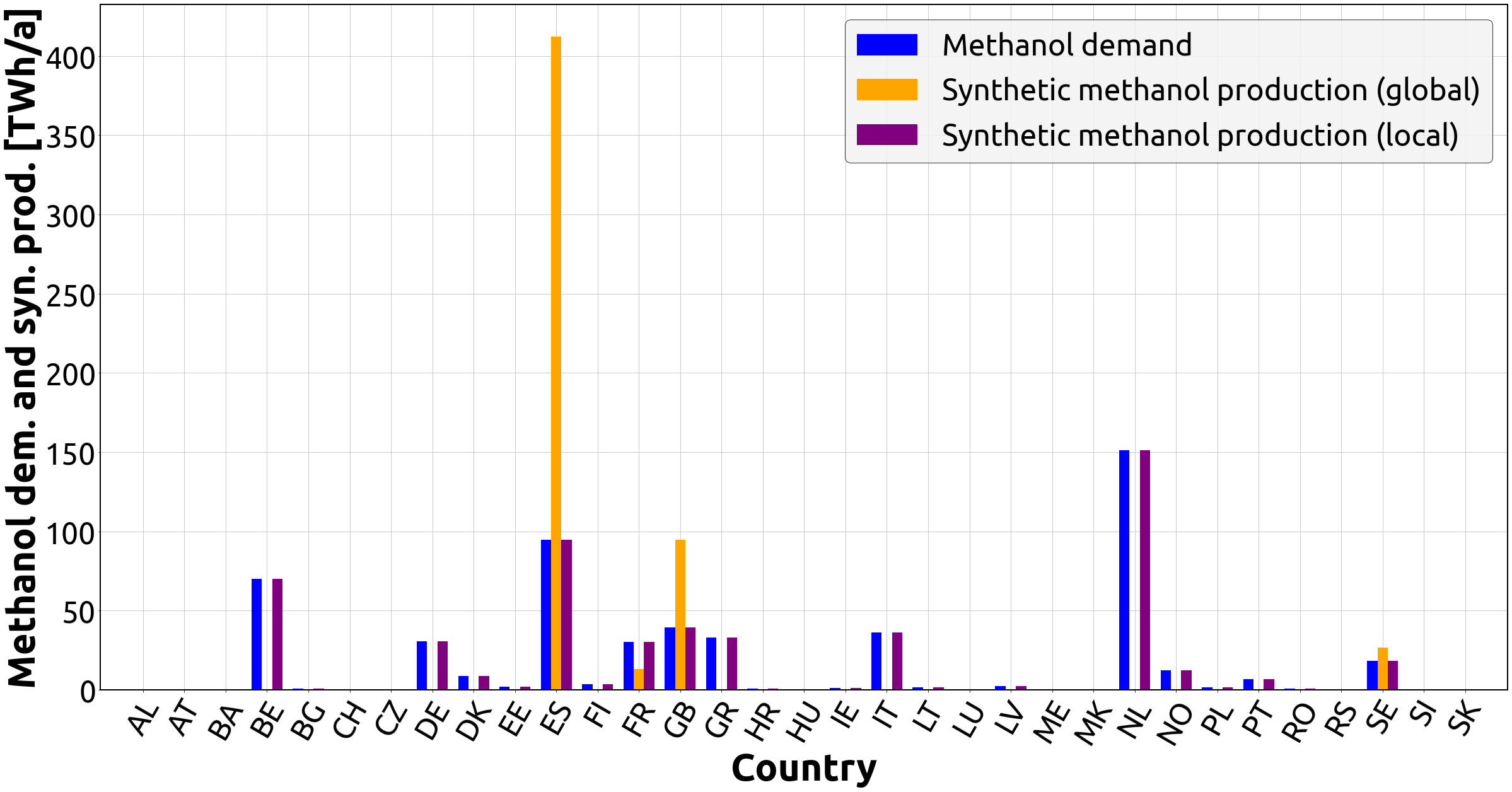}}
    \caption{Synthetic methanol production per node in (a) the global net-zero CO$_2$ emissions scenario and (b) the local net-zero CO$_2$ emissions scenario, and (c) Exogenous methanol demand and synthetic production per country in the global and local net-zero CO$_2$ emissions scenarios. Under a global constraint, Spain and Great Britain produce the bulk of synthetic methanol to satisfy the demand for this product in most of the modelled countries, whereas under local constraints, each country satisfies its own methanol demand with no exchange occurring between them.}
    \label{supplemental:figure_synthetic_methanol_producers_consumers_map_variation}
\end{figure}

We observe a substantial increase in synthetic methanol production in the interior countries under local constraints to meet the needs of their shipping sector, especially in The Netherlands. This is because they can no longer depend on Spain and Great Britain, two major producers of synthetic methanol, to supply this product as it occurs under a global constraint. While these two countries produce the bulk of synthetic methanol to satisfy the demand in most of the remaining countries, under local constraints, each country satisfies its own methanol demand with no exchange occurring between them (Figure \ref{supplemental:figure_synthetic_methanol_producers_consumers_map_variation}).

Regarding the production of synthetic methane gas through the Sabatier reaction, the model does not favour it in the global scenario and further reduces it in the local scenario. The high cost and high heat required by the process make it a challenging option in both scenarios. As a result, boilers consuming gas are replaced by solid biomass boilers and air heat pumps, which offer cleaner and cost-effective heating (Figure \ref{figure_total_system_cost_and_variation}).

While in Germany and Great Britain (both net CO$_2$ emitters in the global scenario), much more CO$_2$ is sequestered underground, in Denmark, Ireland, Italy, Portugal, Latvia, and especially in Greece (all net CO$_2$ absorbers in the global scenario), there is a substantial decrease of CO$_2$ sequestered underground (Figures \ref{figure_spatial_co2_conversion_variation} and \ref{supplemental:figure_co2_sequestration_potential_usage}). This is because net emitter countries capture additional CO$_2$ under local constraints, forcing them to either sequester more CO$_2$ underground or convert more CO$_2$ into synthetic products. Conversely, net absorber countries capture less CO$_2$ mainly from DAC (Figure \ref{figure_spatial_co2_capture_and_variation}) and/or receive less CO$_2$ from neighbouring countries (Figure \ref{figure_co2_flow}), thus are less inclined to sequester CO$_2$ underground. Most net CO$_2$ emitter countries, including the interior countries, do not show any changes in the amount of CO$_2$ sequestered underground, with their potentials being fully utilised in both scenarios. Furthermore, the maximum limit of 200 MtCO$_2$ that can be sequestered underground annually across Europe is fully exhausted in both scenarios.

From a temporal perspective, our results show that both the Sabatier and the methanolisation processes exhibit a seasonal trend, whereas the Fischer-Tropsch process remains almost constant in the two modelled scenarios. The reason for this is that Fischer-Tropsch has high investment cost. Once the model selects it, the process typically operates at full capacity to recover costs. On the other hand, due to lower investment costs, both Sabatier and methanolisation operate (more) flexibly, usually during periods of low electricity prices. Flexible operation of those technologies could be technically attained as described in the literature \cite{BROWN20232414, MOIOLI2024131}. Furthermore, although flexible, operating these two processes does not require very frequent ramping as that observed for H$_2$ electrolysis (Figures \ref{supplemental:figure_temporal_co2_conversion_sequestration} and \ref{supplemental:figure_temporal_h2_production}). Thanks to affordable electricity in the summer, which is essential for powering DAC to capture and provide the necessary CO$_2$ for all the conversion processes, as well as for powering the methanolisation process itself, the production of synthetic methane gas and methanol naturally increases during that season and decreases during the winter. Sequestering CO$_2$ underground also exhibits a seasonal trend in both scenarios, but it is opposite to the first two conversion processes mentioned earlier, with the model using this method more to deal with CO$_2$ during winter and less during summer. This is because CO$_2$ is needed to manufacture synthetic products during the summer, which reduces the amount of CO$_2$ sequestered underground in this period.

\subsection{Transport via CO$_2$ pipelines is the main strategy for carbon exchange amongst countries}
We shift our attention now to investigating the flows of carbon in the model. This involves analysing the flows of CO$_2$ through newly built pipelines but also the flows of energy carriers that contain carbon such as methane gas and solid biomass. We will also analyse the flows of electricity to help understanding where countries obtain the electricity needed for their CO$_2$ capture and conversion activities, as well as the flows of H$_2$ (through the modelled hydrogen network), considering the impact that changes in the locations where synthetic products are manufactured might have on these flows. Although advanced techniques have been developed to trace arbitrary flows in renewable energy systems such as in \cite{HORSCH2018390}, for the sake of simplicity, we analyse flows simply by multiplying the amount of energy or carbon flowing with the physical length of the mean transporting the flow (e.g. CO$_2$ pipeline).

\begin{figure}[!htb]
    \centering
    %\subfloat[]{\label{figure_co2_flow_global}\includegraphics[width = 0.8\linewidth]{./figures/figure_5a.png}}
    \subfloat[]{\label{figure_co2_flow_global}\includegraphics[width = 0.66\linewidth]{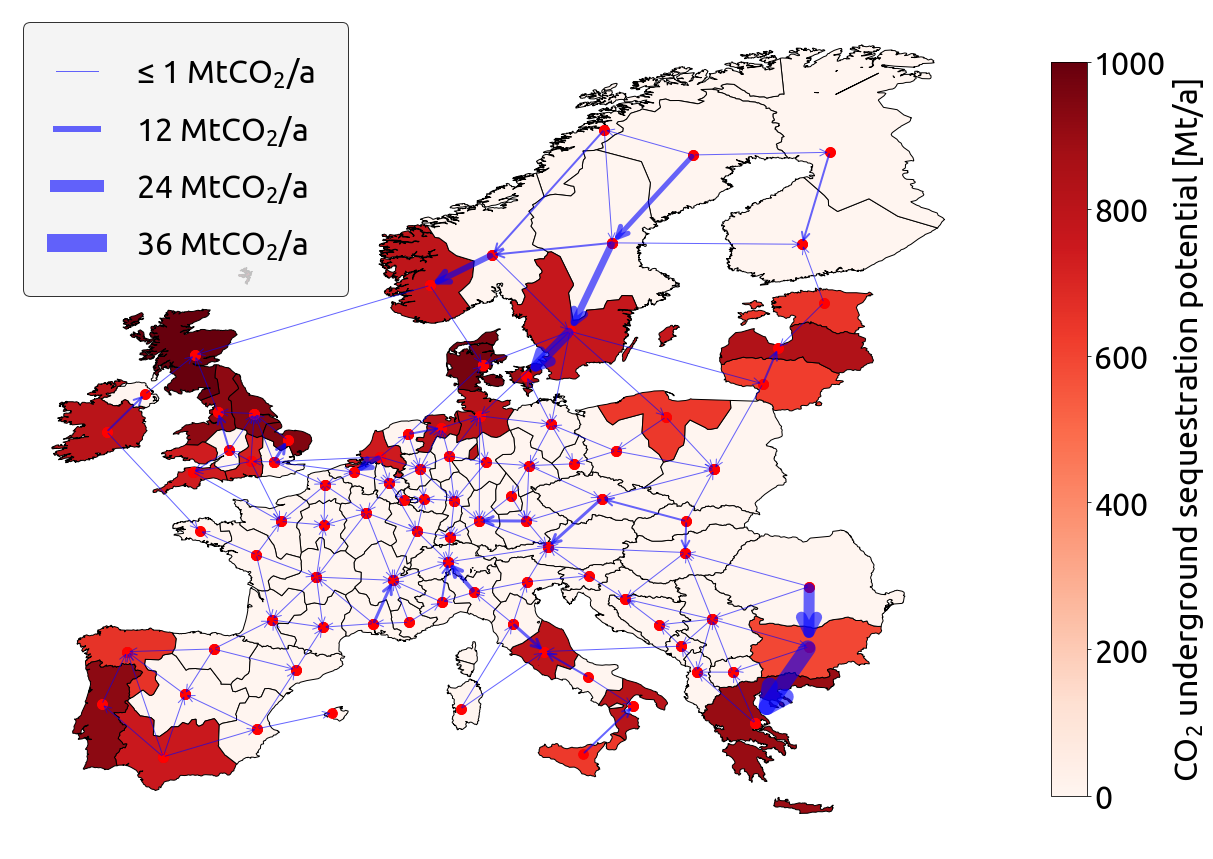}}
    \vspace{20pt}
    %\subfloat[]{\label{figure_co2_flow_local}\includegraphics[width = 0.8\linewidth]{./figures/figure_5b.png}}
    \subfloat[]{\label{figure_co2_flow_local}\includegraphics[width = 0.66\linewidth]{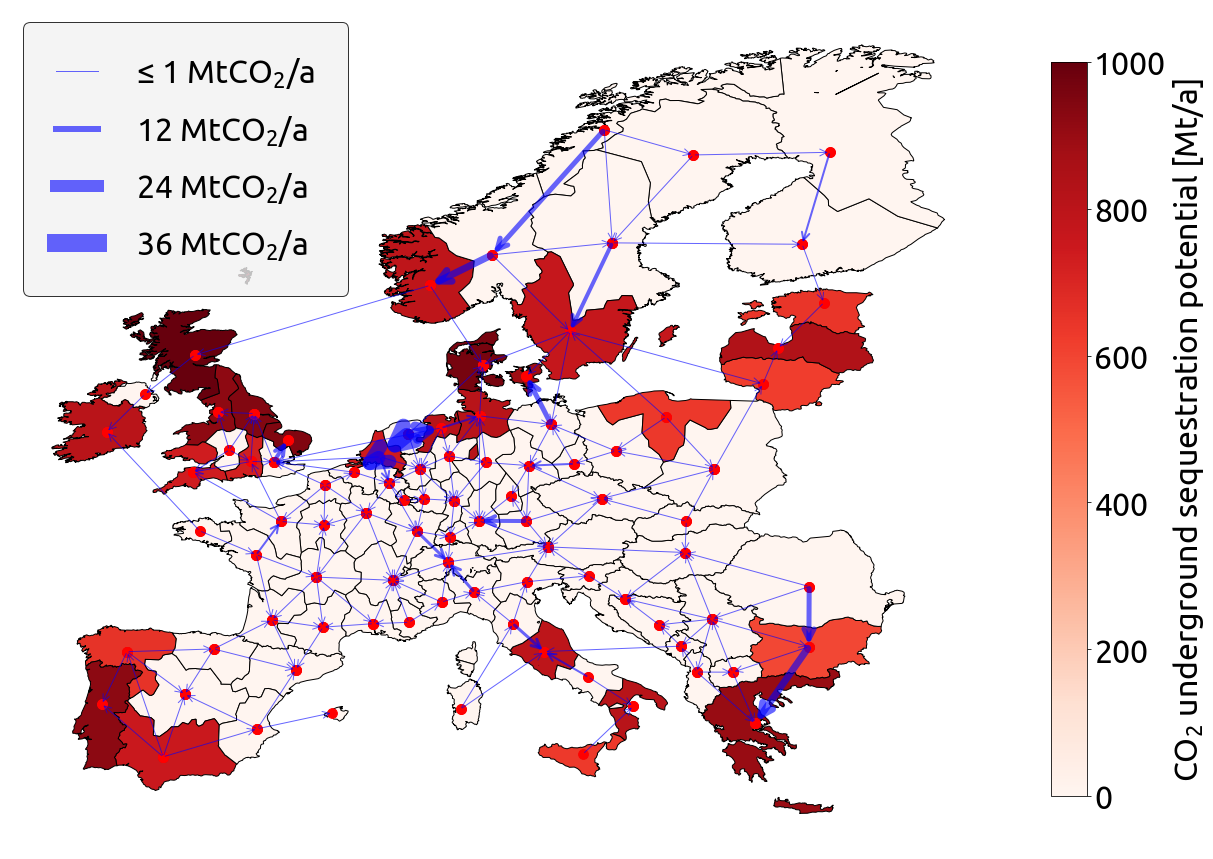}}
    \caption{CO$_2$ flow in (a) the global net-zero CO$_2$ emissions scenario and (b) the local net-zero CO$_2$ emissions scenario. Under local constraints, the flow of CO$_2$ decreases by 30\% on the continent relative to a global constraint. This happens because countries which where net CO$_2$ absorbers under the latter constraint capture less CO$_2$ and consequently send less CO$_2$ to neighbouring countries through pipelines for underground sequestration and/or to produce additional synthetic oil and methanol.}
    \label{figure_co2_flow}
\end{figure}

In the global and local net-zero CO$_2$ emissions scenarios, 124 GtCO$_2$·km/a and 87 GtCO$_2$·km/a are transported amongst and within countries through pipelines, respectively (Figure \ref{figure_co2_flow}). Given that the total captured CO$_2$ amounts to 724 MtCO$_2$/a, transporting 124 GtCO$_2$·km/a can be interpreted as if, on average, 124 MtCO$_2$/a (17\% of the captured CO$_2$) are transported 1000 km.

In both scenarios, the main factor determining the directions and destinations of flows is the assumed potential for sequestering CO$_2$ underground. As illustrated in Figure \ref{supplemental:figure_co2_flow_1000mt_sequestration}, increasing the sequestration potential assumption to 1000 MtCO$_2$/a (from 200 MtCO$_2$/a) across Europe clearly shows CO$_2$ flows converging towards countries with significant underground sequestration potential, such as Great Britain, Denmark, Portugal, Latvia, and Greece. This is in line with the results found in \cite{Hofmann2025}.

\begin{figure}[!htb]
    \centering
    %\subfloat[]{\label{supplemental:figure_co2_sequestration_usage_global}\includegraphics[width = 0.485\linewidth]{./figures/figure_S8a.png}}\hfill
    \subfloat[]{\label{supplemental:figure_co2_sequestration_usage_global}\includegraphics[width = 0.475\linewidth]{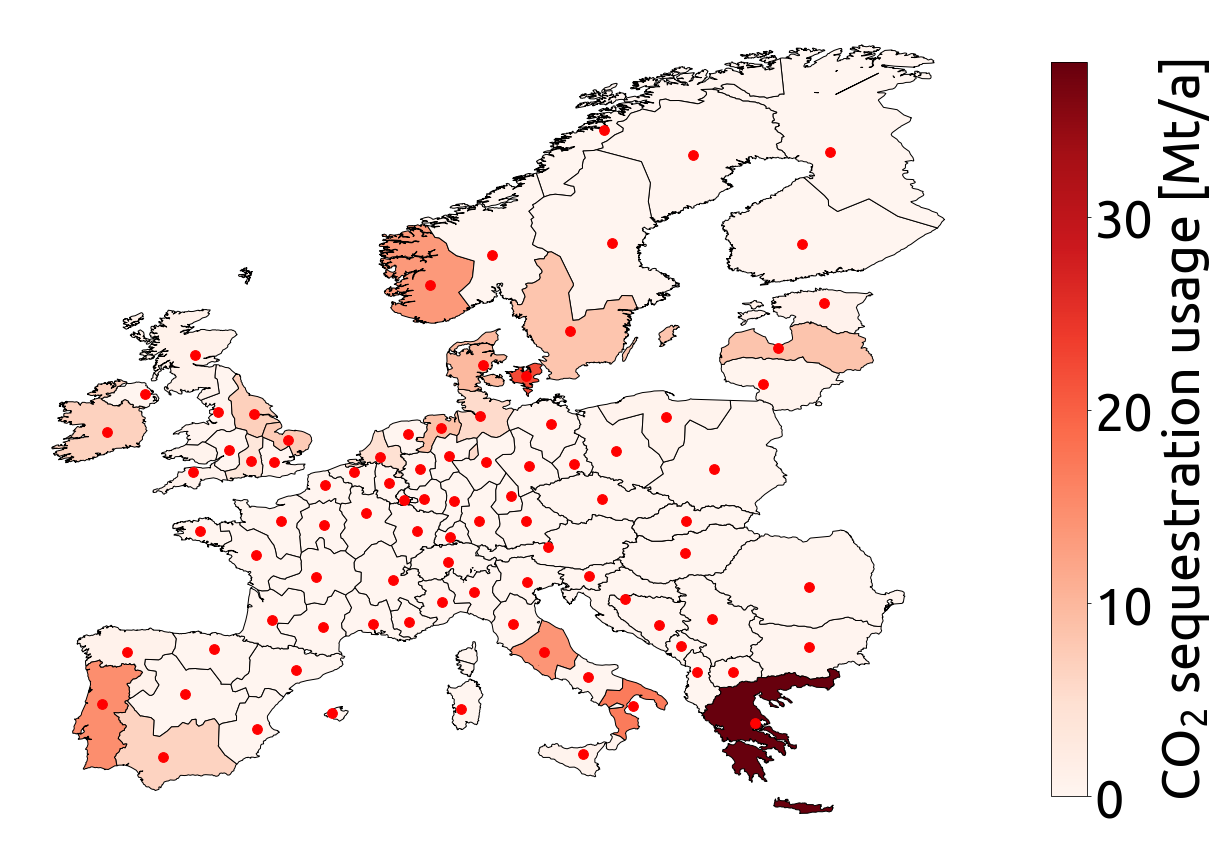}}\hfill
    \subfloat[]{\label{supplemental:figure_co2_sequestration_usage_local}\includegraphics[width = 0.475\linewidth]{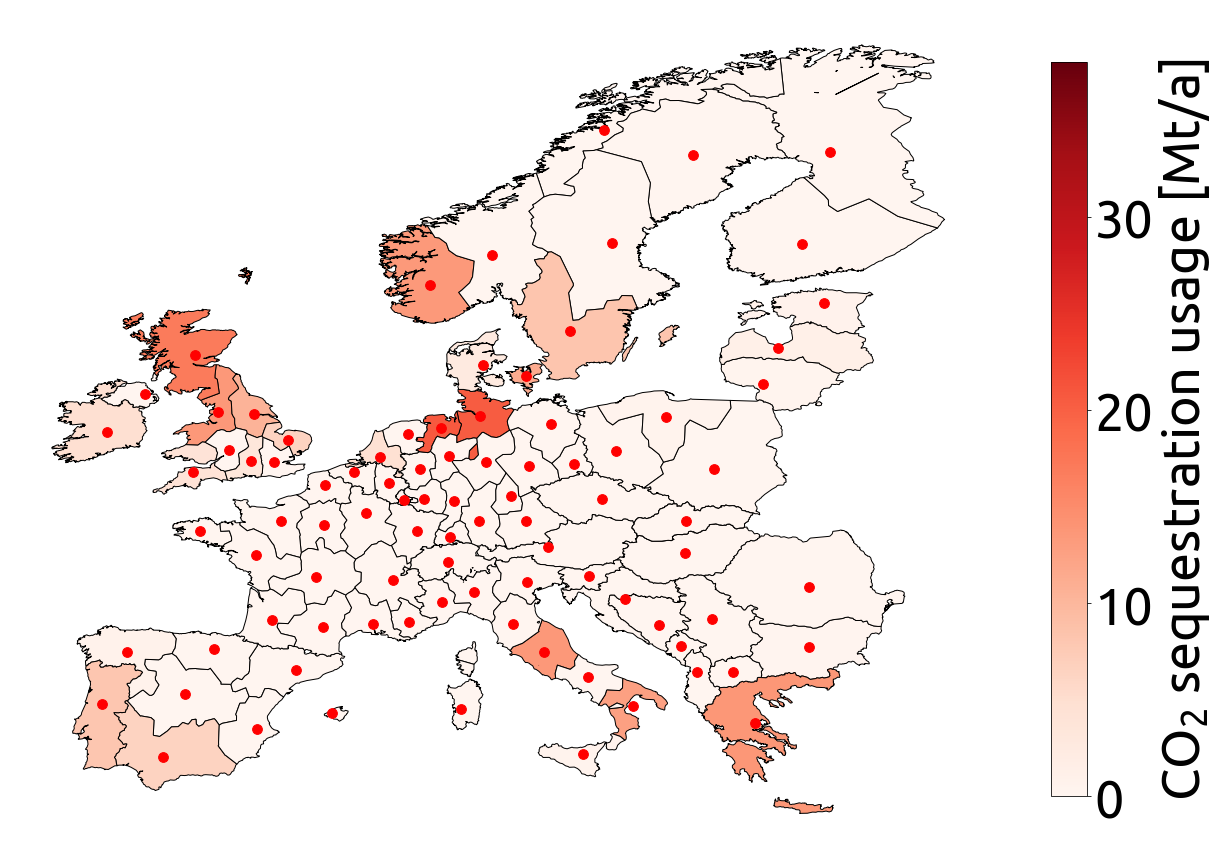}}\par
    %\subfloat[]{\label{supplemental:figure_co2_sequestration_usage_local}\includegraphics[width = 0.485\linewidth]{./figures/figure_S8b.png}}\par
    \vspace{20pt}
    %\subfloat[]{\label{supplemental:figure_co2_sequestration_potential_usage_global_local}\includegraphics[width = 0.96\textwidth]{./figures/figure_S8c.png}}
    \subfloat[]{\label{supplemental:figure_co2_sequestration_potential_usage_global_local}\includegraphics[width = 0.8\textwidth]{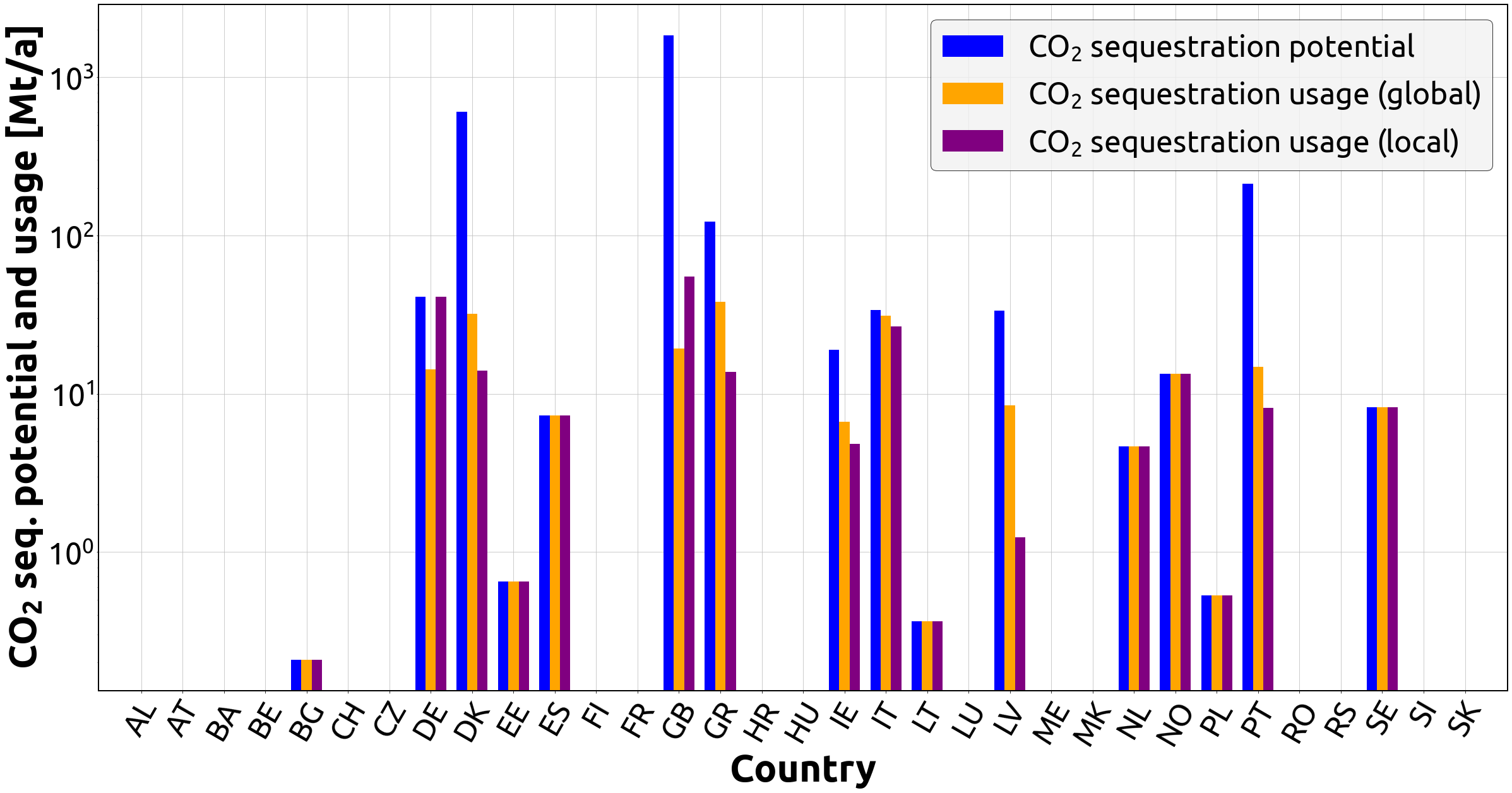}}
    \caption{CO$_2$ underground sequestration usage per node in (a) the global net-zero CO$_2$ emissions scenario and (b) the local net-zero CO$_2$ emissions scenario, and (c) CO$_2$ underground sequestration potential and usage per country in the global and local net-zero CO$_2$ emissions scenarios. Despite the varying levels of sequestration usage amongst countries and between the two scenarios, the conservative sequestration potential limit of 200 MtCO$_2$/a is fully exhausted in both scenarios.}
    \label{supplemental:figure_co2_sequestration_potential_usage}
\end{figure}

Comparing both scenarios, we observe a 30\% (37 GtCO$_2$·km/a) decrease in the flow of CO$_2$ in the local scenario. This is mainly due to Sweden and Romania, both important net absorbers under a global constraint, capturing less CO$_2$ under local constraints. As a result, these countries send less CO$_2$ to Denmark and Greece (via Bulgaria) for underground sequestration, respectively, considering their significant storage capacity (Figure \ref{supplemental:figure_co2_sequestration_potential_usage}). Conversely, The Netherlands, the most important net emitter under a global constraint, captures much more CO$_2$ under local constraints (Figure \ref{figure_spatial_co2_capture_and_variation}). After using its captured CO$_2$ to produce additional synthetic methanol to meet the demand (Figure \ref{figure_spatial_co2_conversion_and_variation}), The Netherlands starts sending excess CO$_2$ to Germany and Belgium. In turn, Germany sequesters the received CO$_2$ underground and Belgium uses it to produce additional synthetic oil and methanol required by local constraints. In addition, northeastern Germany starts sending CO$_2$ to Denmark for underground sequestration. The rise of CO$_2$ flow triggered by The Netherlands and Germany does not offset the previously described decrease in flow resulting in an overall 30\% reduction in CO$_2$ flow.

\subsection{Biomass transport is the second mechanism for exchanging carbon}
\begin{figure}[!htb]
    \centering
    %\subfloat[]{\label{figure_solid_biomass_transport_global}\includegraphics[width = 0.8\linewidth]{./figures/figure_6a.png}}
    \subfloat[]{\label{figure_solid_biomass_transport_global}\includegraphics[width = 0.66\linewidth]{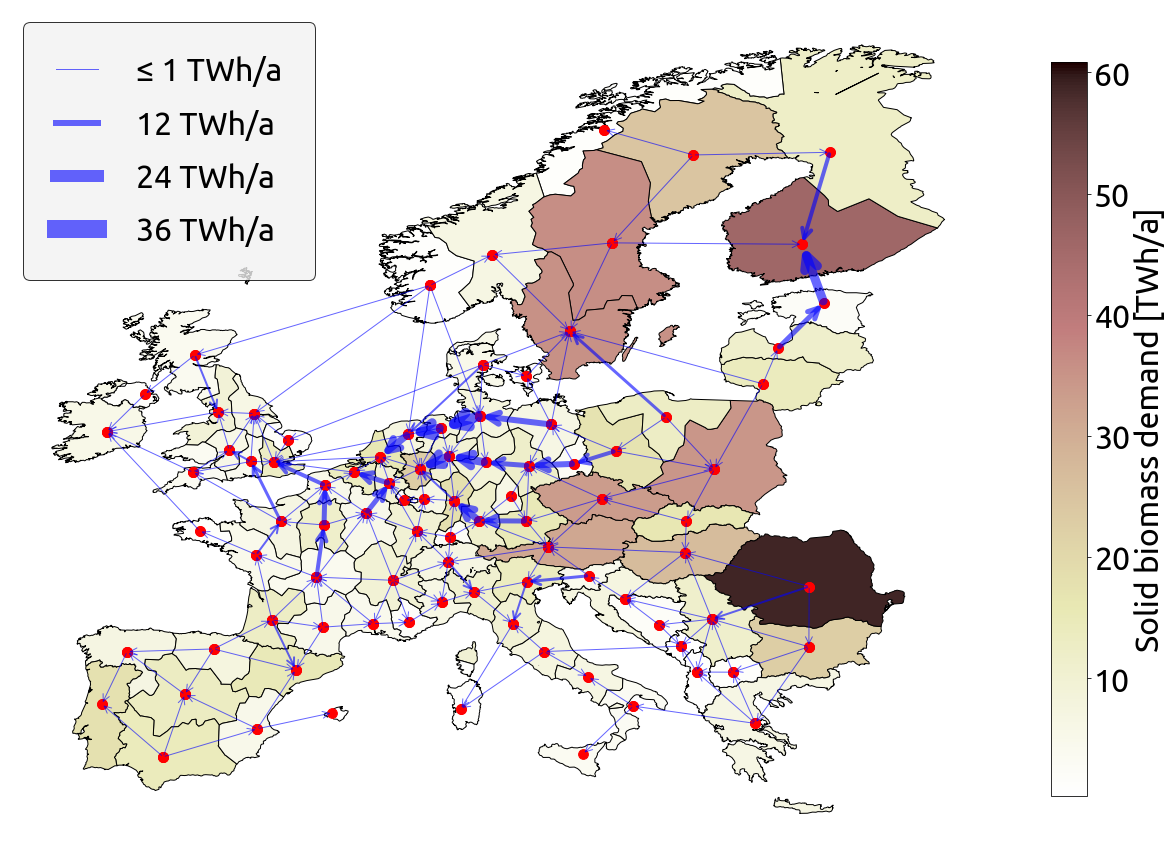}}
    \vspace{20pt}
    %\subfloat[]{\label{figure_solid_biomass_transport_local}\includegraphics[width = 0.8\linewidth]{./figures/figure_6b.png}}
    \subfloat[]{\label{figure_solid_biomass_transport_local}\includegraphics[width = 0.66\linewidth]{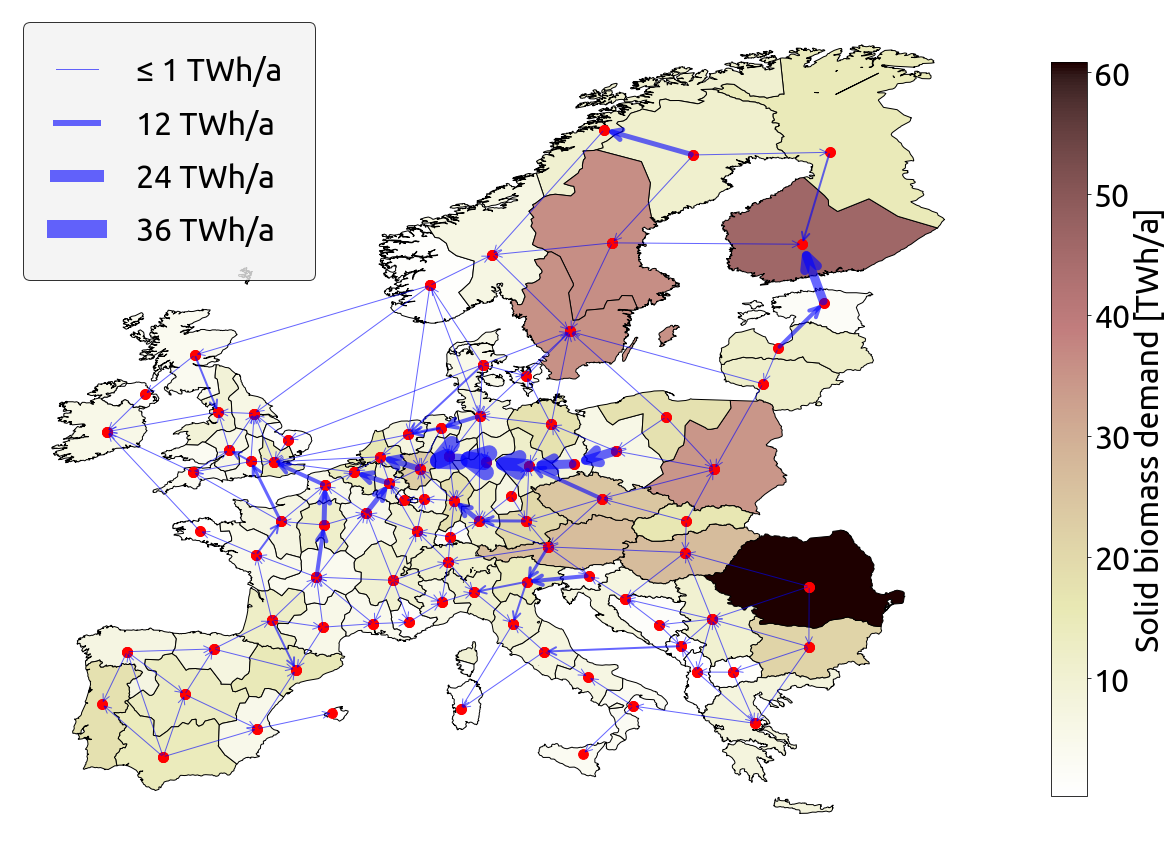}}
    \caption{Solid biomass transportation in (a) the global net-zero CO$_2$ emissions scenario and (b) the local net-zero CO$_2$ emissions scenario. In both scenarios, solid biomass transportation concentrates in central Europe, moving from the East, where there is high solid biomass potential, to the West, where there is high demand for solid biomass but low local potential.}
    \label{figure_solid_biomass_transport}
\end{figure}

Solid biomass can be transported by road across Europe, enabling countries to exchange carbon amongst and within themselves. In the global and local net-zero CO$_2$ emissions scenarios, 87000 TWh·km/a and 92000 TWh·km/a of solid biomass are transported, respectively (Figure \ref{figure_solid_biomass_transport}). Based on the stoichiometric ratios provided in Table \ref{supplemental:table_energy_co2_conversion} and assuming that 50\% of the solid biomass is carbon, this is equivalent to effectively moving 32 GtCO$_2$·km/a in the global scenario and 34 GtCO$_2$·km/a in the local scenario, making this method the second most relevant in distributing carbon across Europe.

In both scenarios, the transportation pattern for solid biomass in Europe involves countries or regions with high demand but low potential for this material, which are supplied by neighbouring countries or regions with a higher potential. This typically occurs from eastern-central to western-central Europe.

\begin{figure}[!htb]
    \centering
    %\subfloat[]{\label{supplemental:figure_solid_biomass_potential_per_node}\includegraphics[width = 0.8\linewidth]{./figures/figure_S10a.png}}
    \subfloat[]{\label{supplemental:figure_solid_biomass_potential_per_node}\includegraphics[width = 0.65\linewidth]{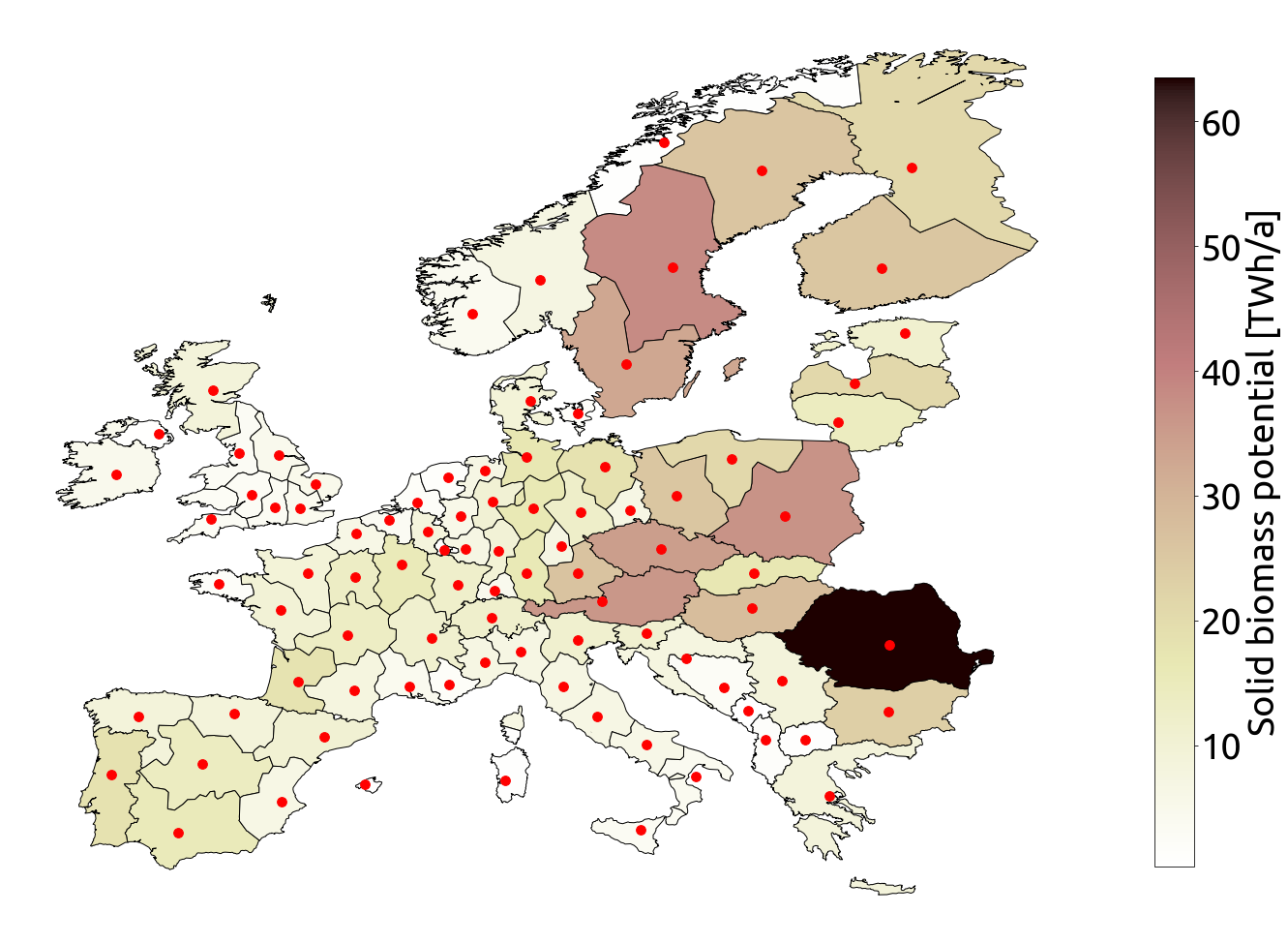}}
    \vspace{20pt}
    %\subfloat[]{\label{supplemental:figure_solid_biomass_potential_per_country}\includegraphics[width = 0.96\linewidth]{./figures/figure_S10b.png}}
    \subfloat[]{\label{supplemental:figure_solid_biomass_potential_per_country}\includegraphics[width = 0.8\linewidth]{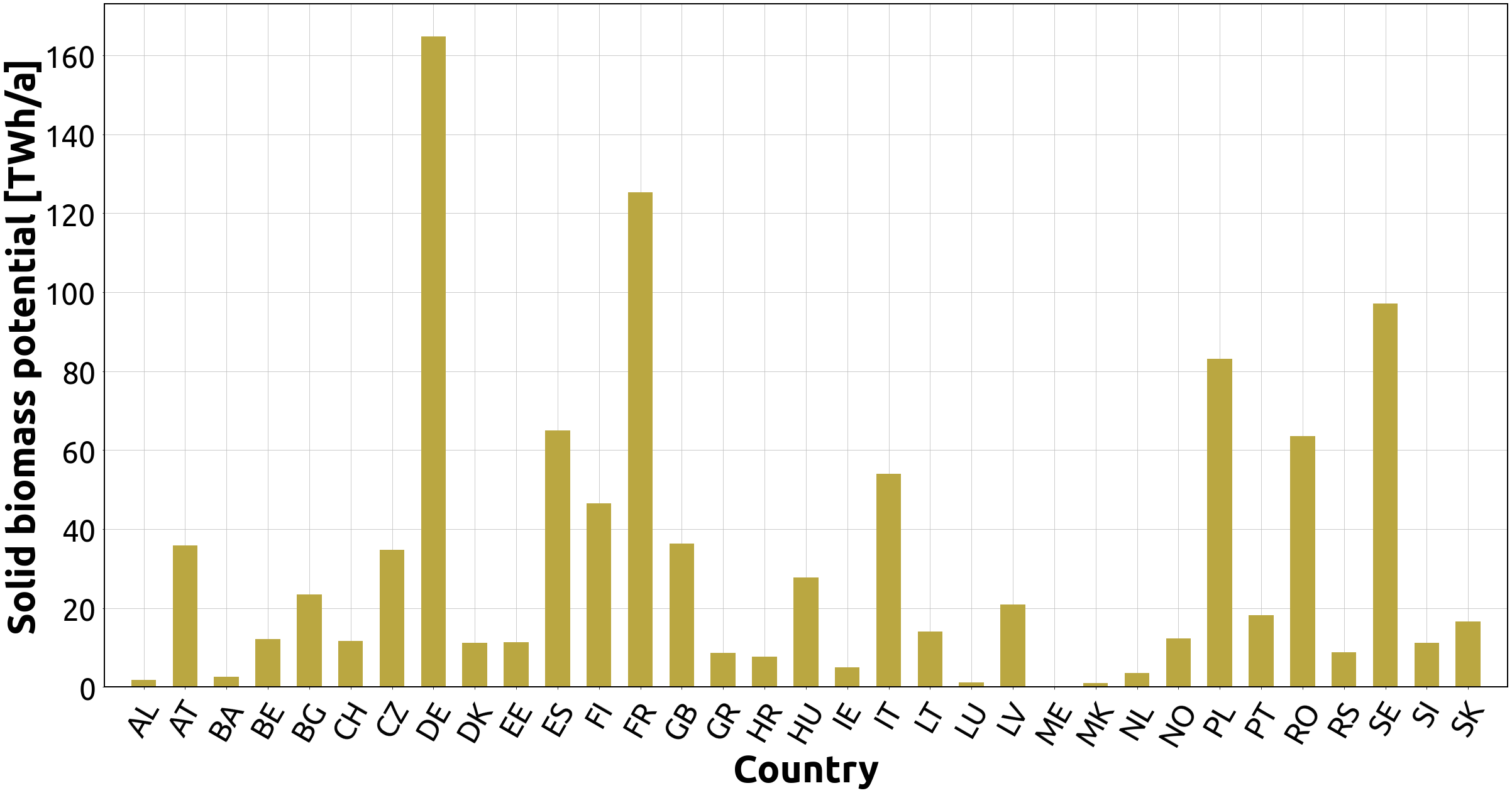}}
    \caption{Solid biomass potential (a) per node and (b) per country. Europe is assumed to have an aggregated potential of 1038 TWh per year of solid biomass. This assumption is based on the medium bioenergy availability profile of the ENSPRESO database, which the model uses in both the global and local net-zero CO$_2$ emissions scenarios.}  \label{supplemental:figure_solid_biomass_potential}
\end{figure}

Under local net-zero CO$_2$ emissions constraints, we observe a rise in the transportation of solid biomass from Poland and the Czech Republic to Germany, and from Sweden to Norway. This is due to the growing use of solid biomass-based CHP units with CC in the importing countries to meet their heat and power requirements (Figures \ref{figure_spatial_co2_capture_and_variation} and \ref{supplemental:figure_demand}). In addition, there are noticeable variations in solid biomass transportation within Germany. The transportation increases from its central regions to its western regions and decreases within its southern regions to meet the different solid biomass demand and potential levels in these regions. In both scenarios, large amounts of solid biomass are transported from France to Great Britain and Belgium, from Germany to The Netherlands, and from northern Finland, Latvia, and Lithuania (via Latvia) to southern Finland. While Great Britain, Belgium, and The Netherlands primarily import solid biomass due to their low potential for this material (Figure \ref{supplemental:figure_solid_biomass_potential}), southern Finland does so due to its high demand for solid biomass.

\subsection{Synthetic and fossil methane gas is the third mechanism for exchanging carbon}
\begin{figure}[!htb]
    \centering
    %\subfloat[]{\label{figure_methane_gas_flow_global}\includegraphics[width = 0.8\linewidth]{./figures/figure_7a.png}}
    \subfloat[]{\label{figure_methane_gas_flow_global}\includegraphics[width = 0.66\linewidth]{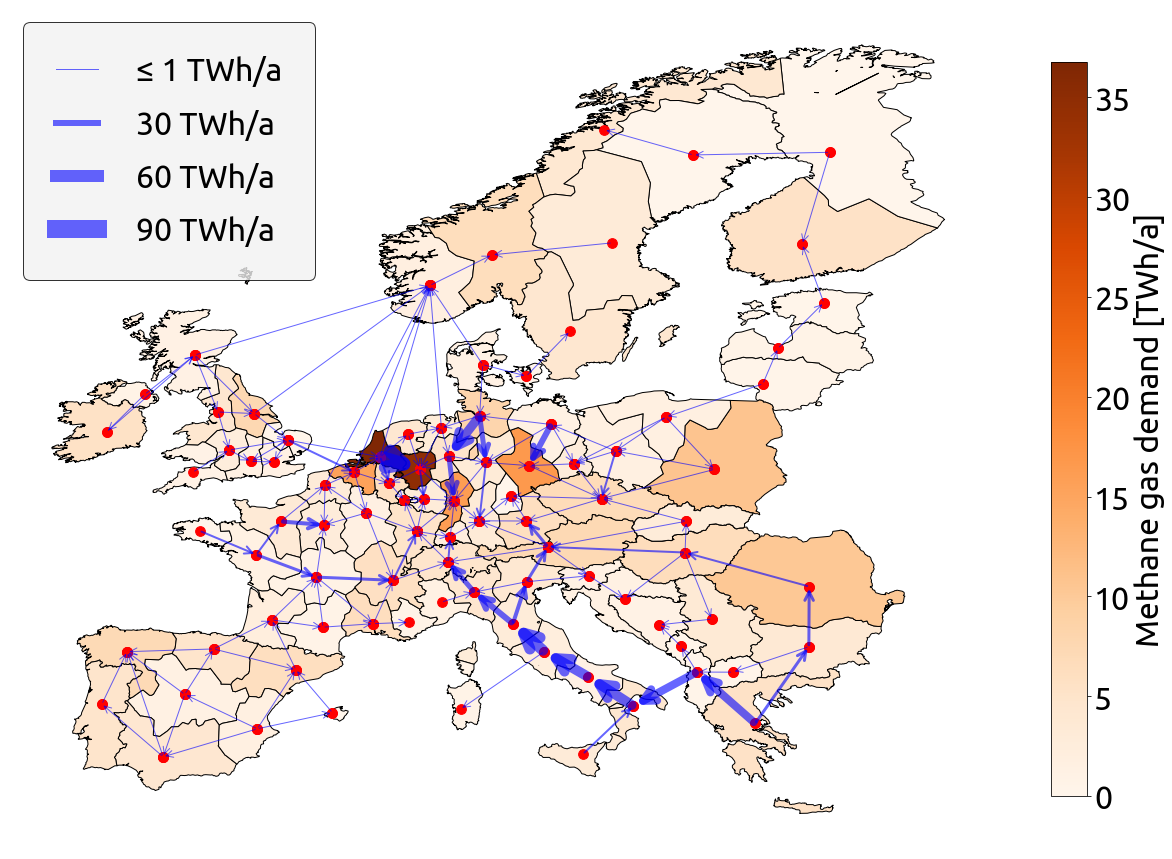}}
    \vspace{20pt}
    %\subfloat[]{\label{figure_methane_gas_flow_local}\includegraphics[width = 0.8\linewidth]{./figures/figure_7b.png}}
    \subfloat[]{\label{figure_methane_gas_flow_local}\includegraphics[width = 0.66\linewidth]{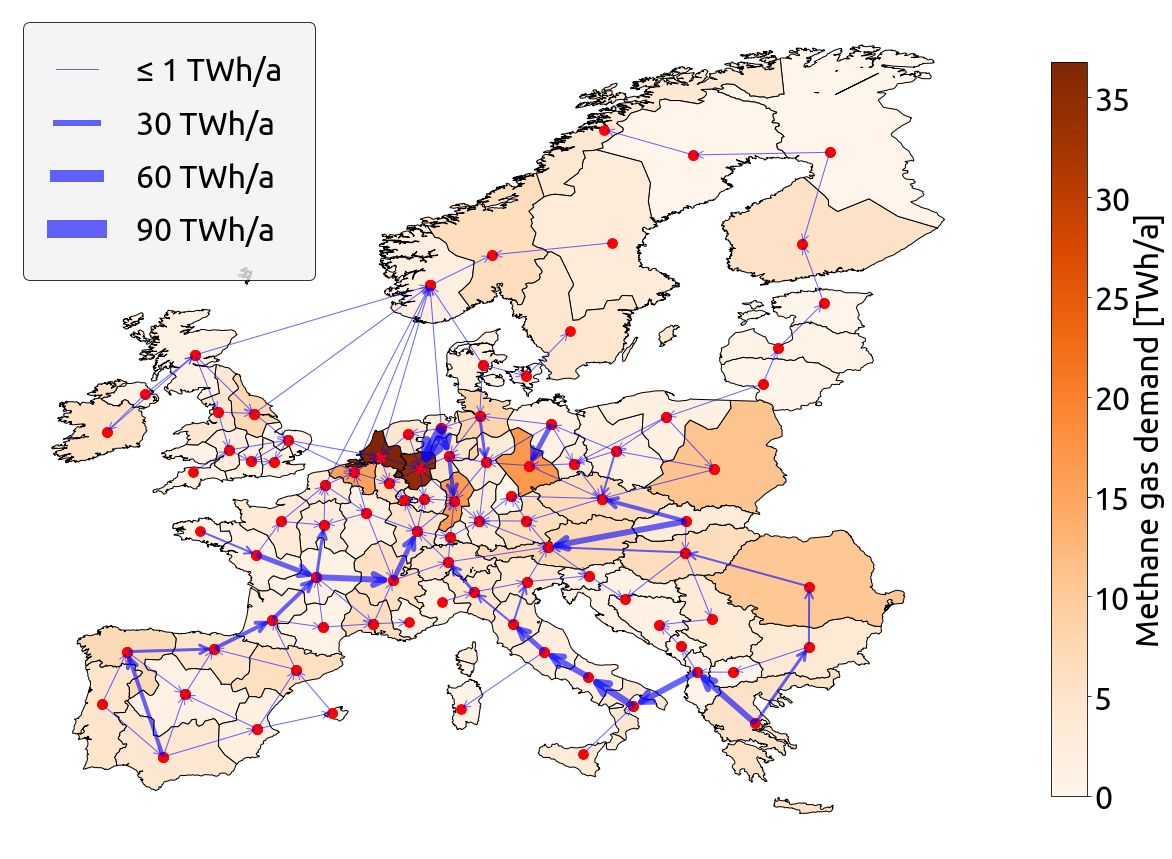}}
    \caption{Methane gas flow in (a) the global net-zero CO$_2$ emissions scenario and (b) the local net-zero CO$_2$ emissions scenario. Most of the flows occur to satisfy northern Italy, Switzerland, and Austria with methane gas imported mainly from Greece (in both scenarios) and Slovakia (in the local scenario). Significant flows also occur within Germany and France.}
    \label{figure_methane_gas_flow}
\end{figure}

Methane gas can be transported through a network of pipelines, allowing countries to exchange this material amongst themselves and within their borders, as well as indirectly transport CO$_2$. In the global and local scenarios aiming for net-zero CO$_2$ emissions, 124000 TWh·km/a and 109000 TWh·km/a of methane gas are transported across the entire continent, respectively (Figure \ref{figure_methane_gas_flow}).

The total methane gas consumption in Europe amounts to 800 TWh/a. This implies that, on average, 16\% of that demand is transported 1000 km. Out of the annual demand, 55\% of methane gas comes from synthetic or biogenic sources. The remaining 45\% has fossil origin and is imported into Europe from liquefied natural gas (LNG) terminals in the North Sea and Mediterranean Sea and via pipelines from Russia and North Africa.

Assuming the stoichiometric ratios provided in Table \ref{supplemental:table_energy_co2_conversion}, the transportation of 124000 TWh·km/a of methane gas effectively results in moving 24 GtCO$_2$·km/a in the global scenario, while 109000 TWh·km/a of methane gas results in moving 21 GtCO$_2$·km/a in the local scenario. Therefore, transporting CO$_2$ in the form of methane gas is the third most relevant method for carbon distribution in Europe.

Important flows of methane gas driven by existing infrastructure occur in the global scenario, including the flow from The Netherlands to Germany and flows from Greece, southern Italy and southern Spain where LNG terminals and entry points for pipelines from northern Africa are located \cite{Neumann2023}. Both flows decrease in significance in the local scenario though. The decrease in the former is connected to Germany requiring less methane gas, while the decrease in the latter is due to countries in central Europe requiring less methane (Figure \ref{supplemental:figure_demand}).

\begin{table}[!htb]
    \renewcommand{\arraystretch}{0.8}% Tighter
    \small
    \centering
    \caption{Energy and carbon flows in the global and local net-zero CO$_2$ emissions scenarios. Carbon is also transported in the form of synthetic oil and methanol (Figures \ref{supplemental:figure_synthetic_oil_producers_consumers_map_variation} and \ref{supplemental:figure_synthetic_methanol_producers_consumers_map_variation}), but our model does not include a detailed representation of the networks for these energy carriers.}
    \label{table_energy_co2_flow}
    \begin{tabular}{*5c}
        \toprule
       & \multicolumn{2}{c}{\textbf{Energy flow [TWh·km/a]}} & \multicolumn{2}{c}{\textbf{Carbon flow [GtCO$_2$·km/a]}} \\
        \midrule
        {}                & Global & Local                                            & Global & Local                                           \\
        CO$_2$            & -         & -                                             & 124    & 87                                              \\
        Solid biomass     & 87000     & 92000                                         & 32     & 34                                              \\
        Methane gas       & 124000    & 109000                                        & 24     & 21                                              \\
        H$_2$             & 518000    & 623000                                        & -      & -                                               \\
        Electricity       & 962000    & 978000                                        & -      & -                                               \\
        \bottomrule
    \end{tabular}
\end{table}

\subsection{Hydrogen is transported to where demand is high and carbon is collected}
H$_2$ is crucial for producing synthetic oil, methane gas, and methanol. It is also used in fuel cells, land transportation, and industry. As such, it is important to understand where H$_2$ is produced and how it is transported across Europe, as well as how these differ between global and local climate-neutral scenarios.

\begin{figure}[!htb]
    \centering
    %\subfloat[]{\label{supplemental:figure_wind_capacity_factors}\includegraphics[width = 0.485\linewidth]{./figures/figure_S9a.png}}\hfill
    \subfloat[]{\label{supplemental:figure_wind_capacity_factors}\includegraphics[width = 0.475\linewidth]{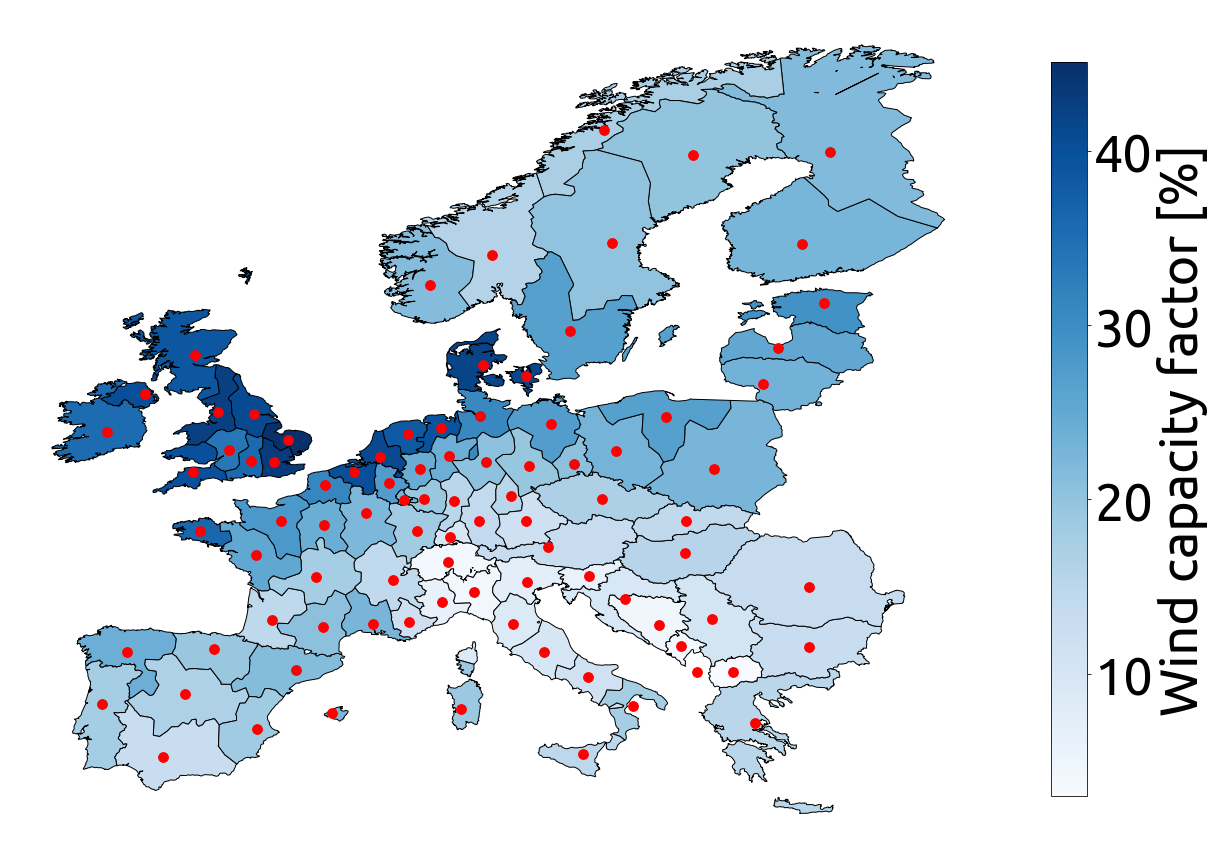}}\hfill
    %\subfloat[]{\label{supplemental:figure_solar_capacity_factors}\includegraphics[width = 0.485\linewidth]{./figures/figure_S9b.png}}\par
    \subfloat[]{\label{supplemental:figure_solar_capacity_factors}\includegraphics[width = 0.475\linewidth]{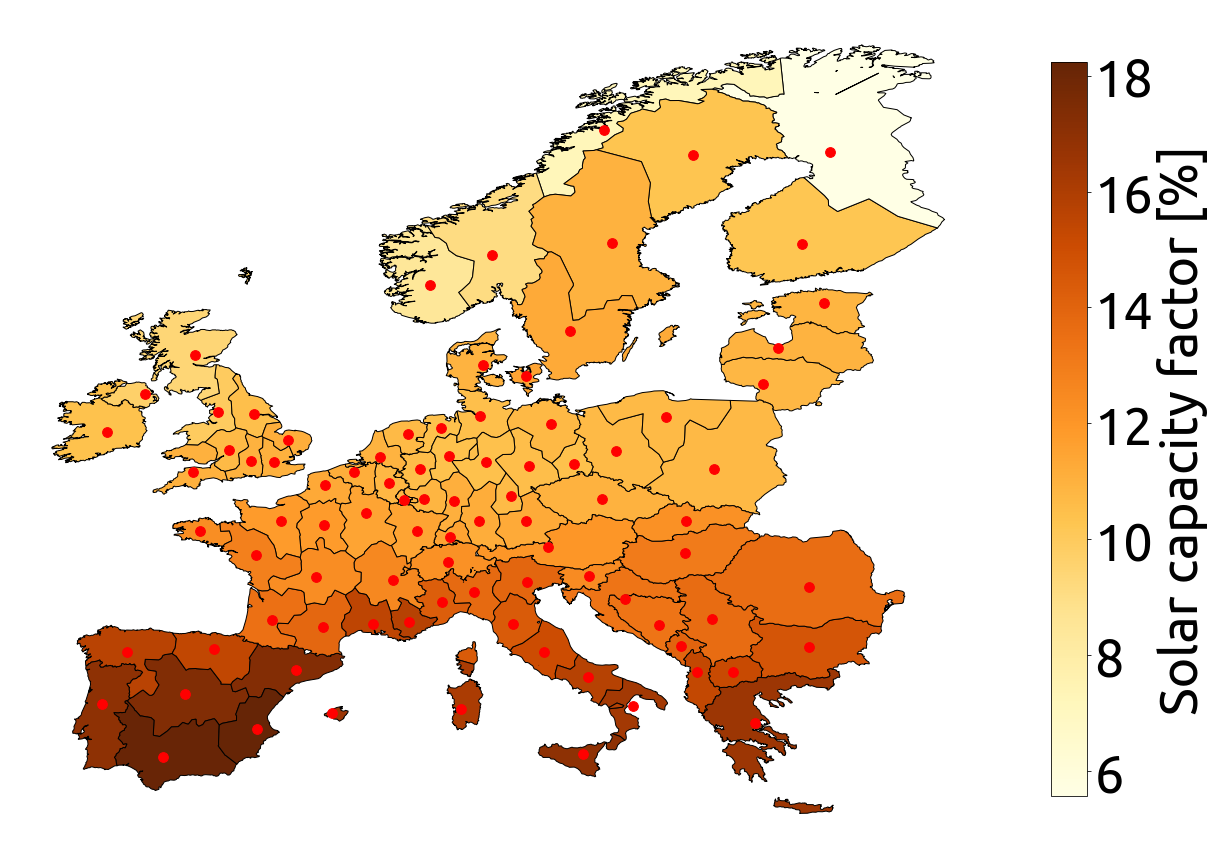}}\par
    \vspace{20pt}
    %\subfloat[]{\label{supplemental:figure_wind_solar_factors_per_country}\includegraphics[width = 0.96\textwidth]{./figures/figure_S9c.png}}
    \subfloat[]{\label{supplemental:figure_wind_solar_factors_per_country}\includegraphics[width = 0.8\textwidth]{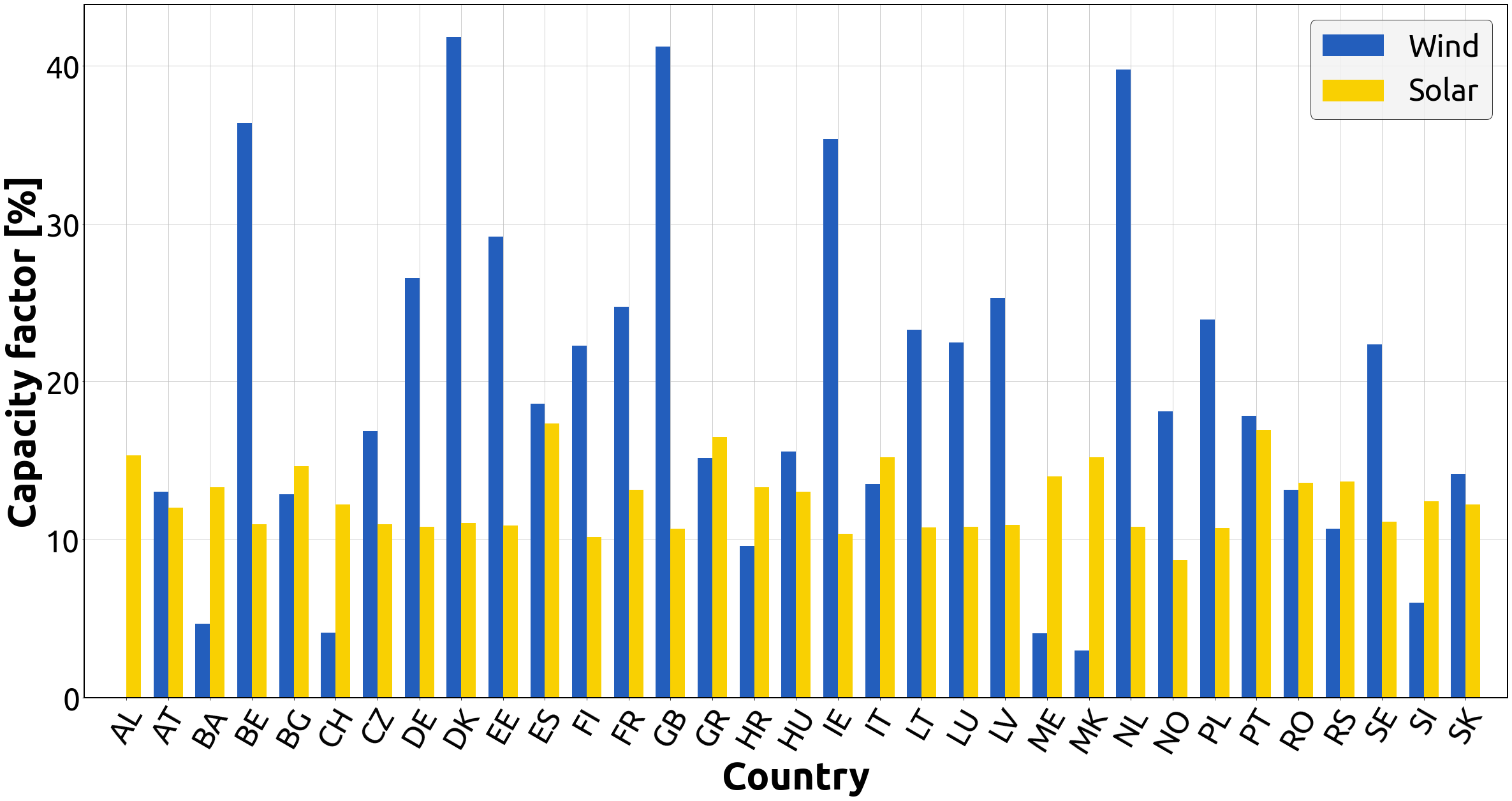}}
    \caption{(a) Onshore wind capacity factors per node and (b) Solar photovoltaic capacity factors per node, and (c) Onshore wind and solar photovoltaic capacity factors per country. The capacity factors for wind are predominantly higher in northern Europe, whereas the capacity factors for solar are predominantly higher in southern Europe.}
    \label{supplemental:figure_renewable_capacity_factors}
\end{figure}

\begin{figure}[!htb]
    \centering
    %\subfloat[]{\label{supplemental:figure_temporal_h2_production_global}\includegraphics[width = 0.96\linewidth]{./figures/figure_S32a.png}}
    \subfloat[]{\label{supplemental:figure_temporal_h2_production_global}\includegraphics[width = 0.8\linewidth]{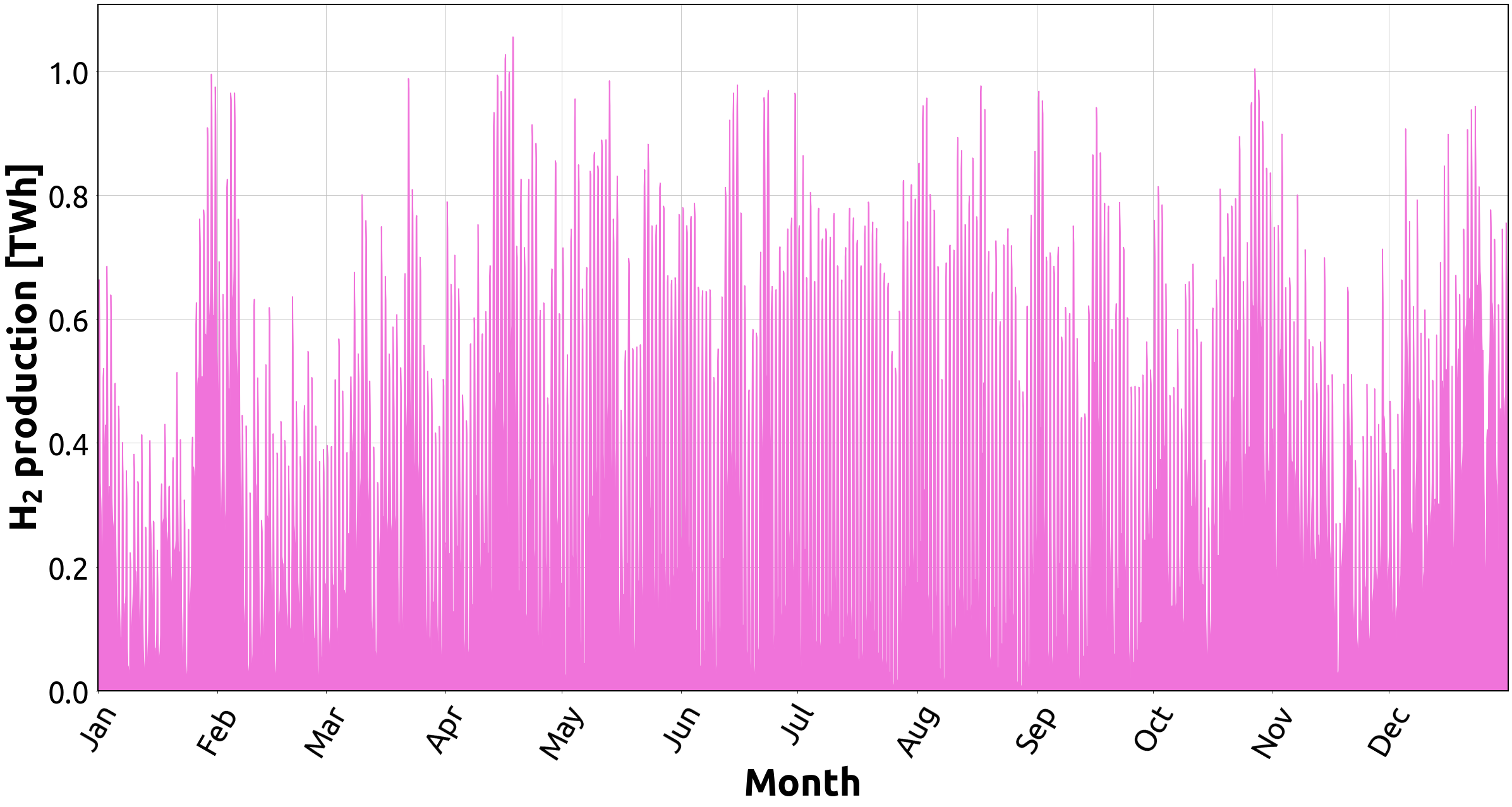}}
    \vspace{20pt}
    %\subfloat[]{\label{supplemental:figure_temporal_h2_production_local}\includegraphics[width = 0.96\linewidth]{./figures/figure_S32b.png}}
    \subfloat[]{\label{supplemental:figure_temporal_h2_production_local}\includegraphics[width = 0.8\linewidth]{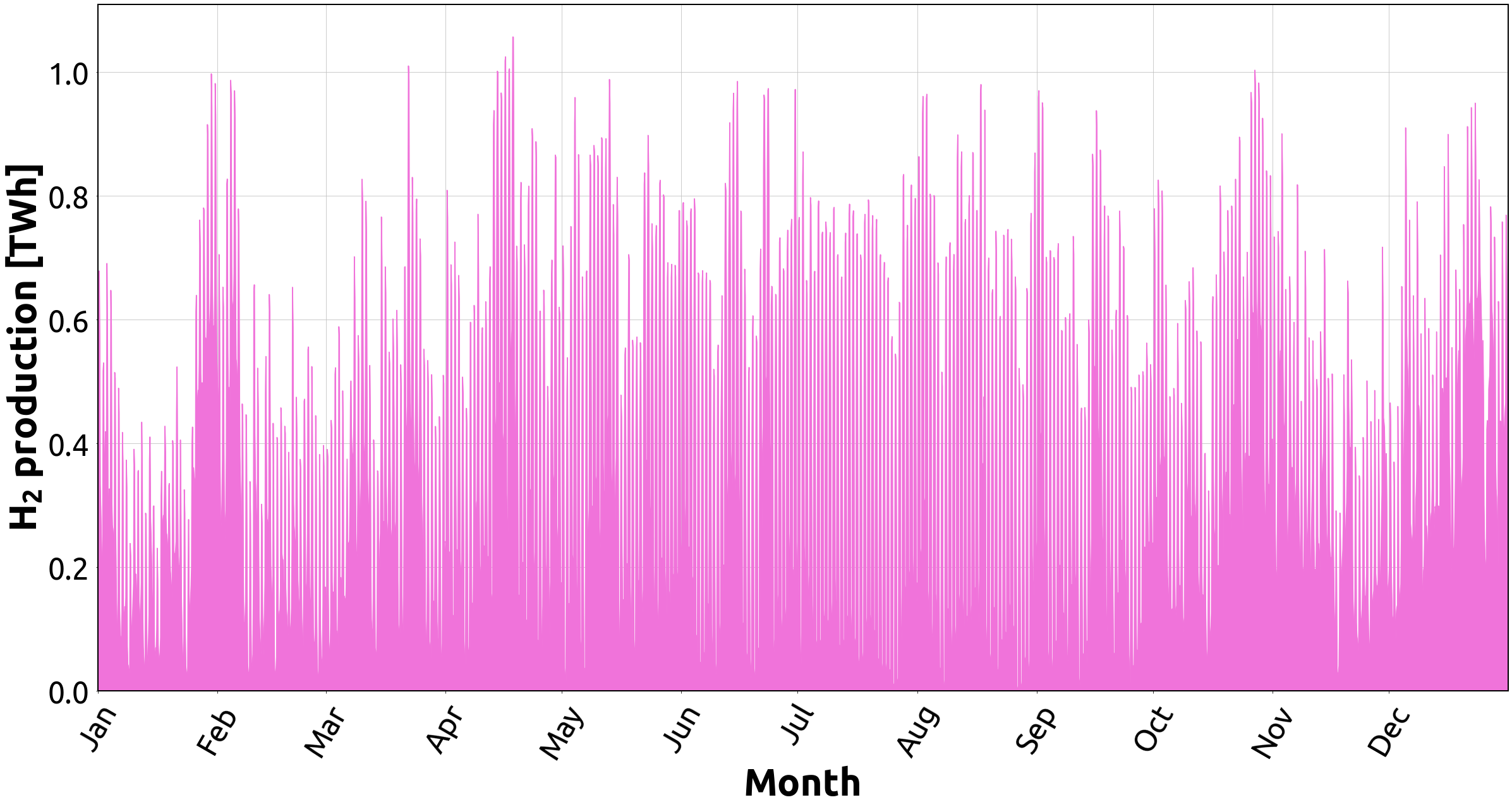}}
    \caption{Temporal H$_2$ production across all of Europe in (a) the global net-zero CO$_2$ emissions scenario and (b) the local net-zero CO$_2$ emissions scenario. In both scenarios, the production of H$_2$ varies seasonally, with higher levels during the summer when there is an abundance of cost-effective energy to power H$_2$ electrolysis compared to winter.}
    \label{supplemental:figure_temporal_h2_production}
\end{figure}

In both scenarios, the flow of H$_2$ is driven by the demand and the cost-effective production in locations with high wind and solar resources (Figure \ref{supplemental:figure_renewable_capacity_factors}). In the global scenario, Great Britain, France, Spain, Italy, and Ireland are the main producers of electrolytic H$_2$
(Figure \ref{supplemental:figure_installed_capacity}), as identified in \cite{WETZEL2023591}. They further increase production levels in the local scenario, in which the interior countries decrease their H$_2$ production (Figures \ref{supplemental:figure_h2_demand_production} and \ref{supplemental:figure_temporal_h2_production}).

\begin{figure}[!htb]
    \centering
    %\subfloat[]{\label{supplemental:figure_h2_flow_global}\includegraphics[width = 0.8\linewidth]{./figures/figure_S28a.png}}
    \subfloat[]{\label{supplemental:figure_h2_flow_global}\includegraphics[width = 0.6\linewidth]{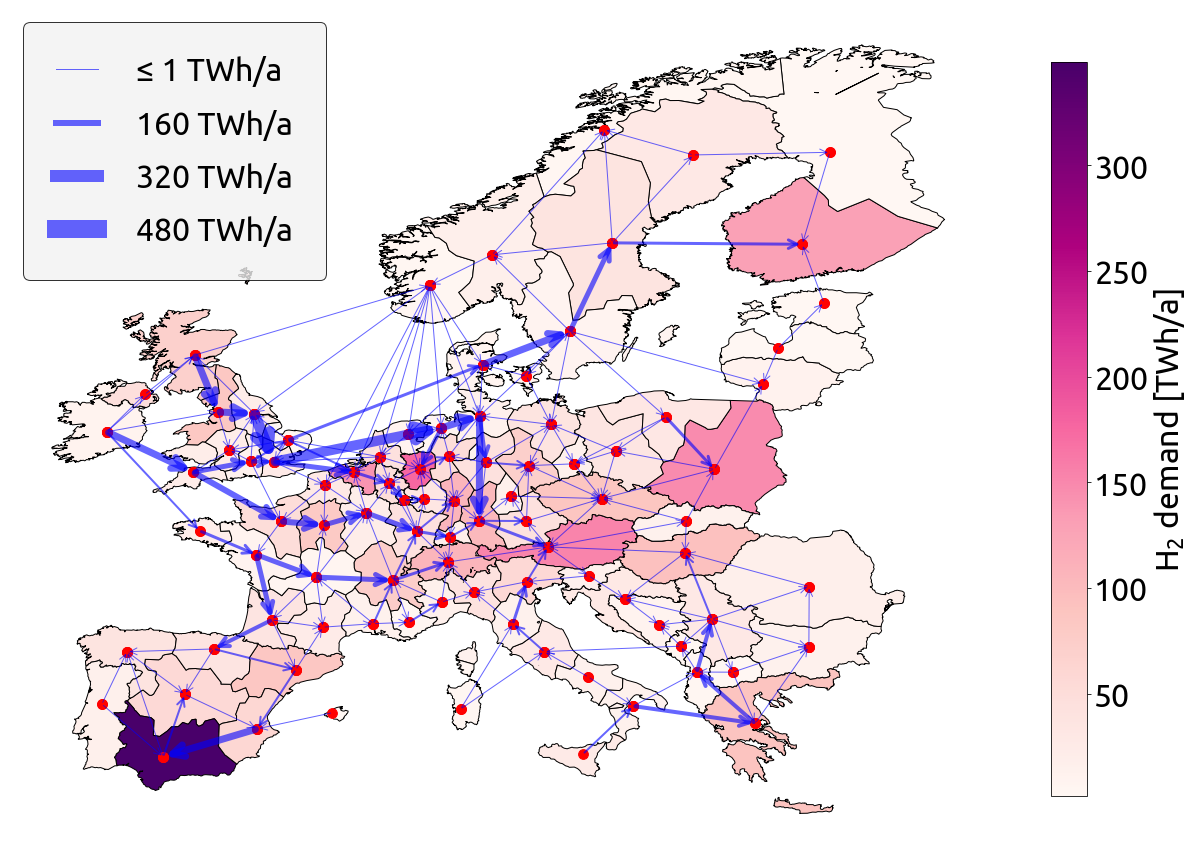}}
    \vspace{20pt}
    %\subfloat[]{\label{supplemental:figure_h2_flow_local}\includegraphics[width = 0.8\linewidth]{./figures/figure_S28b.png}}
    \subfloat[]{\label{supplemental:figure_h2_flow_local}\includegraphics[width = 0.6\linewidth]{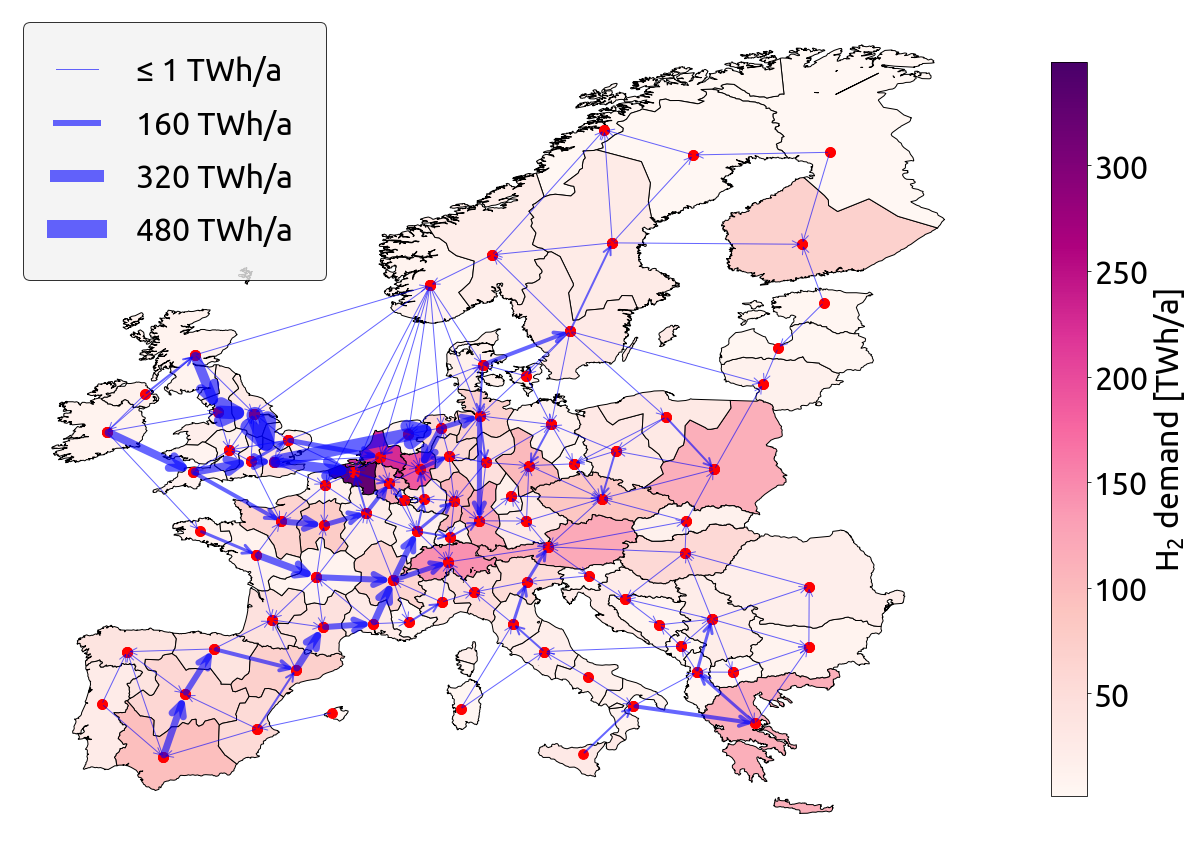}}
    \caption{H$_2$ flow in (a) the global net-zero CO$_2$ emissions scenario and (b) the local net-zero CO$_2$ emissions scenario. The total amount of H$_2$ flowing amongst and within countries increases by 6\% under local constraints relative to the global constraint. Most of this increase is to provide the necessary H$_2$ for the interior countries and Switzerland to meet their industrial needs and produce more synthetic oil and methanol under the local scenario.}
    \label{supplemental:figure_h2_flow}
\end{figure}

Under local constraints, we observe a 20\% increase in the flow of H$_2$ compared to the 518000 TWh·km/a flowing under a global constraint (Figure \ref{supplemental:figure_h2_flow}). This increase can be explained as a side effect of the reduction in CO$_2$ flows. As discussed in \cite{Hofmann2025}, when the carbon flows are lower (in the local scenario), H$_2$ produced in locations with high renewable resources need to be transported to where the carbon is collected to be combined and produce synthetic fuels.

\subsection{Electricity is still the main strategy for energy exchange amongst countries}
Electricity is required in large quantities to enable direct-air-captured CO$_2$ and methanolisation to transform captured CO$_2$ into methanol. It is also vastly needed for producing H$_2$ electrolytically, a key component in manufacturing synthetic products. This emphasises the importance of understanding where electricity is produced and transmitted amongst and within countries for consumption.

\begin{figure}[!htb]
    \centering
    %\subfloat[]{\label{supplemental:figure_electricity_production_map}\includegraphics[width = 0.8\linewidth]{./figures/figure_S18a.png}}
    \subfloat[]{\label{supplemental:figure_electricity_production_map}\includegraphics[width = 0.66\linewidth]{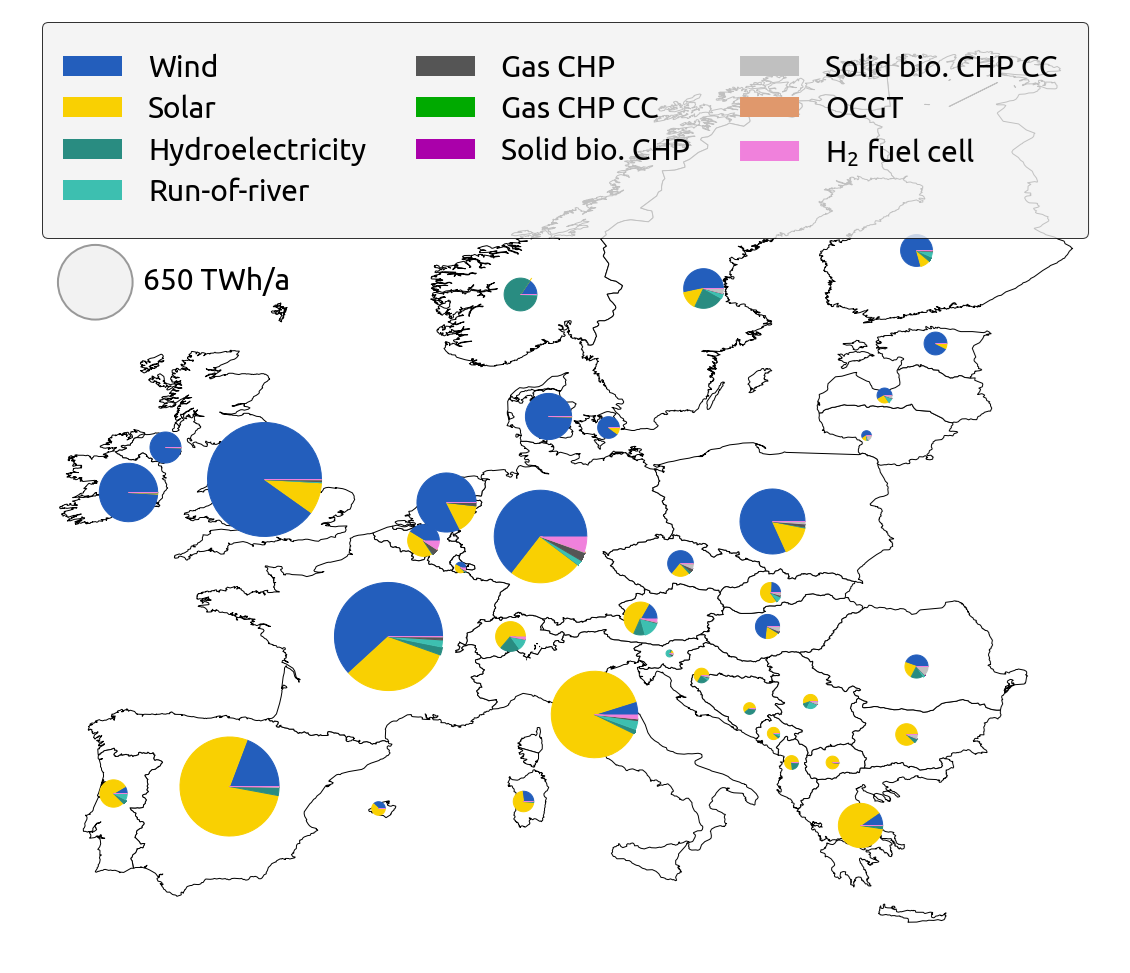}}
    \vspace{20pt}
    %\subfloat[]{\label{supplemental:figure_electricity_production_variation}\includegraphics[width = 0.96\linewidth]{./figures/figure_S18b.png}}
    \subfloat[]{\label{supplemental:figure_electricity_production_variation}\includegraphics[width = 0.8\linewidth]{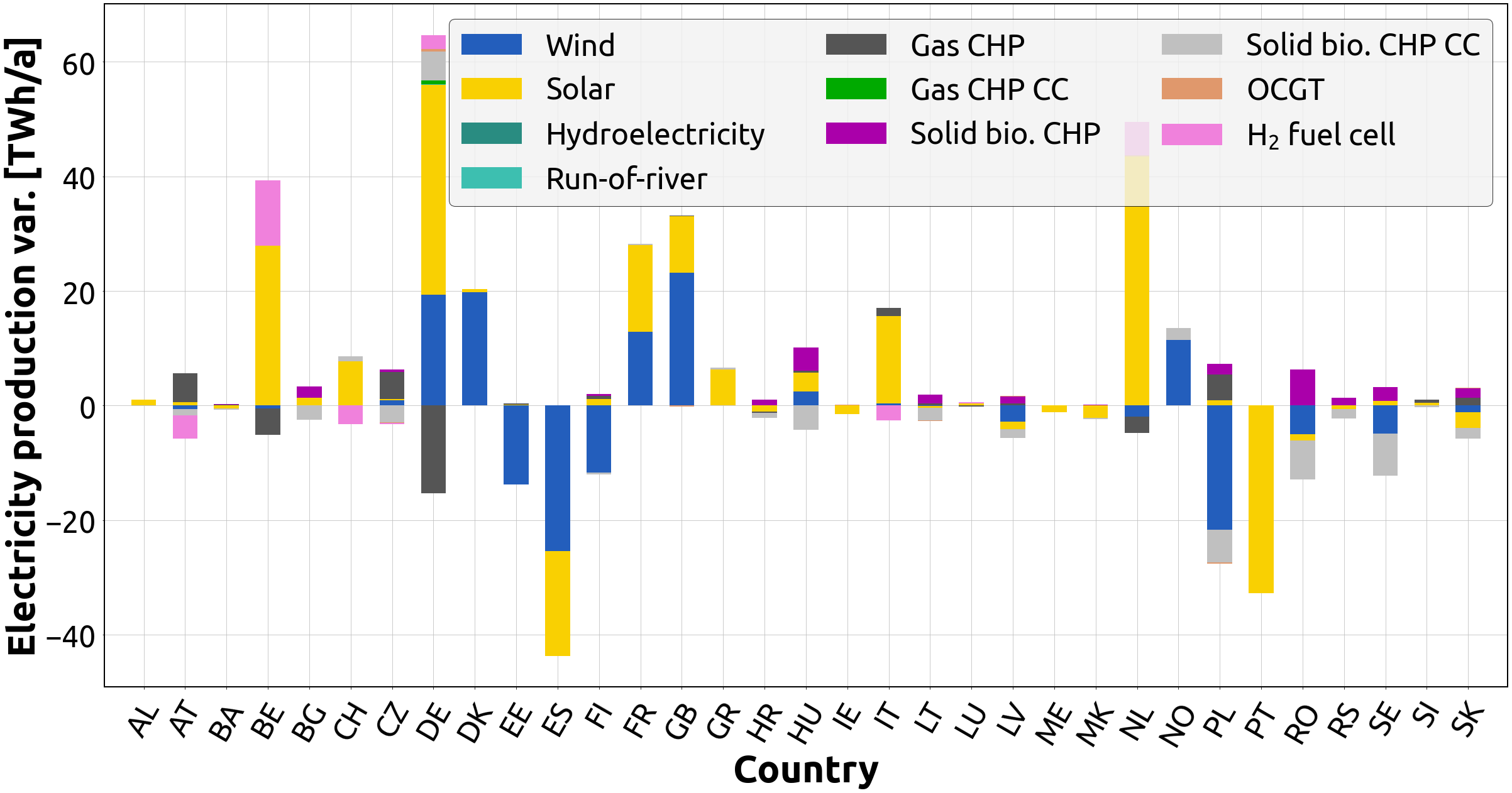}}
    \caption{(a) Electricity production in the global net-zero CO$_2$ emissions scenario and (b) Electricity production variation per country and technology between the global net-zero CO$_2$ emissions scenario and the local net-zero CO$_2$ emissions scenario. Major variations occur in electricity production from renewable sources. While countries that were net CO$_2$ emitters increase solar and wind to produce electricity, net CO$_2$ absorber countries, on the other hand, decrease these. This is because the former countries capture much more CO$_2$ under local constraints using DAC and convert it into methanol, both processes requiring significant amounts of power, while the latter countries relax these (since they have to handle less captured CO$_2$).}
    \label{supplemental:figure_electricity_production}
\end{figure}

\begin{figure}[!htb]
    \centering
    %\subfloat[]{\label{supplemental:figure_installed_capacity_map}\includegraphics[width = 0.8\linewidth]{./figures/figure_S30a.png}}
    \subfloat[]{\label{supplemental:figure_installed_capacity_map}\includegraphics[width = 0.66\linewidth]{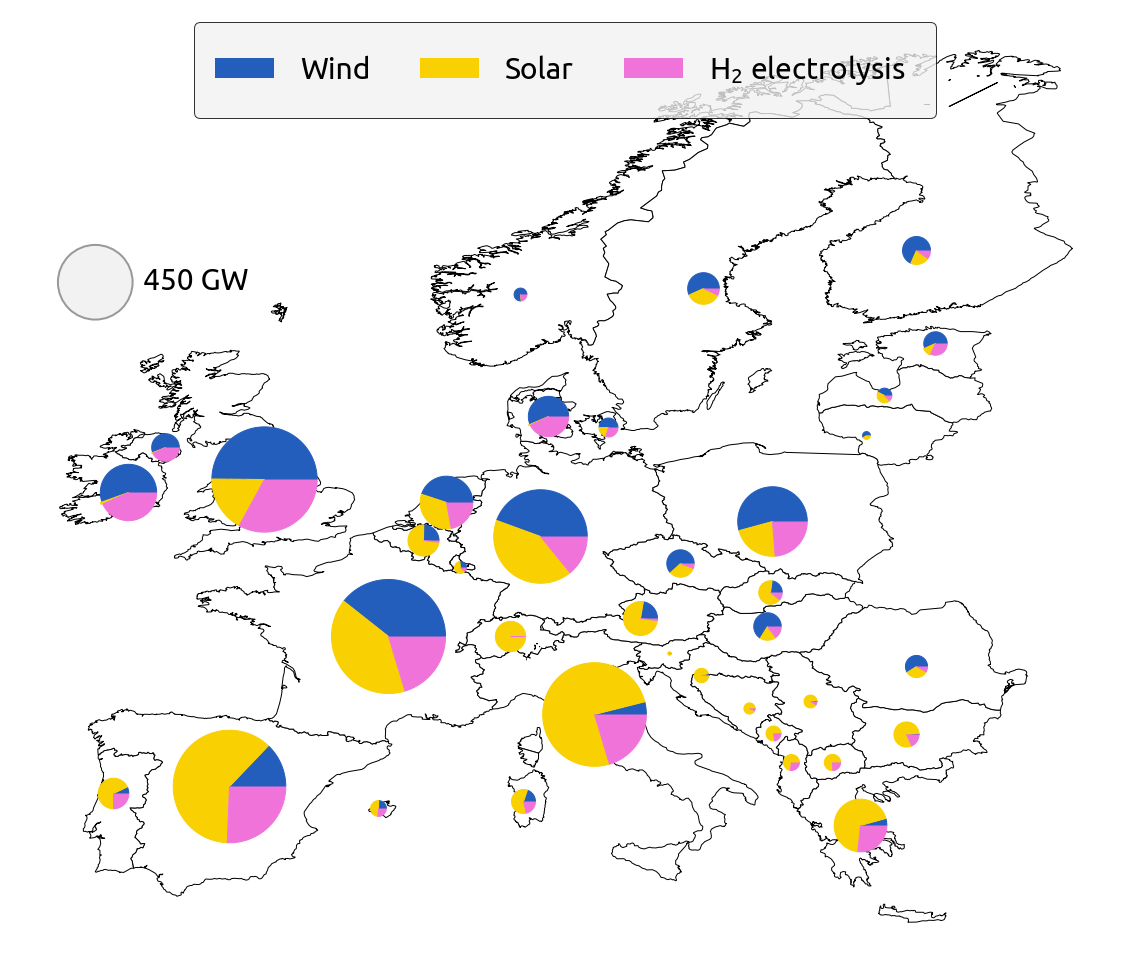}}
    \vspace{20pt}
    %\subfloat[]{\label{supplemental:figure_installed_capacity_variation}\includegraphics[width = 0.96\linewidth]{./figures/figure_S30b.png}}
    \subfloat[]{\label{supplemental:figure_installed_capacity_variation}\includegraphics[width = 0.8\linewidth]{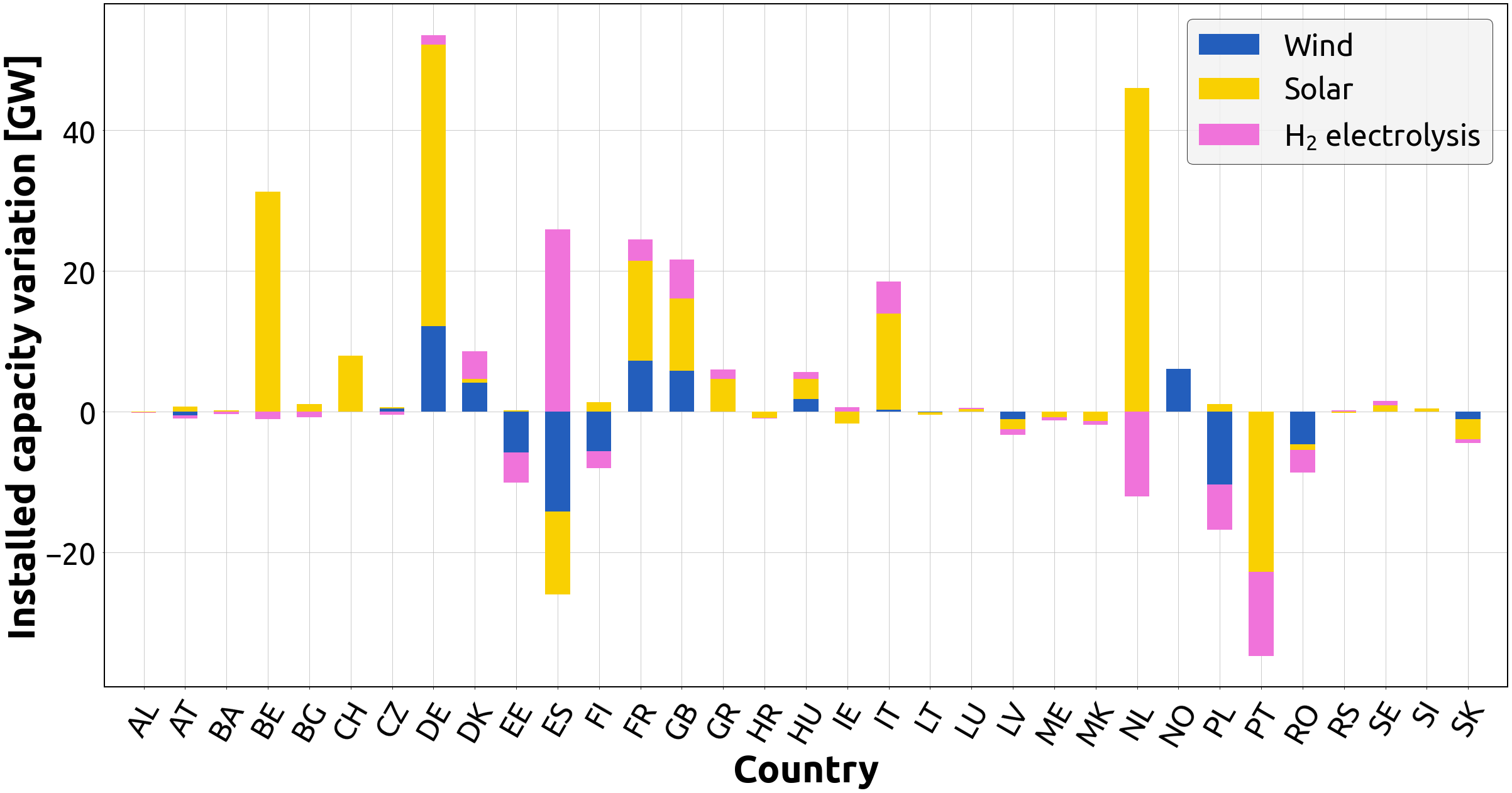}}
    \caption{(a) Wind, solar, and H$_2$ electrolysis installed capacity in the global net-zero CO$_2$ emissions scenario and (b) Wind, solar, and H$_2$ electrolysis installed capacity variation per country between the global net-zero CO$_2$ emissions scenario and the local net-zero CO$_2$ emissions scenario.}
    \label{supplemental:figure_installed_capacity}
\end{figure}

In a climate-neutral scenario, the production of electricity in Europe is mainly determined by the accessibility of cost-effective renewable energy. Although nuclear power plants are an option for producing electricity in our model, they are not chosen because of their high cost. From this perspective, major production is observed in Great Britain, France, Spain, Germany, and Italy. In the global scenario, a total of 10068 TWh of electricity is produced annually, with a slight increase of 1.1\% in the local scenario (Figures \ref{supplemental:figure_installed_capacity} and \ref{supplemental:figure_electricity_production}). The model's intrinsic constraint for the electricity supply to meet the demand at every time step of the selected temporal resolution (i.e. 3 hours) provides stability for the entire system (Equation \ref{supplemental:material_electricity_demand_supply_constraint}), resulting in minimal production differences when comparing the two scenarios.

\begin{figure}[!htb]
    \centering
    %\subfloat[]{\label{supplemental:figure_electricity_flow_global}\includegraphics[width = 0.8\linewidth]{./figures/figure_S29a.png}}
    \subfloat[]{\label{supplemental:figure_electricity_flow_global}\includegraphics[width = 0.6\linewidth]{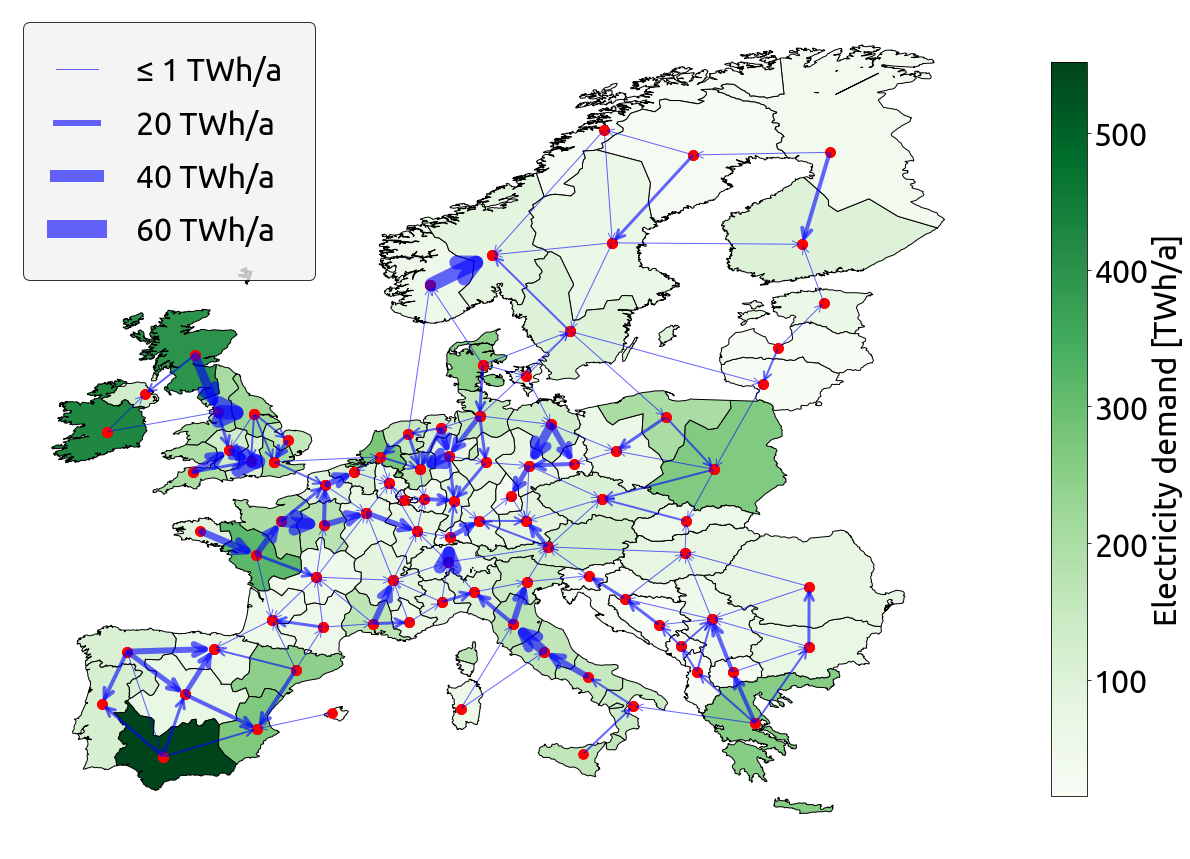}}
    \vspace{20pt}
    %\subfloat[]{\label{supplemental:figure_electricity_flow_local}\includegraphics[width = 0.8\linewidth]{./figures/figure_S29b.png}}
    \subfloat[]{\label{supplemental:figure_electricity_flow_local}\includegraphics[width = 0.6\linewidth]{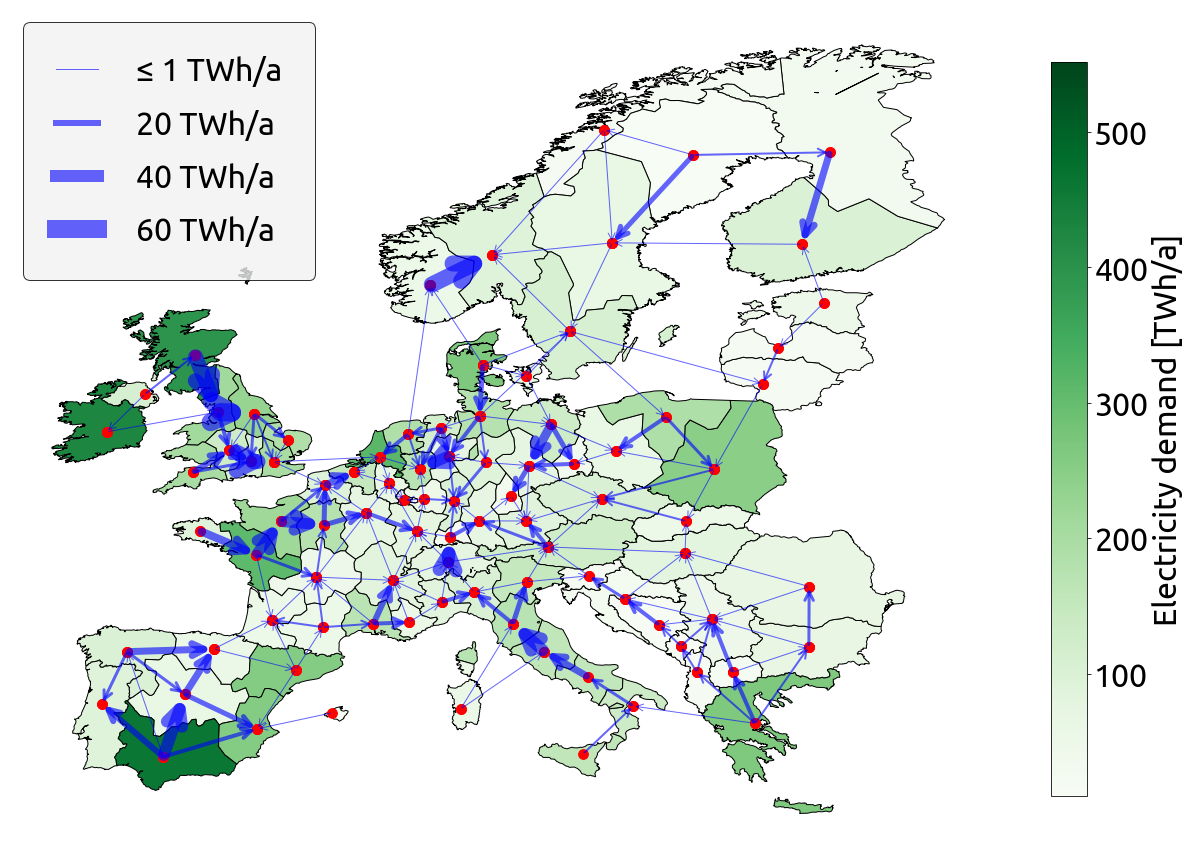}}
    \caption{Electricity flow in (a) the global net-zero CO$_2$ emissions scenario and (b) the local net-zero CO$_2$ emissions scenario. In both scenarios, most of the electricity flows are confined within individual countries. Regions within the same country supply electricity to each other to power DAC and methanolisation, as well as H$_2$ production, with limited electricity exchange occurring between countries.}
    \label{supplemental:figure_electricity_flow}
\end{figure}

\begin{figure}[!htb]
    \centering
    %\subfloat[]{\label{supplemental:figure_h2_production_global_map}\includegraphics[width = 0.485\linewidth]{./figures/figure_S31a.png}}\hfill
    \subfloat[]{\label{supplemental:figure_h2_production_global_map}\includegraphics[width = 0.475\linewidth]{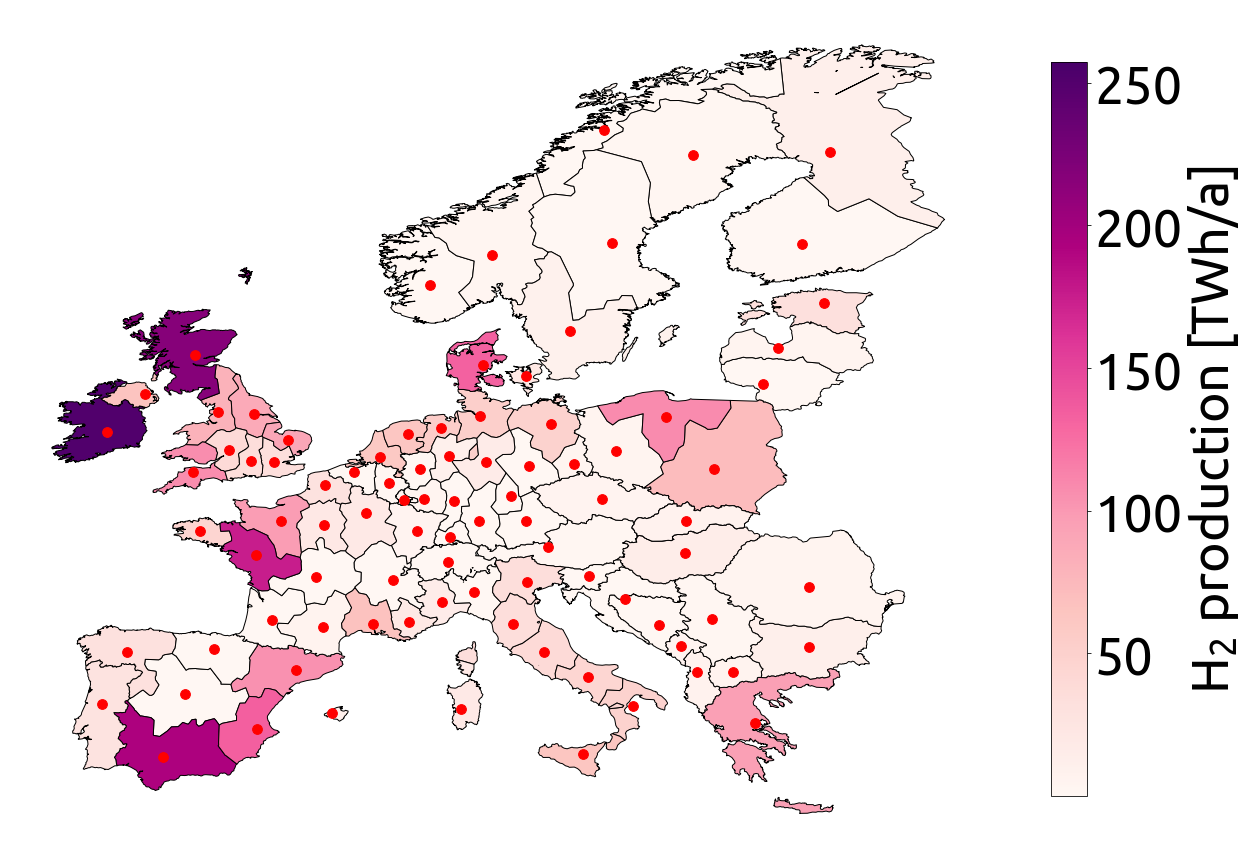}}\hfill
    %\subfloat[]{\label{supplemental:figure_h2_production_local_map}\includegraphics[width = 0.485\linewidth]{./figures/figure_S31b.png}}\par
    \subfloat[]{\label{supplemental:figure_h2_production_local_map}\includegraphics[width = 0.475\linewidth]{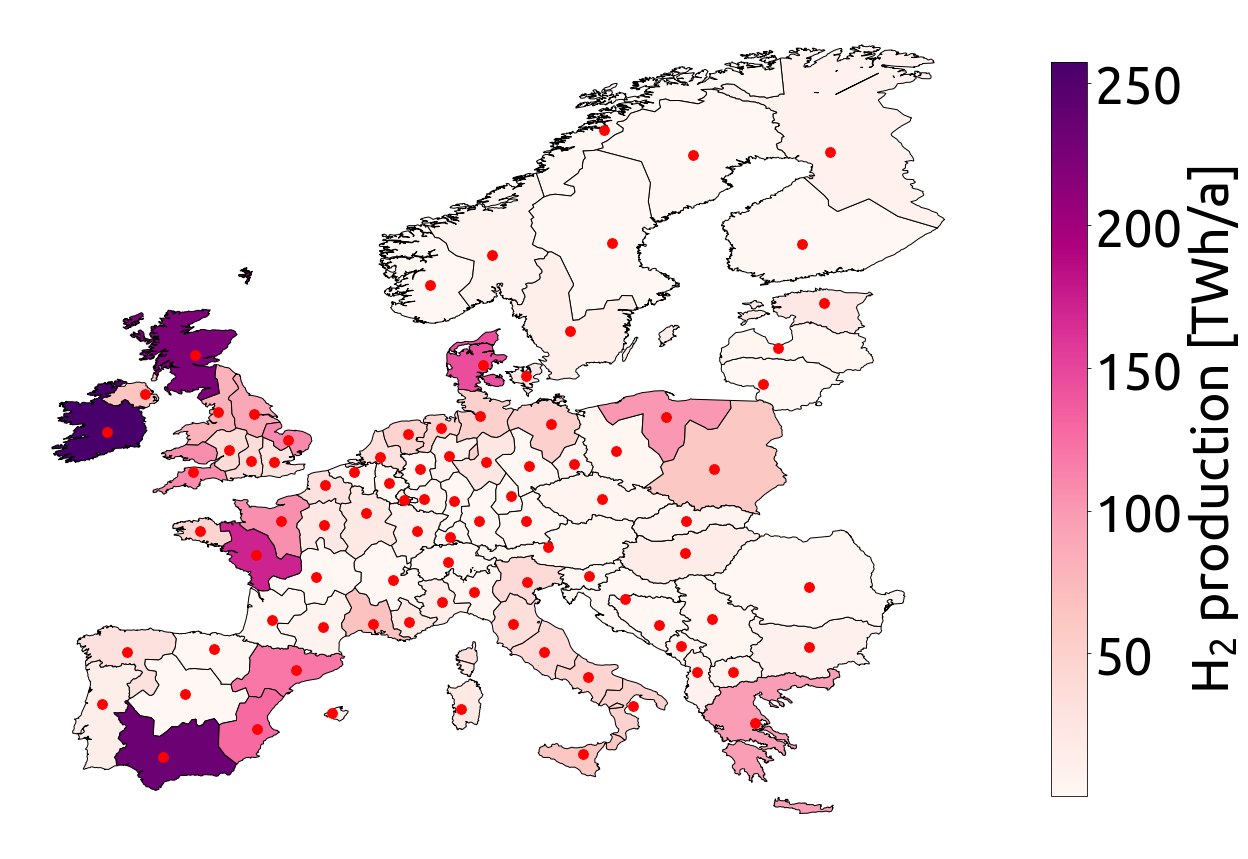}}\par
    \vspace{20pt}
    %\subfloat[]{\label{supplemental:figure_h2_demand_production_variation}\includegraphics[width = 0.96\textwidth]{./figures/figure_S31c.png}}
    \subfloat[]{\label{supplemental:figure_h2_demand_production_variation}\includegraphics[width = 0.8\textwidth]{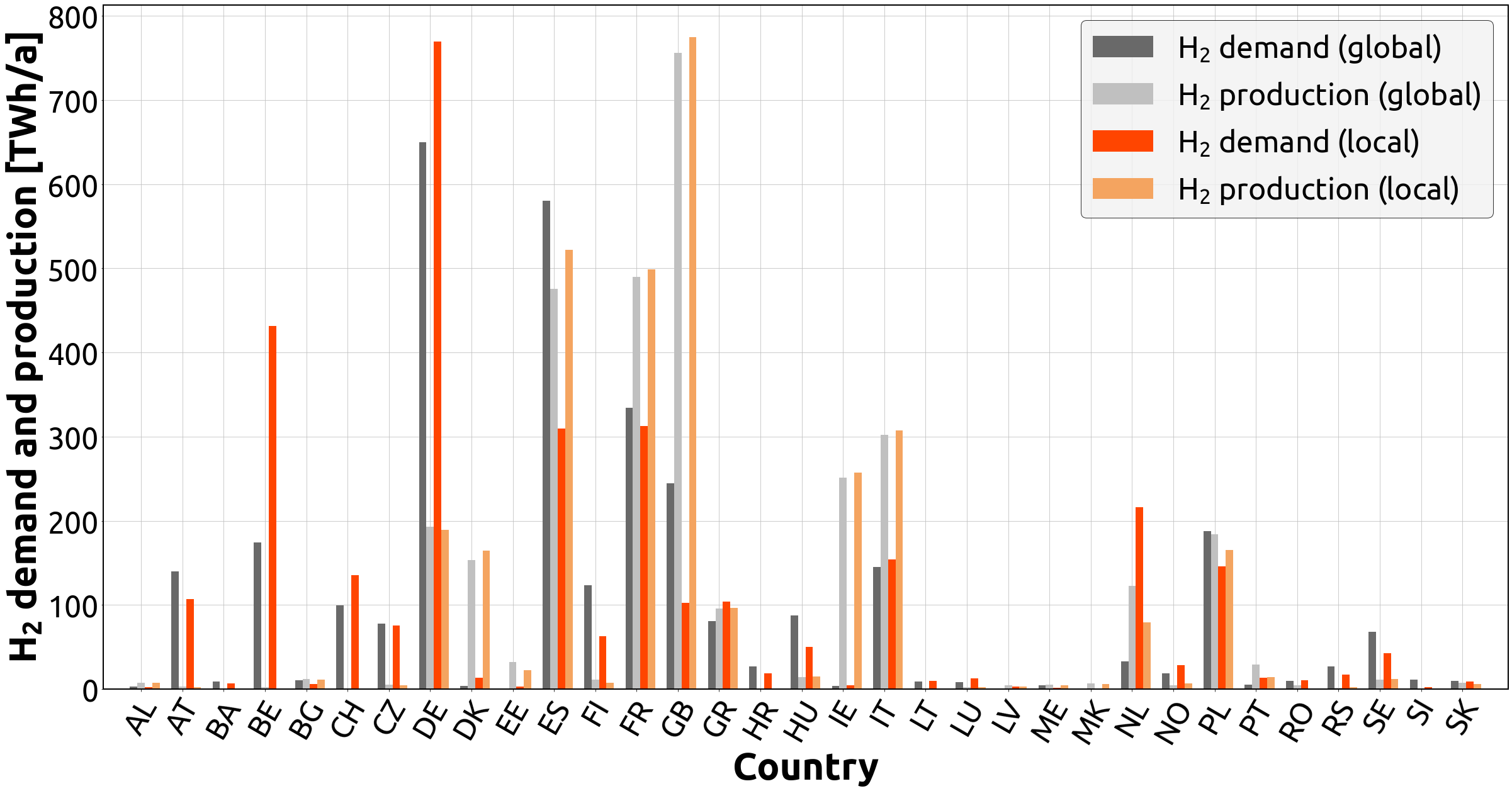}}
    \caption{H$_2$ production per node in (a) the global net-zero CO$_2$ emissions scenario and (b) the local net-zero CO$_2$ emissions scenario, and (c) H$_2$ demand and synthetic production per country in the global and local net-zero CO$_2$ emissions scenarios. Thanks to high renewable energy capacity factors and their relative proximity to the interior countries and Switzerland, Spain, Denmark, and Great Britain  significantly increase H$_2$ production to satisfy a growing demand for H$_2$ in the former countries when under local constraints.}
    \label{supplemental:figure_h2_demand_production}
\end{figure}

For simplicity, we do not allow expansion of transmission capacity beyond today's values. This results in a flow of electricity that amounts to 962000 TWh·km/a in the global scenario and 978000 TWh·km/a in the local scenario (Figure \ref{supplemental:figure_electricity_flow}). In both scenarios, most of the electricity flows within each country, typically those that produce the most H$_2$ in Europe, with minimal cross-border exchange (Figure \ref{supplemental:figure_h2_demand_production}).

\subsection{Similar trends observed for relaxed CO$_2$ targets}

\begin{figure}[!htb]
    \centering
    %\subfloat[]{\label{supplemental:figure_system_cost_sensitivity_analysis_global}\includegraphics[width = 0.96\linewidth]{./figures/figure_S13a.png}}
    \subfloat[]{\label{supplemental:figure_system_cost_sensitivity_analysis_global}\includegraphics[width = 0.8\linewidth]{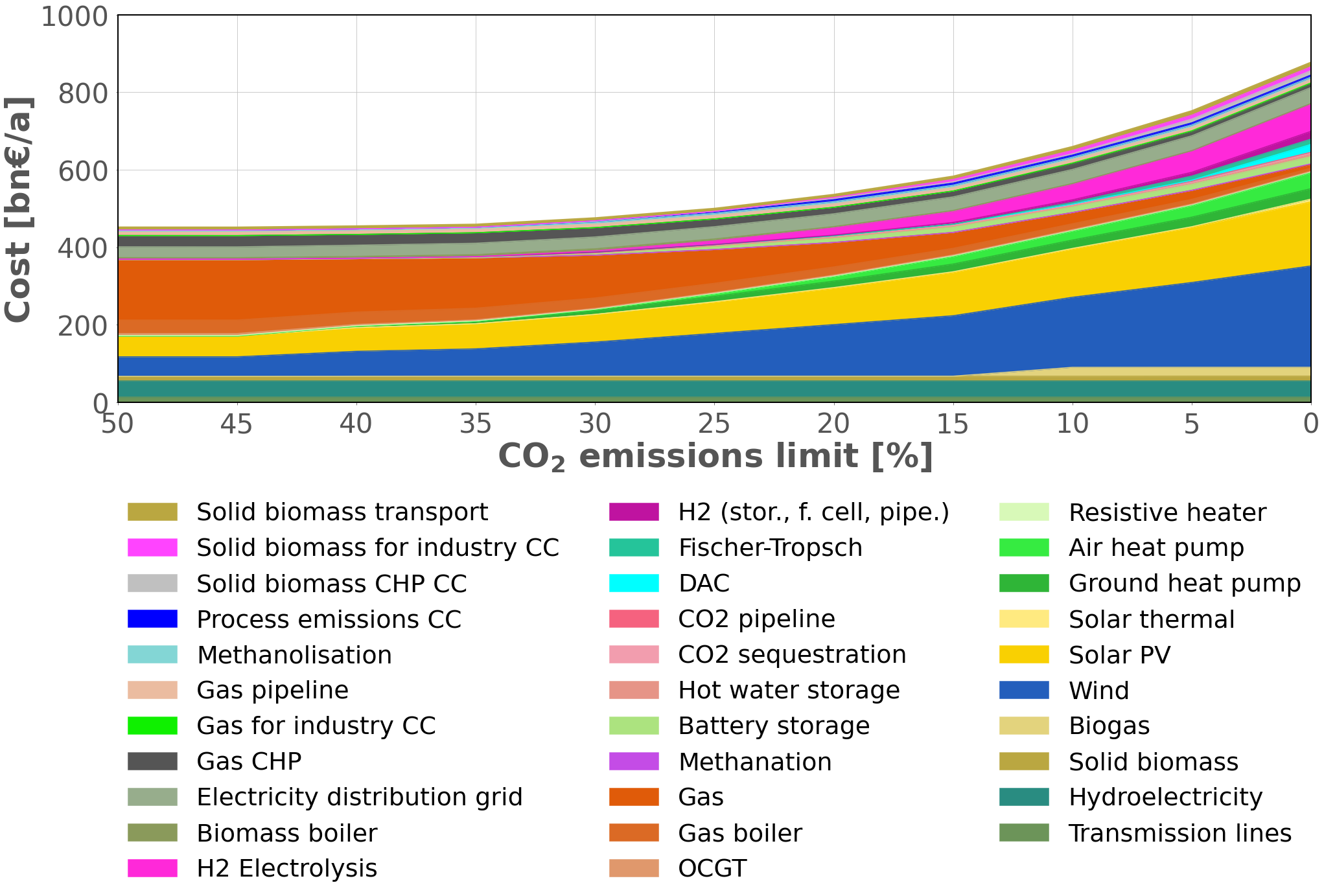}}
    \vspace{20pt}
    %\subfloat[]{\label{supplemental:figure_system_cost_sensitivity_analysis_local}\includegraphics[width = 0.96\linewidth]{./figures/figure_S13b.png}}
    \subfloat[]{\label{supplemental:figure_system_cost_sensitivity_analysis_local}\includegraphics[width = 0.8\linewidth]{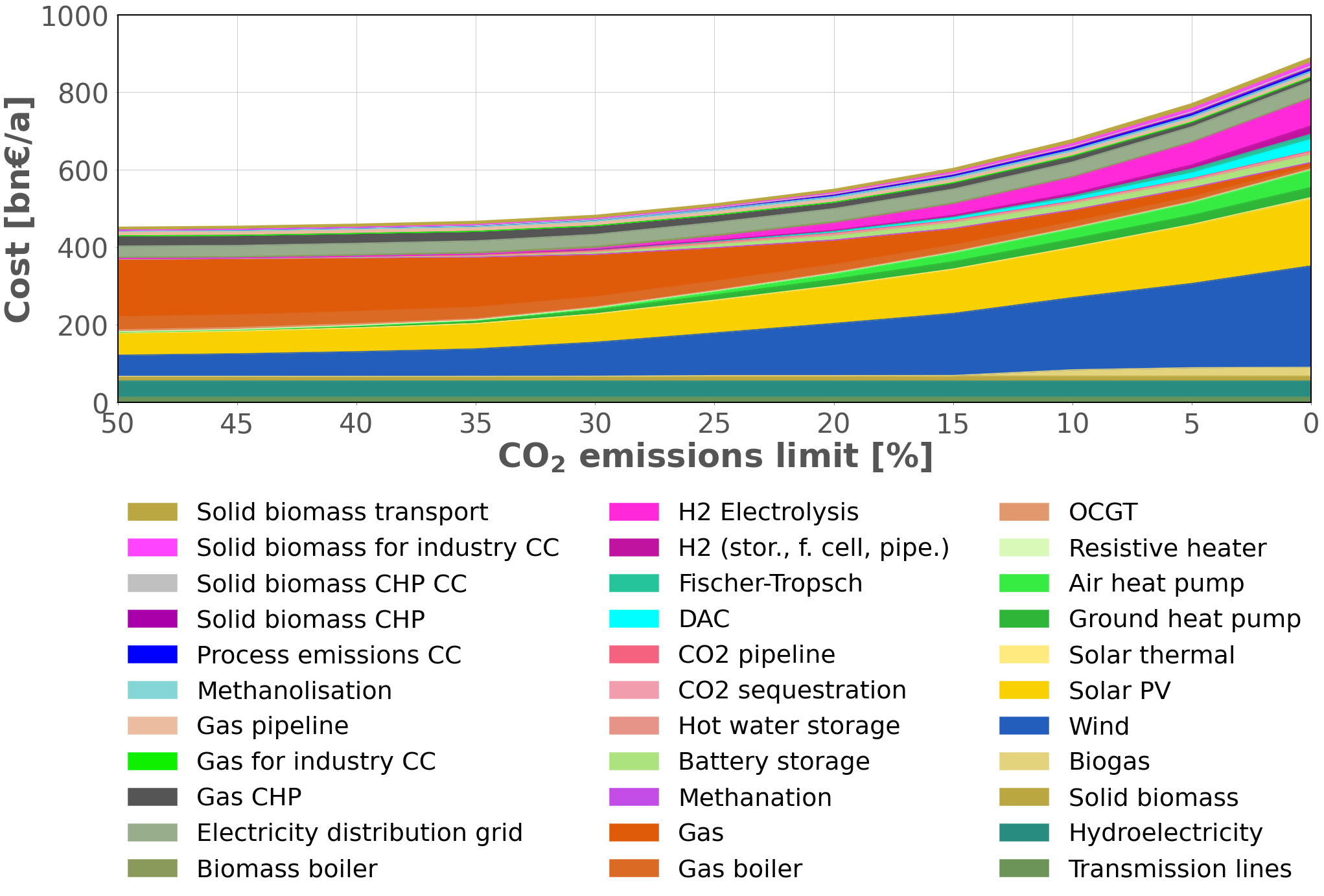}}
    \caption{Total system cost sensitivity analysis under different CO$_2$ emissions limits and respective technology configuration in (a) the global scenario and (b) the local scenario. Until reaching a 35\% limit in CO$_2$ emissions (relative to 1990 levels), the total system cost remains constant since the CO$_2$ constraint is not binding. After that point, the cost increases non-linearly in both scenarios until Europe is fully decarbonised.}  \label{supplemental:figure_system_cost_sensitivity_analysis}
\end{figure}

To understand better the impacts of the CO$_2$ constraints, we conducted a sensitivity analysis to assess how the model reacts under different CO$_2$ emissions limits in terms of cost and technology configuration (Figure \ref{supplemental:figure_system_cost_sensitivity_analysis}). The results show that up to the point where CO$_2$ emissions are reduced to 35\% of 1990 levels, the total system cost stays constant at 540 billion € per year in the globally constrained scenario and 548 billion € per year in the locally constrained scenario. After this point, reducing emissions further increases costs non-linearly in both scenarios until the continent is fully decarbonised. Furthermore, as the limit becomes stricter, the model gradually shifts from using methane gas to renewable technologies and technologies for direct and indirect electrification. The technologies composing each of the modelled scenarios remain the same, regardless of the imposed CO$_2$ emissions limit.

\begin{figure}[!htb]
    \centering
    %\subfloat[]{\label{supplemental:figure_co2_capture_sensitivity_analysis_global}\includegraphics[width = 0.96\linewidth]{./figures/figure_S19a.png}}
    \subfloat[]{\label{supplemental:figure_co2_capture_sensitivity_analysis_global}\includegraphics[width = 0.8\linewidth]{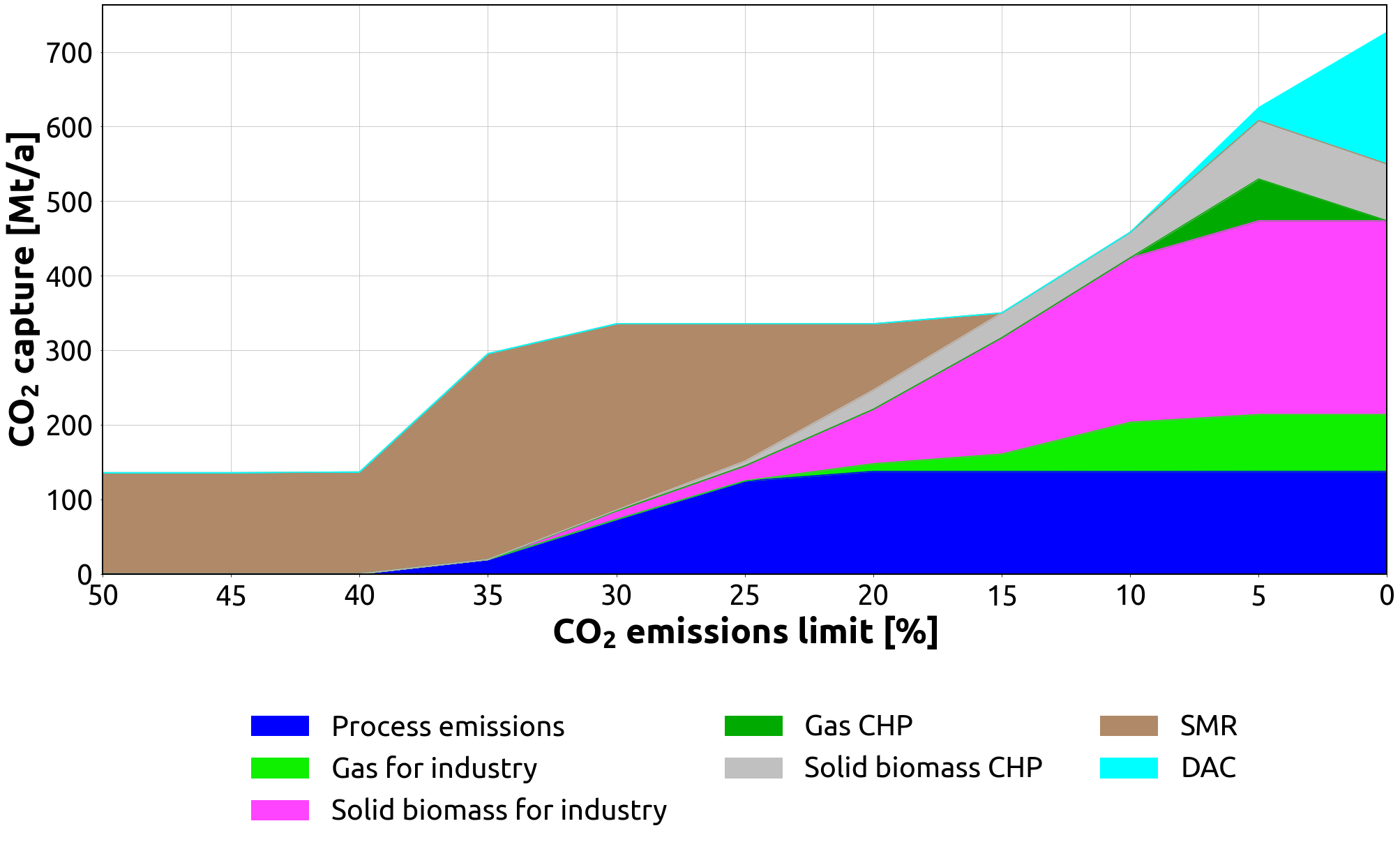}}
    \vspace{20pt}
    %\subfloat[]{\label{supplemental:figure_co2_capture_sensitivity_analysis_local}\includegraphics[width = 0.96\linewidth]{./figures/figure_S19b.png}}
    \subfloat[]{\label{supplemental:figure_co2_capture_sensitivity_analysis_local}\includegraphics[width = 0.8\linewidth]{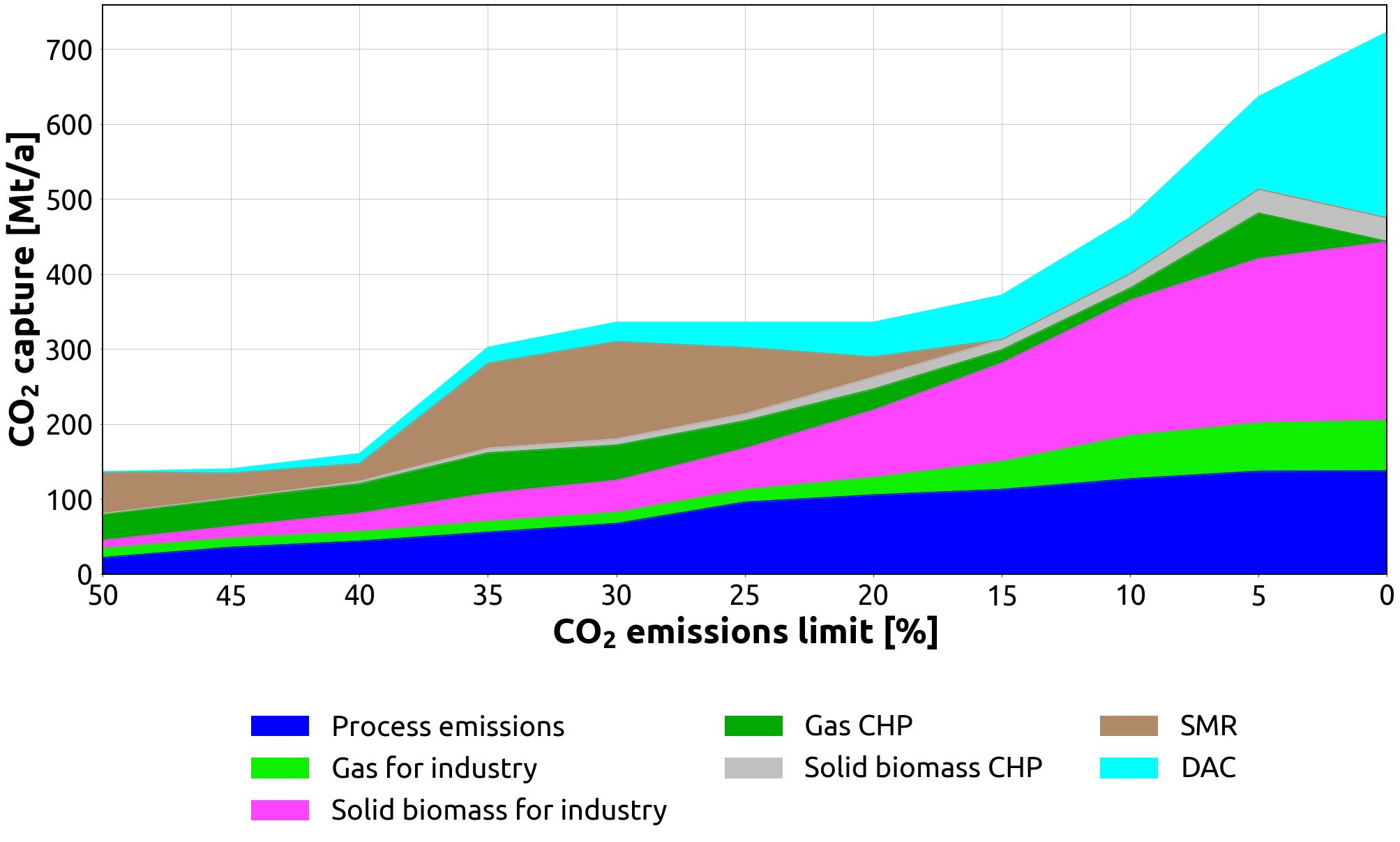}}
    \caption{CO$_2$ capture sensitivity analysis under different CO$_2$ emissions limits in (a) the global scenario and (b) the local scenario. In the local scenario, CO$_2$ captured from DAC and solid biomass used in the industry are adopted for less stringent CO$_2$ targets in the CO$_2$ reduction path compared to the global scenario. In a fully decarbonised Europe, these two technologies combined account for 60\% and 67\% of the total CO$_2$ captured in the global and local scenarios, respectively.}
    \label{supplemental:figure_co2_capture_sensitivity_analysis}
\end{figure}

\begin{figure}[!htb]
    \centering
    %\subfloat[]{\label{supplemental:figure_co2_capture_sensitivity_analysis_global_without_smr}\includegraphics[width = 0.96\linewidth]{./figures/figure_S20a_without_SMR.png}}
    \subfloat[]{\label{supplemental:figure_co2_capture_sensitivity_analysis_global_without_smr}\includegraphics[width = 0.8\linewidth]{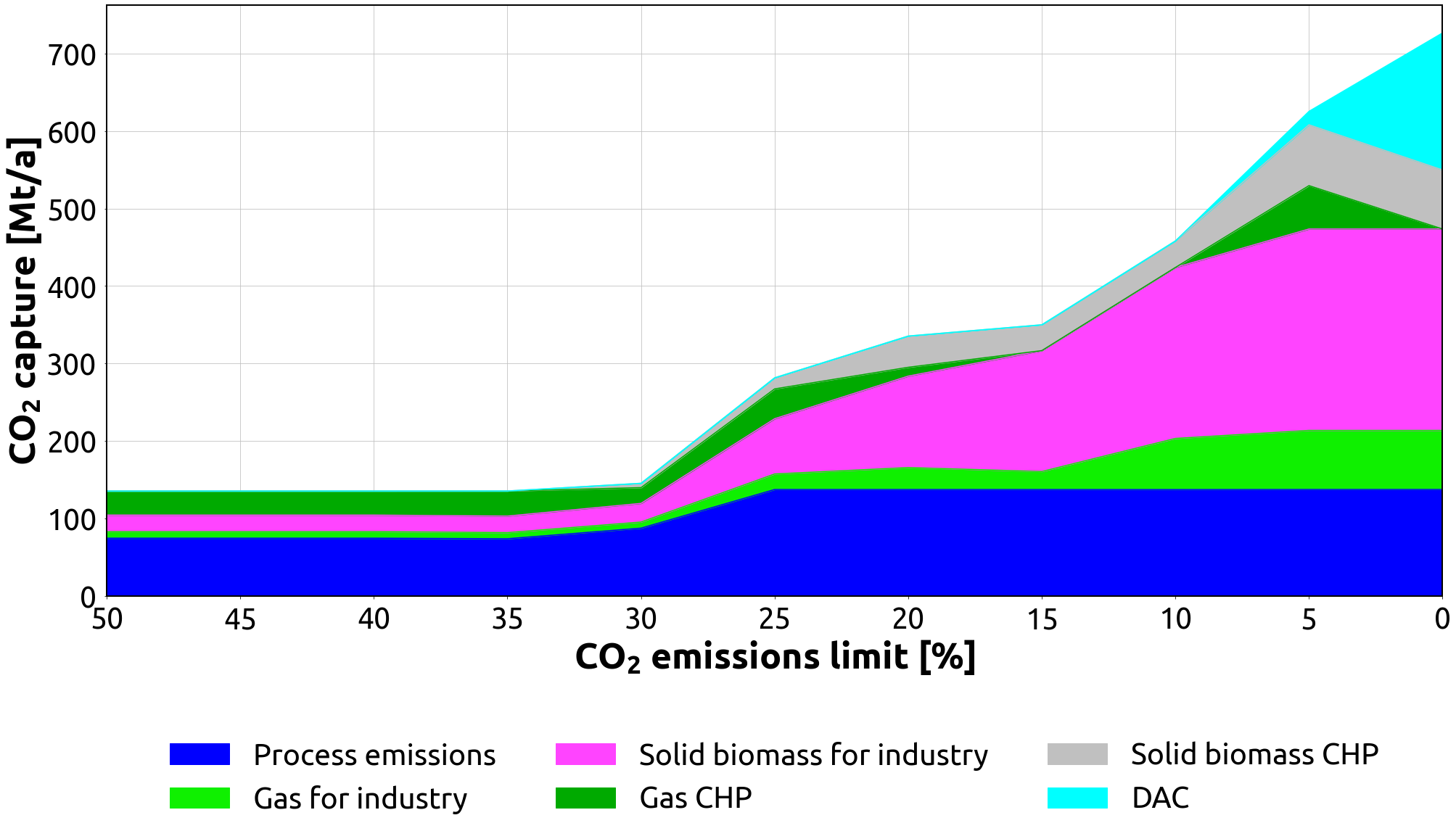}}
    \vspace{20pt}
    %\subfloat[]{\label{supplemental:figure_co2_capture_sensitivity_analysis_local_without_smr}\includegraphics[width = 0.96\linewidth]{./figures/figure_S20b_without_SMR.png}}
    \subfloat[]{\label{supplemental:figure_co2_capture_sensitivity_analysis_local_without_smr}\includegraphics[width = 0.8\linewidth]{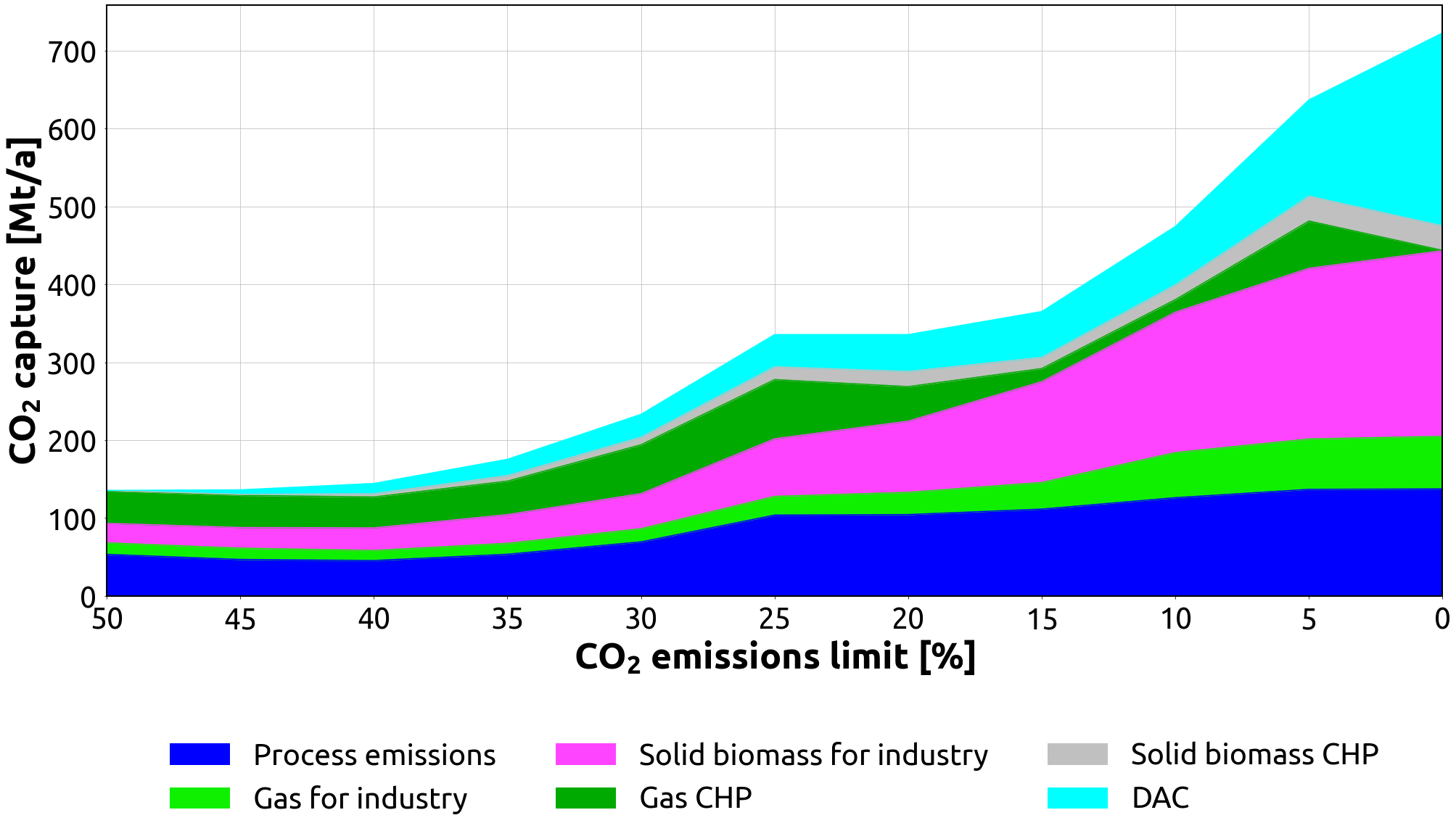}}
    \caption{CO$_2$ capture sensitivity analysis under different CO$_2$ emissions limits without SMR in (a) the global scenario and (b) the local scenario. In both scenarios, a substantial amount of additional CO$_2$ is captured from process emissions and gas-based CHP units to compensate for the CO$_2$ not captured from SMR.}
    \label{supplemental:figure_co2_capture_sensitivity_analysis_without_smr}
\end{figure}

Regarding carbon capture, DAC is adopted at a less stringent CO$_2$ emissions limit (45\%) in the local scenario compared to the global scenario (5\%). Along with CO$_2$ captured from solid biomass used in the industry, which is also adopted at a less stringent limit (50\%), they capture the bulk of CO$_2$ emissions, reaching 60\% and 67\% in the global and local net-zero CO$_2$ emissions scenarios, respectively. Although SMR with CC plays a crucial role in both scenarios for low CO$_2$ emissions reduction targets, as well as gas-based CHP units with CC in the local scenario, they both gradually decrease CO$_2$ capture levels until they become irrelevant for climate-neutral scenarios (Figure \ref{supplemental:figure_co2_capture_sensitivity_analysis}). When SMR with CC is deactivated, additional CO$_2$ is captured from process emissions and gas-based CHP units to compensate for the CO$_2$ not captured from the former (Figure \ref{supplemental:figure_co2_capture_sensitivity_analysis_without_smr}).

\begin{figure}[!htb]
    \centering
    %\subfloat[]{\label{supplemental:figure_co2_conversion_sequestration_sensitivity_analysis_global}\includegraphics[width = 0.96\linewidth]{./figures/figure_S20a.png}}
    \subfloat[]{\label{supplemental:figure_co2_conversion_sequestration_sensitivity_analysis_global}\includegraphics[width = 0.8\linewidth]{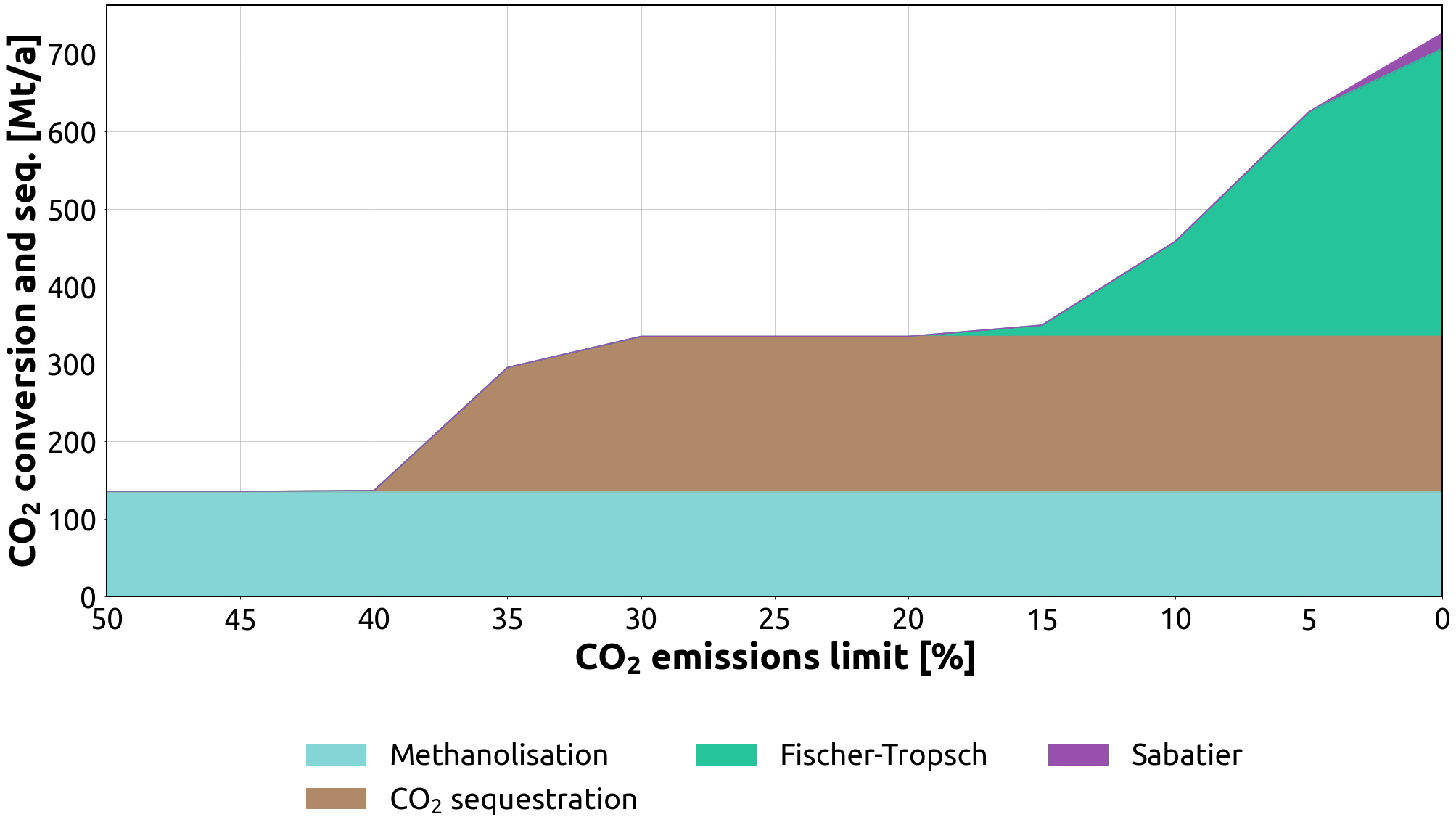}}
    \vspace{20pt}
    %\subfloat[]{\label{supplemental:figure_co2_conversion_sequestration_sensitivity_analysis_local}\includegraphics[width = 0.96\linewidth]{./figures/figure_S20b.png}}
    \subfloat[]{\label{supplemental:figure_co2_conversion_sequestration_sensitivity_analysis_local}\includegraphics[width = 0.8\linewidth]{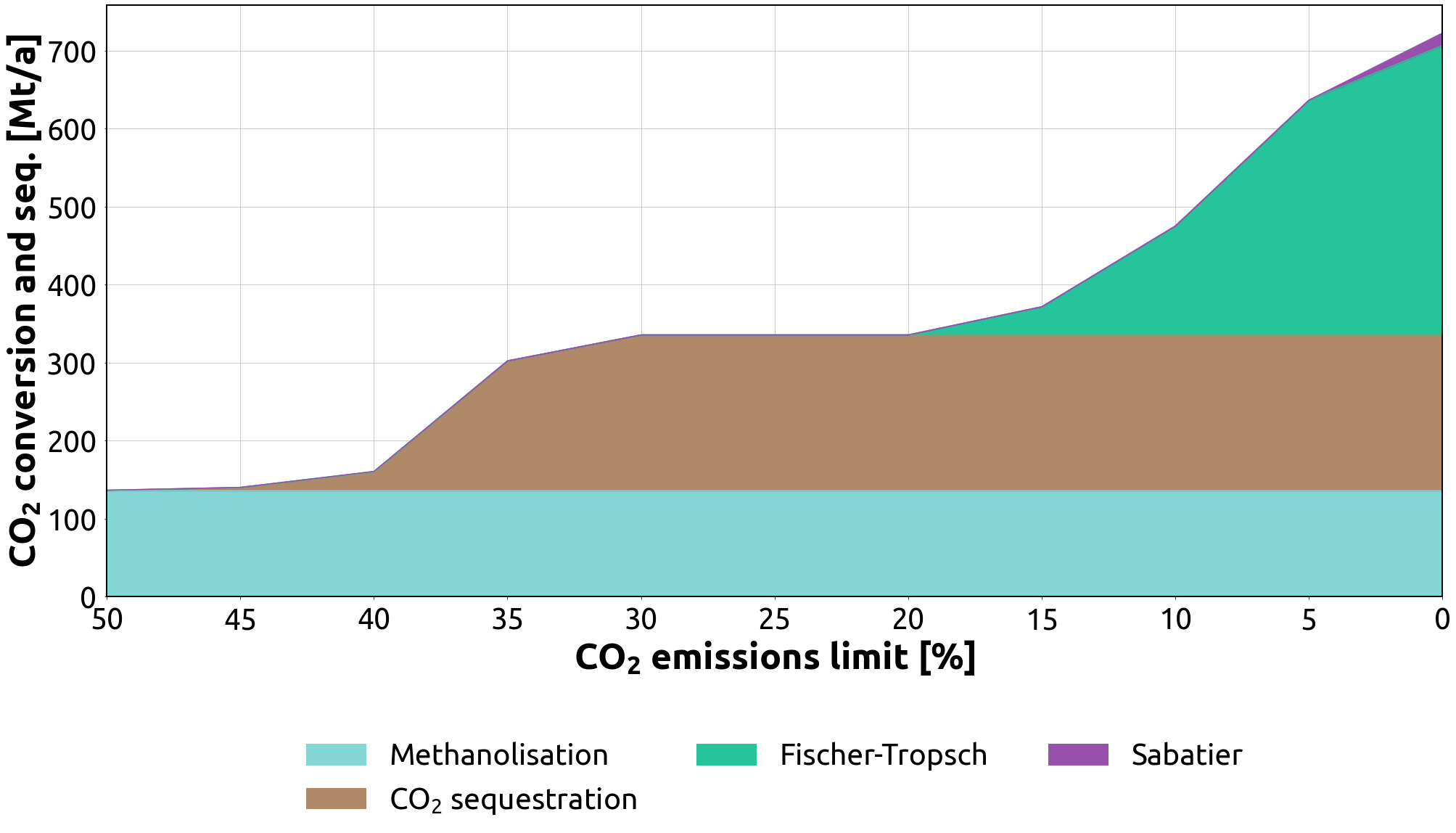}}
    \caption{CO$_2$ conversion and sequestration sensitivity analysis under different CO$_2$ emissions limits in (a) the global scenario and (b) the local scenario. In both scenarios, Fischer-Tropsch and methanolisation play an important role in a fully decarbonised Europe by converting 70\% of all captured CO$_2$ into synthetic oil and methanol. The remaining captured CO$_2$ is sequestered underground to enable negative emissions. As expected, the CO$_2$ conversion and sequestration profile mimics the CO$_2$ capture profile in both scenarios.}
    \label{supplemental:figure_co2_conversion_sequestration_sensitivity_analysis}
\end{figure}

Regarding carbon conversion, regardless of the CO$_2$ target, methanolisation is constantly used in both scenarios since it is the only option to supply the exogenous methanol demand. Sequestering captured CO$_2$ underground in deep salt caverns and depleted hydrocarbon reservoirs starts at 40\% and 45\% of the CO$_2$ emissions limits in global and local scenarios, respectively. This allows for sequestering emissions from various industry processes, as well as enabling negative emissions. Fischer-Tropsch, which is adopted at a more stringent target (20\% CO$_2$ emissions limit in both scenarios), deals with the remaining captured CO$_2$ by converting it into synthetic oil (Figure \ref{supplemental:figure_co2_conversion_sequestration_sensitivity_analysis}).

\section{Limitations of our study}
\begin{figure}[!htb]
    \centering
    %\subfloat[]{\label{supplemental:figure_spatial_co2_capture_global_brownfield}\includegraphics[width = 0.96\linewidth]{./figures/figure_S10a_brownfield.png}}
    \subfloat[]{\label{supplemental:figure_spatial_co2_capture_global_brownfield}\includegraphics[width = 0.66\linewidth]{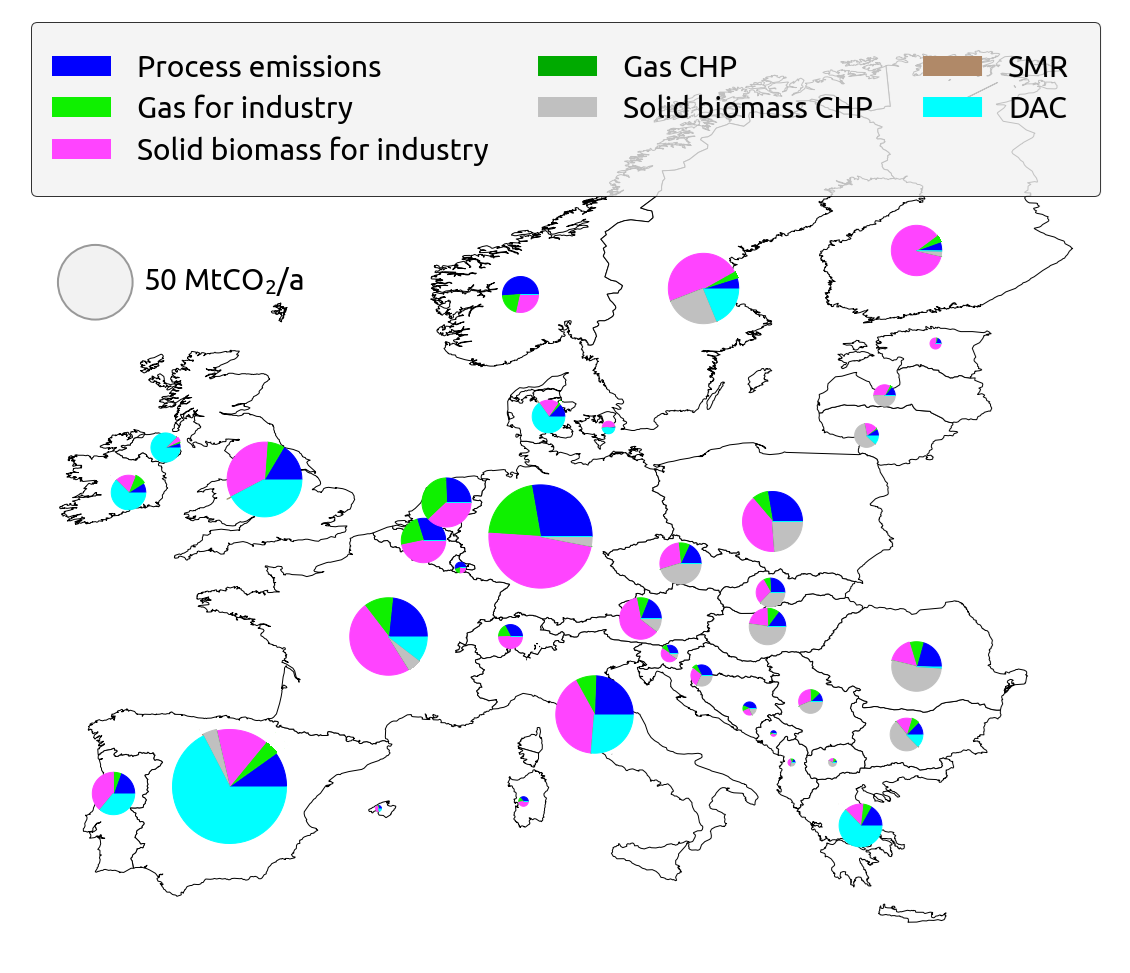}}
    \vspace{20pt}
    %\subfloat[]{\label{supplemental:figure_spatial_co2_capture_variation_brownfield}\includegraphics[width = 0.96\linewidth]{./figures/figure_S10b_brownfield.png}}
    \subfloat[]{\label{supplemental:figure_spatial_co2_capture_variation_brownfield}\includegraphics[width = 0.8\linewidth]{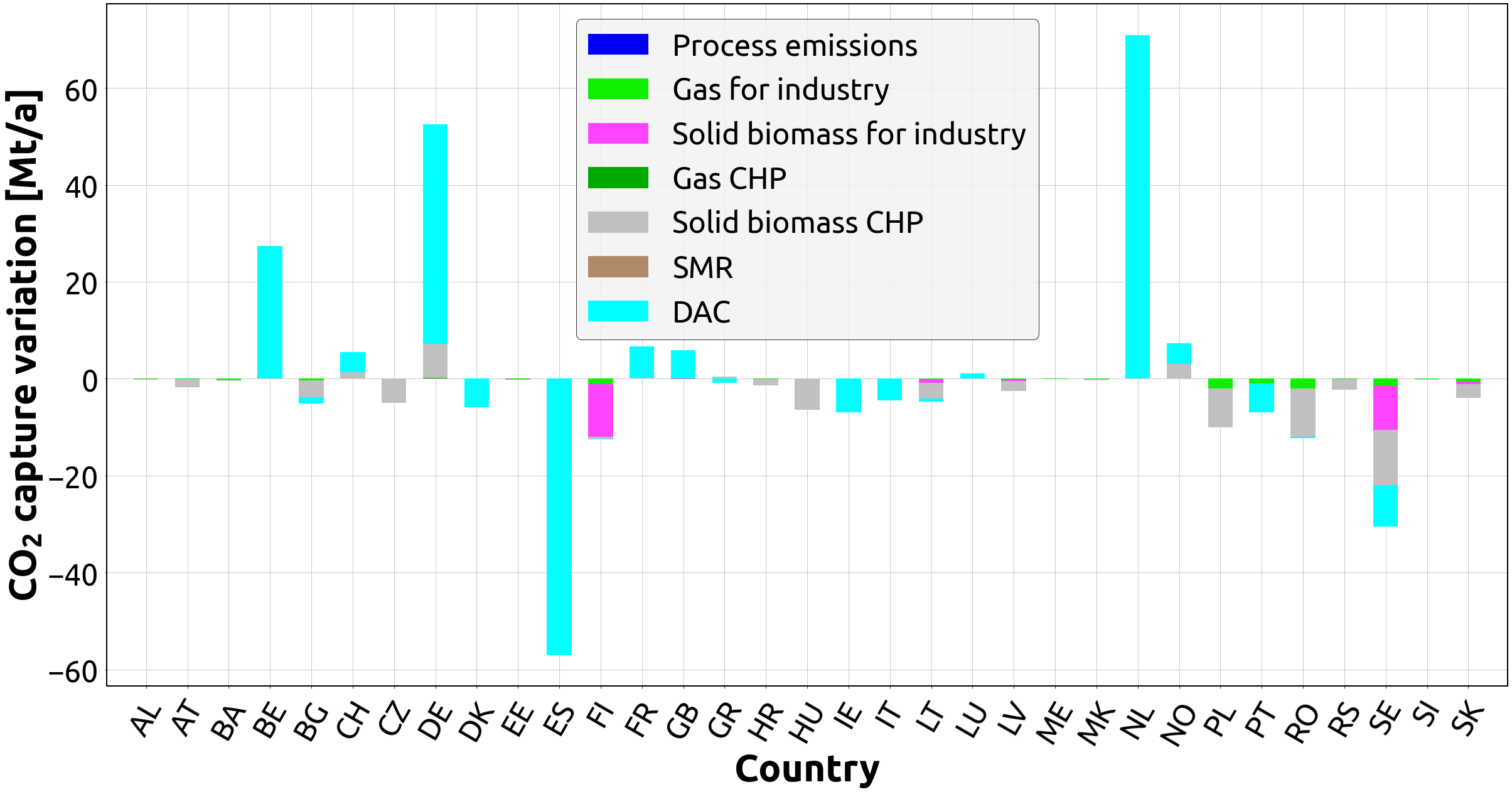}}
    \caption{(a) CO$_2$ capture in the global net-zero CO$_2$ emissions scenario based on a brownfield optimisation and (b) CO$_2$ capture variation per country and technology between the global net-zero CO$_2$ emissions scenario and the local net-zero CO$_2$ emissions scenario based on a brownfield optimisation.}
    \label{supplemental:figure_spatial_co2_capture_and_variation_brownfield}
\end{figure}

\begin{figure}[!htb]
    \centering
    %\subfloat[]{\label{supplemental:figure_spatial_co2_conversion_global_brownfield}\includegraphics[width = 0.96\linewidth]{./figures/figure_S11a_brownfield.png}}
    \subfloat[]{\label{supplemental:figure_spatial_co2_conversion_global_brownfield}\includegraphics[width = 0.66\linewidth]{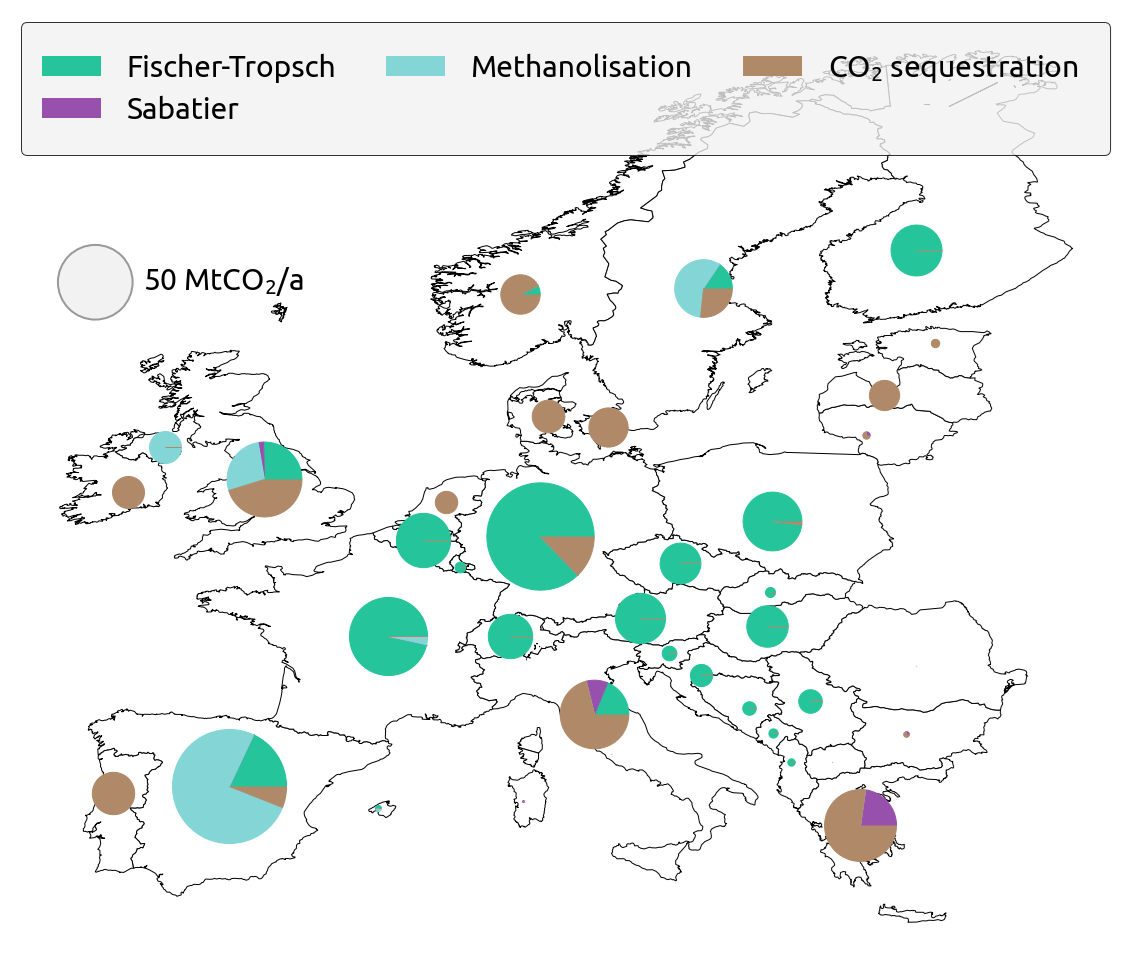}}
    \vspace{20pt}
    %\subfloat[]{\label{supplemental:figure_spatial_co2_conversion_variation_brownfield}\includegraphics[width = 0.96\linewidth]{./figures/figure_S11b_brownfield.png}}
    \subfloat[]{\label{supplemental:figure_spatial_co2_conversion_variation_brownfield}\includegraphics[width = 0.8\linewidth]{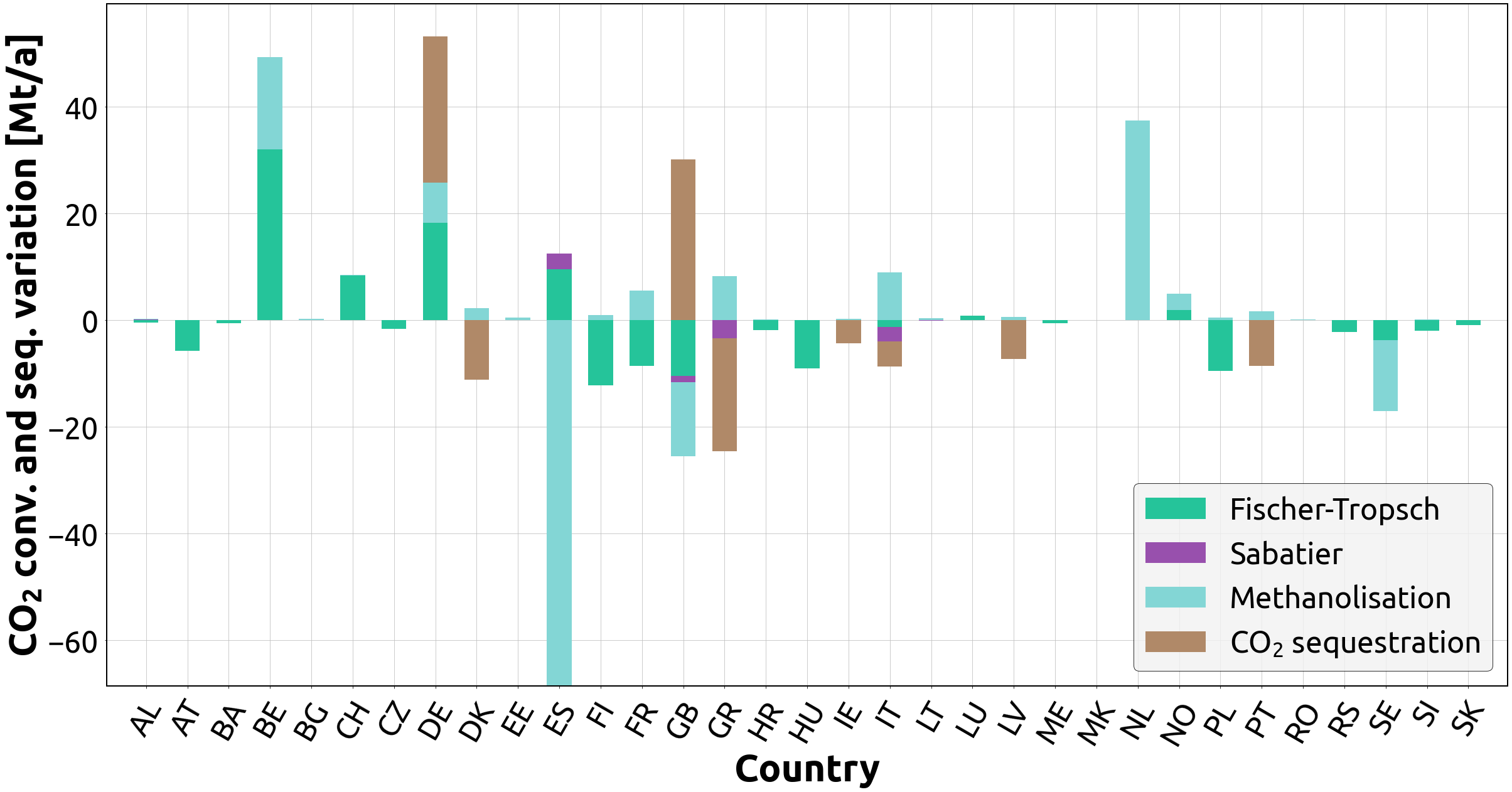}}
    \caption{(a) CO$_2$ conversion and sequestration in the global net-zero CO$_2$ emissions scenario based on a brownfield optimisation and (b) CO$_2$ conversion and sequestration variation per country and technology between the global net-zero CO$_2$ emissions scenario and the local net-zero CO$_2$ emissions scenario based on a brownfield optimisation.}
    \label{supplemental:figure_spatial_co2_conversion_and_variation_brownfield}
\end{figure}

\begin{figure}[!htb]
    \centering
    %\subfloat[]{\label{supplemental:figure_total_system_brownfield}\includegraphics[width = 0.96\linewidth]{./figures/figure_S17a_brownfield.png}}
    \subfloat[]{\label{supplemental:figure_total_system_brownfield}\includegraphics[width = 0.8\linewidth]{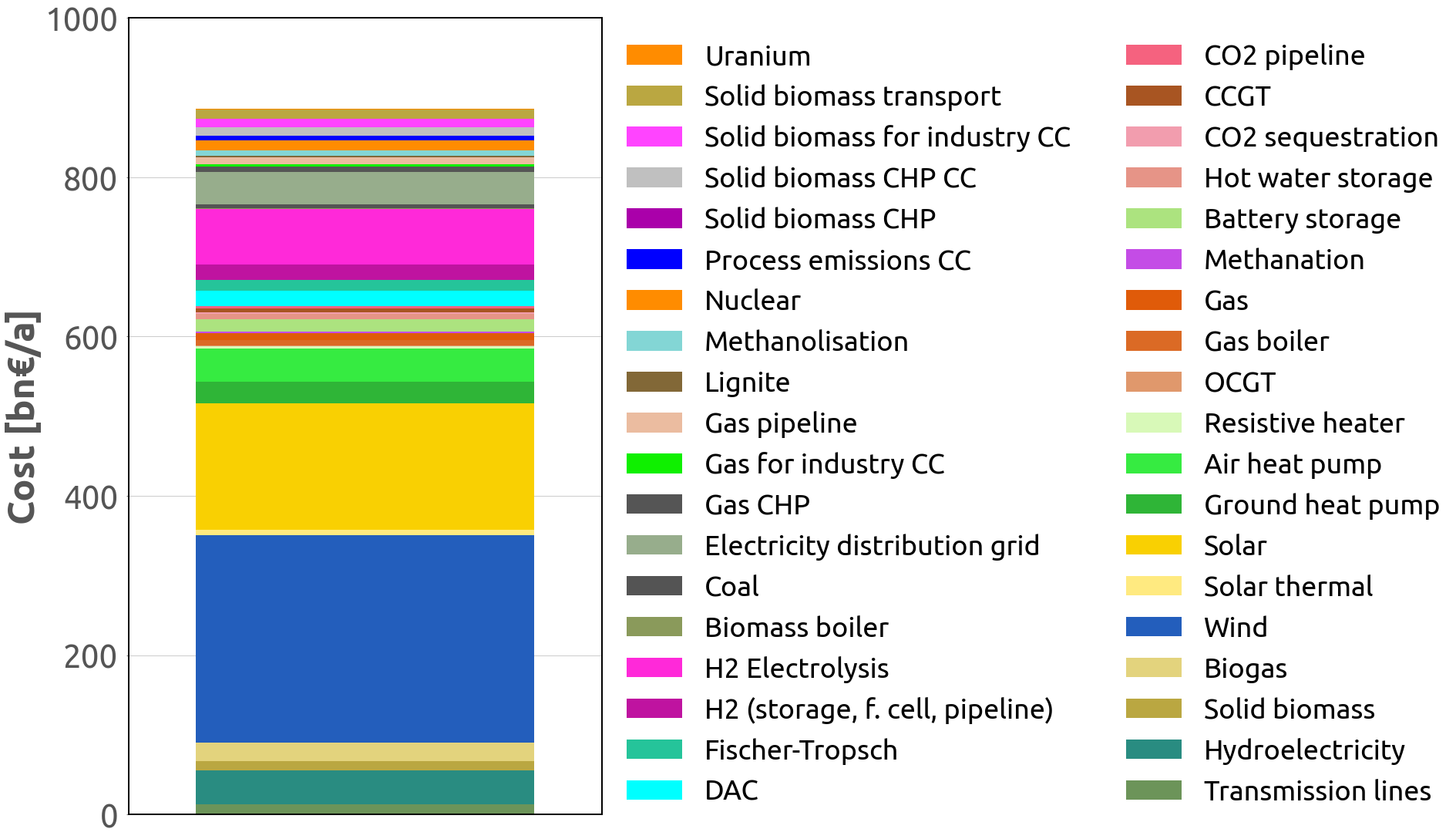}}
    \vspace{20pt}
    %\subfloat[]{\label{supplemental:figure_cost_variations_per_technology_global_brownfield}\includegraphics[width = 0.96\linewidth]{./figures/figure_S17b_brownfield.png}}
    \subfloat[]{\label{supplemental:figure_cost_variations_per_technology_global_brownfield}\includegraphics[width = 0.8\linewidth]{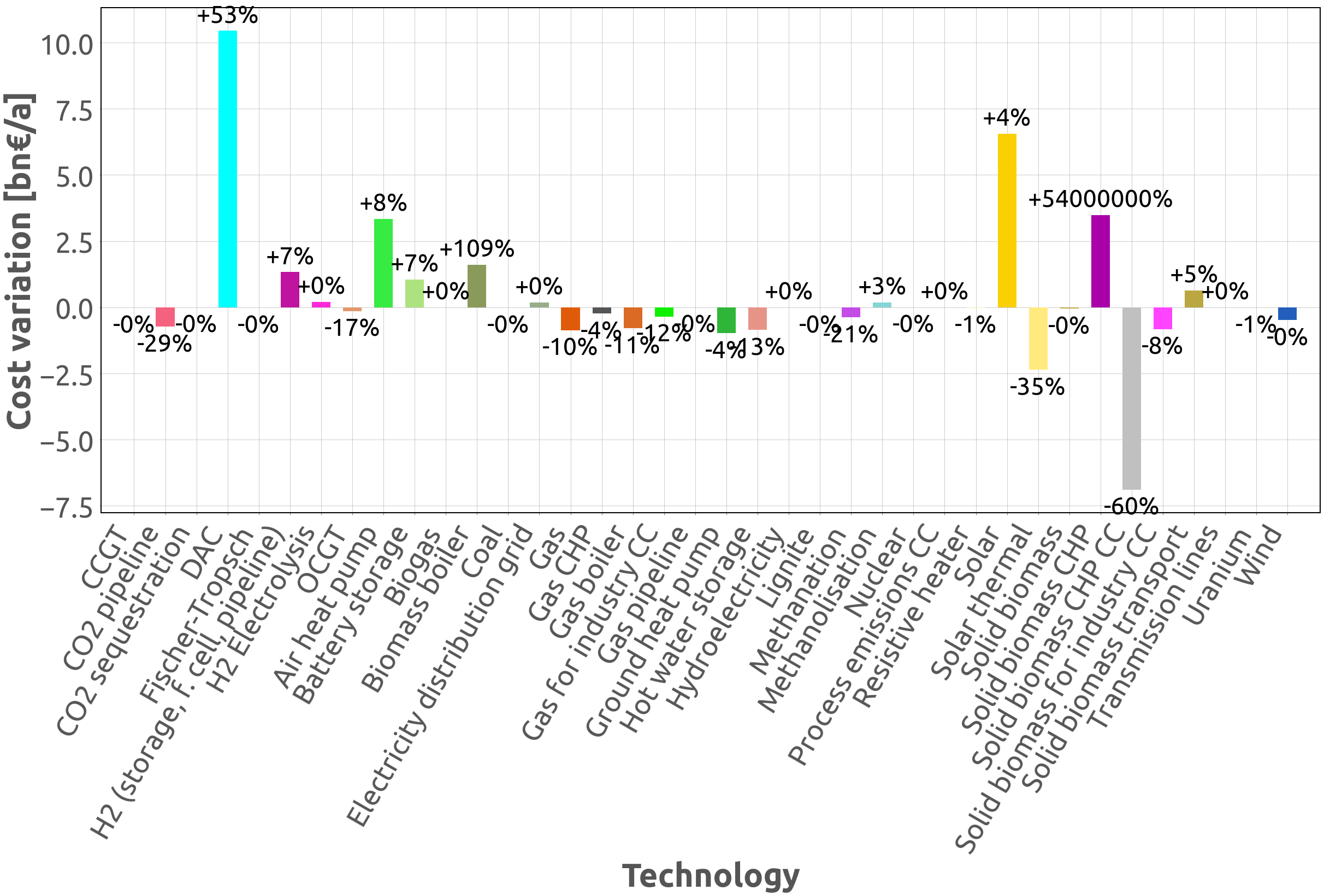}}
    \caption{(a) Total system cost and technology configuration in the global net-zero CO$_2$ emissions scenario based on a brownfield optimisation and (b) Cost variation per technology between the global net-zero CO$_2$ emissions scenario and the local net-zero CO$_2$ emissions scenario based on a brownfield optimisation.} \label{supplemental:figure_total_system_cost_and_variation_global_brownfield}
\end{figure}

In this section, we briefly discuss the limitations of our analysis. First, the model presented in this study is based on an overnight greenfield optimisation. Since we do not model the transition path, we might miss some of the technological lock-ins or path-dependent effects during the system transformation. Our default greenfield optimisation includes existing transmission network and hydropower capacities, but it does not include today's fossil-based capacities which are expected to remain in operation in the year 2050. To check that the impact of this assumption is limited, we ran a sensitivity analysis based on a brownfield optimisation in which existing nuclear, oil, combined cycle gas turbine (CCGT), coal, lignite, geothermal, and biomass generators (whose lifetime extends beyond 2050) are included exogenously. The annual total system cost increased by approximately 1\%, accounting for 886 billion € in the global scenario and 899 billion € in the local scenario (Figure \ref{supplemental:figure_total_system_cost_and_variation_global_brownfield}), mainly due to the cost of existing nuclear and CCGT capacities. The technology portfolio to generate electricity and capture, convert, and sequester CO$_2$ remains identical to the greenfield optimisation (compare Figure \ref{supplemental:figure_spatial_co2_capture_and_variation_brownfield} versus \ref{figure_spatial_co2_capture_and_variation} and \ref{supplemental:figure_spatial_co2_conversion_and_variation_brownfield} versus \ref{figure_spatial_co2_conversion_and_variation}), including the total amount of captured and sequestered CO$_2$.

\begin{figure}[!htb]
    \centering
    %\subfloat[]{\label{supplemental:figure_spatial_co2_capture_global_costs2050}\includegraphics[width = 0.96\linewidth]{./figures/figure_S8a_costs2050.png}}
    \subfloat[]{\label{supplemental:figure_spatial_co2_capture_global_costs2050}\includegraphics[width = 0.66\linewidth]{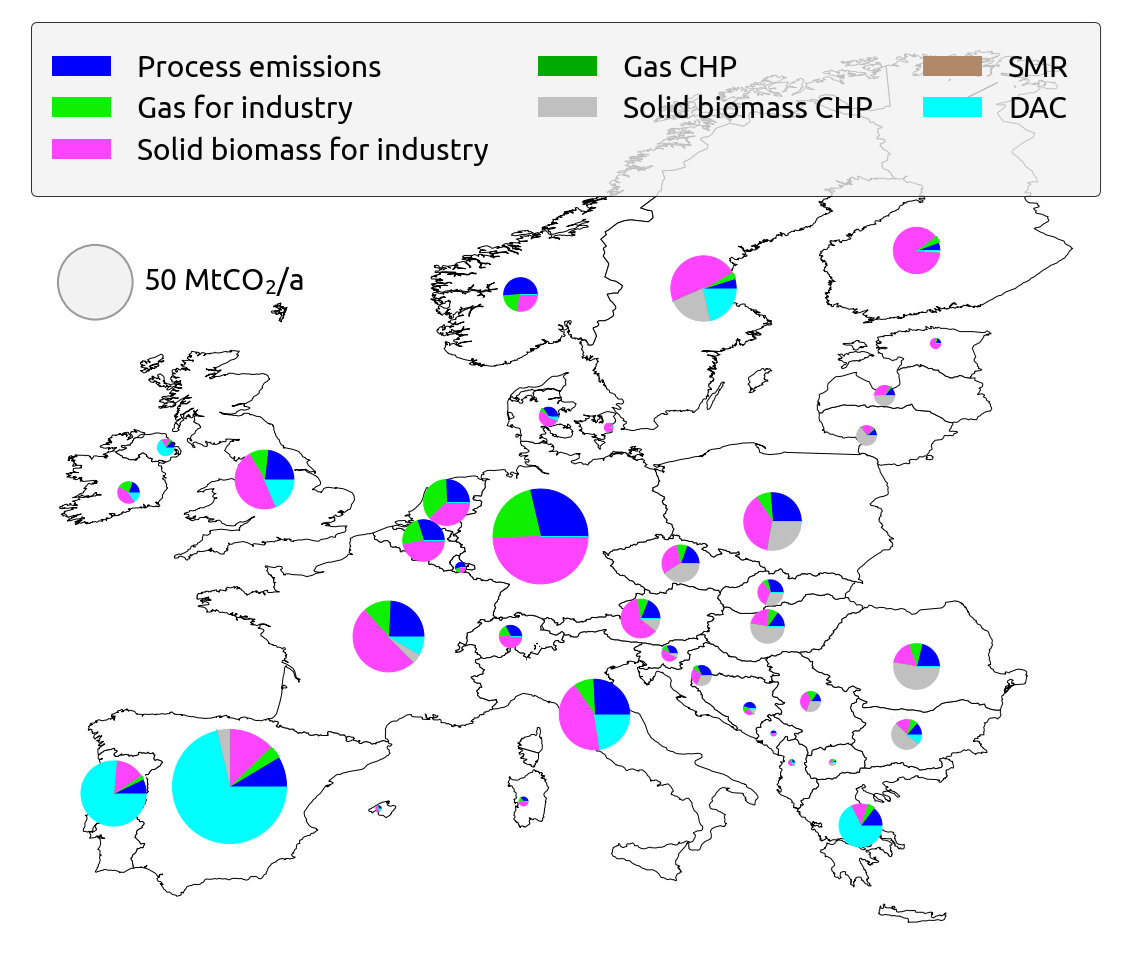}}
    \vspace{20pt}
    %\subfloat[]{\label{supplemental:figure_spatial_co2_capture_variation_costs2050}\includegraphics[width = 0.96\linewidth]{./figures/figure_S8b_costs2050.png}}
    \subfloat[]{\label{supplemental:figure_spatial_co2_capture_variation_costs2050}\includegraphics[width = 0.8\linewidth]{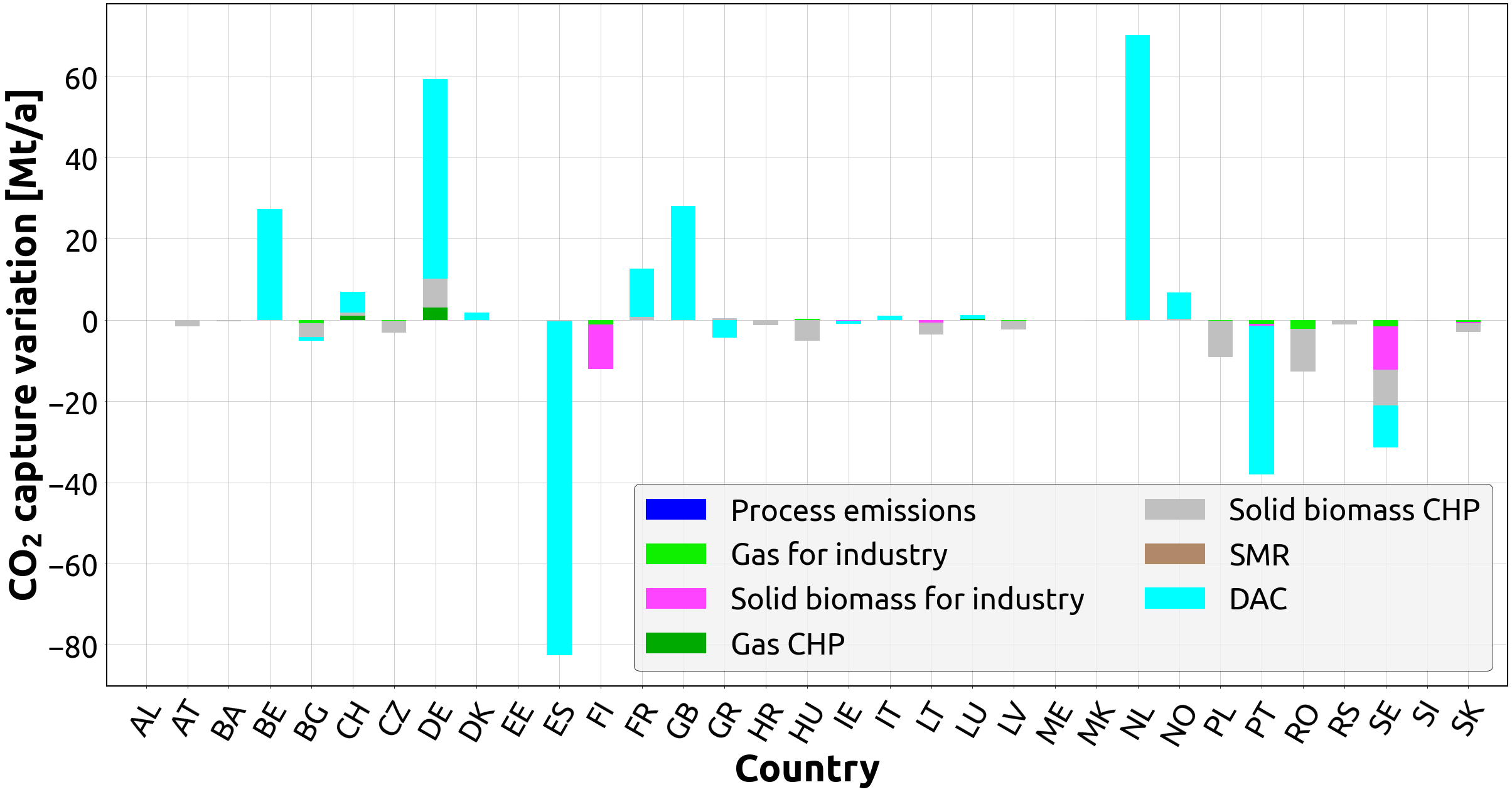}}
    \caption{(a) CO$_2$ capture in the global net-zero CO$_2$ emissions scenario based on costs for the year 2050 and (b) CO$_2$ capture variation per country and technology between the global net-zero CO$_2$ emissions scenario and the local net-zero CO$_2$ emissions scenario based on costs for the year 2050.}
    \label{supplemental:figure_spatial_co2_capture_and_variation_costs2050}
\end{figure}

\begin{figure}[!htb]
    \centering
    %\subfloat[]{\label{supplemental:figure_spatial_co2_conversion_global_costs2050}\includegraphics[width = 0.96\linewidth]{./figures/figure_S9a_costs2050.png}}
    \subfloat[]{\label{supplemental:figure_spatial_co2_conversion_global_costs2050}\includegraphics[width = 0.66\linewidth]{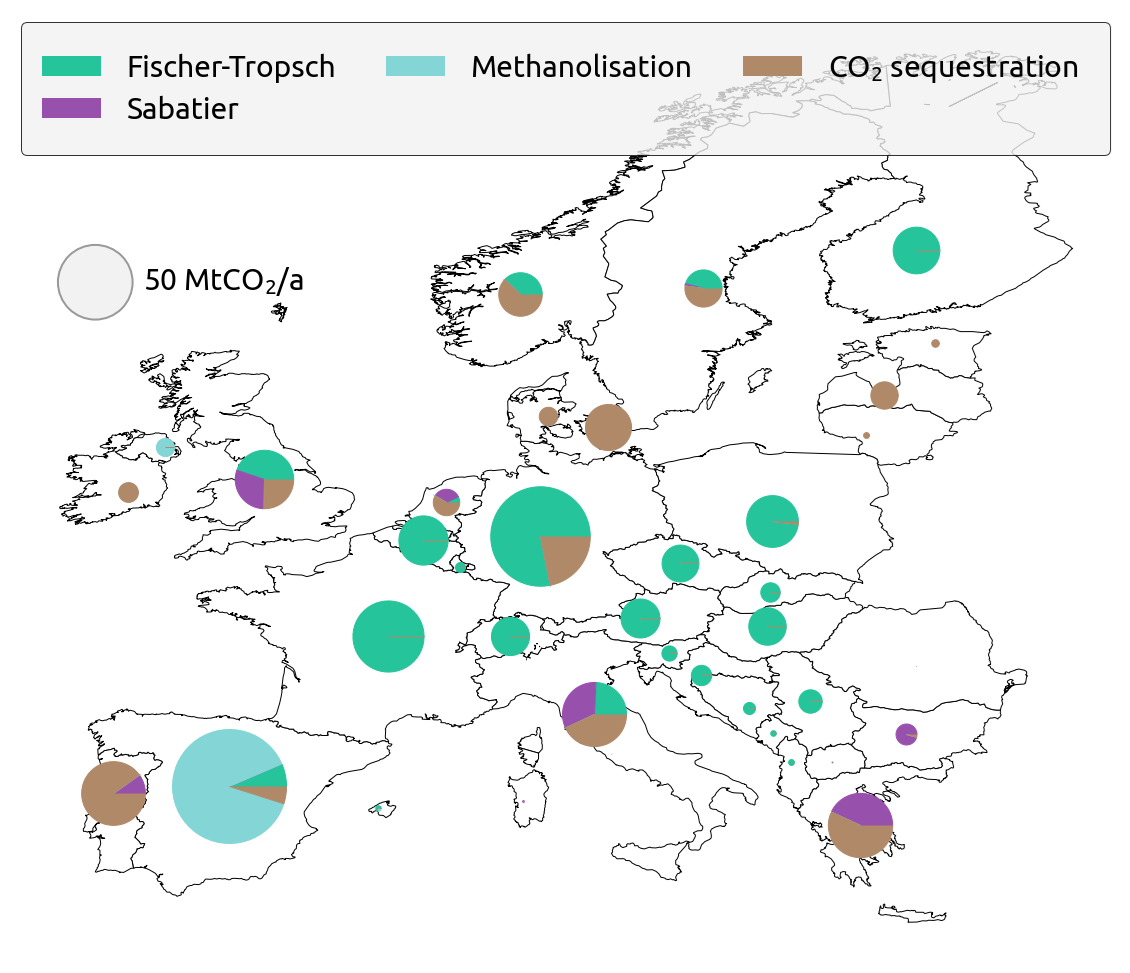}}
    \vspace{20pt}
    %\subfloat[]{\label{supplemental:figure_spatial_co2_conversion_variation_costs2050}\includegraphics[width = 0.96\linewidth]{./figures/figure_S9b_costs2050.png}}
    \subfloat[]{\label{supplemental:figure_spatial_co2_conversion_variation_costs2050}\includegraphics[width = 0.8\linewidth]{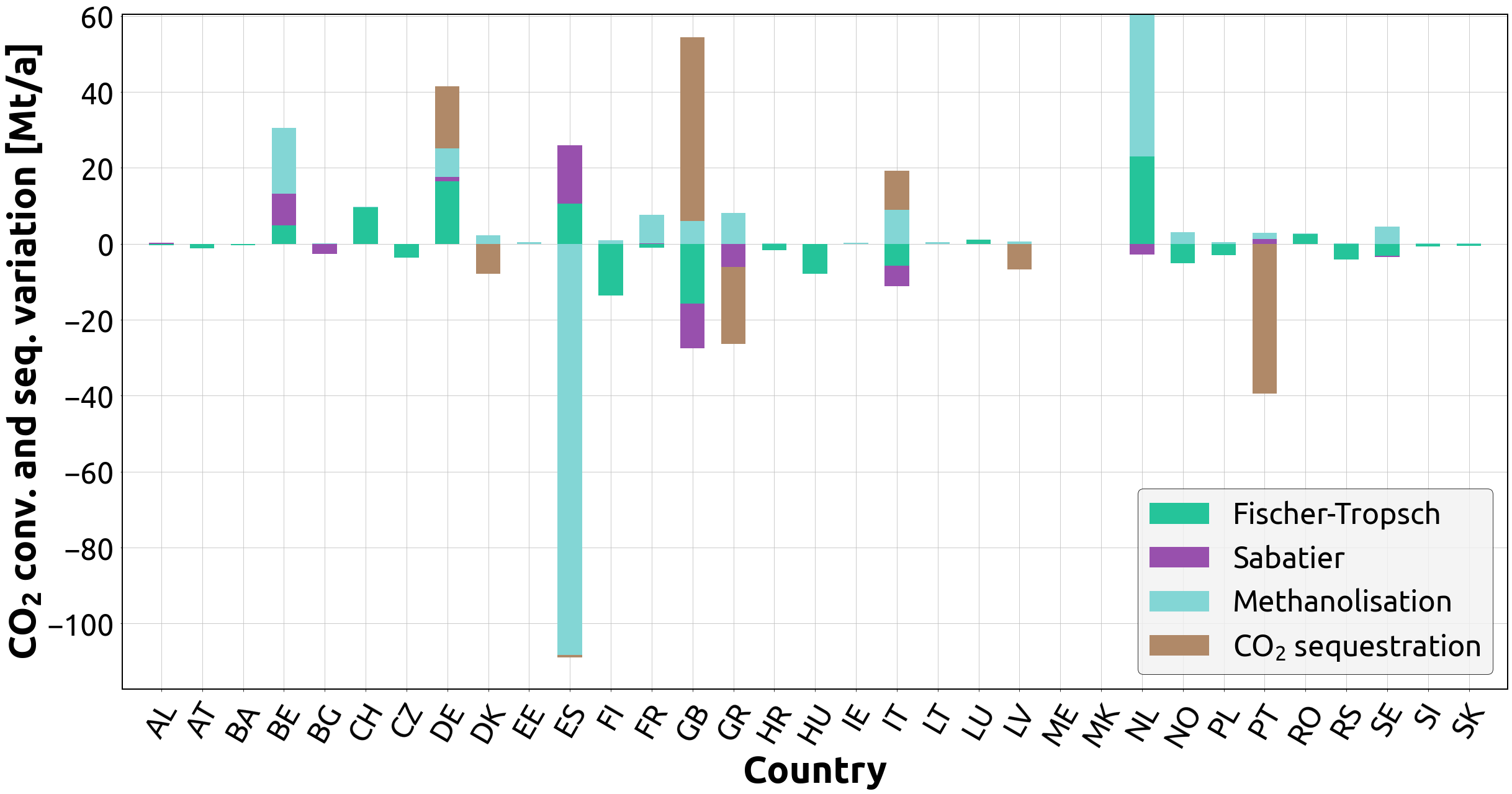}}
    \caption{(a) CO$_2$ conversion and sequestration in the global net-zero CO$_2$ emissions scenario based on costs for the year 2050 and (b) CO$_2$ conversion and sequestration variation per country and technology between the global net-zero CO$_2$ emissions scenario and the local net-zero CO$_2$ emissions scenario based on costs for the year 2050.}
    \label{supplemental:figure_spatial_co2_conversion_and_variation_costs2050}
\end{figure}

\begin{figure}[!htb]
    \centering
    %\subfloat[]{\label{supplemental:figure_total_system_costs2050}\includegraphics[width = 0.96\linewidth]{./figures/figure_S16a_costs2050.png}}
    \subfloat[]{\label{supplemental:figure_total_system_costs2050}\includegraphics[width = 0.8\linewidth]{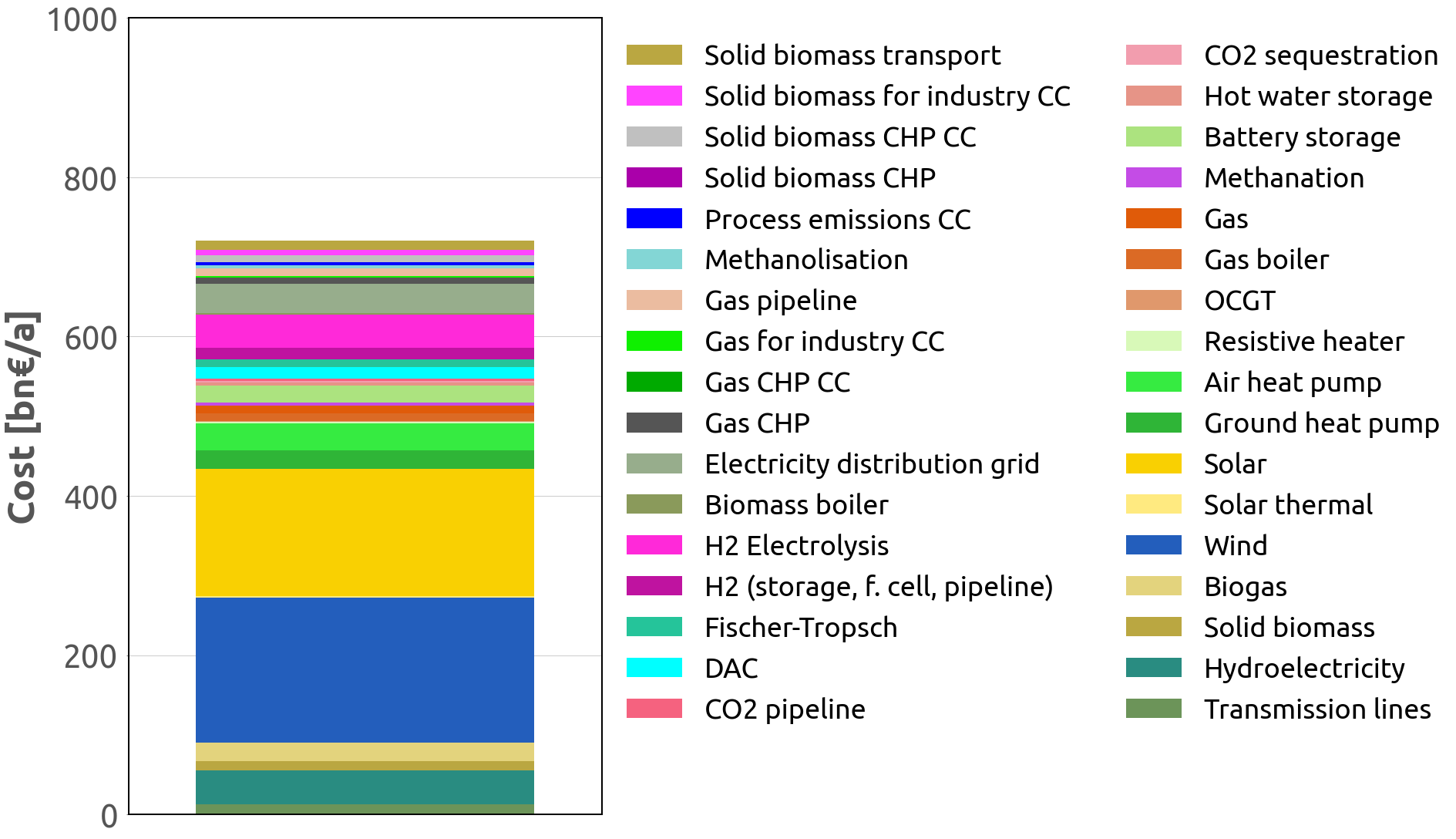}}
    \vspace{20pt}
    %\subfloat[]{\label{supplemental:figure_cost_variations_per_technology_global_costs2050}\includegraphics[width = 0.96\linewidth]{./figures/figure_S16b_costs2050.png}}
    \subfloat[]{\label{supplemental:figure_cost_variations_per_technology_global_costs2050}\includegraphics[width = 0.8\linewidth]{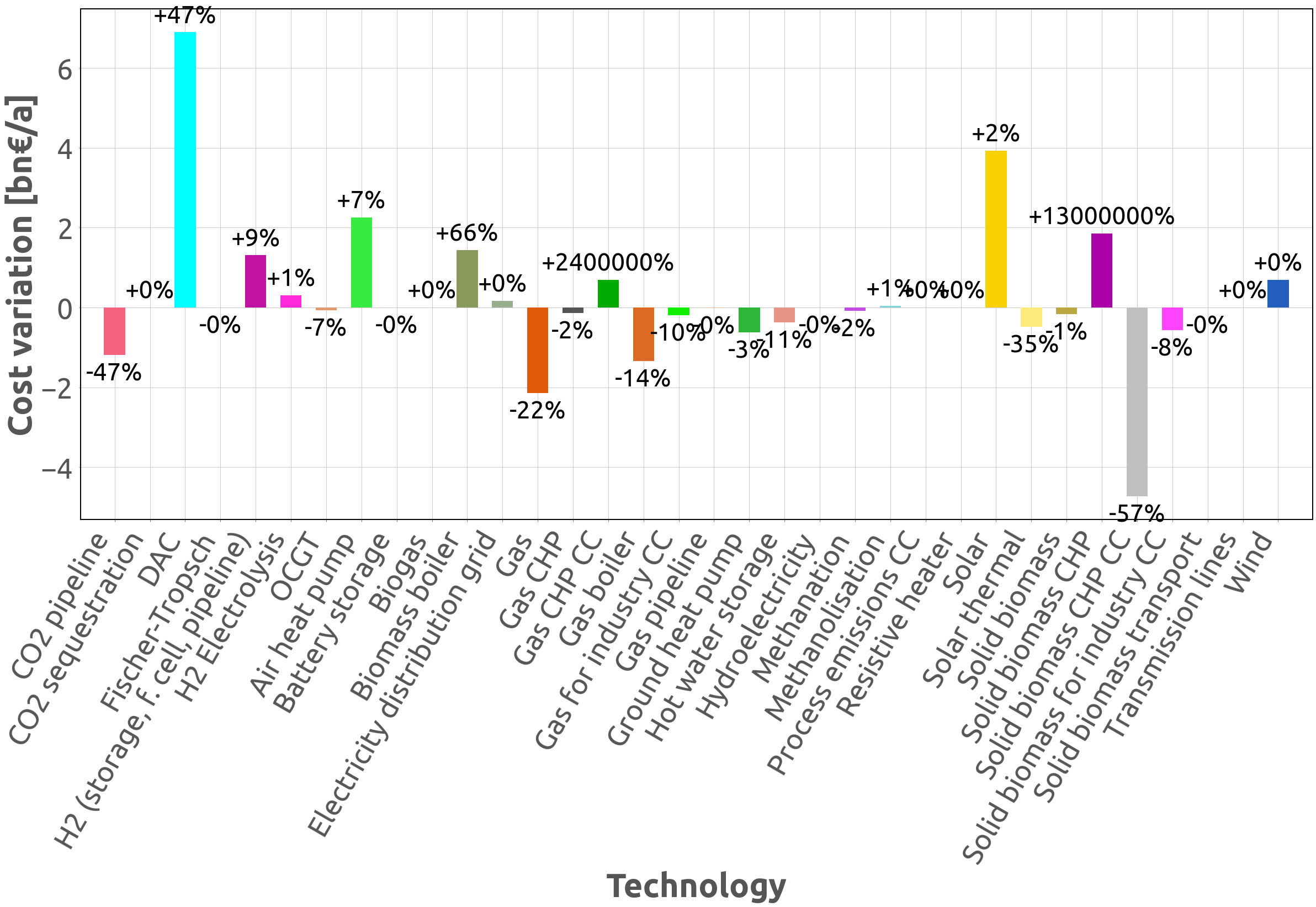}}
    \caption{(a) Total system cost and technology configuration in the global net-zero CO$_2$ emissions scenario based on cost assumptions for the year 2050 and (b) Cost variation per technology between the global net-zero CO$_2$ emissions scenario and the local net-zero CO$_2$ emissions scenario based on cost assumptions for the year 2050.} \label{supplemental:figure_total_system_cost_and_variation_global_costs2050}
\end{figure}

Second, exogenous cost assumptions for different technologies, summarised in Table \ref{supplemental:table_technology_characteristics_assumptions}, might be different when the system reaches carbon neutrality. We do not model technological learning endogenously \cite{Zeyen2023, BEHRENS2024100192} nor consider different costs of capital amongst European countries. Due to the inherent difficulties of projecting a meaningful cost of capital for future years \cite{Bogdanov2019}, we assume it to be uniform across Europe to ensure that the main dynamics determining the optimal location of carbon infrastructure (e.g. access to renewables or the amount of CO$_2$ that needs to be captured in every country) are not masked by future uncertain financial assumptions. To assess the impact of varying cost assumptions, we ran a sensitivity analysis with optimistic cost values, those corresponding to the year 2050 cost projections and included in the Energy System Technology Data version 0.5.0. With optimistic cost assumptions, the annual total system cost decreases by 18\% and amounts to 720 billion € in the global scenario and 728 billion € in the local scenario (Figure \ref{supplemental:figure_total_system_cost_and_variation_global_costs2050}). Wind and H$_2$ electrolysis are the main technologies responsible for this decrease. The amount of captured CO$_2$ increases by 6\%, mostly driven by cheaper DAC. While the spatial layout of technologies capturing CO$_2$ remains unchanged (compare Figure \ref{supplemental:figure_spatial_co2_capture_and_variation_costs2050} versus \ref{figure_spatial_co2_capture_and_variation}), there are minor changes in the layout of technologies converting CO$_2$, and the production of synthetic methane gas through the Sabatier reaction is increased (compare Figure \ref{supplemental:figure_spatial_co2_conversion_and_variation_costs2050} versus \ref{figure_spatial_co2_conversion_and_variation}). This sensitivity analysis proves that our results are robust to different cost assumptions.

\begin{figure}[!htb]
    \centering
    %\subfloat[]{\label{supplemental:figure_spatial_co2_capture_global_weather2010}\includegraphics[width = 0.96\linewidth]{./figures/figure_S12a_weather2010.png}}
    \subfloat[]{\label{supplemental:figure_spatial_co2_capture_global_weather2010}\includegraphics[width = 0.66\linewidth]{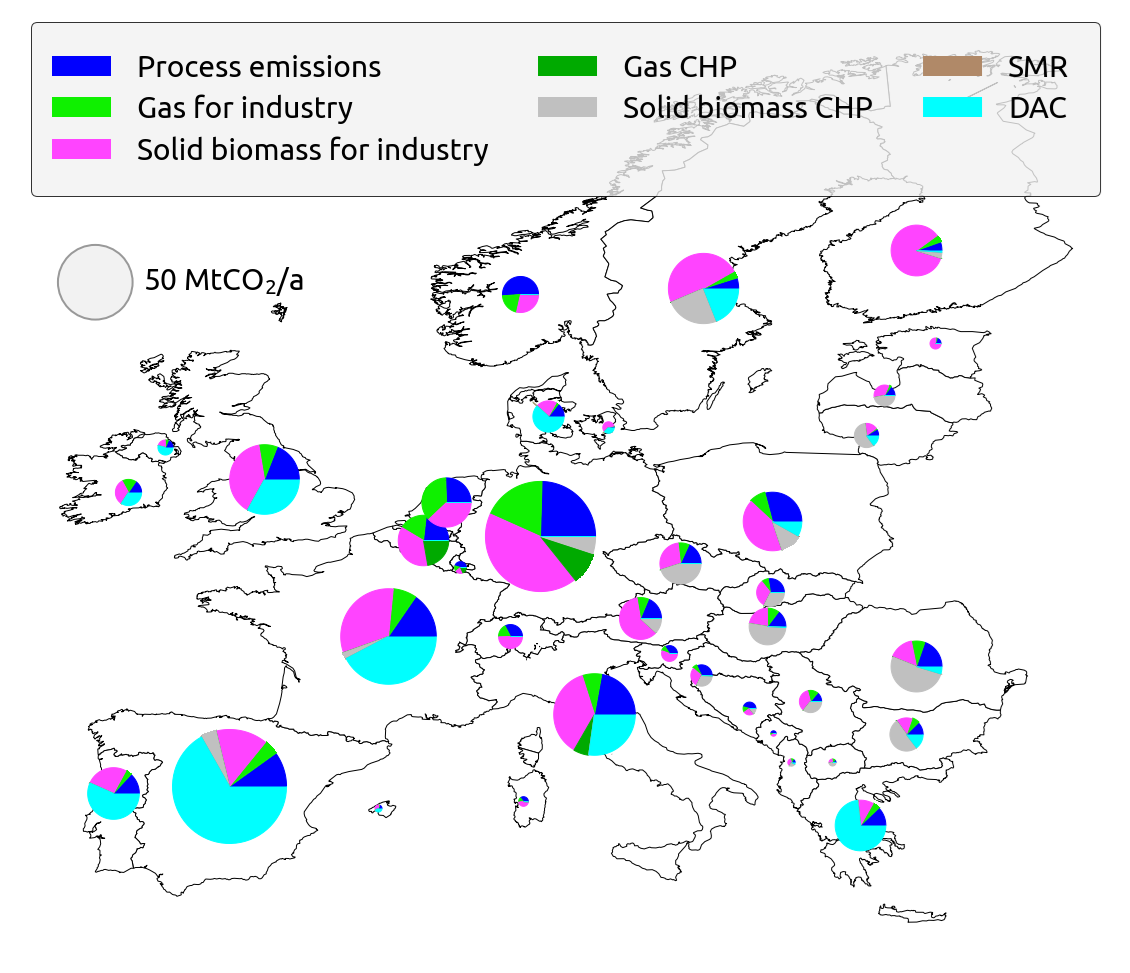}}
    \vspace{20pt}
    %\subfloat[]{\label{supplemental:figure_spatial_co2_capture_variation_weather2010}\includegraphics[width = 0.96\linewidth]{./figures/figure_S12b_weather2010.png}}
    \subfloat[]{\label{supplemental:figure_spatial_co2_capture_variation_weather2010}\includegraphics[width = 0.8\linewidth]{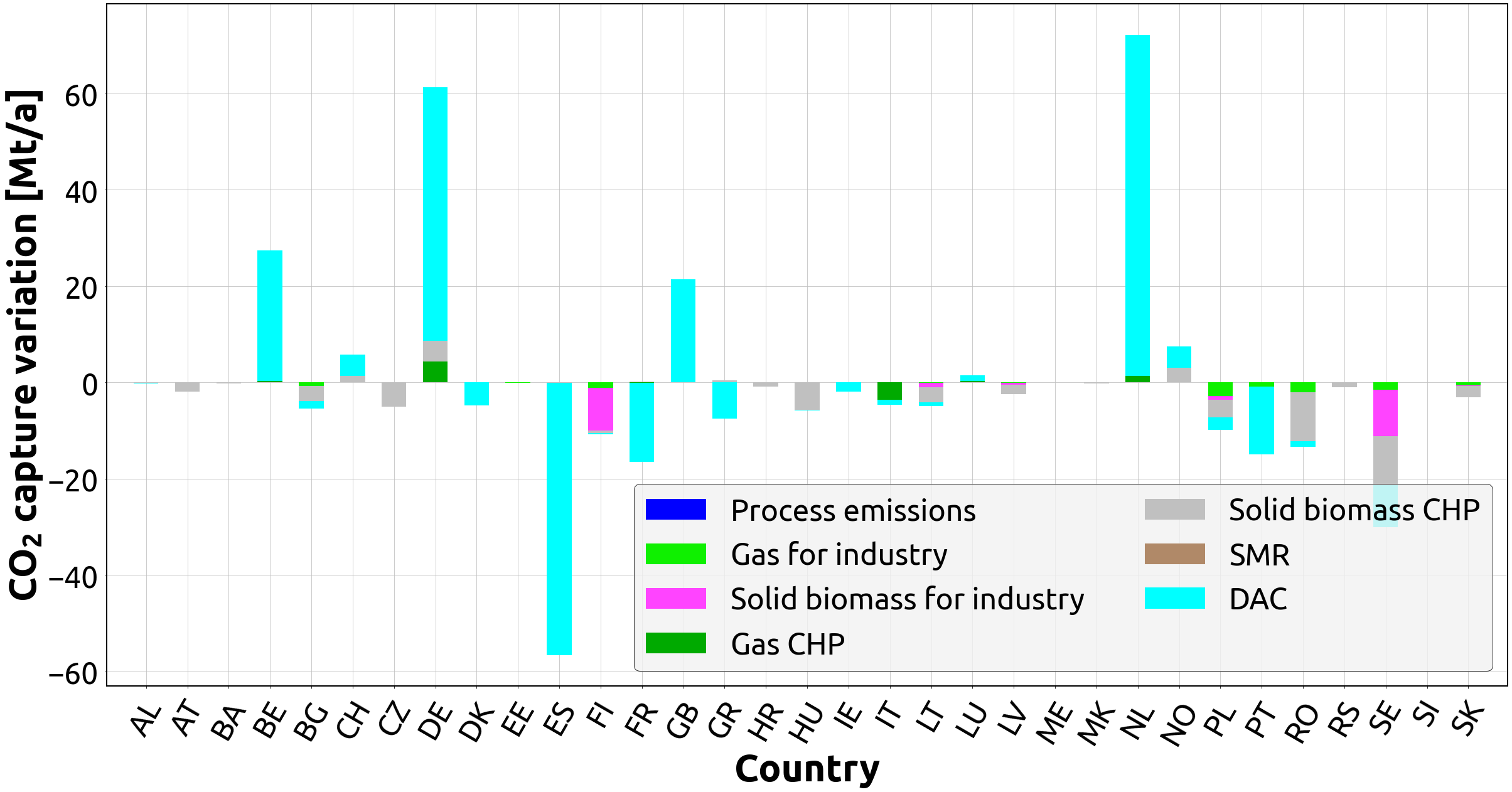}}
    \caption{(a) CO$_2$ capture in the global net-zero CO$_2$ emissions scenario based on the weather year 2010 and (b) CO$_2$ capture variation per country and technology between the global net-zero CO$_2$ emissions scenario and the local net-zero CO$_2$ emissions scenario based on the weather year 2010.}
    \label{supplemental:figure_spatial_co2_capture_and_variation_weather2010}
\end{figure}

\begin{figure}[!htb]
    \centering
    %\subfloat[]{\label{supplemental:figure_spatial_co2_conversion_global_weather2010}\includegraphics[width = 0.96\linewidth]{./figures/figure_S13a_weather2010.png}}
    \subfloat[]{\label{supplemental:figure_spatial_co2_conversion_global_weather2010}\includegraphics[width = 0.66\linewidth]{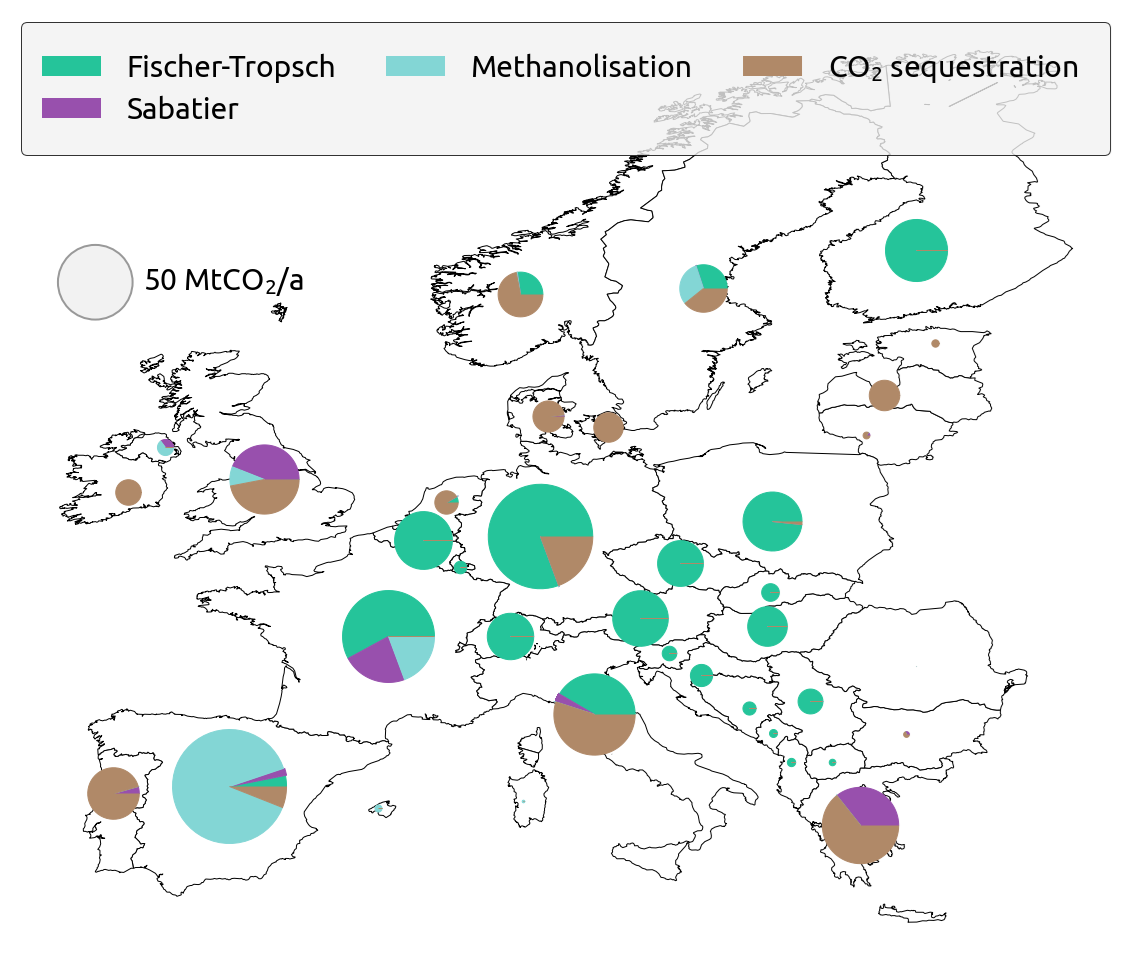}}
    \vspace{20pt}
    %\subfloat[]{\label{supplemental:figure_spatial_co2_conversion_variation_weather2010}\includegraphics[width = 0.96\linewidth]{./figures/figure_S13b_weather2010.png}}
    \subfloat[]{\label{supplemental:figure_spatial_co2_conversion_variation_weather2010}\includegraphics[width = 0.8\linewidth]{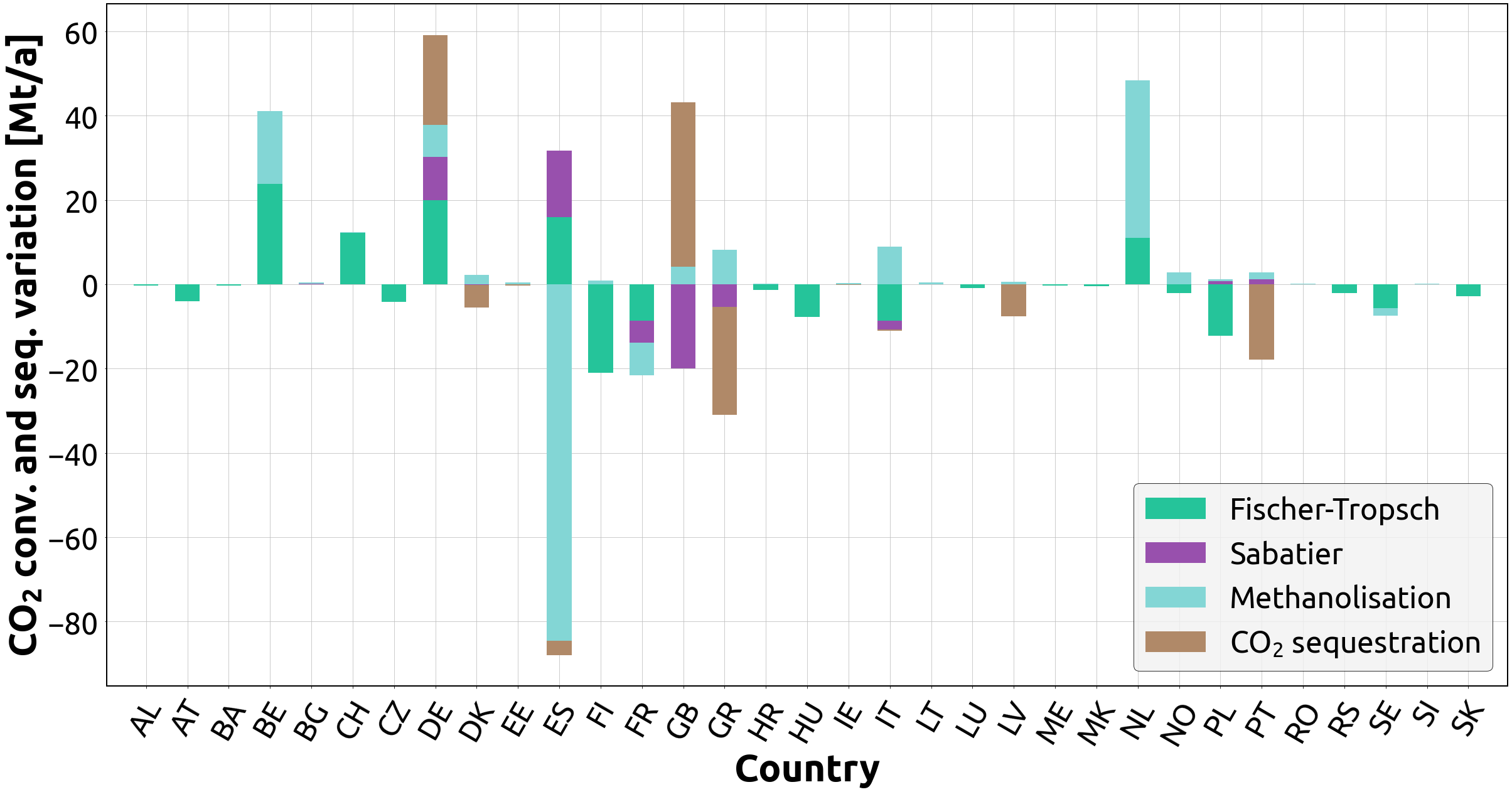}}
    \caption{(a) CO$_2$ conversion and sequestration in the global net-zero CO$_2$ emissions scenario based on the weather year 2010 and (b) CO$_2$ conversion and sequestration variation per country and technology between the global net-zero CO$_2$ emissions scenario and the local net-zero CO$_2$ emissions scenario based on the weather year 2010.}
    \label{supplemental:figure_spatial_co2_conversion_and_variation_weather2010}
\end{figure}

\begin{figure}[!htb]
    \centering
    %\subfloat[]{\label{supplemental:figure_total_system_weather2010}\includegraphics[width = 0.96\linewidth]{./figures/figure_S18a_weather2010.png}}
    \subfloat[]{\label{supplemental:figure_total_system_weather2010}\includegraphics[width = 0.8\linewidth]{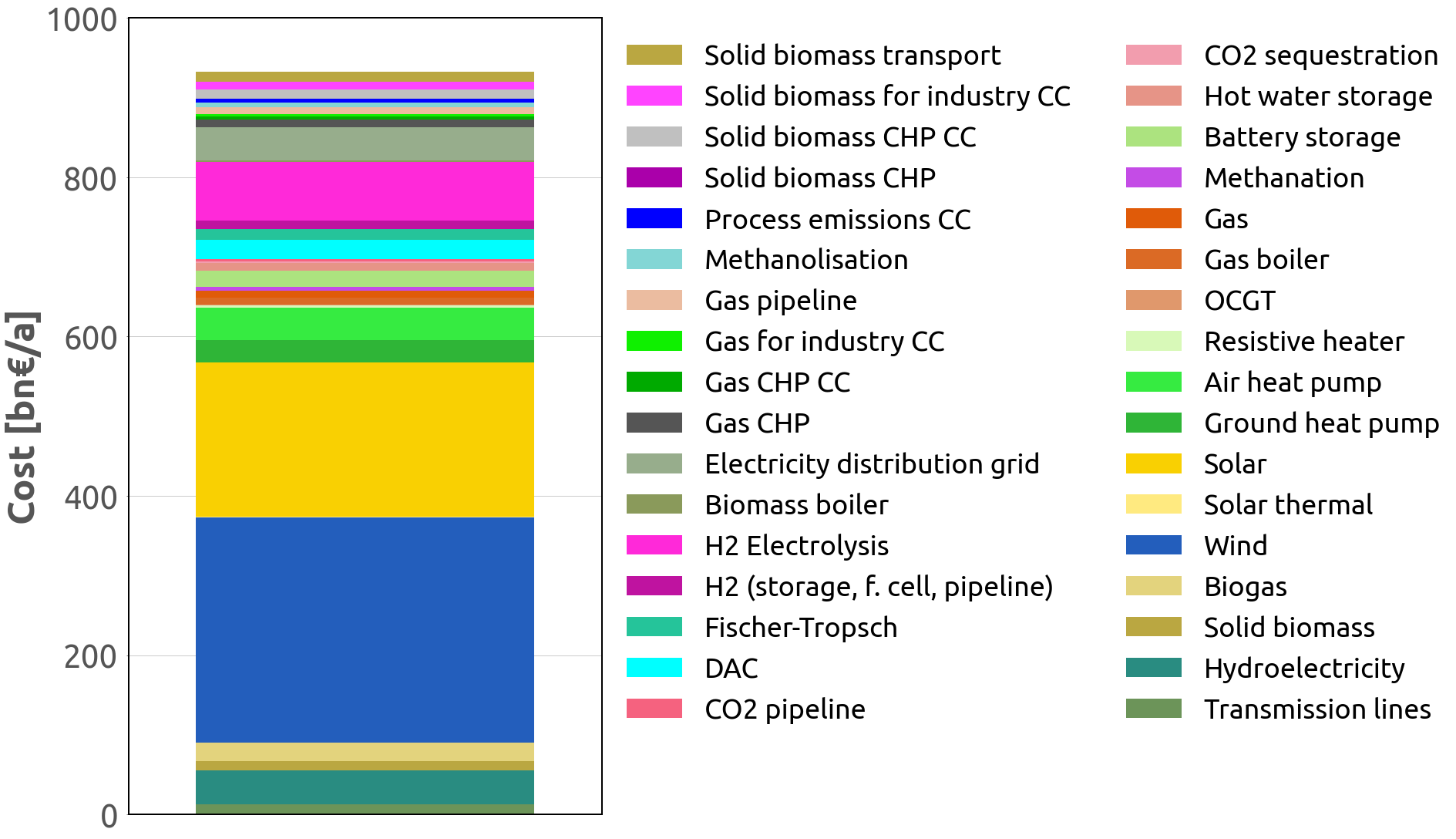}}
    \vspace{20pt}
    %\subfloat[]{\label{supplemental:figure_cost_variations_per_technology_global_weather2010}\includegraphics[width = 0.96\linewidth]{./figures/figure_S18b_weather2010.png}}
    \subfloat[]{\label{supplemental:figure_cost_variations_per_technology_global_weather2010}\includegraphics[width = 0.8\linewidth]{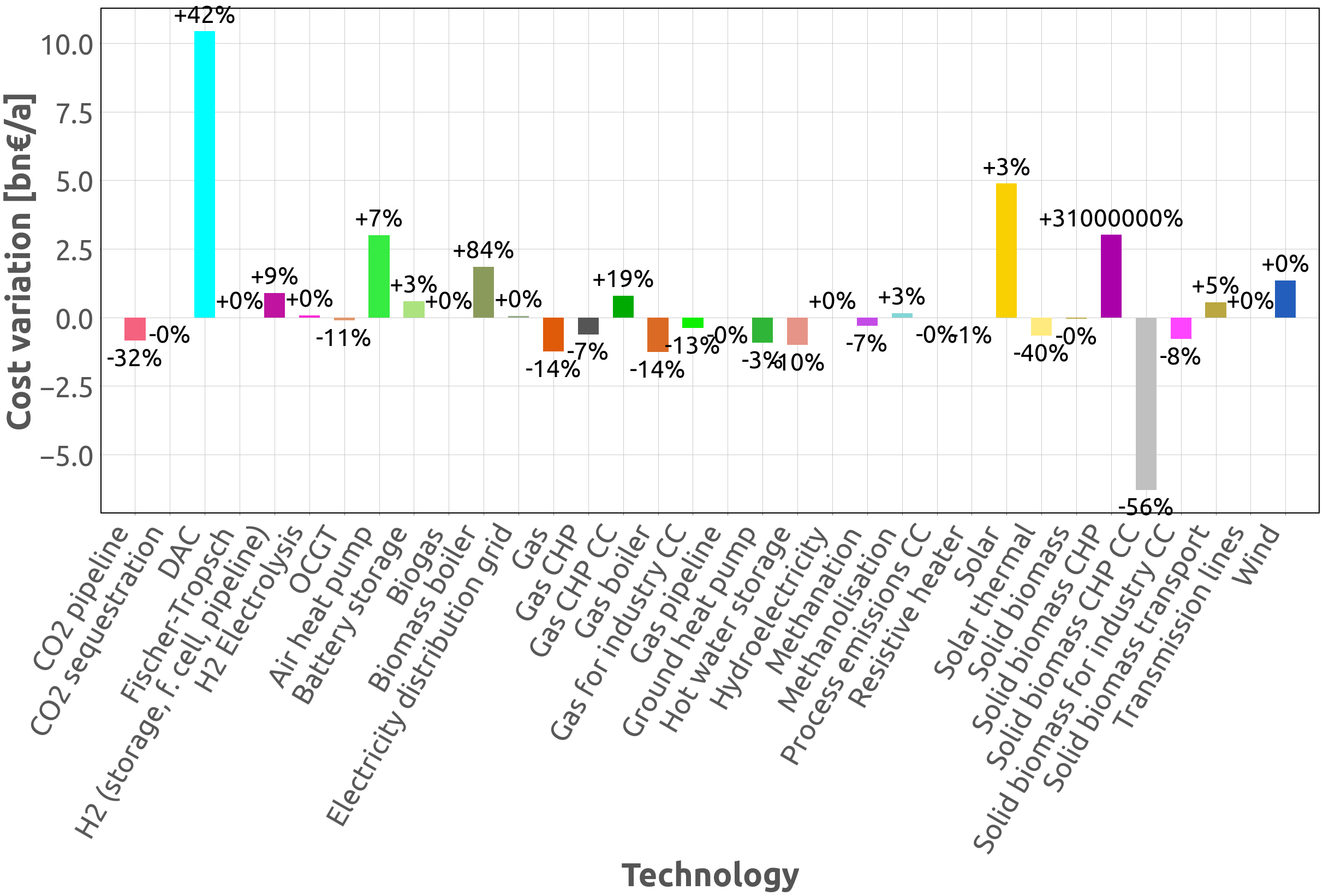}}
    \caption{(a) Total system cost and technology configuration in the global net-zero CO$_2$ emissions scenario based on the weather year 2010 and (b) Cost variation per technology between the global net-zero CO$_2$ emissions scenario and the local net-zero CO$_2$ emissions scenario based on the weather year 2010.} \label{supplemental:figure_total_system_cost_and_variation_global_weather2010}
\end{figure}

\begin{figure}[!htb]
    \centering
    %\subfloat[]{\label{supplemental:figure_temporal_co2_capture_endogenous_global}\includegraphics[width = 0.485\linewidth]{./figures/figure_S25a.png}}\hfill
    \subfloat[]{\label{supplemental:figure_temporal_co2_capture_endogenous_global}\includegraphics[width = 0.415\linewidth]{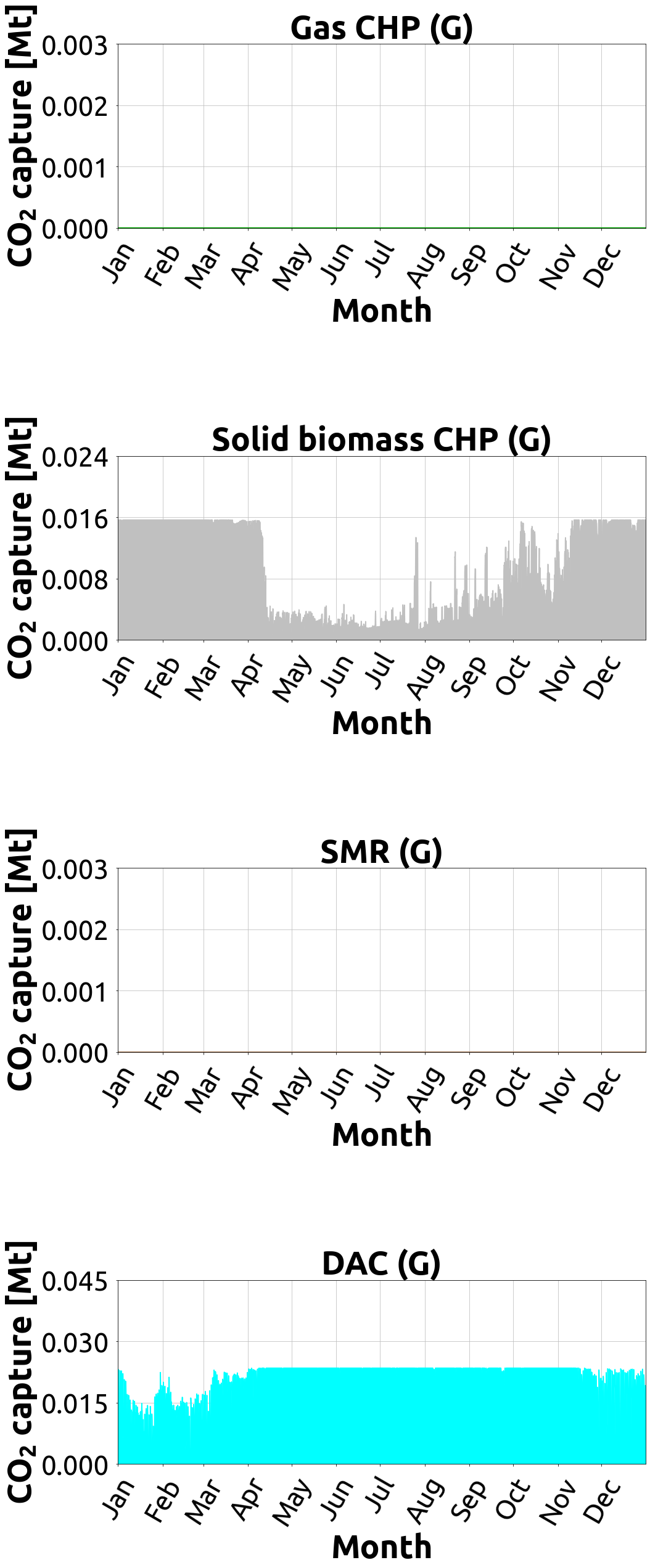}}\hfill
    %\subfloat[]{\label{supplemental:figure_temporal_co2_capture_endogenous_local}\includegraphics[width = 0.485\linewidth]{./figures/figure_S25b.png}}\par
    \subfloat[]{\label{supplemental:figure_temporal_co2_capture_endogenous_local}\includegraphics[width = 0.415\linewidth]{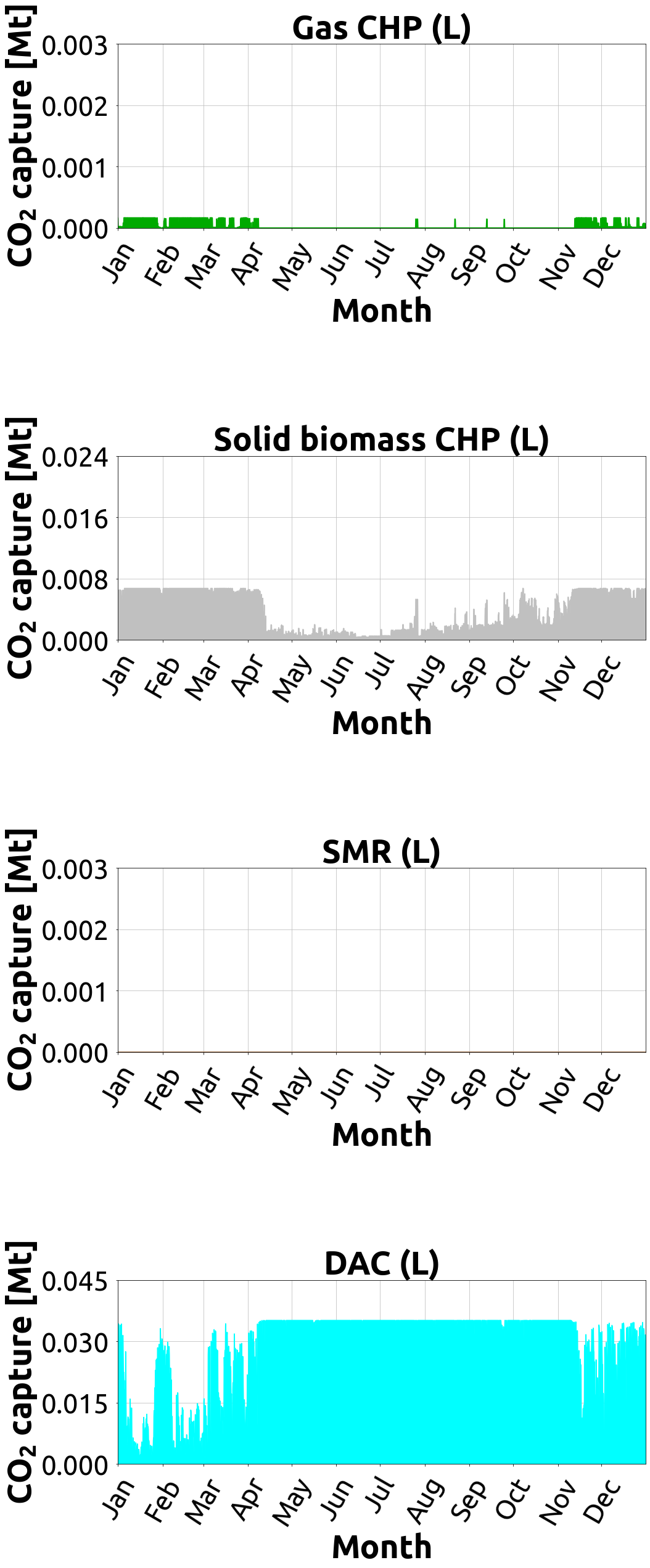}}\par
    \caption{Temporal CO$_2$ capture from endogenous sources across all of Europe based on the weather year 2013 in (a) the global net-zero CO$_2$ emissions scenario and (b) the local net-zero CO$_2$ emissions scenario. Except for the absence of CO$_2$ captured from gas-based CHP units and SMR, which are not cost-competitive, the other two processes of capturing CO$_2$ exhibit a seasonal pattern. CO$_2$ captured from solid biomass-based CHP units increases in winter due to higher heat demand and decreases in summer. In contrast, CO$_2$ captured from DAC decreases in winter and increases in summer due to the availability of low-cost electricity.}
    \label{supplemental:figure_temporal_co2_capture_endogenous}
\end{figure}

\begin{figure}[!htb]
    \centering
    %\subfloat[]{\label{supplemental:figure_temporal_co2_conversion_sequestration_global}\includegraphics[width = 0.485\linewidth]{./figures/figure_S26a.png}}\hfill
    \subfloat[]{\label{supplemental:figure_temporal_co2_conversion_sequestration_global}\includegraphics[width = 0.415\linewidth]{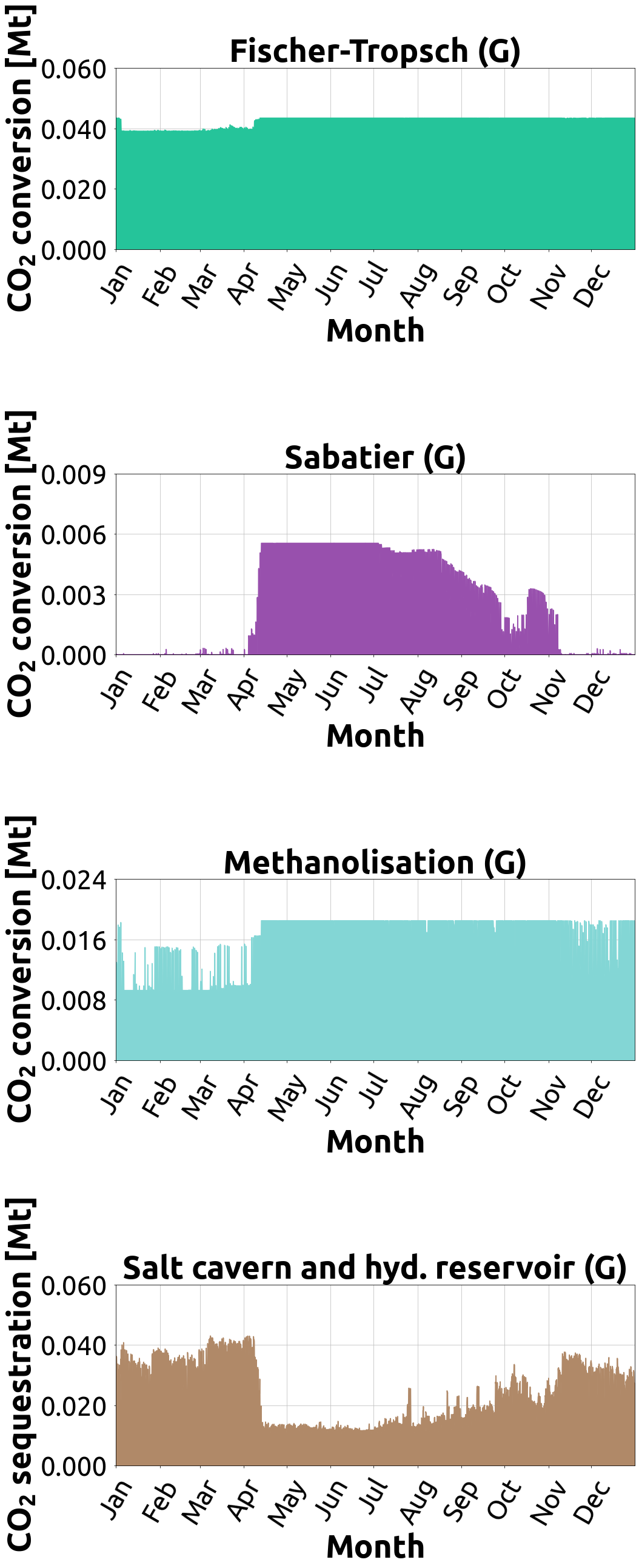}}\hfill
    %\subfloat[]{\label{supplemental:figure_temporal_co2_conversion_sequestration_local}\includegraphics[width = 0.485\linewidth]{./figures/figure_S26b.png}}\par
    \subfloat[]{\label{supplemental:figure_temporal_co2_conversion_sequestration_local}\includegraphics[width = 0.415\linewidth]{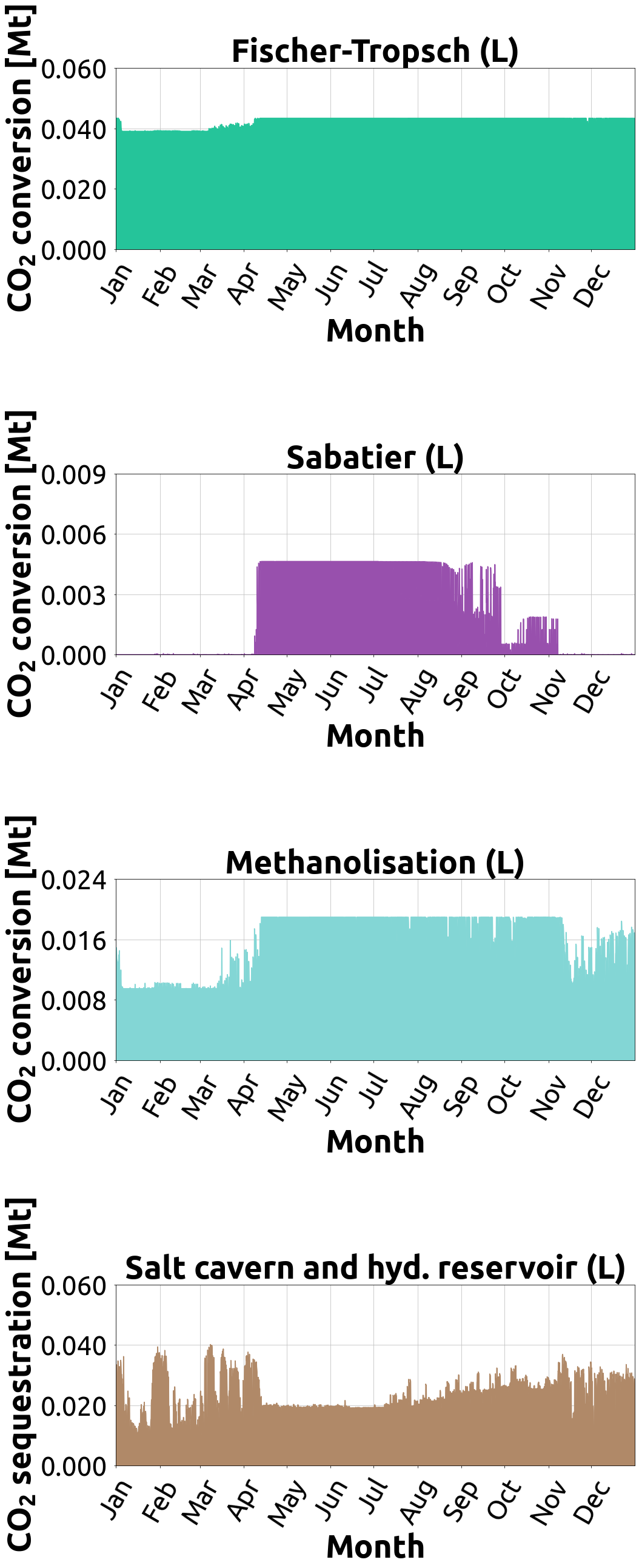}}\par
    \caption{Temporal CO$_2$ conversion and sequestration across all of Europe based on the weather year 2013 in (a) the global net-zero CO$_2$ emissions scenario and (b) the local net-zero CO$_2$ emissions scenario. Although less prominent in Fischer-Tropsch, all CO$_2$ conversion processes exhibit a seasonal pattern. These processes convert more CO$_2$ during summer due to lower electricity and heating prices in this season and convert less during winter. Sequestration of CO$_2$ underground also follows a seasonal pattern, albeit inverted from the one governing the CO$_2$ conversion processes. During winter, more CO$_2$ is sequestered underground, while less is sequestered during summer due to its utilisation in conversion processes.}
    \label{supplemental:figure_temporal_co2_conversion_sequestration}
\end{figure}

\begin{figure}[!htb]
    \centering
    %\subfloat[]{\label{supplemental:figure_temporal_co2_capture_endogenous_global_weather2010}\includegraphics[width = 0.485\linewidth]{./figures/figure_S27a_weather2010.png}}\hfill
    \subfloat[]{\label{supplemental:figure_temporal_co2_capture_endogenous_global_weather2010}\includegraphics[width = 0.415\linewidth]{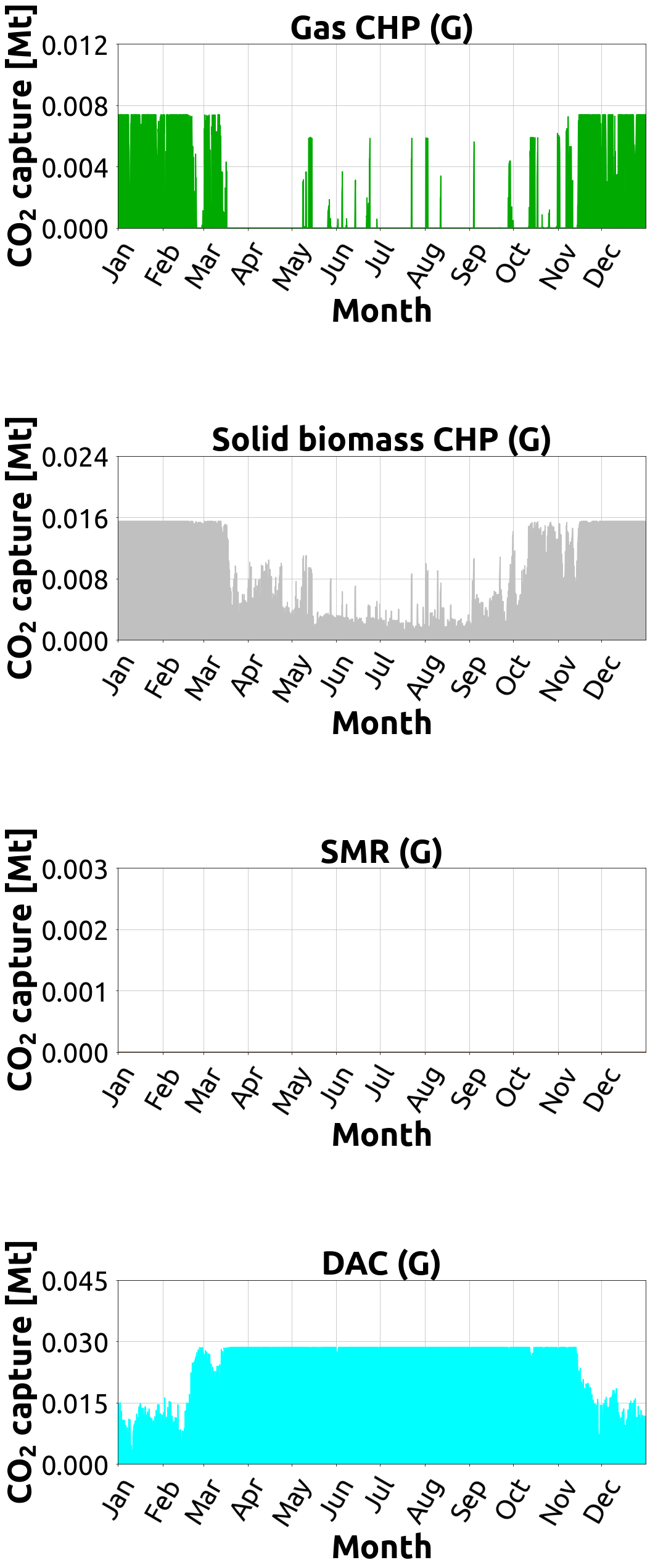}}\hfill
    %\subfloat[]{\label{supplemental:figure_temporal_co2_capture_endogenous_local_weather2010}\includegraphics[width = 0.485\linewidth]{./figures/figure_S27b_weather2010.png}}\par
    \subfloat[]{\label{supplemental:figure_temporal_co2_capture_endogenous_local_weather2010}\includegraphics[width = 0.415\linewidth]{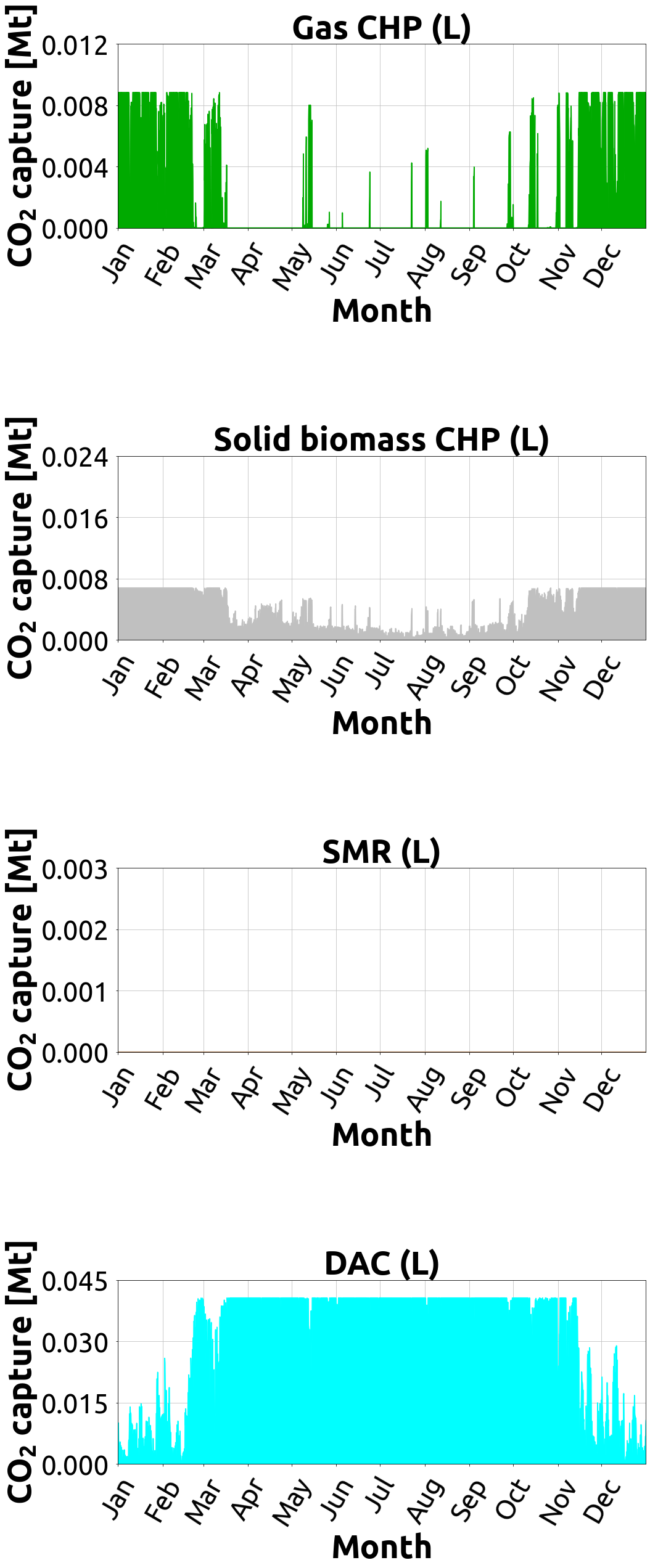}}\par
    \caption{Temporal CO$_2$ capture from endogenous sources across all of Europe based on the weather year 2010 in (a) the global net-zero CO$_2$ emissions scenario and (b) the local net-zero CO$_2$ emissions scenario. Except for the absence of CO$_2$ captured from SMR, which is not cost-competitive, the other three processes of capturing CO$_2$ exhibit a seasonal pattern similar to the pattern of the model based on the weather year 2013 (Figure \ref{supplemental:figure_temporal_co2_capture_endogenous}).}
    \label{supplemental:figure_temporal_co2_capture_endogenous_weather2010}
\end{figure}

\begin{figure}[!htb]
    \centering
    %\subfloat[]{\label{supplemental:figure_temporal_co2_conversion_sequestration_global_weather2010}\includegraphics[width = 0.485\linewidth]{./figures/figure_S28a_weather2010.png}}\hfill
    \subfloat[]{\label{supplemental:figure_temporal_co2_conversion_sequestration_global_weather2010}\includegraphics[width = 0.415\linewidth]{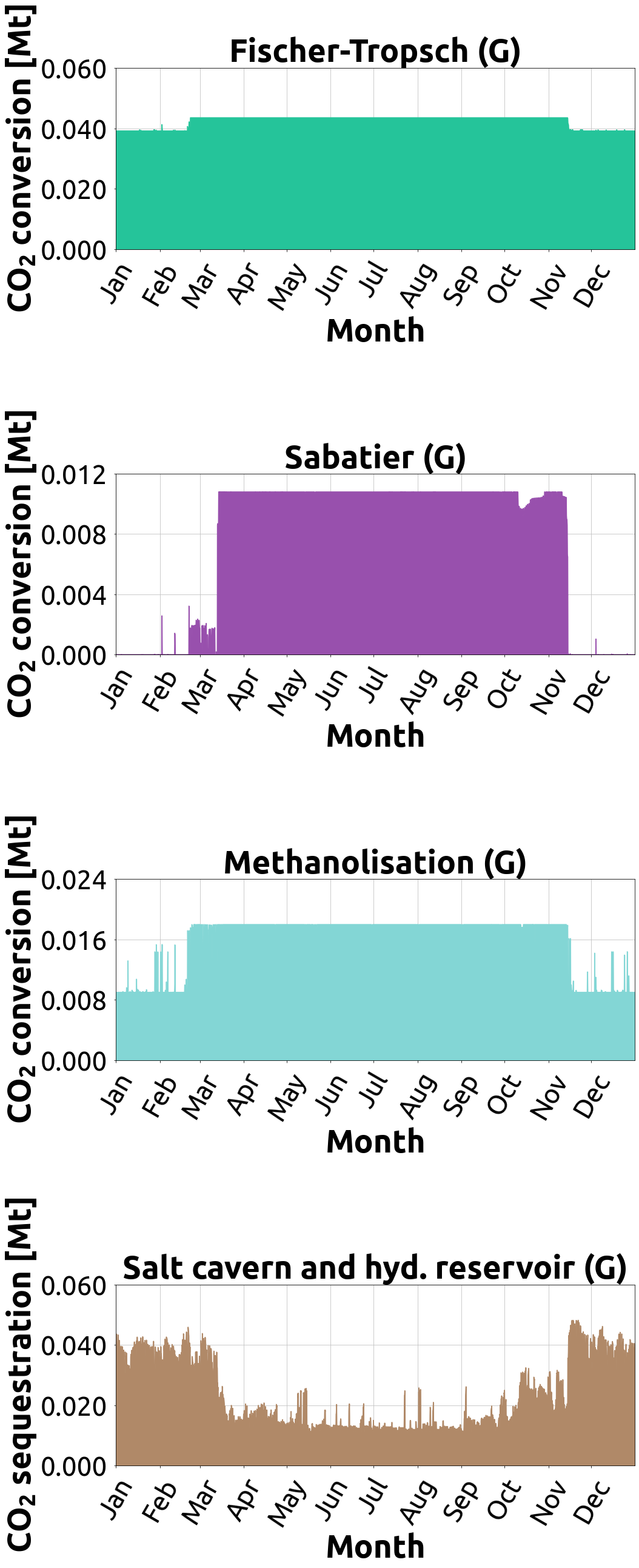}}\hfill
    %\subfloat[]{\label{supplemental:figure_temporal_co2_conversion_sequestration_local_weather2010}\includegraphics[width = 0.485\linewidth]{./figures/figure_S28b_weather2010.png}}\par
    \subfloat[]{\label{supplemental:figure_temporal_co2_conversion_sequestration_local_weather2010}\includegraphics[width = 0.415\linewidth]{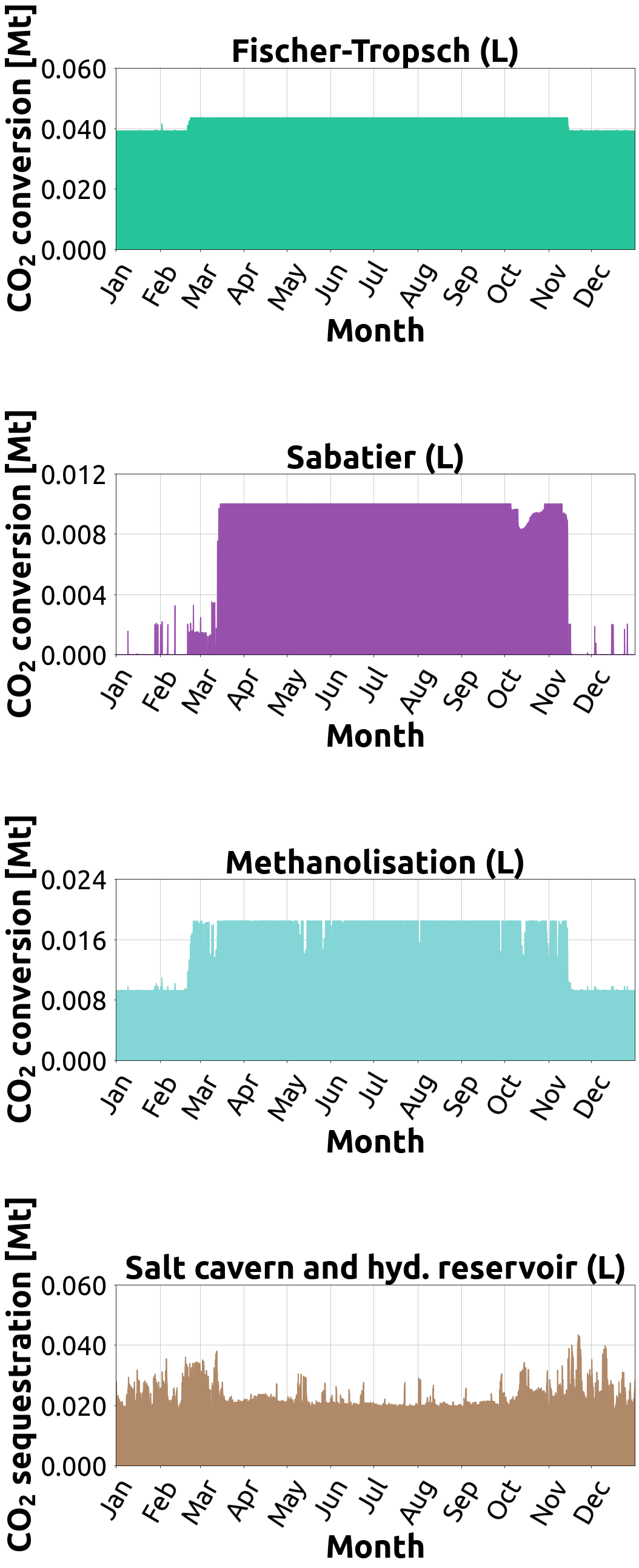}}\par
    \caption{Temporal CO$_2$ conversion and sequestration across all of Europe based on the weather year 2010 in (a) the global net-zero CO$_2$ emissions scenario and (b) the local net-zero CO$_2$ emissions scenario. All the processes of converting and sequestering captured CO$_2$ exhibit a seasonal pattern similar to the pattern of the model based on the weather year 2013 (Figure \ref{supplemental:figure_temporal_co2_conversion_sequestration}).}
    \label{supplemental:figure_temporal_co2_conversion_sequestration_weather2010}
\end{figure}

Third, due to the required computation, our model is based solely on the weather data from 2013, which is known to be an average weather year \cite{Gøtske2024, COKER2020509} to calculate renewable resources. We ran a sensitivity analysis using weather data from 2010, which was found to be the most demanding weather year based on a multi-decadal analysis \cite{Gøtske2024}. The annual total system cost increases by 6.2\% relative to 2013-weather results, amounting to 932 billion € in the global scenario and 945 billion € in the local scenario (Figure \ref{supplemental:figure_total_system_cost_and_variation_global_weather2010}). The cost increase is driven by the additional wind and solar PV capacity needed due to the low wind resources available in 2010. The remaining technology portfolio remains very similar between the two weather years. Captured CO$_2$ increased by 6\% when assuming 2010 weather, mostly due to emissions from gas-based CHP units providing heat during demand peaks. Although CO$_2$ capture levels differ when using weather data from 2013 or 2010, the spatial pattern of CO$_2$ capture remains essentially unchanged (compare Figure \ref{supplemental:figure_spatial_co2_capture_and_variation_weather2010} versus \ref{figure_spatial_co2_capture_and_variation}). The spatial pattern of CO$_2$ conversion and underground sequestration is slightly modified, again favouring the production of synthetic methane gas (compare Figure \ref{supplemental:figure_spatial_co2_conversion_and_variation_weather2010} versus \ref{figure_spatial_co2_conversion_and_variation}). The seasonal operation patterns for CO$_2$ technologies described in Sections \ref{sec_capture} and \ref{sec_conversion}, remain unchanged when using weather data from 2010 (compare Figure \ref{supplemental:figure_temporal_co2_capture_endogenous_weather2010} versus \ref{supplemental:figure_temporal_co2_capture_endogenous} and \ref{supplemental:figure_temporal_co2_conversion_sequestration_weather2010} versus
\ref{supplemental:figure_temporal_co2_conversion_sequestration}).

Fourth, as described in the Methods section, for the sake of simplicity, only the most relevant technologies for capturing CO$_2$ are included in the model, and no expansion of the electricity transmission grid is allowed. H$_2$ pipelines can only be built from scratch (i.e. no retrofitting of gas pipelines into H$_2$ is assumed). For a deeper discussion on the repurposing of gas pipelines into H$_2$ and the complementary role of power and H$_2$ transmission, we refer to \cite{Neumann2023}.

Fifth, energy demands in electricity, heating, transport, and industry (assuming a significant industrial transformation to reduce CO$_2$ emissions) are estimated using current values. Electricity demand is increased endogenously when the model decides to electrify heat.

Finally, the model assumes that Europe produces all the synthetic fuels and solid biomass it needs and does not rely on imports from other regions. Neumann et al. \cite{Neumann2025}, estimated that the total system cost can decrease by 14\% if fuels imports are allowed. Millinger et al. \cite{Millinger2025} estimated that it can be cost-effective to import a large amount of biomass to achieve a climate-neutral Europe, but this might raise questions on the global sustainability of this solution.

\section{Discussion}
In this study, we modelled the European sector-coupled energy system, imposing a net-zero CO$_2$ emissions target either globally or locally in each country. Under a global CO$_2$ target, Germany, Belgium, and The Netherlands (interior countries) become the biggest net CO$_2$ emitters, while Spain, Sweden, Finland, Romania, and Poland (exterior countries) become the biggest net CO$_2$ absorbers. In the locally constrained scenario, the total system cost increases by 1.4\%, relative to the global scenario. From a policy perspective, this means that, with a slight additional cost, climate neutrality could be attained in Europe with a more uniform contribution from different countries than that typically obtained in the literature using a global CO$_2$ limit, i.e., by requiring each country to capture its emitted CO$_2$. This first strategy might achieve better social acceptance, as it uniformly distributes the burden. However, the consequences for individual countries would be more drastic. Interior countries, that were net emitters under a global constraint, see a cost increase of 58\% in Belgium, 39\% in The Netherlands, and 12\% in Germany, while exterior countries, that were absorbers, see significant cost reductions. Since cooperation has been proven to pay off, an alternative policy strategy might consist in aiming at a global CO$_2$ target and establishing a mechanism for financial transfer to compensate for non-uniform contributions from different countries. This requires estimating the extra effort that exterior countries are providing, which can be based on their cost differences for the two scenarios. The second strategy uses a uniform CO$_2$ pricing scheme, similar to the current EU emission trading system (ETS), while the first one would require country-specific carbon pricing.

Our detailed modelling, comprising high spatial and temporal resolution with network representation, closes the existing gap in the literature and has allowed us to quantify the economic consequences of a global or country-specific CO$_2$ constraints. Moreover, it has enabled us to investigate the need for carbon capture, conversion, and transport infrastructure. This is not possible with coarser model resolutions previously used. In the first, we learnt that the optimal location of carbon infrastructure gets impacted by the selected strategy. Under country-specific CO$_2$ limits, interior countries that are net emitters under a global constraint capture much more CO$_2$ mainly through DAC to achieve climate neutrality on their own, as they can no longer rely on other nations with low-cost energy to counterbalance their emissions. Conversely, net absorber countries reduce CO$_2$ capture from DAC, as well as from point sources. When it comes to CO$_2$ conversion, net emitter countries produce more synthetic oil and methanol in a locally constrained scenario, while net absorber countries reduce synthetic products manufacturing due to capturing less CO$_2$ in this scenario. The underground sequestration potential of 200 MtCO$_2$/a is fully utilised in every scenario.

We also learnt that countries are willing to exchange carbon amongst themselves in every scenario. Direct transport of CO$_2$ by pipelines is the most used strategy, and building a CO$_2$ network is cost-effective. However, it only reduces the total system cost by 0.3\% and 0.6\% in the global and local scenario respectively, compared to the same scenario without such a network. These gains are smaller than those attained by building a H$_2$ network, and much smaller than those attained by extending transmission lines in Europe \cite{Neumann2023}. Quantifying the relative cost gains of alternative energy transport infrastructure allows prioritising energy policy to make it more effective. Furthermore, considering that the specific topology of the CO$_2$ network depends on whether CO$_2$ emissions are limited globally or locally, coordination between the two is crucial.

Overall, this study highlights the importance of detailed modelling of the capture, transportation, conversion, and sequestration of CO$_2$ infrastructure. For energy modellers, our results suggest the relevance of considering more distributed CO$_2$ emissions constraints and call for not limiting the analysis to a globally imposed CO$_2$ limit. This makes the results more relevant for each country, describes the potential role that they could play in a collaborative Europe, helps estimate the necessary capacity of the selected technologies, and establishes appropriate intermediate targets to reach those goals.

To address some limitations of this study, future research could include modelling the transition paths instead of a future time step, to capture lock-in and path-dependence effects and focus on the biggest net CO$_2$ emitter and absorber countries, as they influence the most the energy system under net-zero CO$_2$ emissions targets.

\section{Methods}

\subsection{Model}
The present study relies on PyPSA-Eur\cite{HORSCH2018207}, an open-source model based on PyPSA \cite{PyPSA} that is used to optimise the capacity and dispatch of an energy system covering the entire European Network of Transmission System Operators for Electricity (ENTSO-E) area. This encompasses 33 European countries: the EU27 (excluding Malta and Cyprus), Norway, Switzerland, Great Britain, Albania, Bosnia and Herzegovina, Montenegro, Macedonia, and Serbia. PyPSA-Eur comprises the electricity, heating, land transport, shipping, aviation, and industry sectors, including industrial feedstocks. It represents generators, stores, transmission lines, and conversion technologies in different sectors of an interconnected energy system and provides a detailed account of carbon capture, transportation, usage, and storage (CCTUS). Our model assumes long-term market equilibrium, perfect competition and foresight.

At its core, the model uses linear programming to determine the most cost-effective energy system based on a set of constraints, with the most important constraints for this study being the allowable limit for CO$_2$ emissions (Equations \ref{supplemental:material_global_co2_constraint} and \ref{supplemental:material_local_co2_constraint}). Refer to \ref{supplemental:material_mathematical_formulation} for a mathematical formulation of the linear optimisation problem.

The characteristics of the technologies represented in the model, including costs, efficiencies, and lifetimes, are based on the Energy System Technology Data \cite{energy_system_technology_data}, which are derived mainly from data published by the Danish Energy Agency \cite{danish_energy_agency}. These are summarised in Table \ref{supplemental:table_technology_characteristics_assumptions}. In our main scenarios, characteristics and cost assumptions of the technologies from the Energy System Technology Data version 0.5.0 in the year 2030 are chosen because they include cost reduction for renewable and carbon-related technologies, but are less uncertain than those for a more distant year. We implement a sensitivity analysis assuming optimistic costs corresponding to the Energy System Technology Data in the year 2050. We apply cost assumptions and discount rates uniformly across the continent.

\begin{figure}[!htb]
    \centering
    \includegraphics[height = 12.5cm, angle = 90]{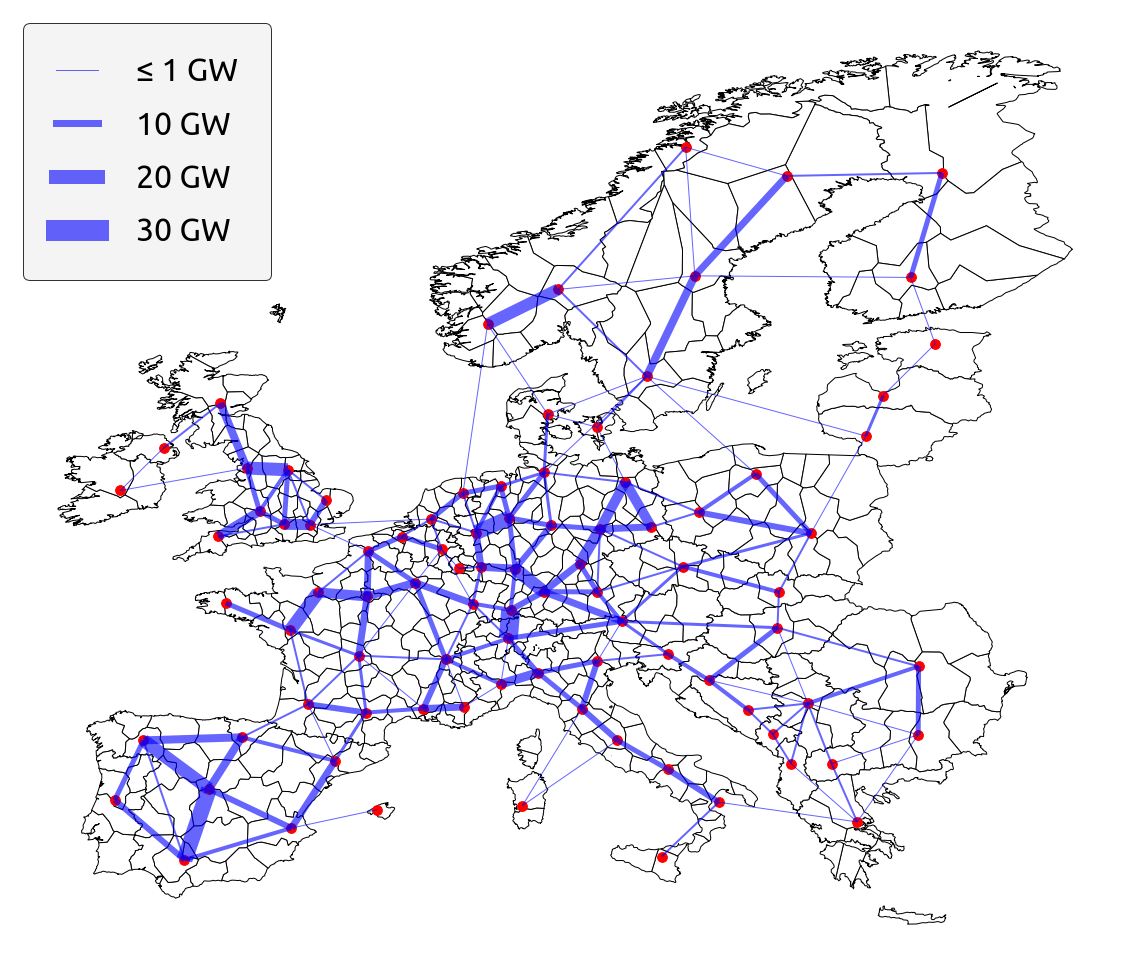}
    \caption{The European energy system spatial layout spanning across 33 countries in both the global and local net-zero CO$_2$ emissions scenarios. It is modelled with a 90 nodes resolution (indicated by red dots) and 370 regions representing renewable resources, with each region associated with one of the nodes. The (blue) lines connecting nodes represent the electricity grid topology, which includes both existing power lines and new ones planned for construction in Europe as per ENTSO-E's TYNDP 2018.}
    \label{supplemental:figure_nodes_regions_electricity_grid}
\end{figure}

\begin{figure}[!htb]
    \centering
    %\subfloat[]{\label{supplemental:figure_temporal_resolution_sensitivity_analysis_global}\includegraphics[width = 0.96\linewidth]{./figures/figure_S5a.png}}
    \subfloat[]{\label{supplemental:figure_temporal_resolution_sensitivity_analysis_global}\includegraphics[width = 0.8\linewidth]{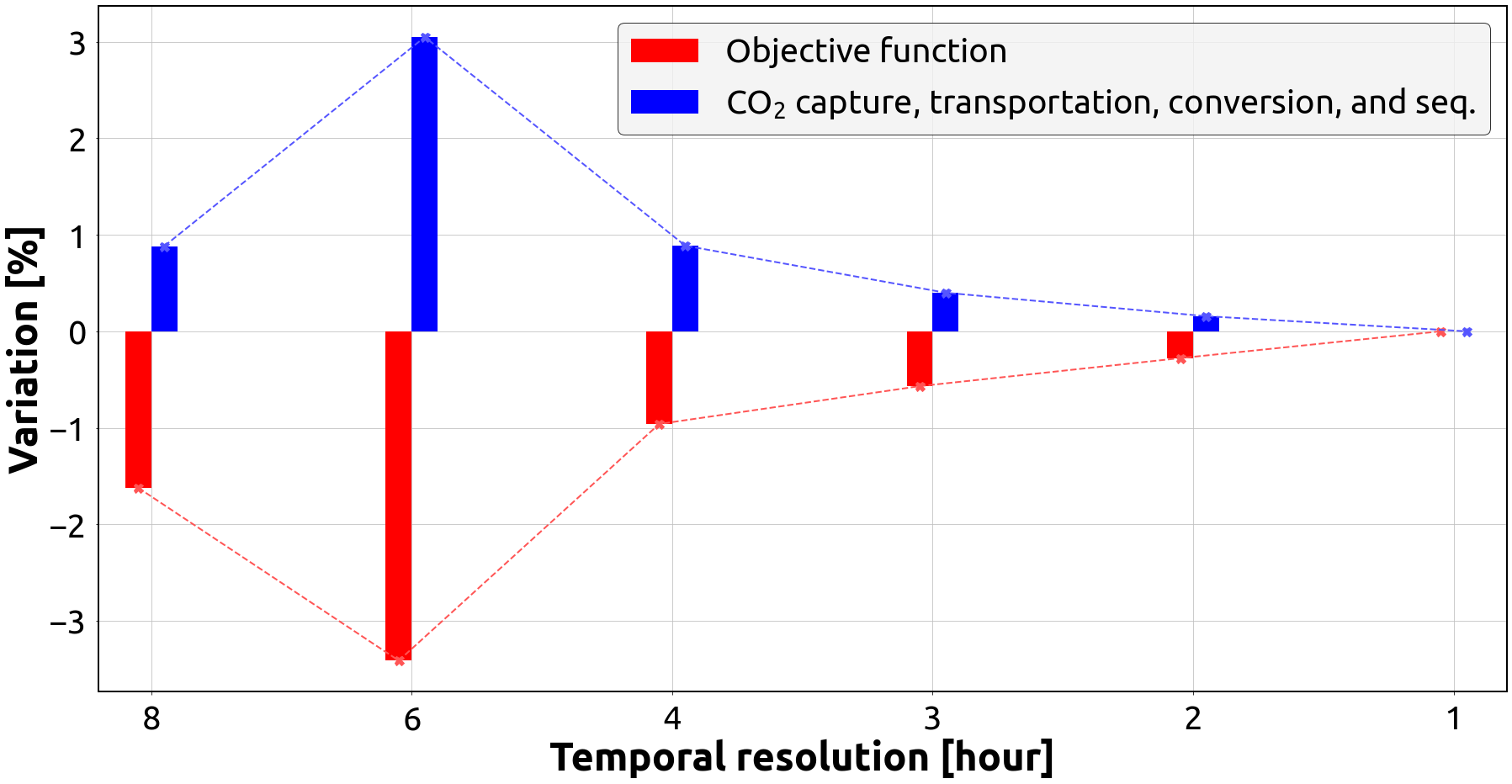}}
    \vspace{20pt}
    %\subfloat[]{\label{supplemental:figure_temporal_resolution_sensitivity_analysis_local}\includegraphics[width = 0.96\linewidth]{./figures/figure_S5b.png}}
    \subfloat[]{\label{supplemental:figure_temporal_resolution_sensitivity_analysis_local}\includegraphics[width = 0.8\linewidth]{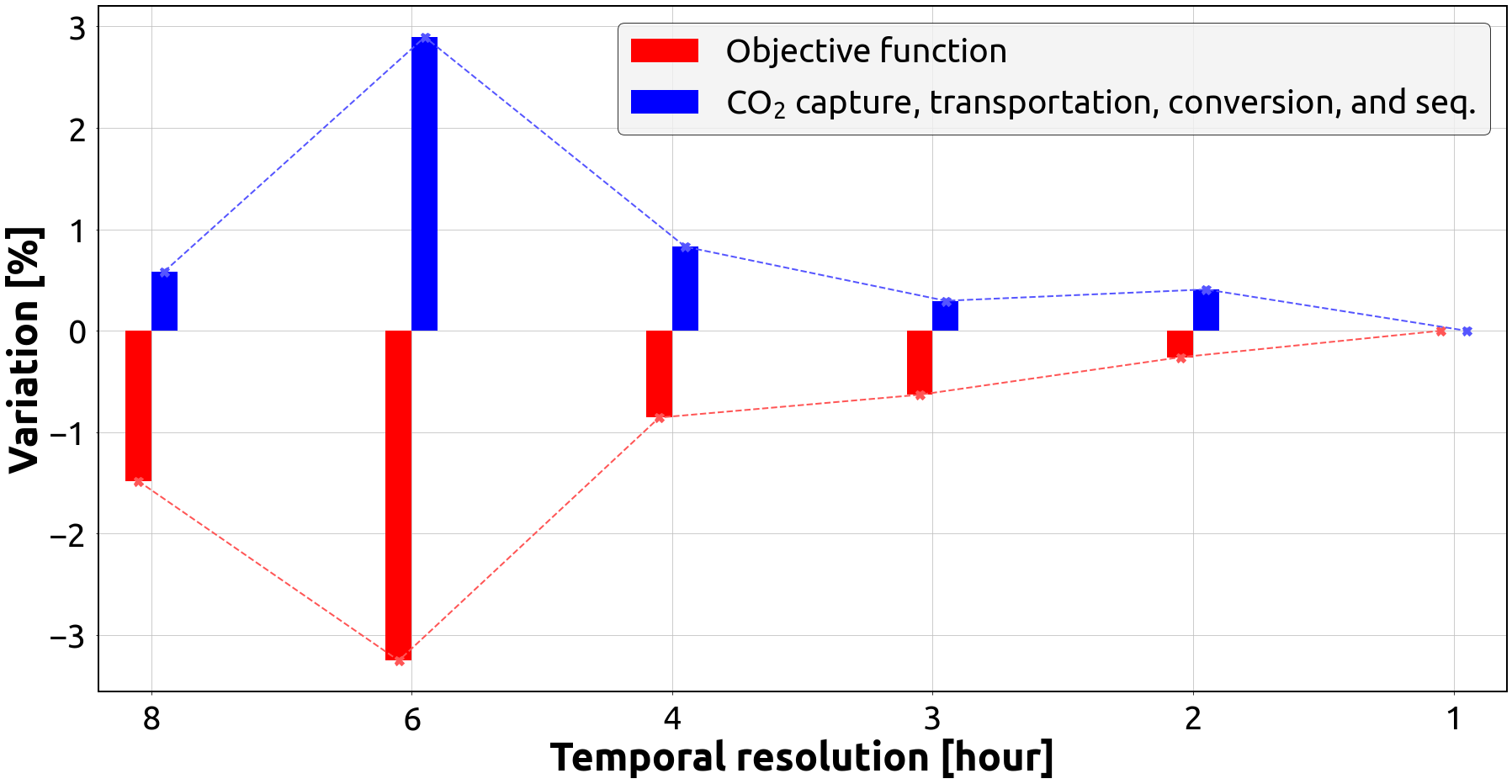}}
    \caption{Sensitivity analysis of the model under different temporal resolutions in (a) the global net-zero CO$_2$ emissions scenario and (b) the local net-zero CO$_2$ emissions scenario. Due to computational limitations, both sensitivity analyses were conducted with an 80-node spatial resolution of the model instead of the 90-node spatial resolution used in our study. The red and blue bars show the variations of the objective function and the aggregated amount of CO$_2$ captured, transported, converted, and sequestered underground, respectively, in comparison to the baseline of one-hour temporal resolution.}
    \label{supplemental:figure_temporal_resolution_sensitivity_analysis}
\end{figure}

The model is set up to represent the European energy system characterised by net-zero CO$_2$ emissions, either globally imposed for the whole of Europe or locally for each of the 33 European countries. It performs an overnight greenfield optimisation of the energy network system, which consists of 90 nodes, 370 regions representing renewable resources (each associated with one of the nodes, allowing a more granular modelling of their distribution), and a 3-hour time resolution for the entire year (Figures \ref{supplemental:figure_nodes_regions_electricity_grid} and \ref{supplemental:figure_temporal_resolution_sensitivity_analysis}). This setup represents well the European landscape and the variations in wind and solar patterns throughout the day, week, and season \cite{PFENNINGER20171}. It also addresses the concerns raised in \cite{horsch_the_role_2017} and \cite{FLEISCHER2020100563} regarding sufficient spatial and temporal resolution, while still being computationally feasible \cite{SCHYSKA20212606} \cite{FrysztackiRechtBrown2022_1000153979}. The renewable energy capacity factors in Europe were calculated using atlite \cite{Hofmann2021} and data from a single year of weather provided by ERA5 \cite{hersbach2018era5}. Here, 2013 was chosen as it represents a year with average wind and solar resources \cite{Gøtske2024, COKER2020509}.

The transmission capacity of the electricity grid is exogenously fixed (i.e. the model is not allowed to increase its capacity) and it includes existing lines as well as new ones planned for Europe, in accordance with ENTSO-E’s Ten-Year Network Development Plan 2018 \cite{entsoe_tyndp_2018} (Figure \ref{supplemental:figure_nodes_regions_electricity_grid}).

\begin{figure}[!htb]
    \centering
    %\subfloat[]{\label{supplemental:figure_demand_map}\includegraphics[width = 0.8\linewidth]{./figures/figure_S21a.png}}
    \subfloat[]{\label{supplemental:figure_demand_map}\includegraphics[width = 0.66\linewidth]{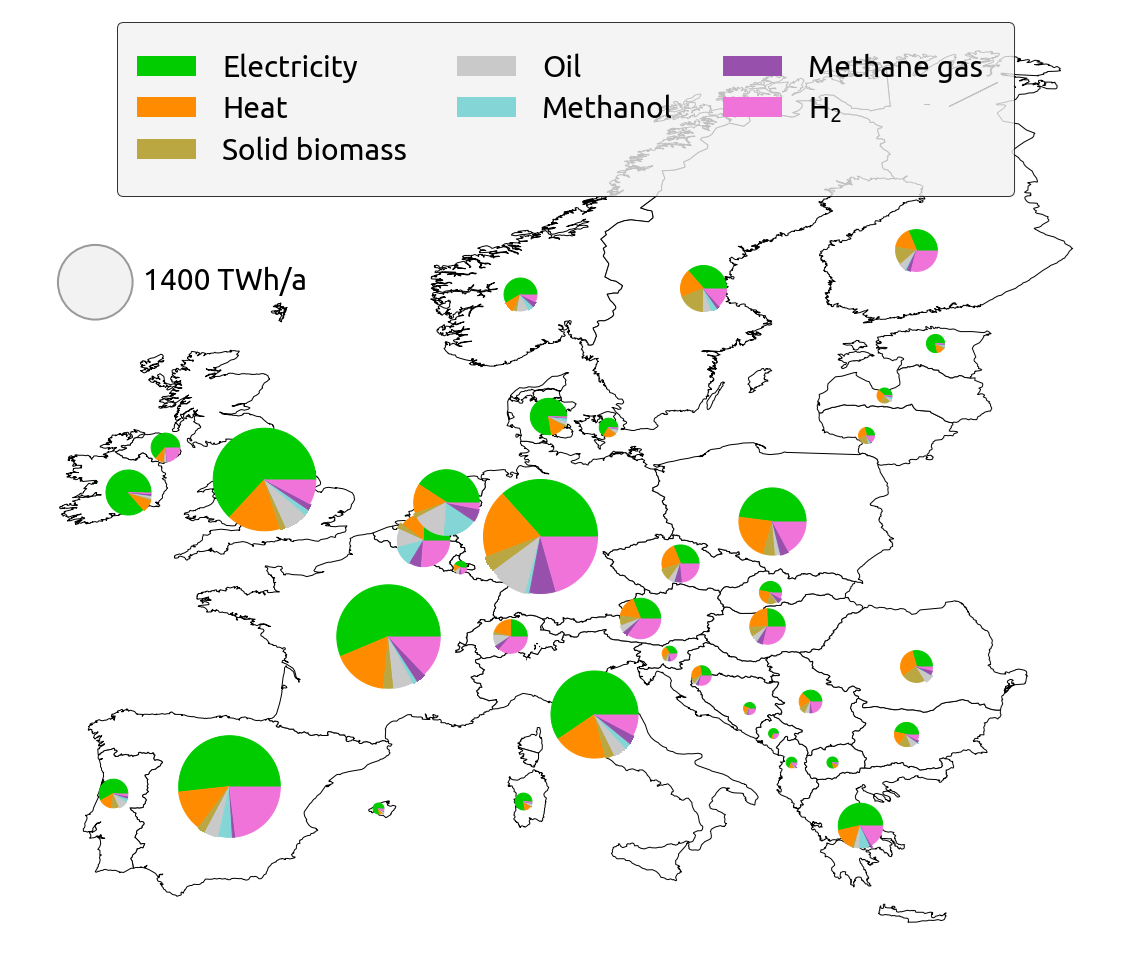}}
    \vspace{20pt}
    %\subfloat[]{\label{supplemental:figure_demand_variation}\includegraphics[width = 0.96\linewidth]{./figures/figure_S21b.png}}
    \subfloat[]{\label{supplemental:figure_demand_variation}\includegraphics[width = 0.8\linewidth]{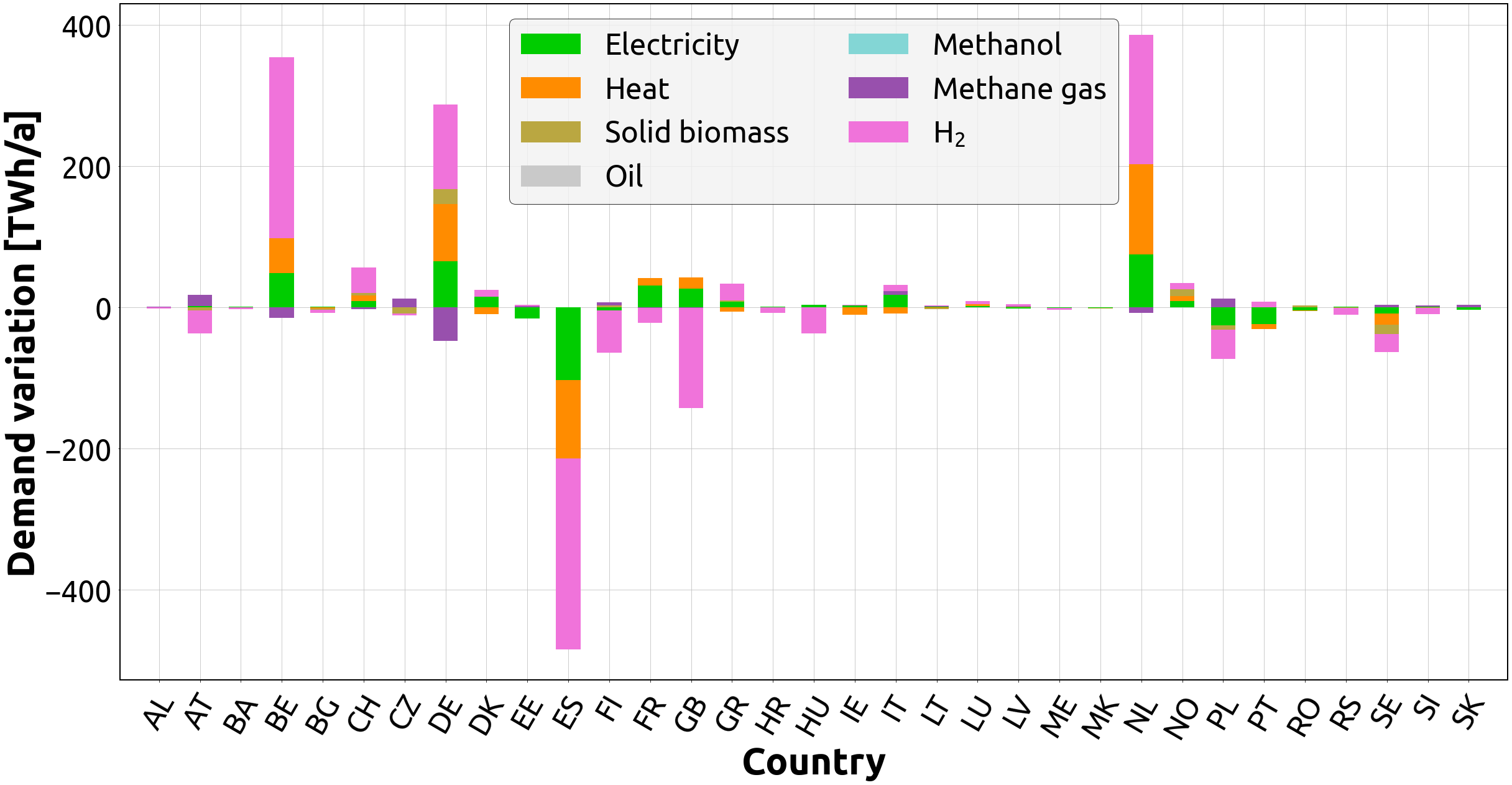}}
    \caption{(a) Electricity, heat, solid biomass, oil, methanol, methane gas, and H$_2$ demand in the global net-zero CO$_2$ emissions scenario and (b) Electricity, heat, solid biomass, oil, methanol, methane gas, and H$_2$ demand variation per country between the global net-zero CO$_2$ emissions scenario and the local net-zero CO$_2$ emissions scenario. In the local scenario, the interior countries significantly increase the demand for electricity, heat, and H$_2$ due to the growth in manufacturing synthetic products, along with the deployment of DAC and air heat pumps.}
    \label{supplemental:figure_demand}
\end{figure}

The demand for electricity (from the industry, agriculture, residential, services, and land transport sectors), heat (from the industry, agriculture, residential, and services sectors), solid biomass (from the industry sector), oil (from the industry, agriculture, and aviation sectors), methanol (from the shipping sector), methane gas (from the industry sector), and H$_2$ (from the industry and land transport sectors) across Europe is exogenously set in the model. The total annual values for these demands are 3526, 3313, 702, 1427, 546, 381, and 471 TWh, respectively (Figure \ref{supplemental:figure_demand}). For electricity, heat, oil, methane gas, and H$_2$, the model endogenously decides the technologies used to produce them and whether they are of fossil origin. The data sources for the demands and the transformation assumed in every industrial sector are based on/described in \cite{Neumann2023} and the JRC Integrated Database of the European Energy Sector (JRC-IDEES) \cite{doi/10.2760/182725}.

\begin{figure}[!htb]
    \centering
    %\subfloat[]{\label{supplemental:figure_methane_gas_network_global}\includegraphics[width = 0.72\linewidth]{./figures/figure_S2a.png}}
    \subfloat[]{\label{supplemental:figure_methane_gas_network_global}\includegraphics[width = 0.6\linewidth]{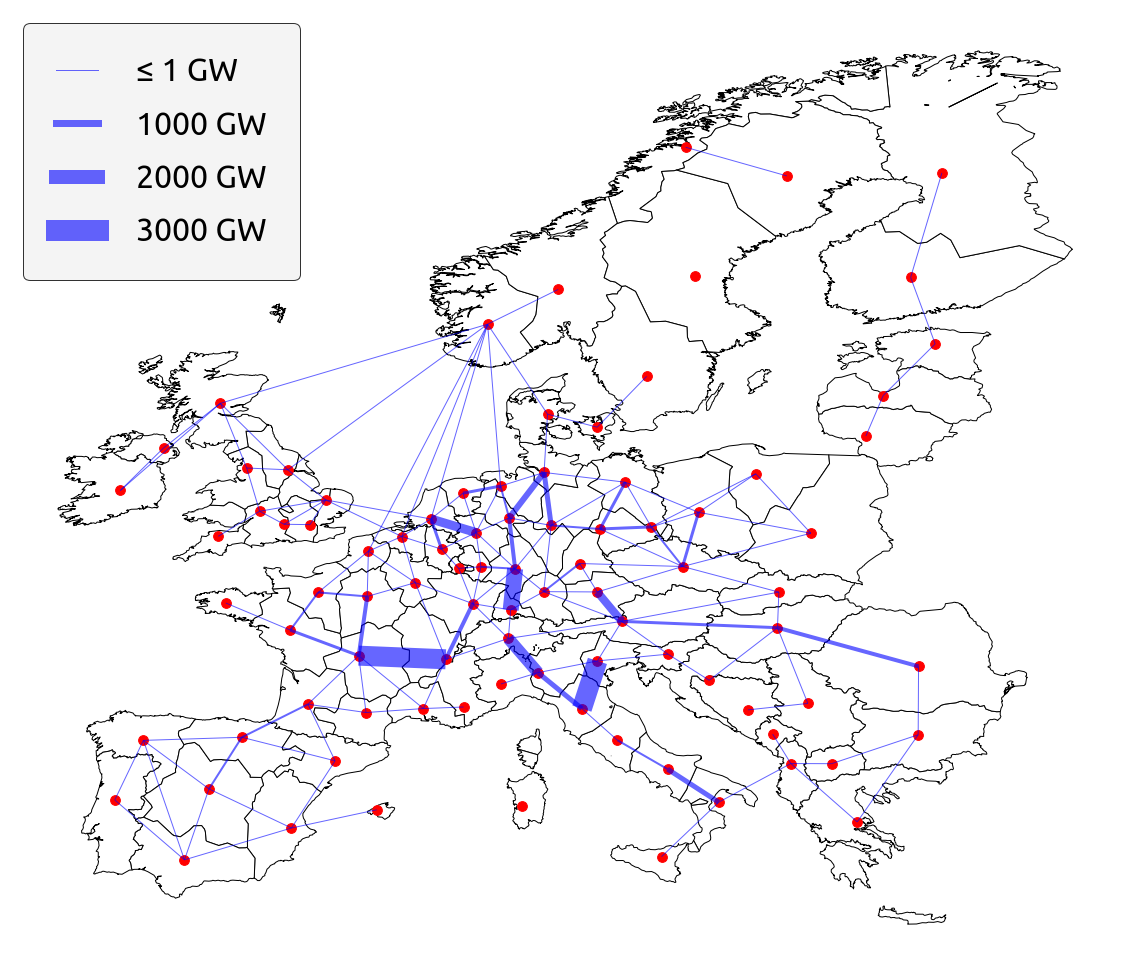}}
    \vspace{20pt}
    %\subfloat[]{\label{supplemental:figure_methane_gas_network_local}\includegraphics[width = 0.72\linewidth]{./figures/figure_S2b.png}}
    \subfloat[]{\label{supplemental:figure_methane_gas_network_local}\includegraphics[width = 0.6\linewidth]{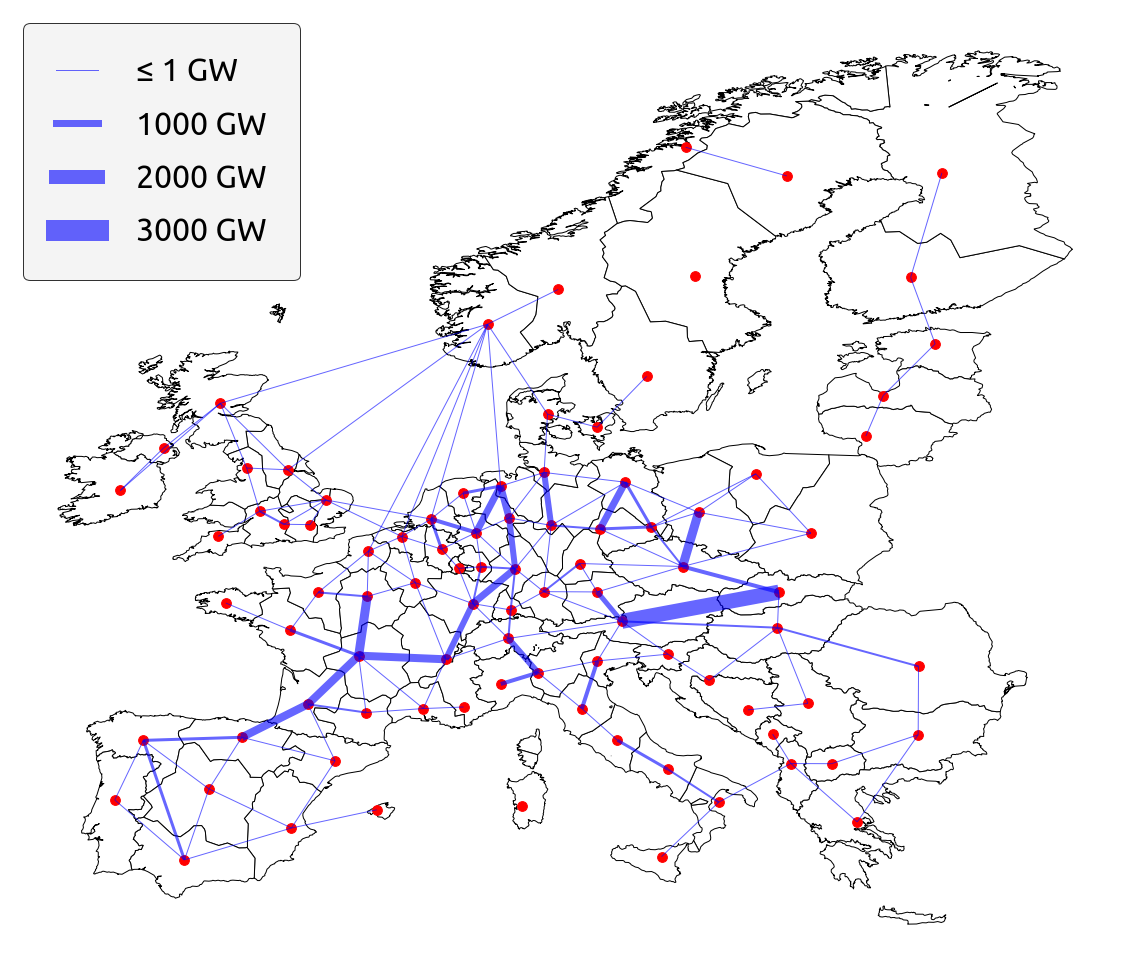}}
    \caption{The European methane gas network topology in (a) the global net-zero CO$_2$ emissions scenario and (b) the local net-zero CO$_2$ emissions scenario. Each pipeline has a specific capacity and is organised according to the SciGRID\_gas project, which forms the basis for transporting methane gas in the model. The model is allowed to expand the capacity of pipelines if it reduces the total system cost.}
    \label{supplemental:figure_methane_gas_network}
\end{figure}

The model allows the transportation of methane gas through pipelines (Figure \ref{supplemental:figure_methane_gas_network}), which are organised and have limited capacities according to the SciGRID\_gas \cite{pluta_scigrid_gas_2022} project. This project represents the existing methane gas network currently installed in Europe. While the model is allowed to expand pipelines’ capacities if it reduces the total system cost, repurposing existing pipelines to transport H$_2$ (instead of methane gas) is not allowed. Furthermore, due to compression losses, these have a transportation efficiency of 99\% per 1000 km.

\begin{figure}[!htb]
    \centering
    %\subfloat[]{\label{supplemental:figure_h2_network_global}\includegraphics[width = 0.72\linewidth]{./figures/figure_S3a.png}}
    \subfloat[]{\label{supplemental:figure_h2_network_global}\includegraphics[width = 0.6\linewidth]{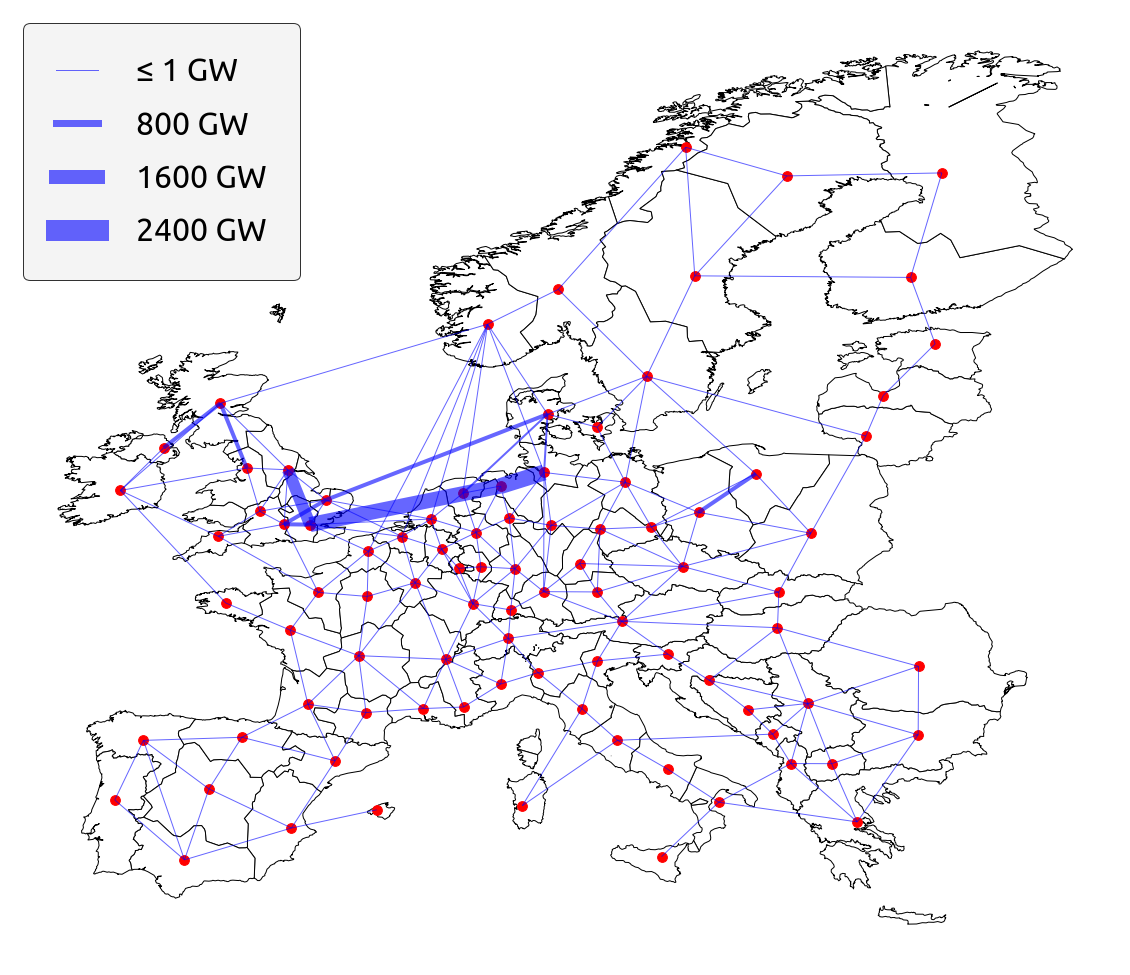}}
    \vspace{20pt}
    %\subfloat[]{\label{supplemental:figure_h2_network_local}\includegraphics[width = 0.72\linewidth]{./figures/figure_S3b.png}}
    \subfloat[]{\label{supplemental:figure_h2_network_local}\includegraphics[width = 0.6\linewidth]{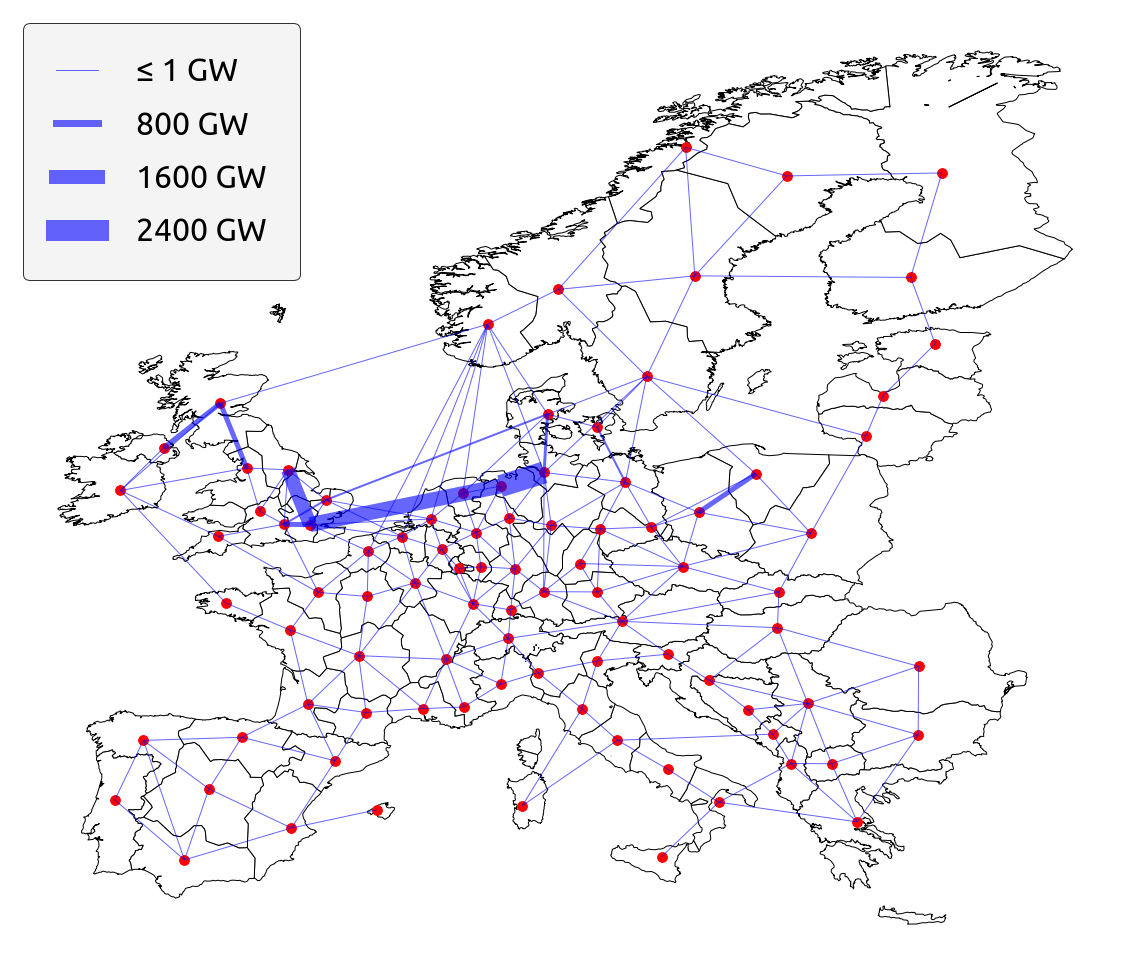}}
    \caption{The European H$_2$ network topology in (a) the global net-zero CO$_2$ emissions scenario and (b) the local net-zero CO$_2$ emissions scenario. The model is allowed to build new H$_2$ pipelines if it reduces the total system cost. These have optimal capacities and are organised according to the electricity grid and methane gas network topologies.}
    \label{supplemental:figure_h2_network}
\end{figure}

The model is allowed to build bidirectional pipelines to transport hydrogen (H$_2$) with optimal capacities also when deemed cost-effective. Due to compression losses, pipelines have a transportation efficiency of 98.2\% per 1000 km. For simplicity, these are spatially arranged based on the existing electricity grid and methane gas network topologies (Figure \ref{supplemental:figure_h2_network}).

\begin{figure}[!htb]
    \centering
    %\subfloat[]{\label{supplemental:figure_co2_network_global}\includegraphics[width = 0.72\linewidth]{./figures/figure_S4a.png}}
    \subfloat[]{\label{supplemental:figure_co2_network_global}\includegraphics[width = 0.6\linewidth]{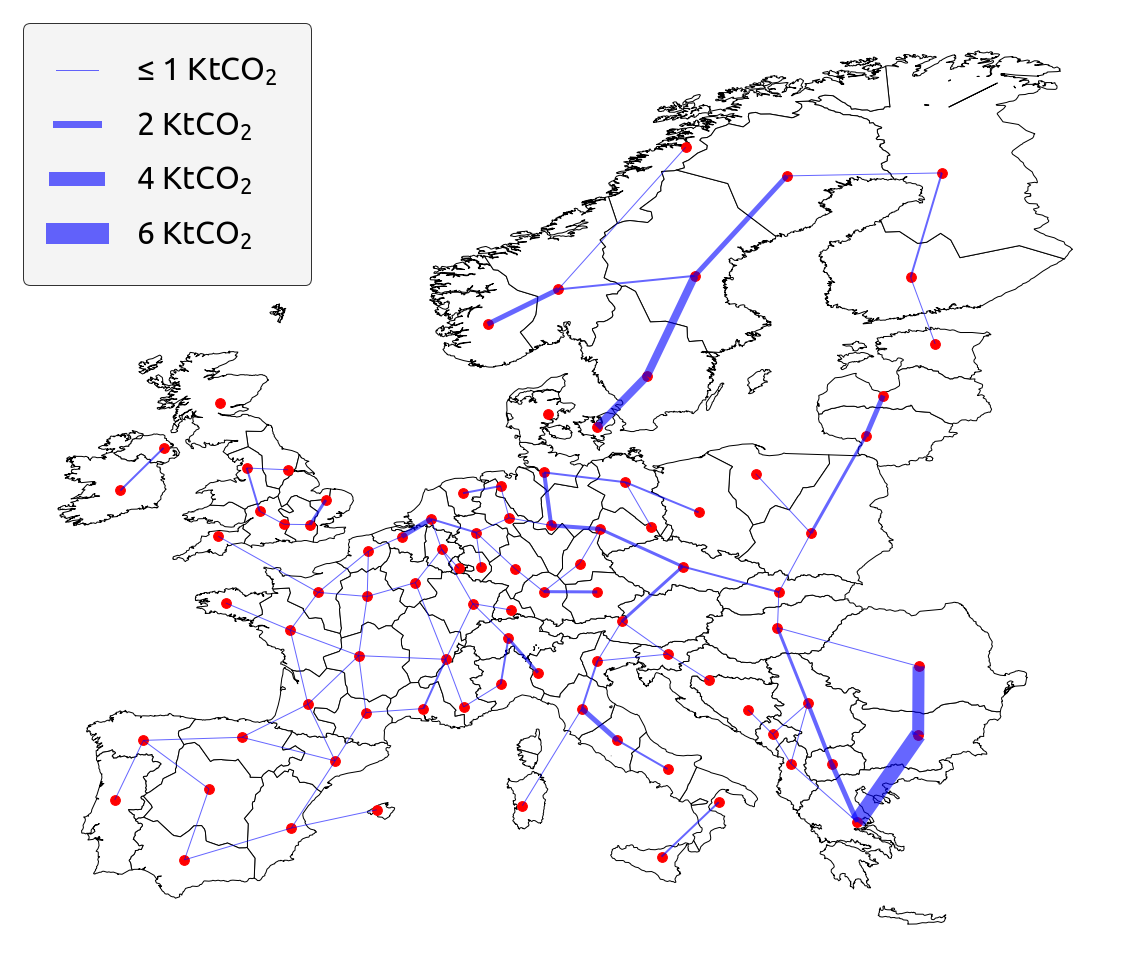}}
    \vspace{20pt}
    %\subfloat[]{\label{supplemental:figure_co2_network_local}\includegraphics[width = 0.72\linewidth]{./figures/figure_S4b.png}}
    \subfloat[]{\label{supplemental:figure_co2_network_local}\includegraphics[width = 0.6\linewidth]{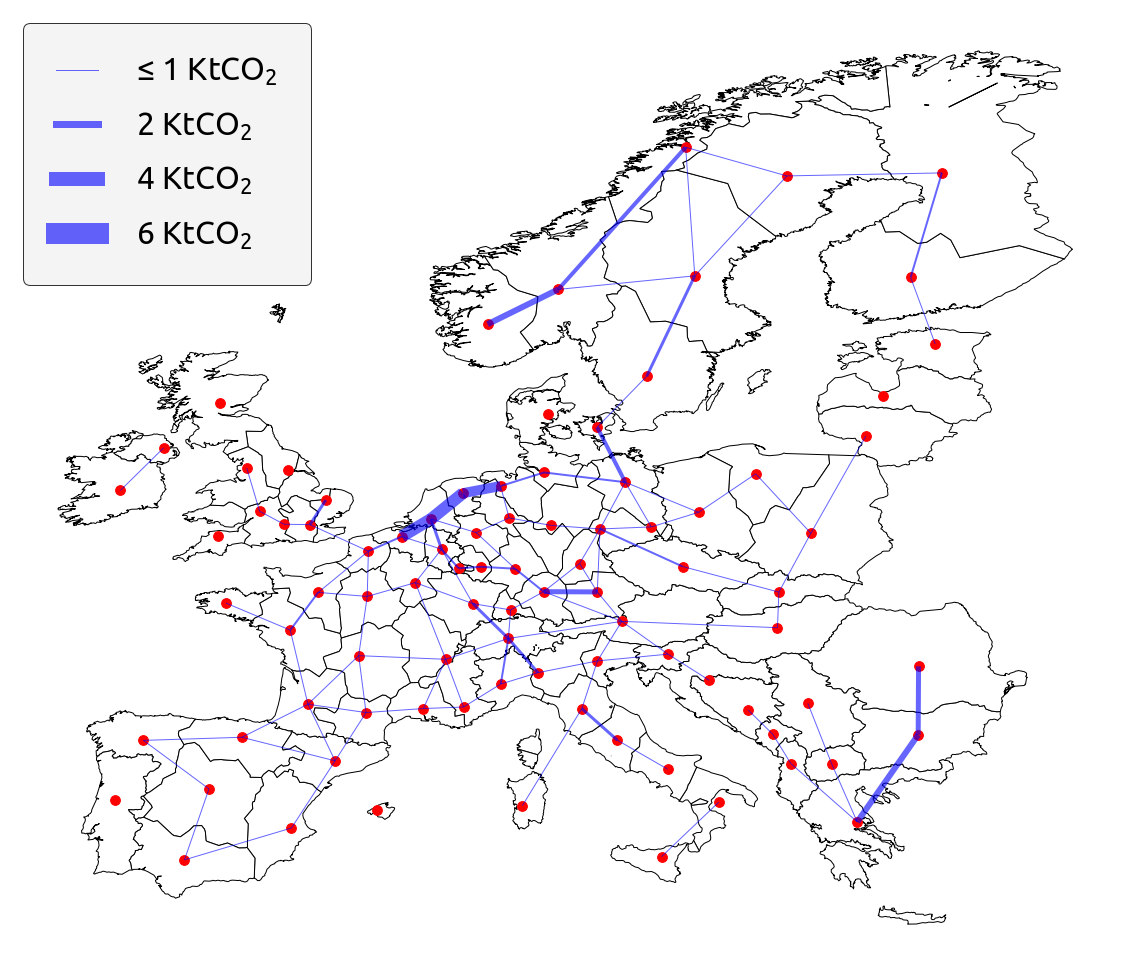}}
    \caption{The European CO$_2$ network topology in (a) the global net-zero CO$_2$ emissions scenario and (b) the local net-zero CO$_2$ emissions scenario. The model is allowed to build new CO$_2$ pipelines if it reduces the total system cost. These have optimal capacities and are organised according to the electricity grid topology.}
    \label{supplemental:figure_co2_network}
\end{figure}

CO$_2$ can be transported through a network of pipelines (Figure \ref{supplemental:figure_co2_network}). The model is allowed to build bidirectional, lossless pipelines when deemed cost-effective with optimal capacities. For simplicity, these are spatially arranged based on the existing electricity grid topology.

In addition, the model allows the transportation of solid biomass amongst and within countries, representing the potential transport by road using trucks. The cost of transportation for each country is based on the JRC-EU-TIMES model (Table \ref{supplemental:table_solid_biomass_transportation_cost}). Other means of transporting solid biomass such as ships and trains are excluded from the model due to the lack of datasets representing their routes.

Oil and methanol, on the other hand, have unlimited transportation capacity amongst countries in the model. The possibility of importing and exporting synthetic fuels to and from Europe is not modelled.

The capacities of run-of-river (ROR), hydroelectricity, and pump hydro storage (PHS) are exogenously set in the model based on existing facilities in Europe.

The model only includes CO$_2$ emissions from energy consumption in the agriculture sector, and it assumes that the rest of CO$_2$ emissions from this sector are offset by the Land Use, Land Use Change, and Forestry (LULUCF) sector.

\subsection{Carbon capture}
In our model, CO$_2$ can be captured from process emissions in industries, gas and solid biomass used in industry, CHP units combusting methane gas or solid biomass, SMR to produce H$_2$, and biomass used to produce biogas, which is then upgraded to methane gas. For the latter, a 90\% capture rate is assumed. In addition, CO$_2$ can be captured using DAC, which is modelled as a low-temperature heat-requiring system. DAC can be installed collocated with industrial demands where the required heat is supplied by air heat pumps, methane gas boilers, and resistive heaters. It can also be installed connected to district-heating system in urban areas. In this case, apart from the mentioned sources, the required heat is supplied by solar thermal, gas and solid biomass-based CHP units, along with the excess heat from the Fischer-Tropsch process and H$_2$ fuel cells.

Although showing promising results, additional methods for capturing CO$_2$, such as enhanced rock weathering, biochar, agroforestry, afforestation/reforestation, soil carbon sequestration, coastal wetland management, ocean alkalinisation, ocean fertilisation and artificial upwelling, and direct ocean capture (DOC) have been excluded from the model \cite{Smith2023, Digdaya2020}. Bioenergy with carbon capture and storage (BECCS) is represented in our model although not explicitly as a technology. This is done by having solid biomass being consumed in industry and CHP plants, both equipped with point-source capture, where the CO$_2$ captured from these is sequestered underground or used to produce synthetic fuels. The model does account for conservative amounts of biomass, i.e., biomass not competing with food crops. First, solid biomass includes agricultural waste, fuelwood residues, secondary forestry residues (woodchips), sawdust, residues from landscape care, and municipal waste. A potential of 1038 TWh per year of solid biomass is assumed for Europe, based on the medium bioenergy availability scenario of the ENSPRESO database \cite{RUIZ2019100379} for the year 2030. This solid biomass can be used to produce heat in the industry or combusted in CHP units, in both cases with or without CC. Biomass import from outside the continent is not allowed. The model also includes biogas, with a potential of 336 TWh/a for Europe, which is upgraded to methane gas.

The amount of process emissions and gas and solid biomass demand in the industry are exogenous to the model, while CHP and DAC capacities are endogenous.

\subsection{Carbon conversion and sequestration}
In the model, CO$_2$ can be converted (used) to manufacture valuable products such as synthetic oil through the Fischer-Tropsch process, synthetic methanol through methanolisation, and synthetic methane gas through the Sabatier reaction. While synthetic oil and methane gas production are endogenous to the model, the production of methanol is fixed since its demand is exogenously set.

\begin{figure}[!htb]
    \centering
    %\subfloat[]{\label{supplemental:figure_co2_flow_1000mt_sequestration_global}\includegraphics[width = 0.8\linewidth]{./figures/figure_S27a.png}}
    \subfloat[]{\label{supplemental:figure_co2_flow_1000mt_sequestration_global}\includegraphics[width = 0.6\linewidth]{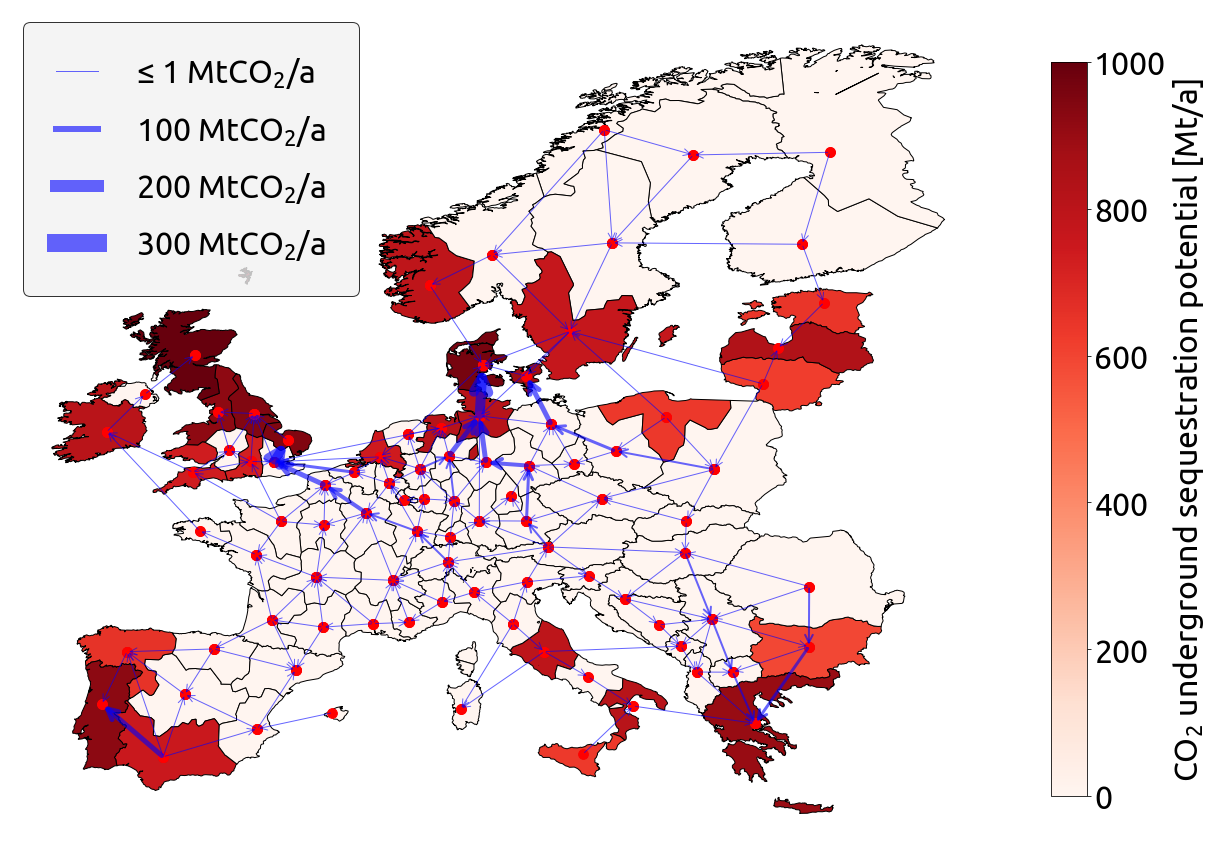}}
    \vspace{20pt}
    %\subfloat[]{\label{supplemental:figure_co2_flow_1000mt_sequestration_local}\includegraphics[width = 0.8\linewidth]{./figures/figure_S27b.png}}
    \subfloat[]{\label{supplemental:figure_co2_flow_1000mt_sequestration_local}\includegraphics[width = 0.6\linewidth]{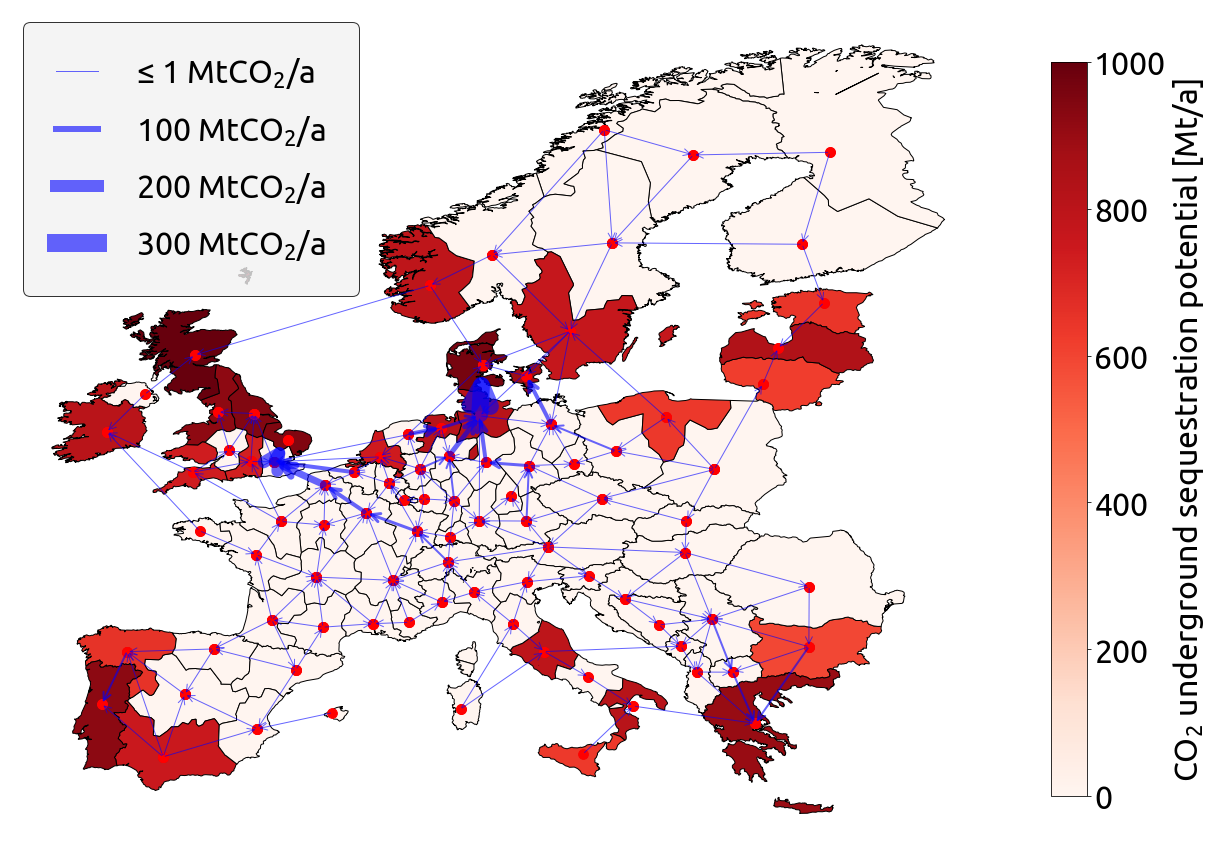}}
    \caption{CO$_2$ flow with a 1000 Mt/a underground sequestration potential in (a) the global net-zero CO$_2$ emissions scenario and (b) the local net-zero CO$_2$ emissions scenario. The main factor determining the directions and destinations of CO$_2$ flows is the assumed potential for sequestering CO$_2$ underground. Increasing the potential assumption to 1000 Mt/a (from 200 Mt/a) across Europe clearly shows CO$_2$ flows converging towards countries with significant underground sequestration potential.}
    \label{supplemental:figure_co2_flow_1000mt_sequestration}
\end{figure}

In addition, CO$_2$ can be sequestered underground in the model. Despite Europe having a huge potential for sequestering CO$_2$ both onshore and offshore, estimated at 117 Gt \cite{european_co2_sequestration_database}, the model restricts this to 2.9 Gt and only considers offshore underground potential, specifically in deep salt caverns and depleted hydrocarbon reservoirs. Table \ref{supplemental:table_co2_sequestration_potential} summarises how this amount is shared amongst countries. Based on these shares, the model optimally determines the amount of CO$_2$ that can potentially be sequestered in each country. Furthermore, since CO$_2$ sequestration is a highly uncertain technology, the model is constrained to a maximum annual sequestration of 200 MtCO$_2$ for the entire continent. This amount allows for sequestering process emissions, which, assuming an industrial transformation, account for 153 MtCO$_2$ per year in Europe, with the remaining 47 MtCO$_2$ enabling negative emissions. A lower amount would make the model unfeasible under a net-zero CO$_2$ emissions constraint. A sensitivity analysis to the assumptions for potential and cost for CO$_2$ underground sequestration can be found in \cite{VICTORIA20221066} and in Figure \ref{supplemental:figure_co2_flow_1000mt_sequestration}.

%\newpage

\section{Experimental procedures}

\subsection{Lead contact}
Further information and requests for resources should be directed to and will be fulfilled by the lead contact, Ricardo Fernandes (ricardo.fernandes@mpe.au.dk).

\subsection{Materials availability}
This study did not generate new unique materials.

\subsection{Data and code availability}
The model used in this study is based on a fork of PyPSA-Eur version 0.8.0, which includes new logic that implements local (national) CO$_2$ constraints. The fork is available at \url{https://github.com/ricnogfer/pypsa-eur}.

The technology assumptions in the model are based on the Energy System Technology Data version 0.5.0 for the year 2030 and are available at \url{https://github.com/PyPSA/technology-data}.

The energy system networks that this study is based on have been deposited in Zenodo and are available at \url{https://doi.org/10.5281/zenodo.12527792}.

\section{Acknowledgments}
R.F. is fully funded by Novo Nordisk CO$_2$ Research Center (CORC) under grant number CORC005. We sincerely thank Sleiman Farah, Alberto Alamia, and Ebbe Kyhl Gøtske, all from the Energy System Modelling Group at Aarhus University, for their scientific insights and technical support.

\section{Author contributions}
Conceptualisation, R.F. and M.V.; Software, R.F. and M.V.; Investigation, R.F.; Project Administration, R.F. and M.V.; Visualisation, R.F.; Resources, R.F. and M.V.; Writing-Original Draft, R.F. and M.V.; Writing-Review \& Editing, R.F. and M.V.; Supervision, M.V. and M.G.

\section{Declaration of interests}
The authors declare no competing interests.

\end{linenumbers}

\clearpage

%\nocite{*}

\bibliography{bibliography.bib}

\clearpage
\onecolumn

\beginsupplement
\pagenumbering{gobble}

\clearpage

\section{Mathematical formulation of the model}\label{supplemental:material_mathematical_formulation}
The model used in our study is based on PyPSA-Eur, which relies on a linear programme to minimise the total annualised cost of the entire energy system in an optimal fashion. Formally, this minimisation is represented by the following objective function:
\begin{equation}\label{supplemental:material_objective_function}
    \min_{G_{n,s},E_{n,s},F_{l},g_{n,s,t}}
    \left[
        \sum_{n,s} c_{n,s} G_{n,s} +
        \sum_{n,s} \hat{c}_{n,s} E_{n,s} +
        \sum_{l} c_l F_l +
        \sum_{n,s,t} o_{n,s,t} g_{n,s,t}
    \right]
\end{equation}
where, for technology $s$ in node $n$, $c_{n,s}$ is the annualised cost for generator and storage power capacity $G_{n,s}$, $\hat{c}_{n,s}$ is the annualised cost for storage energy capacity $E_{n,s}$, $c_l$ is the fixed annualised cost for capacity $F_l$ of link $l$, and $o_{n,s,t}$ are the marginal costs of generation and storage dispatch $g_{n,s,t}$ at time step $t$. In addition, the linear programme consists of several constraints. The most relevant constraints imposed on our model are succinctly described next.
\newline

At every time step of the model's temporal resolution, the demand for electricity must be met by the supply, which can be expressed as follows:
\begin{equation}\label{supplemental:material_electricity_demand_supply_constraint}
    \sum_{s} g_{n,s,t} +
    \sum_{l} \alpha_{n,l,t} f_{l,t} = d_{n,t}
    \hspace{0.15cm} \leftrightarrow \hspace{0.15cm} \lambda_{n,t}
    \hspace{0.5cm} \forall n,t
\end{equation}
where the sum of generation and storage dispatch $g_{n,s,t}$ of technology $s$, added to the sum of power flow $f_{l,t}$ in link $l$ with a direction and efficiency $\alpha_{n,l,t}$, equals demand $d_{n,t}$ in every node $n$ at every time step $t$. The dual value $\lambda_{n,t}$ of this constraint represents the electricity shadow price for node $n$ at time step $t$.
\newline

By default, a model based on PyPSA-Eur is constrained with a global cap (limit) on CO$_2$ emissions. To pursue our study, we extended PyPSA-Eur with new logic to allow for two additional types of constraints on CO$_2$ emissions: local (national) and nodal. In detail, the global constraint applies at a continental level (Europe), requiring that all the nodes of the model comply with a specific cap on CO$_2$ emissions collectively. The local (national) constraint applies at a country level, requiring that (only) the nodes of a given country comply with a specific cap on CO$_2$ emissions assigned to the country collectively, whereas the nodal constraint applies to each node, requiring each one to comply with a specific cap on CO$_2$ emissions assigned to it individually. The constraint representing a model based on a global CO$_2$ emissions cap can be formulated as follows:
\begin{equation}\label{supplemental:material_global_co2_constraint}
    \sum_{n,s,t} \varepsilon_{s} \frac{g_{n,s,t}}{\eta_{s}} \leq CAP_{CO_2}
    \hspace{0.15cm} \leftrightarrow \hspace{0.15cm} \mu_{CO_2}
\end{equation}
where the sum of CO$_2$ emissions $\varepsilon_{s}$ in tonnes per each MWh$_\text{th}$ produced by technology $s$ with efficiency $\eta_{s}$ in all nodes $n$ of the model must be equal to or lower than CO$_2$ limit $CAP_{CO_2}$. The dual value $\mu_{CO_2}$ of this constraint represents the CO$_2$ shadow price for the entire Europe.
\newline

The constraint representing a model based on local (national) CO$_2$ emissions caps can be formulated as follows:
\begin{equation}\label{supplemental:material_local_co2_constraint}
    \sum_{n,s,t} \varepsilon_{s} \frac{g_{n,s,t}}{\eta_{s}} \leq CAP_{CO_2,k}
    \hspace{0.15cm} \leftrightarrow \hspace{0.15cm} \mu_{CO_2,k}
    \hspace{0.5cm} n \in k
    \hspace{0.5cm} \forall k
\end{equation}
where, for every country $k$, the sum of CO$_2$ emissions $\varepsilon_{s}$ in tonnes per each MWh$_\text{th}$ produced by technology $s$ with efficiency $\eta_{s}$ in all nodes $n$ belonging to $k$ must be equal to or lower than CO$_2$ limit $CAP_{CO_2,k}$. The dual value $\mu_{CO_2,k}$ of this constraint represents the CO$_2$ shadow price for country $k$.
\newline

At last, the constraint representing a model based on nodal CO$_2$ emissions caps can be formulated as follows:
\begin{equation}\label{supplemental:material_nodal_co2_constraint}
    \sum_{s,t} \varepsilon_{s} \frac{g_{n,s,t}}{\eta_{s}} \leq CAP_{CO_2,n}
    \hspace{0.15cm} \leftrightarrow \hspace{0.15cm} \mu_{CO_2,n}
    \hspace{0.5cm} \forall n
\end{equation}
where, for every node $n$, CO$_2$ emissions $\varepsilon_{s}$ in tonnes per each MWh$_\text{th}$ produced by technology $s$ with efficiency $\eta_{s}$ in $n$ must be equal to or lower than CO$_2$ limit $CAP_{CO_2,n}$. The dual value $\mu_{CO_2,n}$ of this constraint represents the CO$_2$ shadow price for node $n$.
\newline

\clearpage

\section{Figures}\label{supplemental:figures}

\clearpage

\clearpage

\clearpage

\clearpage

\clearpage

\vspace{20pt}

\clearpage

\clearpage

\clearpage

\clearpage

\clearpage

\clearpage

\clearpage

\clearpage

%\begin{figure}[!htb]
%    \centering
%    \subfloat[]{\label{supplemental:figure_wind_capacity_factors}\includegraphics[width = 0.8\linewidth]{./figures/figure_S9a.png}}
%    \vspace{20pt}
%    \subfloat[]{\label{supplemental:figure_solar_capacity_factors}\includegraphics[width = 0.8\linewidth]{./figures/figure_S9b.png}}
%    \caption{(a) Onshore wind capacity factors across Europe and (b) Solar photovoltaic capacity factors across Europe. The capacity factors for wind are predominantly higher in northern Europe, whereas the capacity factors for solar are predominantly higher in southern Europe.}
%    \label{supplemental:figure_renewable_capacity_factors}
%\end{figure}

\clearpage

\clearpage

\clearpage

\clearpage

\clearpage

\clearpage

\clearpage

\clearpage

\clearpage

\clearpage

\clearpage

\clearpage

\clearpage

\clearpage

\clearpage

\clearpage

\clearpage

%\begin{figure}[!htb]
%    \centering
%    \subfloat[]{\label{supplemental:figure_net_synthetic_oil_producers_consumers_map}\includegraphics[width = 0.8\linewidth]{./figures/figure_S22a.png}}
%    \vspace{20pt}
%    \subfloat[]{\label{supplemental:figure_net_synthetic_oil_producers_consumers_variation}\includegraphics[width = 0.96\linewidth]{./figures/figure_S22b.png}}
%    \caption{(a) Net synthetic oil producers and consumers in the global net-zero CO$_2$ emissions scenario and (b) Net synthetic oil variation per country in the global net-zero CO$_2$ emissions scenario. Countries that produce more synthetic oil than they consume are represented by green bars, while countries that consume more synthetic oil than they produce are represented by red bars. The percentages on top of the green bars indicate how much additional synthetic oil net producer countries manufacture, beyond what is needed to meet their own demand. On the other hand, the percentages at the bottom of the red bars indicate how much synthetic oil net consumer countries do not manufacture to meet their own demand, relying instead on the former countries.}
%    \label{supplemental:figure_net_synthetic_oil_producers_consumers_map_variation}
%\end{figure}

\clearpage

%\begin{figure}[!htb]
%    \centering
%    \subfloat[]{\label{supplemental:figure_net_synthetic_methanol_producers_consumers_map}\includegraphics[width = 0.8\linewidth]{./figures/figure_S23a.png}}
%    \vspace{20pt}
%    \subfloat[]{\label{supplemental:figure_net_synthetic_methanol_producers_consumers_variation}\includegraphics[width = 0.96\linewidth]{./figures/figure_S23b.png}}
%    \caption{(a) Net synthetic methanol producers and consumers in the global net-zero CO$_2$ emissions scenario and (b) Net synthetic methanol variation per country in the global net-zero CO$_2$ emissions scenario. Countries that produce more synthetic methanol than they consume are represented by green bars, while countries that consume more synthetic methanol than they produce are represented by red bars. The percentages on top of the green bars indicate how much additional synthetic methanol net producer countries manufacture, beyond what is needed to meet their own demand. On the other hand, the percentages at the bottom of the red bars indicate how much synthetic methanol net consumer countries do not manufacture to meet their own demand, relying instead on the former countries.}
%    \label{supplemental:figure_net_synthetic_methanol_producers_consumers_map_variation}
%\end{figure}

\clearpage

\clearpage

\clearpage

\clearpage

\clearpage

\clearpage

\clearpage

\clearpage

\clearpage

\clearpage

%\begin{figure}[!htb]
%    \centering
%    \subfloat[]{\label{supplemental:figure_h2_production_global}\includegraphics[width = 0.8\linewidth]%{./figures/figure_S31a.png}}
%    \vspace{20pt}
%    \subfloat[]{\label{supplemental:figure_h2_production_variation}\includegraphics[width = 0.96\linewidth]{./figures/figure_S31b.png}}
%    \caption{(a) H$_2$ production in the global net-zero CO$_2$ emissions scenario and (b) H$_2$ production variation between the global net-zero CO$_2$ emissions scenario and the local net-zero CO$_2$ emissions scenario. Thanks to high renewable energy capacity factors and their relative proximity to the interior countries and Switzerland, Spain, Denmark, and Great Britain  significantly increase H$_2$ production to satisfy a growing demand for H$_2$ in the former countries when under local constraints.}
%    \label{supplemental:figure_h2_production}
%\end{figure}

\clearpage

\clearpage

\section{Tables}\label{supplemental:tables}

\noindent
\small
\renewcommand{\arraystretch}{0.79}% Tighter

\begin{longtable}[H]{p{4.7cm} p{2.65cm} p{2.65cm}}
    \caption{Characteristics assumptions about the main technologies used in PyPSA-Eur are based on the Energy System Technology Data version 0.5.0 for the year 2030. The comprehensive list of technologies is available at \url{https://github.com/PyPSA/technology-data}.}
    \label{supplemental:table_technology_characteristics_assumptions}
    \\
    \toprule
    \textbf{Technology}                  & \textbf{Parameter}                    & \textbf{Value}                              \\
    \midrule
    \multirow{2}{*}
    {Battery storage}                    & \multicolumn{1}{l}{Investment}        & \multicolumn{1}{l}{142 €/kWh}               \\\cline{2-3}
                                         & \multicolumn{1}{l}{Lifetime}          & \multicolumn{1}{l}{25 years}                \\\cline{2-3}

    \hline
    \multirow{5}{*}
    {Biogas}                             & \multicolumn{1}{l}{Investment}        & \multicolumn{1}{l}{1539.62 €/kW}            \\\cline{2-3}
                                         & \multicolumn{1}{l}{Lifetime}          & \multicolumn{1}{l}{20 years}                \\\cline{2-3}
                                         & \multicolumn{1}{l}{Efficiency}        & \multicolumn{1}{l}{1 per unit}              \\\cline{2-3}
                                         & \multicolumn{1}{l}{FOM}               & \multicolumn{1}{l}{12.84\%/year}            \\\cline{2-3}
                                         & \multicolumn{1}{l}{Capture rate}      & \multicolumn{1}{l}{0.98 per unit}           \\\cline{2-3}

    \hline
    \multirow{4}{*}
    {Biomass boiler}                     & \multicolumn{1}{l}{Investment}        & \multicolumn{1}{l}{649.3 €/kW$_{th}$}       \\\cline{2-3}
                                         & \multicolumn{1}{l}{Lifetime}          & \multicolumn{1}{l}{20 years}                \\\cline{2-3}
                                         & \multicolumn{1}{l}{Efficiency}        & \multicolumn{1}{l}{0.86 per unit}           \\\cline{2-3}
                                         & \multicolumn{1}{l}{FOM}               & \multicolumn{1}{l}{7.49\%/year}             \\\cline{2-3}

    \hline
    \multirow{3}{*}
    {CO$_2$ pipeline}                    & \multicolumn{1}{l}{Investment}        & \multicolumn{1}{l}{2000 €/tCO$_2$/h/km}     \\\cline{2-3}
                                         & \multicolumn{1}{l}{Lifetime}          & \multicolumn{1}{l}{50 years}                \\\cline{2-3}
                                         & \multicolumn{1}{l}{FOM}               & \multicolumn{1}{l}{0.9\%/year}              \\\cline{2-3}

    \hline
    \multirow{1}{*}
    {CO$_2$ sequestration}               & \multicolumn{1}{l}{VOM}               & \multicolumn{1}{l}{10 €/tCO$_2$}            \\\cline{2-3}

    \hline
    \multirow{5}{*}
    {Central air heat pump}              & \multicolumn{1}{l}{Investment}        & \multicolumn{1}{l}{856.25 €/kW$_{th}$}      \\\cline{2-3}
                                         & \multicolumn{1}{l}{Lifetime}          & \multicolumn{1}{l}{25 years}                \\\cline{2-3}
                                         & \multicolumn{1}{l}{Efficiency}        & \multicolumn{1}{l}{3.6 per unit}            \\\cline{2-3}
                                         & \multicolumn{1}{l}{FOM}               & \multicolumn{1}{l}{0.23\%/year}             \\\cline{2-3}
                                         & \multicolumn{1}{l}{VOM}               & \multicolumn{1}{l}{2.51 €/MWh$_{th}$}       \\\cline{2-3}

    \hline
    \multirow{5}{*}
    {Central gas boiler}                 & \multicolumn{1}{l}{Investment}        & \multicolumn{1}{l}{50 €/kW$_{th}$}          \\\cline{2-3}
                                         & \multicolumn{1}{l}{Lifetime}          & \multicolumn{1}{l}{25 years}                \\\cline{2-3}
                                         & \multicolumn{1}{l}{Efficiency}        & \multicolumn{1}{l}{1.04 per unit}           \\\cline{2-3}
                                         & \multicolumn{1}{l}{FOM}               & \multicolumn{1}{l}{3.8\%/year}              \\\cline{2-3}
                                         & \multicolumn{1}{l}{VOM}               & \multicolumn{1}{l}{1 €/MWh$_{th}$}          \\\cline{2-3}

    \hline
    \multirow{5}{*}
    {Central ground heat pump}           & \multicolumn{1}{l}{Investment}        & \multicolumn{1}{l}{507.6 €/kW$_{th}$}       \\\cline{2-3}
                                         & \multicolumn{1}{l}{Lifetime}          & \multicolumn{1}{l}{25 years}                \\\cline{2-3}
                                         & \multicolumn{1}{l}{Efficiency}        & \multicolumn{1}{l}{1.73 per unit}           \\\cline{2-3}
                                         & \multicolumn{1}{l}{FOM}               & \multicolumn{1}{l}{0.39\%/year}             \\\cline{2-3}
                                         & \multicolumn{1}{l}{VOM}               & \multicolumn{1}{l}{1.25 €/MWh$_{th}$}       \\\cline{2-3}

    \hline
    \multirow{5}{*}
    {Central resistive heater}           & \multicolumn{1}{l}{Investment}        & \multicolumn{1}{l}{60 €/kW$_{th}$}          \\\cline{2-3}
                                         & \multicolumn{1}{l}{Lifetime}          & \multicolumn{1}{l}{20 years}                \\\cline{2-3}
                                         & \multicolumn{1}{l}{Efficiency}        & \multicolumn{1}{l}{0.99 per unit}           \\\cline{2-3}
                                         & \multicolumn{1}{l}{FOM}               & \multicolumn{1}{l}{1.7\%/year}              \\\cline{2-3}
                                         & \multicolumn{1}{l}{VOM}               & \multicolumn{1}{l}{1 €/MWh$_{th}$}          \\\cline{2-3}

    \hline
    \multirow{3}{*}
    {Central solar thermal}              & \multicolumn{1}{l}{Investment}        & \multicolumn{1}{l}{140000 €/1000m$^2$}       \\\cline{2-3}
                                         & \multicolumn{1}{l}{Lifetime}          & \multicolumn{1}{l}{20 years}                \\\cline{2-3}
                                         & \multicolumn{1}{l}{FOM}               & \multicolumn{1}{l}{1.4\%/year}              \\\cline{2-3}

    \hline
    \multirow{3}{*}
    {Central water tank storage}         & \multicolumn{1}{l}{Investment}        & \multicolumn{1}{l}{0.54 €/kWh}              \\\cline{2-3}
                                         & \multicolumn{1}{l}{Lifetime}          & \multicolumn{1}{l}{25 years}                \\\cline{2-3}
                                         & \multicolumn{1}{l}{FOM}               & \multicolumn{1}{l}{0.55\%/year}             \\\cline{2-3}

    \hline
    \multirow{4}{*}
    {Decentral air heat pump}            & \multicolumn{1}{l}{Investment}        & \multicolumn{1}{l}{850 €/kW$_{th}$}         \\\cline{2-3}
                                         & \multicolumn{1}{l}{Lifetime}          & \multicolumn{1}{l}{18 years}                \\\cline{2-3}
                                         & \multicolumn{1}{l}{Efficiency}        & \multicolumn{1}{l}{3.6 per unit}            \\\cline{2-3}
                                         & \multicolumn{1}{l}{FOM}               & \multicolumn{1}{l}{3\%/year}                \\\cline{2-3}

    \hline
    \multirow{4}{*}
    {Decentral gas boiler}               & \multicolumn{1}{l}{Investment}        & \multicolumn{1}{l}{296.82 €/kW$_{th}$}      \\\cline{2-3}
                                         & \multicolumn{1}{l}{Lifetime}          & \multicolumn{1}{l}{20 years}                \\\cline{2-3}
                                         & \multicolumn{1}{l}{Efficiency}        & \multicolumn{1}{l}{0.98 per unit}           \\\cline{2-3}
                                         & \multicolumn{1}{l}{FOM}               & \multicolumn{1}{l}{6.69\%/year}             \\\cline{2-3}

    \hline
    \multirow{4}{*}
    {Decentral ground heat pump}         & \multicolumn{1}{l}{Investment}        & \multicolumn{1}{l}{1400 €/kW$_{th}$}        \\\cline{2-3}
                                         & \multicolumn{1}{l}{Lifetime}          & \multicolumn{1}{l}{20 years}                \\\cline{2-3}
                                         & \multicolumn{1}{l}{Efficiency}        & \multicolumn{1}{l}{3.9 per unit}            \\\cline{2-3}
                                         & \multicolumn{1}{l}{FOM}               & \multicolumn{1}{l}{1.82\%/year}             \\\cline{2-3}

    \hline
    \multirow{4}{*}
    {Decentral resistive heater}         & \multicolumn{1}{l}{Investment}        & \multicolumn{1}{l}{100 €/kW$_{th}$}         \\\cline{2-3}
                                         & \multicolumn{1}{l}{Lifetime}          & \multicolumn{1}{l}{20 years}                \\\cline{2-3}
                                         & \multicolumn{1}{l}{Efficiency}        & \multicolumn{1}{l}{0.9 per unit}            \\\cline{2-3}
                                         & \multicolumn{1}{l}{FOM}               & \multicolumn{1}{l}{2\%/year}                \\\cline{2-3}

    \hline
    \multirow{3}{*}
    {Decentral solar thermal}            & \multicolumn{1}{l}{Investment}        & \multicolumn{1}{l}{270000 €/1000m$^2$}      \\\cline{2-3}
                                         & \multicolumn{1}{l}{Lifetime}          & \multicolumn{1}{l}{20 years}                \\\cline{2-3}
                                         & \multicolumn{1}{l}{FOM}               & \multicolumn{1}{l}{1.3\%/year}              \\\cline{2-3}

    \hline
    \multirow{3}{*}
    {Decentral water tank storage}       & \multicolumn{1}{l}{Investment}        & \multicolumn{1}{l}{18.38 €/kWh}             \\\cline{2-3}
                                         & \multicolumn{1}{l}{Lifetime}          & \multicolumn{1}{l}{20 years}                \\\cline{2-3}
                                         & \multicolumn{1}{l}{FOM}               & \multicolumn{1}{l}{1\%/year}                \\\cline{2-3}

    \hline
    \multirow{6}{*}
    {Direct air capture}                 & \multicolumn{1}{l}{Investment}        & \multicolumn{1}{l}{6000000 €/tCO$_2$/h}     \\\cline{2-3}
                                         & \multicolumn{1}{l}{Lifetime}          & \multicolumn{1}{l}{20 years}                \\\cline{2-3}
                                         & \multicolumn{1}{l}{FOM}               & \multicolumn{1}{l}{4.95\%/year}             \\\cline{2-3}
                                         & \multicolumn{1}{l}{Electricity input} & \multicolumn{1}{l}{0.32 MWh/tCO$_2$}        \\\cline{2-3}
                                         & \multicolumn{1}{l}{Heat input}        & \multicolumn{1}{l}{2 MWh/tCO$_2$}           \\\cline{2-3}
                                         & \multicolumn{1}{l}{Heat output}       & \multicolumn{1}{l}{1 MWh/tCO$_2$}           \\\cline{2-3}

    \hline
    \multirow{3}{*}
    {Electricity distribution grid}      & \multicolumn{1}{l}{Investment}        & \multicolumn{1}{l}{500 €/kW}                \\\cline{2-3}
                                         & \multicolumn{1}{l}{Lifetime}          & \multicolumn{1}{l}{40 years}                \\\cline{2-3}
                                         & \multicolumn{1}{l}{FOM}               & \multicolumn{1}{l}{2\%/year}                \\\cline{2-3}

    \hline
    \multirow{5}{*}
    {Fischer-Tropsch}                    & \multicolumn{1}{l}{Investment}        & \multicolumn{1}{l}{650711.26 €/MW$_{FT}$}   \\\cline{2-3}
                                         & \multicolumn{1}{l}{Lifetime}          & \multicolumn{1}{l}{20 years}                \\\cline{2-3}
                                         & \multicolumn{1}{l}{Efficiency}        & \multicolumn{1}{l}{0.8 per unit}            \\\cline{2-3}
                                         & \multicolumn{1}{l}{FOM}               & \multicolumn{1}{l}{3\%/year}                \\\cline{2-3}
                                         & \multicolumn{1}{l}{Capture rate}      & \multicolumn{1}{l}{0.98 per unit}           \\\cline{2-3}

    \hline
    \multirow{5}{*}
    {Gas CHP}                            & \multicolumn{1}{l}{Investment}        & \multicolumn{1}{l}{560 €/kW}                \\\cline{2-3}
                                         & \multicolumn{1}{l}{Lifetime}          & \multicolumn{1}{l}{25 years}                \\\cline{2-3}
                                         & \multicolumn{1}{l}{Efficiency}        & \multicolumn{1}{l}{0.41 per unit}           \\\cline{2-3}
                                         & \multicolumn{1}{l}{FOM}               & \multicolumn{1}{l}{3.32\%/year}             \\\cline{2-3}
                                         & \multicolumn{1}{l}{VOM}               & \multicolumn{1}{l}{4.2 €/MWh}               \\\cline{2-3}

    \hline
    \multirow{10}{*}
    {Gas CHP CC}                         & \multicolumn{1}{l}{Investment}        & \multicolumn{1}{l}{560 €/kW}                \\\cline{2-3}
                                         & \multicolumn{1}{l}{Lifetime}          & \multicolumn{1}{l}{25 years}                \\\cline{2-3}
                                         & \multicolumn{1}{l}{Efficiency}        & \multicolumn{1}{l}{0.41 per unit}           \\\cline{2-3}
                                         & \multicolumn{1}{l}{FOM}               & \multicolumn{1}{l}{3.32\%/year}             \\\cline{2-3}
                                         & \multicolumn{1}{l}{VOM}               & \multicolumn{1}{l}{4.2 €/MWh}               \\\cline{2-3}
                                         & \multicolumn{1}{l}{Capture rate}      & \multicolumn{1}{l}{0.9 per unit}            \\\cline{2-3}
                                         & \multicolumn{1}{l}{CO$_2$ intensity}  & \multicolumn{1}{l}{0.2 tCO$_2$/MWh$_{th}$}  \\\cline{2-3}
                                         & \multicolumn{1}{l}{Electricity input} & \multicolumn{1}{l}{0.02 MWh/tCO$_2$}        \\\cline{2-3}
                                         & \multicolumn{1}{l}{Heat input}        & \multicolumn{1}{l}{0.72 MWh/tCO$_2$}        \\\cline{2-3}
                                         & \multicolumn{1}{l}{Heat output}       & \multicolumn{1}{l}{0.72 MWh/tCO$_2$}        \\\cline{2-3}

    \hline
    \multirow{5}{*}
    {Gas for industry CC}                & \multicolumn{1}{l}{Investment}        & \multicolumn{1}{l}{2600000 €/tCO$_2$/h}     \\\cline{2-3}
                                         & \multicolumn{1}{l}{Lifetime}          & \multicolumn{1}{l}{25 years}                \\\cline{2-3}
                                         & \multicolumn{1}{l}{FOM}               & \multicolumn{1}{l}{3\%/year}                \\\cline{2-3}
                                         & \multicolumn{1}{l}{Capture rate}      & \multicolumn{1}{l}{0.9 per unit}            \\\cline{2-3}
                                         & \multicolumn{1}{l}{CO$_2$ intensity}  & \multicolumn{1}{l}{0.2 tCO$_2$/MWh$_{th}$}  \\\cline{2-3}

    \hline
    \multirow{3}{*}
    {Gas storage}                        & \multicolumn{1}{l}{Investment}        & \multicolumn{1}{l}{0.03 €/kWh}              \\\cline{2-3}
                                         & \multicolumn{1}{l}{Lifetime}          & \multicolumn{1}{l}{100 years}               \\\cline{2-3}
                                         & \multicolumn{1}{l}{FOM}               & \multicolumn{1}{l}{3.59\%/year}             \\\cline{2-3}

    \hline
    \multirow{4}{*}
    {H$_2$ electrolysis}                 & \multicolumn{1}{l}{Investment}        & \multicolumn{1}{l}{450 €/kW$_{el}$}          \\\cline{2-3}
                                         & \multicolumn{1}{l}{Lifetime}          & \multicolumn{1}{l}{30 years}                \\\cline{2-3}
                                         & \multicolumn{1}{l}{Efficiency}        & \multicolumn{1}{l}{0.68 per unit}           \\\cline{2-3}
                                         & \multicolumn{1}{l}{FOM}               & \multicolumn{1}{l}{2\%/year}                \\\cline{2-3}

    \hline
    \multirow{4}{*}
    {H$_2$ fuel cell}                    & \multicolumn{1}{l}{Investment}        & \multicolumn{1}{l}{1100 €/kW$_{el}$}         \\\cline{2-3}
                                         & \multicolumn{1}{l}{Lifetime}          & \multicolumn{1}{l}{10 years}                \\\cline{2-3}
                                         & \multicolumn{1}{l}{Efficiency}        & \multicolumn{1}{l}{0.5 per unit}            \\\cline{2-3}
                                         & \multicolumn{1}{l}{FOM}               & \multicolumn{1}{l}{5\%/year}                \\\cline{2-3}

    \hline
    \multirow{3}{*}
    {H$_2$ pipeline}                     & \multicolumn{1}{l}{Investment}        & \multicolumn{1}{l}{267 €/MW/km}             \\\cline{2-3}
                                         & \multicolumn{1}{l}{Lifetime}          & \multicolumn{1}{l}{40 years}                \\\cline{2-3}
                                         & \multicolumn{1}{l}{FOM}               & \multicolumn{1}{l}{3\%/year}                \\\cline{2-3}

    \hline
    \multirow{3}{*}
    {H$_2$ storage tank type 1 + compr.} & \multicolumn{1}{l}{Investment}        & \multicolumn{1}{l}{44.91 €/kWh}             \\\cline{2-3}
                                         & \multicolumn{1}{l}{Lifetime}          & \multicolumn{1}{l}{30 years}                \\\cline{2-3}
                                         & \multicolumn{1}{l}{FOM}               & \multicolumn{1}{l}{1.11\%/year}             \\\cline{2-3}

    \hline
    \multirow{3}{*}
    {H$_2$ storage underground}          & \multicolumn{1}{l}{Investment}        & \multicolumn{1}{l}{2 €/kWh}                 \\\cline{2-3}
                                         & \multicolumn{1}{l}{Lifetime}          & \multicolumn{1}{l}{100 years}               \\\cline{2-3}
                                         & \multicolumn{1}{l}{FOM}               & \multicolumn{1}{l}{0\%/year}                \\\cline{2-3}
                                         & \multicolumn{1}{l}{VOM}               & \multicolumn{1}{l}{0 €/MWh}                 \\\cline{2-3}

    \hline
    \multirow{3}{*}
    {HVAC overhead}                      & \multicolumn{1}{l}{Investment}        & \multicolumn{1}{l}{432.97 €/MW/km}          \\\cline{2-3}
                                         & \multicolumn{1}{l}{Lifetime}          & \multicolumn{1}{l}{40 years}                \\\cline{2-3}
                                         & \multicolumn{1}{l}{FOM}               & \multicolumn{1}{l}{2\%/year}                \\\cline{2-3}

    \hline
    \multirow{3}{*}
    {HVDC overhead}                      & \multicolumn{1}{l}{Investment}        & \multicolumn{1}{l}{432.97 €/MW/km}          \\\cline{2-3}
                                         & \multicolumn{1}{l}{Lifetime}          & \multicolumn{1}{l}{40 years}                \\\cline{2-3}
                                         & \multicolumn{1}{l}{FOM}               & \multicolumn{1}{l}{2\%/year}                \\\cline{2-3}

    \hline
    \multirow{3}{*}
    {HVDC submarine}                     & \multicolumn{1}{l}{Investment}        & \multicolumn{1}{l}{471.16 €/MW/km}          \\\cline{2-3}
                                         & \multicolumn{1}{l}{Lifetime}          & \multicolumn{1}{l}{40 years}                \\\cline{2-3}
                                         & \multicolumn{1}{l}{FOM}               & \multicolumn{1}{l}{0.35\%/year}             \\\cline{2-3}

    \hline
    \multirow{4}{*}
    {Hydroelectricity}                   & \multicolumn{1}{l}{Investment}        & \multicolumn{1}{l}{2208.16 €/kW$_{el}$}     \\\cline{2-3}
                                         & \multicolumn{1}{l}{Lifetime}          & \multicolumn{1}{l}{80 years}                \\\cline{2-3}
                                         & \multicolumn{1}{l}{Efficiency}        & \multicolumn{1}{l}{0.9 per unit}            \\\cline{2-3}
                                         & \multicolumn{1}{l}{FOM}               & \multicolumn{1}{l}{1\%/year}                \\\cline{2-3}

    \hline
    \multirow{5}{*}
    {Methanation (Sabatier)}             & \multicolumn{1}{l}{Investment}        & \multicolumn{1}{l}{628.6 €/MW$_{CH_4}$}     \\\cline{2-3}
                                         & \multicolumn{1}{l}{Lifetime}          & \multicolumn{1}{l}{20 years}                \\\cline{2-3}
                                         & \multicolumn{1}{l}{Efficiency}        & \multicolumn{1}{l}{0.8 per unit}            \\\cline{2-3}
                                         & \multicolumn{1}{l}{FOM}               & \multicolumn{1}{l}{3\%/year}                \\\cline{2-3}
                                         & \multicolumn{1}{l}{Capture rate}      & \multicolumn{1}{l}{0.98 per unit}           \\\cline{2-3}

    \hline
    \multirow{3}{*}
    {Methane gas pipeline}               & \multicolumn{1}{l}{Investment}        & \multicolumn{1}{l}{79 €/MW/km}              \\\cline{2-3}
                                         & \multicolumn{1}{l}{Lifetime}          & \multicolumn{1}{l}{50 years}                \\\cline{2-3}
                                         & \multicolumn{1}{l}{FOM}               & \multicolumn{1}{l}{1.5\%/year}              \\\cline{2-3}

    \hline
    \multirow{3}{*}
    {Methanolisation}                    & \multicolumn{1}{l}{Investment}        & \multicolumn{1}{l}{650711.26 €/MW$_{MeOH}$} \\\cline{2-3}
                                         & \multicolumn{1}{l}{Lifetime}          & \multicolumn{1}{l}{20 years}                \\\cline{2-3}
                                         & \multicolumn{1}{l}{FOM}               & \multicolumn{1}{l}{3\%/year}                \\\cline{2-3}

    \hline
    \multirow{3}{*}
    {Nuclear}                            & \multicolumn{1}{l}{Investment}        & \multicolumn{1}{l}{7940.45 €/kW$_{el}$}     \\\cline{2-3}
                                         & \multicolumn{1}{l}{Lifetime}          & \multicolumn{1}{l}{40 years}                \\\cline{2-3}
                                         & \multicolumn{1}{l}{Efficiency}        & \multicolumn{1}{l}{0.33 per unit}           \\\cline{2-3}
                                         & \multicolumn{1}{l}{FOM}               & \multicolumn{1}{l}{1.4\%/year}              \\\cline{2-3}
                                         & \multicolumn{1}{l}{VOM}               & \multicolumn{1}{l}{3.5 €/MWh$_{el}$}        \\\cline{2-3}

    \hline
    \multirow{5}{*}
    {OCGT}                               & \multicolumn{1}{l}{Investment}        & \multicolumn{1}{l}{435.24 €/kW}             \\\cline{2-3}
                                         & \multicolumn{1}{l}{Lifetime}          & \multicolumn{1}{l}{25 years}                \\\cline{2-3}
                                         & \multicolumn{1}{l}{Efficiency}        & \multicolumn{1}{l}{0.41 per unit}           \\\cline{2-3}
                                         & \multicolumn{1}{l}{FOM}               & \multicolumn{1}{l}{1.78\%/year}             \\\cline{2-3}
                                         & \multicolumn{1}{l}{VOM}               & \multicolumn{1}{l}{4.5 €/MWh}               \\\cline{2-3}

    \hline
    \multirow{4}{*}
    {Offshore wind}                      & \multicolumn{1}{l}{Investment}        & \multicolumn{1}{l}{1523.55 €/kW$_{el}$}     \\\cline{2-3}
                                         & \multicolumn{1}{l}{Lifetime}          & \multicolumn{1}{l}{30 years}                \\\cline{2-3}
                                         & \multicolumn{1}{l}{FOM}               & \multicolumn{1}{l}{2.32\%/year}             \\\cline{2-3}
                                         & \multicolumn{1}{l}{VOM}               & \multicolumn{1}{l}{0.02 €/MWh$_{el}$}       \\\cline{2-3}

    \hline
    \multirow{4}{*}
    {Onshore wind}                       & \multicolumn{1}{l}{Investment}        & \multicolumn{1}{l}{1035.56 €/kW}            \\\cline{2-3}
                                         & \multicolumn{1}{l}{Lifetime}          & \multicolumn{1}{l}{30 years}                \\\cline{2-3}
                                         & \multicolumn{1}{l}{FOM}               & \multicolumn{1}{l}{1.22\%/year}             \\\cline{2-3}
                                         & \multicolumn{1}{l}{VOM}               & \multicolumn{1}{l}{1.35 €/MWh}              \\\cline{2-3}

    \hline
    \multirow{4}{*}
    {Process emissions CC}               & \multicolumn{1}{l}{Investment}        & \multicolumn{1}{l}{2600000 €/tCO$_2$/h}     \\\cline{2-3}
                                         & \multicolumn{1}{l}{Lifetime}          & \multicolumn{1}{l}{25 years}                \\\cline{2-3}
                                         & \multicolumn{1}{l}{FOM}               & \multicolumn{1}{l}{3\%/year}                \\\cline{2-3}
                                         & \multicolumn{1}{l}{Capture rate}      & \multicolumn{1}{l}{0.9 per unit}            \\\cline{2-3}

    \hline
    \multirow{4}{*}
    {SMR}                                & \multicolumn{1}{l}{Investment}        & \multicolumn{1}{l}{493470.4 €/MW$_{CH_4}$}  \\\cline{2-3}
                                         & \multicolumn{1}{l}{Lifetime}          & \multicolumn{1}{l}{30 years}                \\\cline{2-3}
                                         & \multicolumn{1}{l}{Efficiency}        & \multicolumn{1}{l}{0.76 per unit}           \\\cline{2-3}
                                         & \multicolumn{1}{l}{FOM}               & \multicolumn{1}{l}{5\%/year}                \\\cline{2-3}

    \hline
    \multirow{5}{*}
    {SMR CC}                             & \multicolumn{1}{l}{Investment}        & \multicolumn{1}{l}{572425.66 €/MW$_{CH_4}$} \\\cline{2-3}
                                         & \multicolumn{1}{l}{Lifetime}          & \multicolumn{1}{l}{30 years}                \\\cline{2-3}
                                         & \multicolumn{1}{l}{Efficiency}        & \multicolumn{1}{l}{0.69 per unit}           \\\cline{2-3}
                                         & \multicolumn{1}{l}{FOM}               & \multicolumn{1}{l}{5\%/year}                \\\cline{2-3}
                                         & \multicolumn{1}{l}{Capture rate}      & \multicolumn{1}{l}{0.9 per unit}            \\\cline{2-3}

    \hline
    \multirow{4}{*}
    {Solar}                              & \multicolumn{1}{l}{Investment}        & \multicolumn{1}{l}{492.11 €/kW$_{el}$}      \\\cline{2-3}
                                         & \multicolumn{1}{l}{Lifetime}          & \multicolumn{1}{l}{40 years}                \\\cline{2-3}
                                         & \multicolumn{1}{l}{FOM}               & \multicolumn{1}{l}{1.95\%/year}             \\\cline{2-3}
                                         & \multicolumn{1}{l}{VOM}               & \multicolumn{1}{l}{0.01 €/MWh$_{el}$}       \\\cline{2-3}

    \hline
    \multirow{4}{*}
    {Solid biomass}                      & \multicolumn{1}{l}{Investment}        & \multicolumn{1}{l}{2209 €/kW$_{el}$}        \\\cline{2-3}
                                         & \multicolumn{1}{l}{Lifetime}          & \multicolumn{1}{l}{30 years}                \\\cline{2-3}
                                         & \multicolumn{1}{l}{Efficiency}        & \multicolumn{1}{l}{0.47 per unit}           \\\cline{2-3}
                                         & \multicolumn{1}{l}{FOM}               & \multicolumn{1}{l}{4.53\%/year}             \\\cline{2-3}

    \hline
    \multirow{5}{*}
    {Solid biomass CHP}                  & \multicolumn{1}{l}{Investment}        & \multicolumn{1}{l}{3349.49 €/kW$_{el}$}     \\\cline{2-3}
                                         & \multicolumn{1}{l}{Lifetime}          & \multicolumn{1}{l}{25 years}                \\\cline{2-3}
                                         & \multicolumn{1}{l}{Efficiency}        & \multicolumn{1}{l}{0.27 per unit}           \\\cline{2-3}
                                         & \multicolumn{1}{l}{FOM}               & \multicolumn{1}{l}{2.87\%/year}             \\\cline{2-3}
                                         & \multicolumn{1}{l}{VOM}               & \multicolumn{1}{l}{4.58 €/MWh$_{el}$}       \\\cline{2-3}

    \hline
    \multirow{8}{*}
    {Solid biomass CHP CC}               & \multicolumn{1}{l}{Investment}        & \multicolumn{1}{l}{2700000 €/tCO$_2$/h}     \\\cline{2-3}
                                         & \multicolumn{1}{l}{Lifetime}          & \multicolumn{1}{l}{25 years}                \\\cline{2-3}
                                         & \multicolumn{1}{l}{FOM}               & \multicolumn{1}{l}{3\%/year}                \\\cline{2-3}
                                         & \multicolumn{1}{l}{Capture rate}      & \multicolumn{1}{l}{0.9 per unit}            \\\cline{2-3}
                                         & \multicolumn{1}{l}{CO$_2$ intensity}  & \multicolumn{1}{l}{0.2 tCO$_2$/MWh$_{th}$}  \\\cline{2-3}
                                         & \multicolumn{1}{l}{Electricity input} & \multicolumn{1}{l}{0.02 MWh/tCO$_2$}        \\\cline{2-3}
                                         & \multicolumn{1}{l}{Heat input}        & \multicolumn{1}{l}{0.72 MWh/tCO$_2$}        \\\cline{2-3}
                                         & \multicolumn{1}{l}{Heat output}       & \multicolumn{1}{l}{0.72 MWh/tCO$_2$}        \\\cline{2-3}

    \hline
    \multirow{5}{*}
    {Solid biomass for industry CC}      & \multicolumn{1}{l}{Investment}        & \multicolumn{1}{l}{2600000 €/tCO$_2$/h}     \\\cline{2-3}
                                         & \multicolumn{1}{l}{Lifetime}          & \multicolumn{1}{l}{25 years}                \\\cline{2-3}
                                         & \multicolumn{1}{l}{FOM}               & \multicolumn{1}{l}{3\%/year}                \\\cline{2-3}
                                         & \multicolumn{1}{l}{Capture rate}      & \multicolumn{1}{l}{0.9 per unit}            \\\cline{2-3}
                                         & \multicolumn{1}{l}{CO$_2$ intensity}  & \multicolumn{1}{l}{0.37 tCO$_2$/MWh$_{th}$} \\\cline{2-3}
    \bottomrule
\end{longtable}

\newpage

\begin{table}[H]
    \small
    \centering
    \caption{Solid biomass transportation cost per country. The costs are based on the JRC European TIMES Energy System Model (JRC-EU-TIMES) and are available at \url{https://data.jrc.ec.europa.eu/collection/id-00287}.}
    \label{supplemental:table_solid_biomass_transportation_cost}
    \begin{tabular}{p{4cm} p{1.6cm} p{3.4cm}}
        \toprule
        \textbf{Country}                & \textbf{ISO} & \textbf{Cost [€/km/MWh]} \\
        \midrule
        Albania                         & AL           & 0.0552                   \\
        Austria                         & AT           & 0.1365                   \\
        Bosnia and Herzegovina          & BA           & 0.0594                   \\
        Belgium                         & BE           & 0.1406                   \\
        Bulgaria                        & BG           & 0.0635                   \\
        Switzerland                     & CH           & 0.1771                   \\
        Czech Republic                  & CZ           & 0.0875                   \\
        Germany                         & DE           & 0.1333                   \\
        Denmark                         & DK           & 0.1906                   \\
        Estonia                         & EE           & 0.0865                   \\
        Spain                           & ES           & 0.1198                   \\
        Finland                         & FI           & 0.1490                   \\
        France                          & FR           & 0.1427                   \\
        Great Britain                   & GB           & 0.1438                   \\
        Greece                          & GR           & 0.1115                   \\
        Croatia                         & HR           & 0.0802                   \\
        Hungary                         & HU           & 0.0729                   \\
        Ireland                         & IE           & 0.1333                   \\
        Italy                           & IT           & 0.1323                   \\
        Lithuania                       & LT           & 0.0740                   \\
        Luxembourg                      & LU           & 0.1458                   \\
        Latvia                          & LV           & 0.0813                   \\
        Montenegro                      & ME           & 0.0604                   \\
        North Macedonia                 & MK           & 0.0510                   \\
        The Netherlands                 & NL           & 0.1458                   \\
        Norway                          & NO           & 0.1615                   \\
        Poland                          & PL           & 0.0781                   \\
        Portugal                        & PT           & 0.0927                   \\
        Romania                         & RO           & 0.0625                   \\
        Serbia                          & RS           & 0.0573                   \\
        Sweden                          & SE           & 0.1615                   \\
        Slovenia                        & SI           & 0.0979                   \\
        Slovakia                        & SK           & 0.0833                   \\
        \bottomrule
    \end{tabular}
\end{table}

\newpage

\begin{table}[H]
    \small
    \centering
    \caption{CO$_2$ underground sequestration potential per country. The potentials are based on the European CO$_2$ storage database (CO2StoP) and are available at \url{https://energy.ec.europa.eu/publications/assessment-co2-storage-potential-europe-co2stop_en}. The model limits the aggregated CO$_2$ sequestration potential across Europe to 2.9 Gt and only considers the offshore underground potential of deep salt caverns and depleted hydrocarbon reservoirs.}
    \label{supplemental:table_co2_sequestration_potential}
    \begin{tabular}{p{4cm} p{1.6cm} p{3.4cm}}
        \toprule
        \textbf{Country}                & \textbf{ISO} & \textbf{Potential [MtCO$_2$/a]} \\
        \midrule
        Albania                         & AL           & 0                               \\
        Austria                         & AT           & 0                               \\
        Bosnia and Herzegovina          & BA           & 0                               \\
        Belgium                         & BE           & 0                               \\
        Bulgaria                        & BG           & 0.2                             \\
        Switzerland                     & CH           & 0                               \\
        Czech Republic                  & CZ           & 0                               \\
        Germany                         & DE           & 41.1                            \\
        Denmark                         & DK           & 603.6                           \\
        Estonia                         & EE           & 0.7                             \\
        Spain                           & ES           & 7.3                             \\
        Finland                         & FI           & 0                               \\
        France                          & FR           & 0                               \\
        Great Britain                   & GB           & 1836.4                          \\
        Greece                          & GR           & 122                             \\
        Croatia                         & HR           & 0                               \\
        Hungary                         & HU           & 0                               \\
        Ireland                         & IE           & 18.9                            \\
        Italy                           & IT           & 33.8                            \\
        Lithuania                       & LT           & 0.4                             \\
        Luxembourg                      & LU           & 0                               \\
        Latvia                          & LV           & 33.4                            \\
        Montenegro                      & ME           & 0                               \\
        North Macedonia                 & MK           & 0                               \\
        The Netherlands                 & NL           & 4.7                             \\
        Norway                          & NO           & 13.4                            \\
        Poland                          & PL           & 0.5                             \\
        Portugal                        & PT           & 212.2                           \\
        Romania                         & RO           & 0                               \\
        Serbia                          & RS           & 0                               \\
        Sweden                          & SE           & 8.2                             \\
        Slovenia                        & SI           & 0                               \\
        Slovakia                        & SK           & 0                               \\
        \bottomrule
    \end{tabular}
\end{table}

\newpage

\newpage

\begin{table}[H]
    \small
    \centering
    \caption{Total system cost per country in the global net-zero CO$_2$ emissions scenario. The variation refers to the increase or decrease (in percentage) of this cost compared with its counterpart in the local net-zero CO$_2$ emissions scenario. Typically, under local constraints, countries that were net CO$_2$ emitters under a global constraint have positive variations (i.e. higher costs), while countries that were net CO$_2$ absorbers under the same constraint have negative variations (i.e. lower costs).}
    \label{supplemental:table_total_system_cost_per_country}
    \begin{tabular}{p{3.8cm} p{1.4cm} p{2.8cm} p{2.2cm}}
        \toprule
        \textbf{Country}       & \textbf{ISO} & \textbf{Cost [billion €/a]} & \textbf{Variation [\%]} \\
        \midrule
        Albania                & AL           & 1.84                        & -3                      \\
        Austria                & AT           & 14.31                       & -2.7                    \\
        Bosnia and Herzegovina & BA           & 1.63                        & -2.5                    \\
        Belgium                & BE           & 14.21                       & +57.8                   \\
        Bulgaria               & BG           & 6.07                        & -6.6                    \\
        Switzerland            & CH           & 11.3                        & +14.3                   \\
        Czech Republic         & CZ           & 12.16                       & -3.5                    \\
        Germany                & DE           & 107.29                      & +12.5                   \\
        Denmark                & DK           & 19.81                       & -0.6                    \\
        Estonia                & EE           & 4.45                        & -18.9                   \\
        Spain                  & ES           & 98.54                       & -15.7                   \\
        Finland                & FI           & 12.96                       & -12.3                   \\
        France                 & FR           & 114.8                       & +2.1                    \\
        Great Britain          & GB           & 115.27                      & +1.8                    \\
        Greece                 & GR           & 21.68                       & -1.2                    \\
        Croatia                & HR           & 3.06                        & -8.4                    \\
        Hungary                & HU           & 12.06                       & -3.3                    \\
        Ireland                & IE           & 23.59                       & -3.7                    \\
        Italy                  & IT           & 86.7                        & +0.2                    \\
        Lithuania              & LT           & 2.38                        & -16.5                   \\
        Luxembourg             & LU           & 1.81                        & +13.2                   \\
        Latvia                 & LV           & 3.23                        & -13.5                   \\
        Montenegro             & ME           & 1.41                        & -6.9                    \\
        North Macedonia        & MK           & 1.67                        & -9.6                    \\
        The Netherlands        & NL           & 35.27                       & +39                     \\
        Norway                 & NO           & 9.45                        & +14.9                   \\
        Poland                 & PL           & 51.86                       & -5.1                    \\
        Portugal               & PT           & 8.62                        & -26.8                   \\
        Romania                & RO           & 11.58                       & -13.5                   \\
        Serbia                 & RS           & 3.85                        & -8.1                    \\
        Sweden                 & SE           & 18.5                        & -15.1                   \\
        Slovenia               & SI           & 1.44                        & -6.2                    \\
        Slovakia               & SK           & 6.33                        & +6.5                    \\
        \bottomrule
    \end{tabular}
\end{table}

\newpage

\begin{table}[H]
    \small
    \centering
    \caption{CO$_2$ shadow price per country in the local net-zero CO$_2$ emissions scenario. The prices are the result of solving the model and are referred to as dual values associated with each constraint in a linear programming problem. As a reference, the CO$_2$ shadow price for all of Europe is set at 540 € per tonne of CO$_2$ in the global net-zero CO$_2$ emissions scenario.}
    \label{supplemental:table_co2_shadow_price}
    \begin{tabular}{p{4cm} p{1.6cm} p{3.4cm}}
        \toprule
        \textbf{Country}       & \textbf{ISO} & \textbf{Price [€/tCO$_2$]} \\
        \midrule
        Albania                & AL           & 455                        \\
        Austria                & AT           & 474                        \\
        Bosnia and Herzegovina & BA           & 454                        \\
        Belgium                & BE           & 584                        \\
        Bulgaria               & BG           & 458                        \\
        Switzerland            & CH           & 571                        \\
        Czech Republic         & CZ           & 476                        \\
        Germany                & DE           & 569                        \\
        Denmark                & DK           & 535                        \\
        Estonia                & EE           & 412                        \\
        Spain                  & ES           & 534                        \\
        Finland                & FI           & 435                        \\
        France                 & FR           & 548                        \\
        Great Britain          & GB           & 542                        \\
        Greece                 & GR           & 538                        \\
        Croatia                & HR           & 479                        \\
        Hungary                & HU           & 480                        \\
        Ireland                & IE           & 530                        \\
        Italy                  & IT           & 536                        \\
        Lithuania              & LT           & 405                        \\
        Luxembourg             & LU           & 579                        \\
        Latvia                 & LV           & 402                        \\
        Montenegro             & ME           & 456                        \\
        North Macedonia        & MK           & 436                        \\
        The Netherlands        & NL           & 575                        \\
        Norway                 & NO           & 544                        \\
        Poland                 & PL           & 461                        \\
        Portugal               & PT           & 446                        \\
        Romania                & RO           & 460                        \\
        Serbia                 & RS           & 467                        \\
        Sweden                 & SE           & 436                        \\
        Slovenia               & SI           & 477                        \\
        Slovakia               & SK           & 426                        \\
        \bottomrule
    \end{tabular}
\end{table}

\newpage

\begin{table}[H]
    \small
    \centering
    \caption{Conversion from MWh of energy to tonnes of CO$_2$ content based on stoichiometric reactions and thermal content of energy carriers. A factor of 3.6 is used to convert MWh to gigajoules (GJ). The molecular mass of CO$_2$ is 44.01 g/mol. Solid biomass is assumed to be dry and composed of 50\% carbon (C).}
    \label{supplemental:table_energy_co2_conversion}
    \begin{tabular}{p{2.5cm} p{3.1cm} p{3.1cm} p{3.8cm}}
        \toprule
        \textbf{Energy carrier} & \textbf{GJ per tonne of energy carrier \newline(a)} & \textbf{Material molecular mass [g/mol]\newline(b)} & \textbf{Tonnes of CO$_2$ per MWh of energy carrier\newline3.6 / (a) * 44.01 / (b)} \\
        \midrule
        Solid biomass           & 18                                                  & 12.01 (C)                                           & 0.3664                                                                             \\
        Methane gas             & 50                                                  & 16.04 (CH$_4$)                                      & 0.1976                                                                             \\
        Oil                     & 44                                                  & 14 (C$_{12}$H$_{23}$)                               & 0.2572                                                                             \\
        Methanol                & 22.7                                                & 32.04 (CH$_3$OH)                                    & 0.2178                                                                             \\
        \bottomrule
    \end{tabular}
\end{table}

\end{document}